\documentclass[aps,prx,twocolumn,longbibliography]{revtex4-2}
\usepackage[utf8]{inputenc}
\usepackage[T1]{fontenc}

\usepackage{physics} 
\usepackage{amsmath}
\usepackage{amssymb}

\DeclareMathOperator{\sgn}{sgn}

\usepackage{silence}
\usepackage{graphicx}
\usepackage{color}
\usepackage[
colorlinks=true,
linkcolor=red,
citecolor=blue,
urlcolor=blue
]{hyperref}
\usepackage{orcidlink}

\begin{document}

\title{Unifying description of competing chiral and nematic superconducting states in twisted bilayer graphene}

\author{Lucas Baldo\,\orcidlink{0009-0002-7612-8521}}
\affiliation{Department of Physics and Astronomy, Uppsala University, Box 516, SE-752 37 Uppsala, Sweden}

\author{Patric Holmvall\,\orcidlink{0000-0002-1866-2788}}
\affiliation{Department of Physics and Astronomy, Uppsala University, Box 516, SE-752 37 Uppsala, Sweden}

\author{Annica M. Black-Schaffer\,\orcidlink{0000-0002-4726-5247}}
\affiliation{Department of Physics and Astronomy, Uppsala University, Box 516, SE-752 37 Uppsala, Sweden}

\date{\today}

\begin{abstract}
	We reveal a striking correspondence between electron- and phonon-driven pairing in twisted bilayer graphene (TBG) by mapping an atomistic electronically driven pairing model onto an effective inter-valley, intra-Chern description, originally proposed for phonon-mediated superconductivity. Within the unified framework of intra-Chern pairing, we then analyze the competition between nematic and chiral $d$-wave superconducting states. The latter corresponds to the extreme Chern-polarized limit and hosts unpaired flat bands within the superconducting gap, which generally disfavors it relative to the nematic states. Crucially, nematic order is locally preferred at each momentum, but the optimal nematic directions are incompatible across the Brillouin zone due to the broken rotation symmetry. This momentum-space frustration still enables a chiral ground state at large fillings or weak interactions in a symmetric parent state. Including heterostrain or inter-valley coherence further stabilizes the nematic ground state and sets the nematic direction. Under heterostrain, this direction shows a double winding with the strain axis, revealing the $d$-wave nature of the pairing, while inter-valley coherence makes it filling dependent. Our results thereby provide a unified understanding of superconductivity in TBG, in which electron- and phonon-mediated pairing cooperate, and clarify the microscopic origin of the competition between the chiral and nematic superconducting states.
\end{abstract}

\maketitle

\section{Introduction}
\label{sec:introduction}

Despite intense experimental \cite{Cao2018b, Yankowitz2019, Lu2019, Codecido2019, Polshyn2019, Cao2020, Jaoui2022, Saito2020, Stepanov2020, Stepanov2021, Liu2021, Cao2021a, Oh2021, RodanLegrain2021, DiBattista2022, Nuckolls2023, Tian2023, Tanaka2025} efforts, the  microscopic mechanism underlying superconductivity and the superconducting pairing symmetry in magic-angle twisted bilayer graphene (TBG) remain unsettled. While measured critical magnetic fields near the Pauli limit are compatible with predominantly spin-singlet pairing \cite{Cao2018b}, other observations---such as short coherence lengths \cite{Cao2018b, Lu2019}, $T$-linear resistivity \cite{Polshyn2019, Cao2020, Jaoui2022}, rotation-symmetry breaking \cite{Cao2021a}, nodal quasiparticles \cite{Oh2021, DiBattista2022}, inter-valley coherence \cite{Nuckolls2023}, and signatures of quantum geometry \cite{Tian2023, Tanaka2025}---point towards unconventional superconductivity. These signatures, together with a proximity of the superconducting domes to correlated insulating states, have motivated numerous proposals of purely electronic pairing mechanisms \cite{Po2018b, Isobe2018c, Lin2018, Kennes2018, Fidrysiak2018, Liu2018, You2019d, Fang2019, Lin2019, Gonzalez2019, Gu2020, Sharma2020, Fischer2021, Khalaf2021, WangY2021, Fernandes2021c, Lothman2022, Khalaf2022, MacCari2023, Islam2023}. However, an initially observed robustness of superconductivity under Coulomb screening \cite{Saito2020, Stepanov2020, Liu2021} suggested that superconducting and insulating orders may have distinct, possibly even competing, microscopic origins \cite{Chou2019}. Phonon-mediated pairing has also been considered, but initial predictions favored chiral, and thus rotation-symmetry preserving, superconductivity \cite{Wu2018, Wu2019, Lian2019, Wu2019c, Shavit2021c, Cea2021}, at odds with current experiments \cite{Cao2021a, Oh2021, DiBattista2022}, which point to a nematic superconducting state breaking rotation symmetry and with an anisotropic gap, including nodal points. Theoretical models stabilizing such nematic order have been put forth, but initially relied on electronic mechanisms \cite{Kozii2019, Fernandes2021c, Lothman2022}.

\begin{figure} [b]
	\includegraphics[width=\columnwidth]{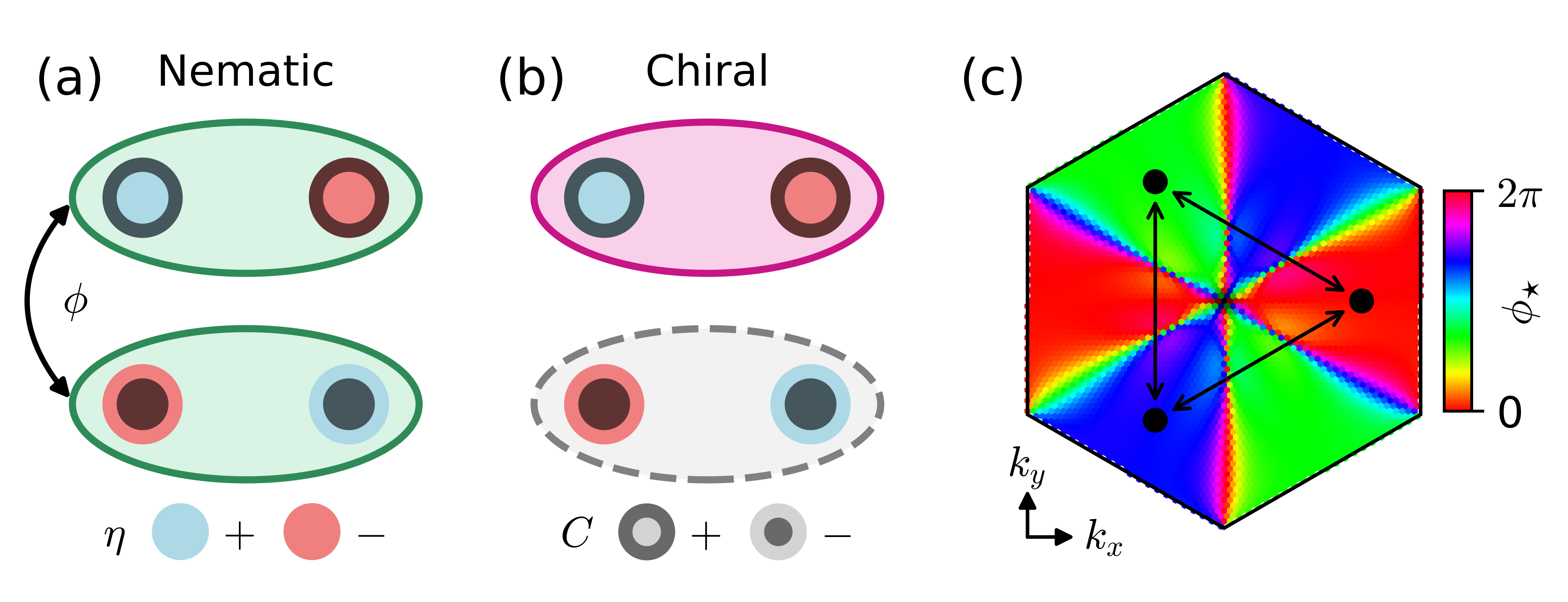}
	\caption{Pairing between opposite valleys $\eta=\pm$ (blue/red) and in the same Chern sector $C=\pm$ (dark rim/center). (a) In the nematic state pairing is finite in both Chern sectors, with relative phase $\phi$ setting the nematic direction. (b) In the chiral state one sector remains unpaired (dashed gray), lowering condensation energy. (c) For the nematic state, the preferred direction $\phi_{\star}$ (colormap) varies with momentum (numerical solution within atomistic model, see Sec.~\ref{sec:atomistic}). Values at $C_{3z}$-related momenta (black markers) are incompatible, leading to frustration.}
	\label{fig:introduction:diagram}
\end{figure}

Recent evidence for strong coupling to monolayer $K$-phonons \cite{Chen2024} has renewed the focus on phonon-mediated pairing in TBG. In particular, such phonons can generate a nematic superconducting state in the $d$-wave spin-singlet channel for sufficiently flat bands \cite{Liu2024}. There, the low-energy pairing is inter-valley and intra-Chern, as illustrated schematically in Fig.~\ref{fig:introduction:diagram}(a,b), where blue/red denote the two valleys $\eta=\pm$ and different gray scales on center/rim represent the two Chern sectors $C = \pm$. Pairing correlations (ellipses) thus occur only between flat bands in opposite valleys and with the same Chern number \cite{Tarnopolsky2019, Wang2021}. The nematic state corresponds to pairing distributed across both Chern sectors, see Fig.~\ref{fig:introduction:diagram}(a), with a relative phase $\phi$ determining the nematic orientation, whereas the chiral state concentrates pairing within a single Chern sector (b), leaving two Chern bands unpaired (dashed gray). Crucially, the kinetic energy hybridizes the Chern bands within each valley, such that the intra-Chern pairing model naturally incorporates inter-eigenband pairing terms. Recent works have specifically emphasized the importance of such pairing in TBG \cite{Christos2023, Putzer2025}, also recently shown to limit the energetic costs of the nodal structure of the nematic state \cite{Liu2025}. 
Importantly, however, these results still do not answer why and when the nematic state is actually favored.

Speaking against phonon-generated superconductivity in TBG is the fact that $K$-phonons alone favor an extended $s$-wave inter-Chern order over the $d$-wave intra-Chern order
, which is inconsistent with observed symmetry breaking and nodal states. Recently it has been argued that superconductivity in TBG cannot be explained by phonon mediation alone, but electronic repulsion is needed to stabilize the $d$-wave channel, as well as control a transition between nematic and chiral ground states \cite{Wagner2024, Wang2024, Wang2025}. Here it is the on-site Hubbard repulsion $U$ at each carbon atom that plays a decisive role. Interestingly, in purely electronic models of superconductivity such $U$ generates spin fluctuations that mediate an effective nearest-neighbor attraction. On the square lattice the same mechanism famously yields $d_{x^2-y^2}$-wave pairing, as in the cuprate superconductors \cite{Scalapino1995, Lee2006}. In sufficiently doped monolayer graphene it has instead been predicted to give rise to chiral $d$-wave pairing \cite{BlackSchaffer2007, Nandkishore2012, Nandkishore2012b, Kiesel2012, Wang2012, BlackSchaffer2014a, BlackSchaffer2014b} and similar results have also recently been reported in a modified dice lattice hosting a flat band \cite{Kakoi2025}. In TBG, $d$-wave superconductivity also emerges as the leading instability up to a critical interaction strength $U_c$, beyond which magnetic order sets in \cite{Fischer2021}, but, importantly, the superconducting ground state has numerically been mostly found to be nematic \cite{Julku2020, Fischer2021, Lothman2022}. A possible transition to chiral pairing at weak interactions was proposed \cite{Su2018, Fischer2021}, but, again, with no clear understanding of the competition between the nematic and chiral states. More broadly, any similarities in the superconducting state between electron- and phonon-mediated pairing remain fully unexplained. 

In this work we establish a key direct connection between existing electron- and phonon-driven models of superconductivity in TBG. Explicitly, we establish the same $d$-wave inter-valley and intra-Chern pairing in Fig.~\ref{fig:introduction:diagram}(a,b) for electron-driven superconductivity mediated by spin fluctuations. We obtain these results by first solving fully self-consistently for superconductivity for an effective attraction, of realistic strength, from spin fluctuations acting between nearest-neighbor carbon atoms and then projecting onto the moir\'e flat-band subspace and its Chern basis. While we focus on the simplest description of electron-driven pairing compatible with $d$-wave superconductivity, we argue that our results hold for other forms of inter-valley, inter-sublattice pairing, due to the fundamental relation between valley, sublattice, and Chern number of the moir\'e flat bands. Importantly, the low-energy pairing structure we derive for electron-driven pairing in TBG is the same as that previously derived for phonon-driven pairing. Since both interactions are likely present, we expect them to reinforce each other rather than compete. The relevant question for superconductivity in TBG may therefore not be which mechanism is responsible, but how the two cooperate.

As a second main result, we establish both a conceptual and quantitative understanding of $d$-wave superconductivity in TBG, especially why and when nematic $d$-wave is favored over chiral $d$-wave superconductivity. This is also a remarkable result, as chiral pairing is usually assumed to be favorable whenever allowed by symmetry, e.g.~on three- or sixfold symmetric lattices. This generic assumption is well-founded and easy to understand in single-band models, where chiral $d$-wave superconductivity hosts a full gap, while all nematic (i.e.~real-valued) $d$-wave solutions have a nodal gap, thus decreasing the condensation energy. 
This energy argument holds in doped monolayer graphene \cite{BlackSchaffer2007, Nandkishore2012, Nandkishore2012b, Kiesel2012, Wang2012, BlackSchaffer2014a, BlackSchaffer2014b}, even if it also hosts some inter-eigenband pairing. However, here we show that in TBG the inter-eigenband pairing becomes crucial, such that for intra-Chern pairing it is equally large as the (standard) intra-eigenband term. We identify that the chiral state is then forced to be fully Chern polarized in order to preserve its rotation symmetry. This results in an unpaired Chern band that often appears within the superconducting gap, thereby reducing the condensation energy of the chiral state even more than the nodal states suppress the condensation energy in the nematic state. We quantify this finding by establishing that for each individual momentum in the moir\'e Brillouin zone, the nematic state is always energetically favored over the chiral state. However, the condensation energy also depends on the nematic direction $\phi$, and the preferred nematic direction, $\phi_\star$, varies across the Brillouin zone, as shown in Fig.~\ref{fig:introduction:diagram}(c). Importantly, preferred directions for $C_{3z}$-related momenta (black markers) are mutually incompatible, leading to momentum-space frustration. This frustration becomes more prominent as the energy splitting of the normal-state moir\'e bands $\tilde{\epsilon}$ increases compared to the effective attractive interaction $\tilde{J}$, such that for $\tilde{\epsilon} \gtrsim \tilde{J}$ a chiral ground state may still emerge. This explains why a chiral state can appear at weak interactions and we also establish that it can be driven by band filling. Finally, we investigate the effects of symmetry-broken parent states on intra-Chern pairing, by considering lattice relaxation, heterostrain, and inter-valley coherence. The latter two favor the nematic state, with the nematic orientation set by the heterostrain axis or, in the presence of inter-valley coherence, varying with filling. We also establish how a nematic-to-chiral transition can still be facilitated by a finite Chern polarization in the normal state.

\section{Electron-driven superconductivity from atomistic modeling}
\label{sec:atomistic}
TBG at the magic angle has about $10^4$ carbon atoms in each moir\'e unit cell. Each carbon atom contributes one $p_z$ electron to the electronic structure of monolayer graphene. Electronic interactions are strong already in the monolayer where they lead for example to logarithmic Coulomb corrections to the Fermi velocity \cite{Gonzalez1994, Sheehy2007, Elias2011, Lee2026}, although ordering is absent due to the vanishing density of states (DOS) at the Fermi level. Also the on-site Hubbard $U$ is known to be large in the $p_z$ orbitals \cite{Wehling2011, Schler2013, Rosner2015, DLee2025}.
In the monolayer, functional renormalization group and quantum Monte Carlo studies have long predicted that such local repulsion leads to superconducting instabilities when the Fermi level is tuned to regions of sufficiently high DOS \cite{Honerkamp2008, Ma2011, BlackSchaffer2014b}, where a chiral $d$-wave ground state is favored \cite{Uchoa2007, BlackSchaffer2007, Honerkamp2008, Pathak2010, Nandkishore2012, Nandkishore2012b, Kiesel2012, Wang2012, BlackSchaffer2014a}. While these predictions have not been experimentally realized in monolayer graphene, as typical doping ranges yield too small DOS, they point to an underlying pairing tendency that is expected to persist in other graphene systems with enhanced DOS.

In fact, random phase approximation (RPA) analysis has shown that the repulsive Hubbard interaction leads to an effectively attractive interaction in non-local spin-singlet channels, dominated by nearest-neighbor attraction \cite{Fischer2021}. There, and in contrast with the monolayer, nematic $d$-wave superconductivity was found to be favored instead in certain regimes \cite{Su2018, Julku2020, Fischer2021, Lothman2022}, although its energetic competition with the chiral state remains poorly understood. Following these results, we here use an effective model for spin-singlet pairing between nearest-neighbor carbon atoms. 
Importantly, pairing on nearest neighbors in real space gives the lowest non-trivial reciprocal space harmonics. Since these are the states with the lowest number of nodes, they are energetically favorable so including longer-ranged pairing is expected to at most yield quantitative corrections.

\subsection{Atomistic model}
\label{sec:atomistic:background}

\begin{figure} 
	\includegraphics[width=\columnwidth]{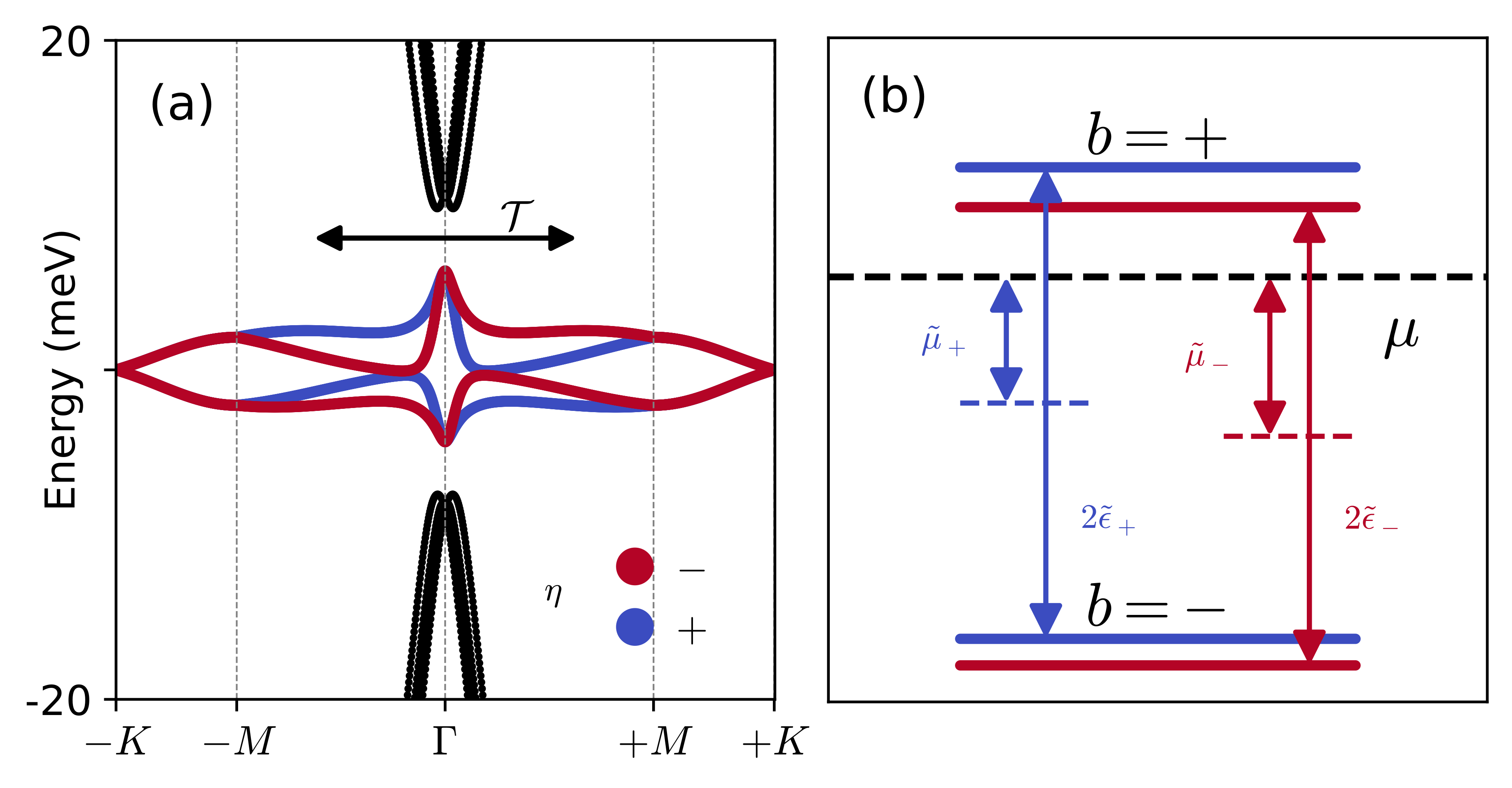}
	\caption{Spectrum of normal-state TBG. (a) Low-energy band structure with the moir\'e flat bands highlighted. States at opposite momenta and valley $\eta = \pm$ (blue/red) are related by spinless time-reversal symmetry $\mathcal{T}$. (b) Band energies are parametrized in terms of the momentum-dependent splitting $\tilde{\epsilon}_{\eta}$ between levels of opposite eigenband $b$, and their offset $\tilde{\mu}_{\eta}$ from the chemical potential $\mu$ (dashed black). }
	\label{fig:atomistic:bands}
\end{figure}

To capture atomic-scale correlations, we treat the $p_z$ orbital of each carbon atom explicitly. We consider commensurate structures, in particular type~I \cite{Zou2018b}, corresponding to a twist axis passing through lattice sites in both layers. The resulting point group is $D_3$, generated by a threefold rotation $C_{3z}$ about this axis and a twofold rotation $C_{2y}$ about an in-plane axis parallel to one of the moir\'e lattice vectors. We focus on a structure with twist angle $\theta \approx 1.2^\circ$, which is close to the first magic angle, and provides a good compromise between numerical feasibility and band flatness.

The symmetric normal state is described by real-valued hopping amplitudes parametrized from \emph{ab initio} calculations \cite{TramblydeLaissardiere2010}. The normal-state Hamiltonian $\hat{h}_0(\mathbf{k})$ is obtained by Fourier transforming these hoppings according to the moir\'e periodicity and including a chemical potential $\mu$ to regulate filling, for more details see Supplemental Material (SM). Diagonalization of $\hat{h}_0$ yields the full bare band structure of TBG, at the center of which lie four isolated bands with extremely small bandwidth, referred to as the (moir\'e) flat bands. We illustrate this in Fig.~\ref{fig:atomistic:bands}(a), where the flat bands (colored) are well separated from the rest of the spectrum (black). The colors blue and red denote valley quantum numbers $\eta=\pm$, respectively \cite{Ramires2018, Colomes2018, Ramires2019}. This labeling reflects an approximate $U(1)$ valley symmetry arising from the suppression of inter-valley scattering between the layers. 

The real-valuedness of the hopping elements further endows the system with an exact spinless time-reversal symmetry $\mathcal{T}$, expressed as $\hat{h}_0^*(\mathbf{k}) = \hat{h}_0(-\mathbf{k})$. Within each valley sector $\eta$, the two flat bands are generically separated in energy, touching only at the moir\'e Dirac points. As a result, the band index $b=\pm$, referring to the upper and lower flat bands, respectively, constitutes a good quantum number known as the \emph{eigenband} \cite{Liu2025}. At a given momentum $\mathbf{k}$, the four flat bands thus acquire distinct energies $\epsilon_{b,\eta}(\mathbf{k})$, as summarized in Fig.~\ref{fig:atomistic:bands}(b). The quantities $\tilde{\epsilon}$ and $\tilde{\mu}$ are defined later. Importantly, while $\mathcal{T}$ preserves the eigenband index, it relates states in opposite valleys, such that ${\epsilon}_{b, \eta}(\mathbf{k}) = {\epsilon}_{b,-\eta}(-\mathbf{k})$, as is evident in Fig.~\ref{fig:atomistic:bands}(a). For our initial results we use this normal state, so as to highlight the symmetry breaking by the superconducting order. In Sec.~\ref{sec:parentstate} we extend this treatment by also considering symmetry-broken normal states as the parent state of superconductivity to understand how superconductivity responds to such symmetry breaking.

To capture spin-singlet $d$-wave pairing efficiently, in agreement with earlier results \cite{Su2018, Julku2020, Fischer2021, Lothman2022} and as argued above, we focus on nearest-neighbor spin-singlet pairing. We thereby assume an effective attractive interaction $J$ between nearest-neighbor sites $\langle i,j \rangle$ in each graphene layer and perform a mean-field decoupling in the Cooper channel. Assuming further that moir\'e translation symmetry is preserved, we obtain the mean-field Hamiltonian \cite{Julku2020, Fischer2021, Lothman2022}
\begin{align} \label{eq:atomistic:Hamiltonian}
	\hat{H}
	&= \sum_{\mathbf{k}} \left[ \hat{H}_0 + \sum_{\langle i,j \rangle} \left( \hat{\Delta}_{ij} \hat{s}_{ij}^\dagger + \mathrm{H.c.} + \frac{|\hat{\Delta}_{ij}|^2}{J}  \right) \right],
\end{align}
where $\sum_{\langle i,j \rangle}$ runs over nearest-neighbor bonds, each counted once. The kinetic term is $\hat{H}_0(\mathbf{k}) = \sum_{\sigma} \sum_{i,j} [\hat{h}_0(\mathbf{k})]_{ij} \hat{c}^\dagger_{i \mathbf{k} \sigma} \hat{c}_{j \mathbf{k} \sigma}$, with $\hat{c}^\dagger_{i \mathbf{k} \sigma}$ the Bloch electron creation operator for site $i$, momentum $\mathbf{k}$ and spin $\sigma$, while $\hat{s}_{ij}^\dagger(\mathbf{k}) = \hat{c}^\dagger_{i \mathbf{k} \uparrow} \hat{c}^\dagger_{j -\mathbf{k} \downarrow} - \hat{c}^\dagger_{i \mathbf{k} \downarrow} \hat{c}^\dagger_{j -\mathbf{k} \uparrow}$ creates a singlet pair. The pairing matrix $\hat{\Delta}(\mathbf{k})$ has nonzero elements defined independently on each carbon--carbon bond. For bonds within a moir\'e unit cell, these elements are given by
\begin{align} \label{eq:atomistic:gap}
	\hat{\Delta}_{ij}(\mathbf{k})
	&= -\frac{J}{N_k} \sum_{\mathbf{k}^\prime} \langle \hat{s}_{ij}(\mathbf{k}^\prime) \rangle .
\end{align}
For bonds crossing the moir\'e unit-cell boundary, Bloch phase factors are present; see the SM for details. Expectation values in Eq.~\eqref{eq:atomistic:gap} are evaluated in the ground state of Eq.~\eqref{eq:atomistic:Hamiltonian}, which establishes a self-consistency problem. With the interaction $J$ and doping $\mu$ as the only free parameters, we can either solve self-consistently for the ground state at any temperature or we can linearize the gap equation Eq.~\eqref{eq:atomistic:gap} to identify the pairing configurations $\hat{\Delta}$ with the highest $T_c$. See Ref.~\cite{Lothman2022} for details on the numerical implementations.

\subsection{Symmetry classification of \texorpdfstring{$d$}{d}-wave pairing states}
\label{sec:atomistic:symmetry}

\begin{figure} 
	\includegraphics[width=\columnwidth]{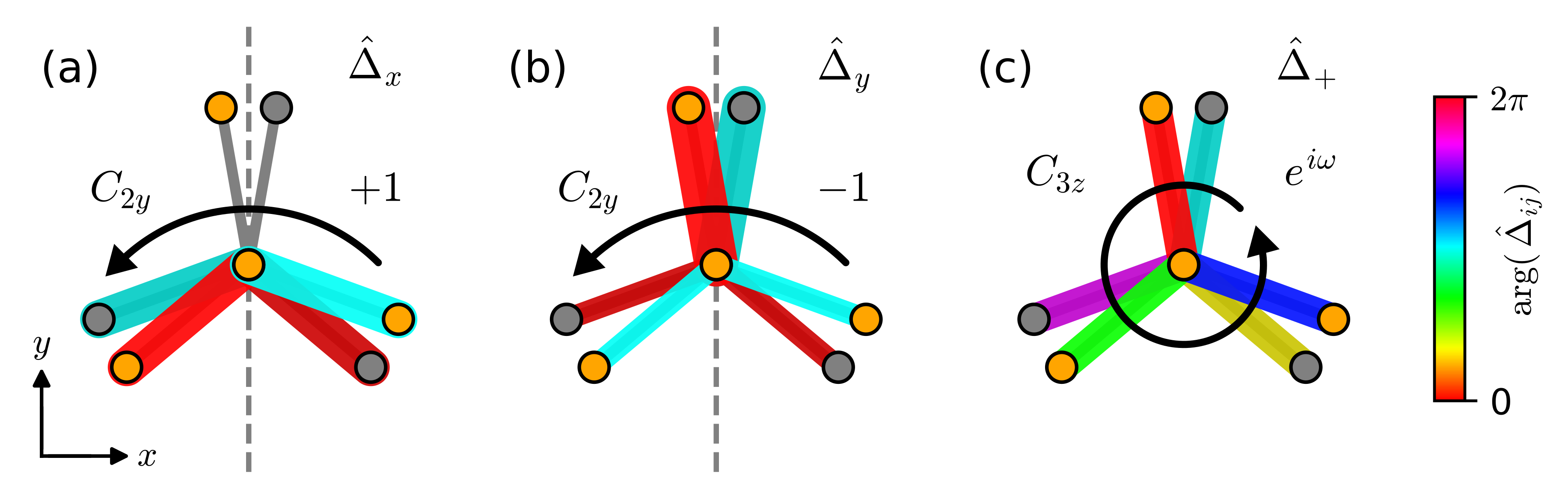}
	\caption{Local $d$-wave pairing symmetry at the center of the $AA$ region. Orange (gray) dots and light (dark) colored lines denote respectively sites and bonds in the top (bottom) layer. Central sites of both layers are perfectly aligned. Bond line width (hue) represents the pairing amplitude $|\hat{\Delta}_{ij}|$ (phase $\arg (\hat{\Delta}_{ij})$). In all cases, the order parameter is $\pi$-shifted between the layers. Nematic states (a) $\hat{\Delta}_{x}$ and (b) $\hat{\Delta}_{y}$ have eigenvalues $+1$ and $-1$, respectively, under $C_{2y}$ transformation (gray dashed line as rotation axis), which swaps the layers. The chiral state (c) $\hat{\Delta}_{+}$ has eigenvalue $e^{i \omega}$, with $\omega=2\pi/3$, under in-plane rotations $C_{3z}$.}
	\label{fig:atomistic:symmetry}
\end{figure}

The linearized gap equation shows that, over a broad parameter range, the solutions with highest $T_c$ correspond to $d$-wave pairing states forming a two-dimensional $E$ irreducible representation (irrep) of the lattice point group $D_3$. Individual configurations within the $E$ manifold can be classified by their transformation properties under the symmetry generators. The pairing is also strongly modulated on the moir\'e scale, with the strongest pairing occurring in the $AA$ regions of the moir\'e unit cell. The central site of the $AA$ region is a symmetry center, making the symmetries of each configuration most apparent there. We schematically illustrate this in Fig.~\ref{fig:atomistic:symmetry}, where sites in the top and bottom layers are indicated by orange and gray dots, respectively. Bond thickness encodes the pairing amplitude and color hue represents its phase (thin gray bonds indicate vanishing pairing). Bonds from the bottom layer are represented in a darker shade. In all cases, the solutions exhibit a $\pi$-phase shift between the layers, a feature known to be essential for their stabilization \cite{Wu2019, Fischer2021, Lothman2022}, attributed to a significant layer counterflow velocity \cite{Bistritzer2011a}.

The two fixed points of the $C_{2y}$ transformation within the $E$ manifold correspond to so-called \emph{nematic} states in Fig.~\ref{fig:atomistic:symmetry}(a,b). Their nematicity spontaneously breaks $C_{3z}$ symmetry and is reflected in a preferred direction of pairing, which we use to label them. The state $\hat{\Delta}_{x}$ in Fig.~\ref{fig:atomistic:symmetry}(a) is an eigenstate of $C_{2y}$ with eigenvalue $+1$, such that pairing vanishes on bonds most nearly parallel to the $y$ direction. By contrast, the state $\hat{\Delta}_{y}$  in Fig.~\ref{fig:atomistic:symmetry}(b) exhibits its strongest pairing along the $y$ direction and transforms with eigenvalue $-1$ under $C_{2y}$. Importantly, the states $\hat{\Delta}_{x,y}$ are real-valued, and all linear combinations of them with real-valued coefficients are also nematic configurations. The nematic states resemble the nodal $d$-wave configurations of single-band superconductors. However, as we will later show, the nematic states in TBG host substantial inter-eigenband pairing components. Instead considering the fixed points of the $C_{3z}$ rotation, and thus states respecting the in-plane rotation symmetry of TBG, gives the \emph{chiral} states $\hat{\Delta}_{\chi}$, with $\chi = \pm$, see Fig.~\ref{fig:atomistic:symmetry}(c). These have eigenvalues $e^{i \chi \omega}$ of $C_{3z}$, where $\omega = 2\pi/3$, reflecting the angular momentum of the Cooper pairs through the chirality $\chi=\pm$ \cite{Sigrist1991, Kallin2016}. Thus, these states spontaneously break both the twofold rotation $C_{2y}$ and spinless time-reversal symmetry $\mathcal{T}$. The latter implies that their pairing matrix cannot be made real-valued by any global gauge choice. Instead, the operations $C_{2y}$ and $\mathcal{T}$ relate states of opposite chirality.

Up to a global $U(1)$ phase, any state $\hat{\Delta}$ in the $E$ manifold can be expressed in the chiral basis as
\begin{align} 
\label{eq:atomistic:parametrization}
	\hat{\Delta}
	&= \Delta_0 \left( \hat{\Delta}_{+} \cos \tfrac{\alpha}{2} \, e^{-i \tfrac{\phi}{2}}
	+ \hat{\Delta}_{-} \sin \tfrac{\alpha}{2} \, e^{+i \tfrac{\phi}{2}} \right),
\end{align}
where we define the norm of an order parameter configuration as $|\hat{\Delta}|^2 = \sum_{\langle i,j \rangle} |\hat{\Delta}_{ij}|^2$, with the pairing amplitude $\Delta_0 = |\hat{\Delta}| > 0$, and normalize the chiral basis states such that $|\hat{\Delta}_{\chi}| = 1$. The angles $\alpha \in [0,\pi]$ and $\phi \in [0,2\pi]$ parametrize a sphere. The two chiral states $\hat{\Delta}_{\chi}$ are at the poles $\alpha = 0,\pi$, while states along the equator, $\alpha = \pi/2$, are real-valued and with a preferred pairing direction and are thus nematic, with $\phi = 0$ corresponding to $\hat{\Delta}_{x}$ and $\phi = \pi$ to $\hat{\Delta}_{y}$.
At $T_c$, all configurations $(\alpha, \phi)$ in Eq.~\eqref{eq:atomistic:parametrization} are degenerate due to the linearity of the gap equation. 

\subsection{Competition between nematic and chiral superconductivity}
\label{sec:atomistic:transition}

\begin{figure}
	\includegraphics[width=\columnwidth]{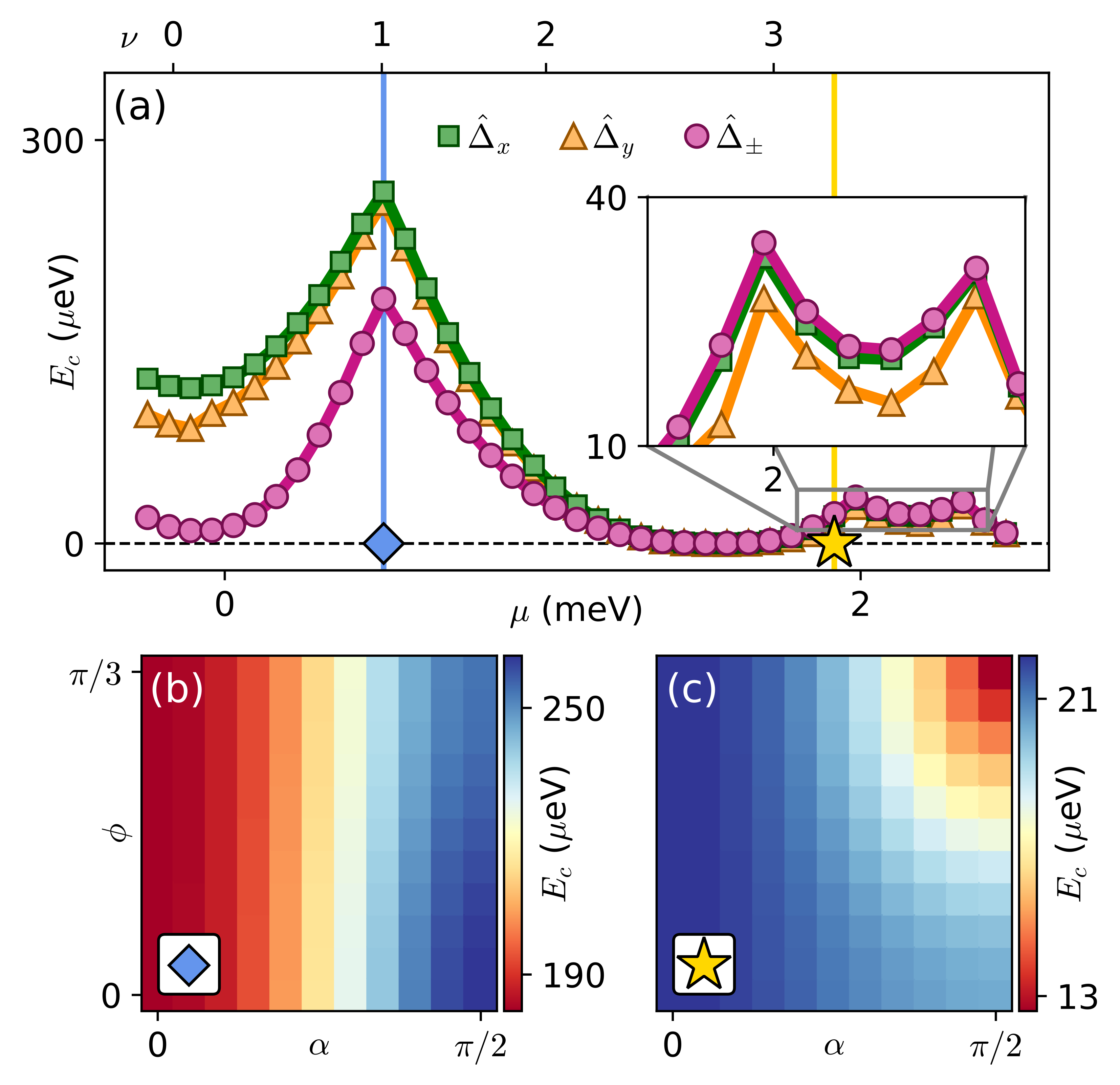}
	\caption{Condensation energy $E_c(\Delta_\star, \alpha, \phi)$ for nearest-neighbor pairing with $J\approx1.35$~eV. (a) Nematic $\hat{\Delta}_x$ (green squares), $\hat{\Delta}_y$ (orange triangles), and chiral $\hat{\Delta}_{\pm}$ (magenta circles) states as a function of doping $\mu$ measured from the Dirac point energy. The normal-state filling factor $\nu$ shown at the top was calculated with a higher $k$-point sampling and is for reference only. Inset shows small region with chiral ground state. (b,c) $E_c$ for different pairing configurations $(\alpha, \phi)$ at (b) $\mu\approx0.5$~meV (blue line/diamond in (a)) and (c) $\mu\approx2$~meV (yellow line/star in (a)). At low doping (b) the nematic state $(\alpha,\phi)=(\pi/2,0)$ ($\hat{\Delta}_x$) is the ground state, while the chiral state $\alpha=0$ ($\hat{\Delta}_{+}$) wins at large doping (c). }
	\label{fig:atomistic:freeenergy}
\end{figure}

Below $T_c$ nonlinear effects emerge and a specific configuration becomes energetically favored. We find this ground state by self-consistently solving Eqs.~\eqref{eq:atomistic:Hamiltonian} and~\eqref{eq:atomistic:gap} within the leading manifold Eq.~\eqref{eq:atomistic:parametrization}~\footnote{The basis matrices in Eq.~\eqref{eq:atomistic:parametrization} may in principle change below $T_c$, including nucleating a finite extended $s$-wave component. However, we find no such evidence, even at $T=0$ or with changing doping. We attribute this stability to the fact that the $N$ leading solutions to the linearized gap equation (with at least $N>10$) have only local $d$-wave components. We therefore always use basis matrices extracted at $\mu \approx 0.5$~meV, corresponding to the positive energy van Hove peak.}.
We note that our model is motivated by RPA analysis, where the attractive interaction $J$ is derived starting from a bare on-site Hubbard repulsion $U$ \cite{Fischer2021}. At the same time, \emph{Ab initio} estimates for $U$ vary substantially with factors such as screening \cite{Wehling2011, Schler2013, Rosner2015, DLee2025}, which in turn depend on the experimental setup \cite{Saito2020, Stepanov2020, Liu2021}. $J$ should therefore be viewed as a tuning parameter. For concreteness, we here use $J\approx1.35$~eV, corresponding to a $T_c$ of a few kelvin. This is of the same order as the nearest-neighbor attraction in the RPA analysis above, whose strength is ultimately set by $U$ \cite{Fischer2021}. Note that this corresponds to an attraction of only $0.6$~meV on the moir\'e scale (see SM). The same analysis yielded other, longer-ranged attractive channels, but which are subleading and which we here discard as an approximation in favor of computational feasibility. We expect that their inclusion would likely reduce the $J$ value needed to obtain the same $T_c$. Previous studies already pointed out the robustness of the manifold of solutions in Eq.~\eqref{eq:atomistic:parametrization} with respect to $J$ at $T_c$ \cite{Lothman2022}. At zero temperature, we find that $J$ essentially sets the energy scale for the quasiparticle energies and chemical potential in the phase transition between different superconducting ground states which we discuss below.

Since no pairing configuration preserves all normal-state symmetries, further spontaneous symmetry breaking necessarily occurs. However, in the absence of external perturbations, physically distinct states related by the broken symmetries still remain degenerate. Consequently, the nematic and chiral states are threefold and twofold degenerate, respectively. Generally, we can compare different configurations in Eq.~\eqref{eq:atomistic:parametrization} by extracting the superconducting condensation energy $E_c(\Delta_0, \alpha,\phi)$, calculated as the free energy $F$ gain relative to the normal state,
\begin{align} 
\label{eq:atomistic:condensation}
	E_c(\Delta_0, \alpha,\phi) = F(0) - F(\hat{\Delta}).
\end{align}
Superconducting phases are characterized by $E_c > 0$, with the ground state maximizing $E_c$. To ensure a fair comparison between different pairing symmetries, for each $(\alpha,\phi)$ we obtain the optimal amplitude $\Delta_\star$ maximizing $E_c$ by projecting onto the corresponding normalized $\hat{\Delta}$ at each iteration of the self-consistency loop.

In Fig.~\ref{fig:atomistic:freeenergy}(a) we show the condensation energy $E_c(\Delta_\star, \alpha, \phi)$ as a function of chemical potential $\mu$ for the chiral states $\hat{\Delta}_{\pm}$ (magenta) and the nematic states $\hat{\Delta}_{x}$, $\hat{\Delta}_{y}$ (green, orange). Computational constraints limit us to a coarse $4\times4$ sampling of momentum space (similar to patch models \cite{Chichinadze2020}). Finer sampling is expected to possibly shift the condensation peaks in $\mu$ ($\nu$), but our analysis in the SM indicates that the conclusions hold for denser sampling. For reference, we also show the filling factor $\nu$ calculated at a higher $k$-point sampling. All values of $\nu$ in this work are calculated in the normal state and measured from charge neutrality. We find that sizable $E_c$ generally correlates with the sampled normal-state DOS, peaking at $\mu \approx 0.5$~meV (blue diamond), corresponding here to $\nu\approx 1$, where nematic states are strongly favored.  As $\mu$ increases, the condensation energy decreases, until around $\mu\approx 2$~meV (yellow star), where instead the chiral state acquires a slight energetic advantage. We note that the precise value of $\mu$ (and $\nu$) corresponding to the transition depends on the choice of the interaction strength $J$, with weaker interactions leading to transitions at smaller $\mu$ ($\nu$). Although we here focus on the parameter regime $\nu > 0$, we note that an approximate particle-hole symmetry of the symmetric parent state we consider here results in an approximately mirrored condensation energy landscape for $\nu < 0$. We also note that the appearance of finite superconductivity at integer fillings is a consequence of using a symmetric parent state. In Sec.~\ref{sec:parentstate} we for example show how the inclusion of inter-valley coherence in the parent state leads to suppression of superconductivity at $\nu=-2$.


In Figs.~\ref{fig:atomistic:freeenergy}(b,c) we consider the condensation energy landscape $E_c(\alpha,\phi)$ across the entire $E$ manifold at two different doping values. At $\mu \approx 0.5$~meV the ground state is the nematic state $\hat{\Delta}_{x}$, given by $(\alpha,\phi)=(\pi/2,0)$ (and other $C_{3z}$-related states), while at $\mu \approx 2$~meV, it shifts to the chiral states $\hat{\Delta}_{\pm}$, given by $(\alpha,\phi)=(\{0,\pi\},\phi)$.

Our results illustrate a strong preference for the nematic state in the considered regime of interaction strength and twist angle, as also found earlier \cite{Fischer2021, Lothman2022}, but also show that a nematic-to-chiral transition occurs at large doping. Similar transitions have been predicted previously, driven either by changes in interaction strength \cite{Su2018, Kozii2019, Chichinadze2020, Julku2020, Fischer2021, Wang2024} or by external fields \cite{Yu2021}. Here, by contrast, the transition is induced by a change in doping.

\subsection{Spectrum and sublattice-valley polarization}
\label{sec:atomistic:polarization}

\begin{figure} 
	\includegraphics[width=\columnwidth]{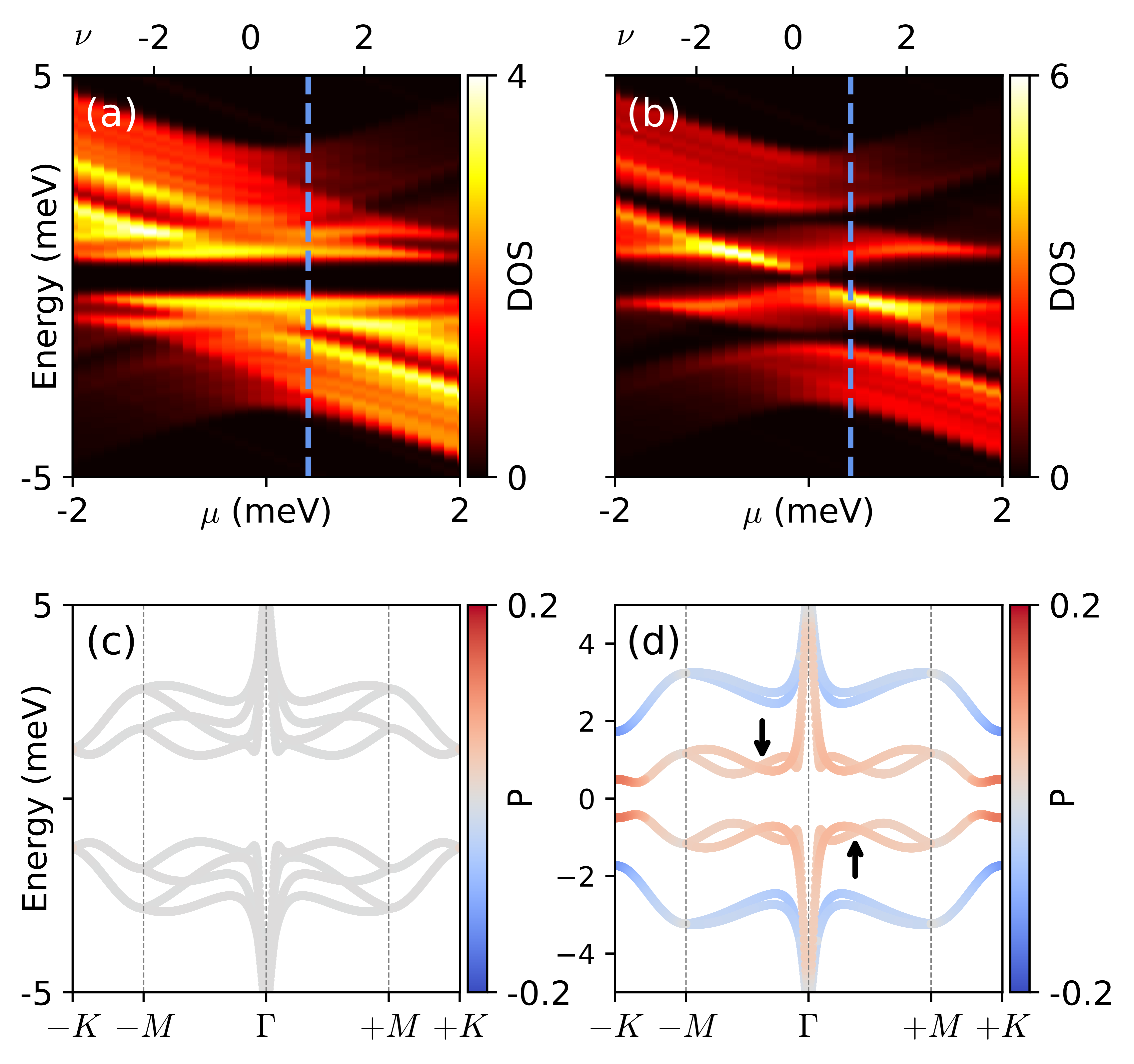}
	\caption{Spectrum of nematic $\hat{\Delta}_{x}$ (left) and chiral $\hat{\Delta}_{-}$ (right) states for nearest-neighbor pairing at fixed pairing strength $\Delta_0 \approx 49$~meV, corresponding to the self-consistent pairing strength for the nematic state at $\mu \approx 0.5$~meV in Fig.~\ref{fig:atomistic:freeenergy}, with an averaged bond pairing amplitude $\langle |\hat{\Delta}_{ij}| \rangle \approx 0.2$~meV. (a,b) Electronic DOS versus chemical potential $\mu$. Filling factor $\nu$ is shown for reference. (c,d) Band structure along high-symmetry lines at $\mu\approx0.5$~meV (blue dashed lines in (a,b)) with color according to sublattice-valley polarization $P$.}
	\label{fig:atomistic:polarization}
\end{figure}

In order to gain further insight into the competition between nematic and chiral states, we investigate their spectral properties in Fig.~\ref{fig:atomistic:polarization} for the nematic $\hat{\Delta}_{x}$ (left) and chiral $\hat{\Delta}_{-}$ (right) states. Top panels show the electronic DOS as a function of chemical potential $\mu$. The nematic state in Fig.~\ref{fig:atomistic:polarization}(a) exhibits an apparent spectral gap for all $\mu$, whereas the chiral state in Fig.~\ref{fig:atomistic:polarization}(b) displays pronounced low-energy features near $\mu=0$. We provide more detail by plotting the band structure at $\mu\approx0.5$~meV (dashed blue) in the bottom panels.
The nematic state is generally gapped along the high-symmetry lines, but still exhibits nodes away from these momenta (see SM). These nodes result from interference between kinetic and pairing terms and their locations in the moir\'e Brillouin zone are therefore not symmetry-constrained. Away from the nodal points, the bands rapidly open a sizable gap, which explains the apparent gap in Fig.~\ref{fig:atomistic:polarization}(a). This indicates that the nodal points have limited impact on the condensation energy, paralleling an argument for nematic superconductivity from electron-phonon coupling \cite{Liu2025}.
By contrast, the in-gap features of the chiral state have a more pronounced effect. Figure~\ref{fig:atomistic:polarization}(d) shows that these features correspond to an inner subset of flat bands (see black arrows), separated from the outer bands. The outer-band energies scale with the pairing strength $\Delta_0$, whereas the inner bands respond to the chemical potential, even crossing zero energy as $\mu$ is tuned across charge neutrality. This behavior suggests the existence of moir\'e flat bands not engaged in pairing for the chiral state.

We also investigate the wavefunctions of the nematic and chiral states. For this, we first compute the valley number $\eta=\pm$ of each energy eigenstate using the TBG valley operator \cite{Ramires2018, Colomes2018, Ramires2019} generalized to the Bogoliubov--de Gennes (BdG) framework. For all pairing configurations considered, we find that the wavefunctions of each band carry opposite values of $\eta$ in the particle and hole sectors, indicating pairing is inter-valley only. This means that, although we only assumed the Cooper pairs to have zero net momentum on the moir\'e scale, we find that this property is also present at the atomic scale. 
We also extract the sublattice polarization $\langle \hat{n}_A - \hat{n}_B \rangle$, where $\hat{n}_{A,B}$ is the number operator on sites of the denoted sublattice. We find a clear valley-dependent polarization for the chiral states, while for the nematic states this quantity is negligible.
The combined effect is captured by the sublattice-valley polarization ${P}$ 
\begin{align} \label{eq:atomistic:polarization}
	{P} =  \eta \, \langle \hat{n}_A - \hat{n}_B \rangle
\end{align}
which we use to color-code the bands in Fig.~\ref{fig:atomistic:polarization}(c,d). While this quantity is vanishing in the nematic state, the inner (outer) bands of the chiral state take positive (negative) values of ${P}$.

To summarize, electron-driven spin-singlet superconductivity in the atomic lattice yields a nematic $d$-wave state, notably more stable than the chiral $d$-wave state. Based on analyzing their energy spectra, we find that the chiral state is hampered by leaving two flat bands effectively unpaired. This gives a larger energy disadvantage for pairing than the nodal quasiparticles present in the nematic state.  However, it is still possible to transition to a chiral ground state at large doping. Understanding these behaviors requires an effective low-energy description, which we uncover and explore in the remainder of this work.

\section{Mapping to minimal Chern model} 
\label{sec:mapping}

To better understand electron-driven superconductivity, we first project the order-parameter configuration $\hat{\Delta}$, consisting of the pairing amplitudes on every nearest-neighbor carbon-carbon bond, onto the four moir\'e flat bands. Motivated by the vanishing kinetic energy in this subspace, we then adopt a basis that diagonalizes the pairing terms, which as we uncover is the so-called Chern basis.

\subsection{Projection onto moir\'e flat bands}
\label{sec:mapping:projection}
To project onto the moir\'e flat bands we first numerically extract the eigenvalues and eigenvectors of these bands from the atomistic normal-state Hamiltonian $\hat{h}_0$. At each momentum $\mathbf{k}$ we compute the sign $\eta$ of the expectation value of the valley operator \cite{Ramires2018, Colomes2018, Ramires2019}. For a fixed $\eta$, the top and bottom bands are labeled by the eigenband number $b=\pm$ \cite{Liu2025}. The flat-band energies and eigenvectors are thus labeled as $\epsilon_{b \eta}(\mathbf{k})$ and $\mid b~\eta~\mathbf{k}\, \rangle$, with the normal-state Hamiltonian matrix elements
\begin{align}
	h^{\eta \eta^\prime}_{bb^\prime} (\mathbf{k})
	&= \langle b ~ \eta ~ \mathbf{k}\mid  \hat{h}_0(\mathbf{k}) \mid b^\prime ~ \eta^\prime ~ \mathbf{k} \rangle.
\end{align}
Inter-valley scattering ($\eta\neq\eta^\prime$) is vanishing, such that
\begin{align}
	h^{\eta \eta^\prime}_{bb^\prime} (\mathbf{k})
	&= \delta_{\eta \eta^\prime} \, [h_{\eta}]_{b b^\prime},
	\quad 
	[h_{\eta}]_{b b^\prime} =  \delta_{b b^\prime} \, (\epsilon_{b \eta} (\mathbf{k}) - \mu).
\end{align}
Likewise, the pairing between electron bands with indices $(b,\eta)$ and hole bands with $(b^\prime, \eta^\prime)$ is obtained by projecting the pairing matrix onto the same flat-band eigenstates. Since the normal-state Hamiltonian appears in the hole sector as $-\hat{h}_0^*(-\mathbf{k})$, the associated eigenstates are $\mid b^\prime ~ \eta^\prime ~ (-\mathbf{k}) \, \rangle^*$. We then make use of the fact that complex conjugation acts as $\eta \rightarrow -\eta$ and $\mathbf{k} \rightarrow -\mathbf{k}$ to arrive at
\begin{align}
	{\Delta}^{\eta \eta^\prime}_{b b^\prime}(\mathbf{k})
	&= \langle\,  b ~ \eta ~ \mathbf{k}\mid  \hat{\Delta}(\mathbf{k}) \mid b^\prime ~ (-\eta^\prime) ~ \mathbf{k}\, \rangle,
\end{align}
where spin-singlet symmetry is implied. See SM for an explicit derivation. We here note that, as the wavefunctions $\mid b~\eta~\mathbf{k}\, \rangle$ are obtained numerically, the global phase of each eigenstate is uncontrolled, resulting in projected pairing matrices with random phases. We address this by a gauge-fixing procedure, resulting in pairing matrices locally smooth in reciprocal space. We choose the gauge preserving the real-valuedness of the nematic pairing matrices.

Analyzing the flat-band-projected pairing matrices of the nematic and chiral states we find that intra-valley components are always vanishing, see SM for an account of all components. Further, due to the spin-singlet anti-symmetry, the pairing matrix is symmetric in the remaining degrees of freedom, which means the two inter-valley sectors are related by ${\Delta}_{bb^\prime}^{+-}(\mathbf{k}) = {\Delta}_{b^\prime b}^{-+}(-\mathbf{k})$. For simplicity, we henceforth focus on the single sector ${\Delta}_{bb^\prime}^{+-}$ and suppress the valley index. Remarkably, we find that both intra-eigenband terms ($b=b^\prime$) and inter-eigenband terms ($b\neq b^\prime$) are present and with similar magnitude. We next examine their momentum space structure. 

\begin{figure} 
	\includegraphics[width=\columnwidth]{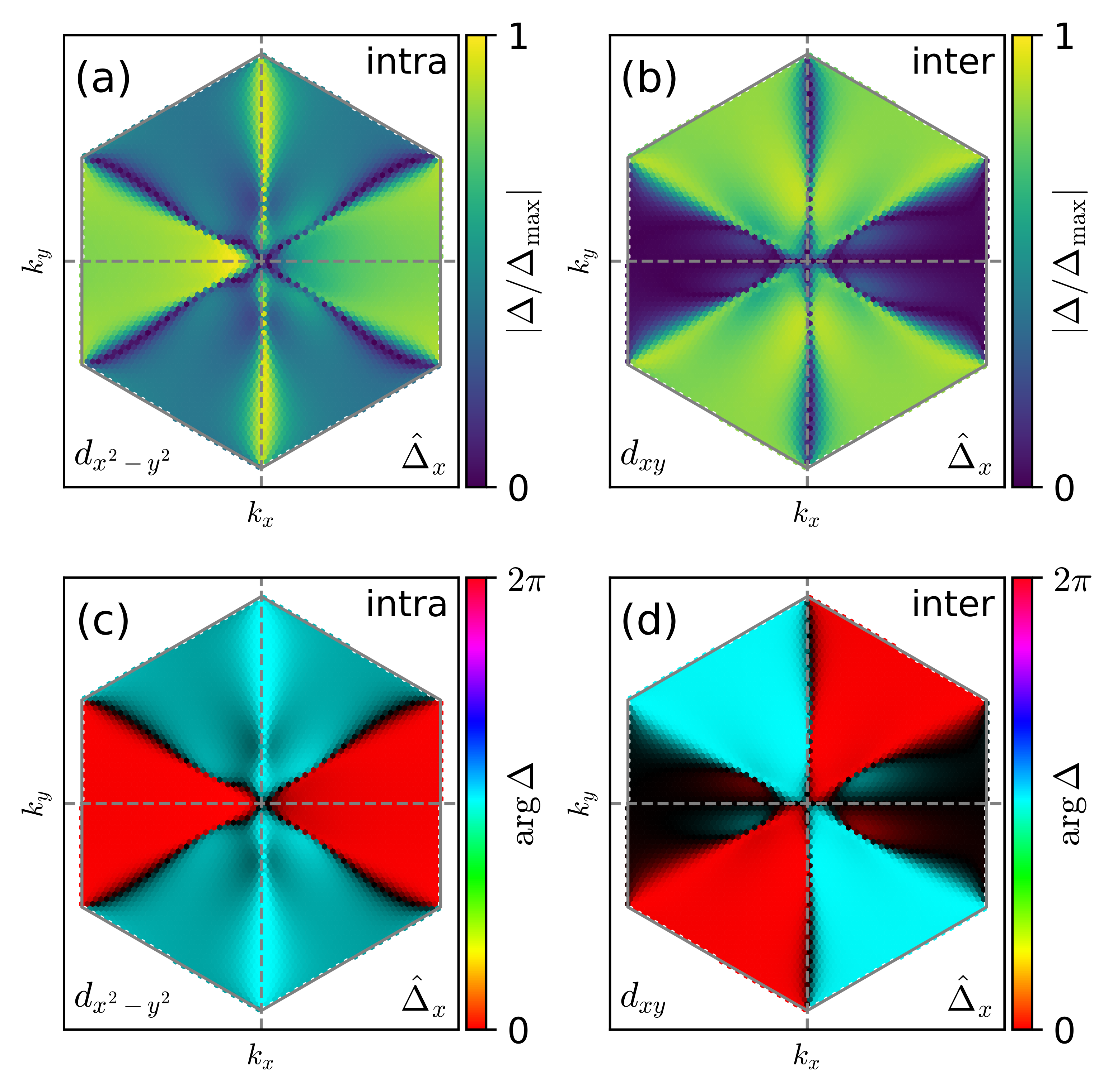}
	\caption{Amplitude (top) and phase (bottom) of the intra- ($\Delta_{++}$, left) and inter-eigenband ($\Delta_{-+}$, right) components of the nematic state $\hat{\Delta}_{x}$ projected onto the moir\'e flat bands in the first moir\'e Brillouin zone. Each component is identified with the form factor $d_{x^2-y^2}$ or $d_{xy}$ approximately satisfying the same symmetries, which become exact when accounting for valley conjugation. }
	\label{fig:mapping:nematic}
\end{figure}

For the nematic state $\hat{\Delta}_{x}$, we illustrate both the intra- and inter-eigenband components in Fig.~\ref{fig:mapping:nematic}. Panels (a,c) depict the amplitude and phase, respectively, of the intra-eigenband component $\Delta_{++}$, which shows an approximate even symmetry under $k_{x,y}$ mirror transformations (dashed gray lines). Panels (b,d) show the inter-eigenband component $\Delta_{-+}$, which is instead odd under these mirror operations. The amplitude plots share a common normalization $\Delta_{\text{max}}$, corresponding to the maximum over both momentum and band space, such that intra- and inter-eigenband terms can be compared. We find that $\Delta_{++} \approx \Delta_{--} \approx \tfrac{1}{2}\, d_{x^2-y^2}$ and $\Delta_{-+} \approx -\Delta_{+-} \approx \tfrac{1}{2}\, d_{xy}$, where $d_{x^2-y^2}$ and $d_{xy}$ are basis functions of the $E$ irreducible representation of the crystal group $D_{3}$ and with a prefactor introduced for later convenience. A discussion of the exact and approximate symmetries satisfied by the pairing matrix elements is found in the SM. We therefore obtain the approximate submatrix in eigenband space
\begin{align}
	{\Delta}_{x}
	&= \frac{1}{2} \begin{pmatrix} d_{x^2 - y^2} & - d_{xy} \\  d_{xy} & d_{x^2 - y^2} \end{pmatrix}.
\end{align}
Performing the same analysis for $\hat{\Delta}_{y}$ yields 
\begin{align}
	{\Delta}_{y}
	&= \frac{1}{2} \begin{pmatrix} d_{xy} & d_{x^2 - y^2} \\ - d_{x^2 - y^2} & d_{xy} \end{pmatrix}.
\end{align}
Then, the projected pairing matrix of the chiral states $\hat{\Delta}_{\pm}$ can be computed by recalling Eq.~\eqref{eq:atomistic:parametrization} and defining $d_{\pm} = (d_{x^2-y^2} \pm i d_{xy})/\sqrt{2}$, yielding
\begin{align} \label{eq:mapping:chiral_band}
	{\Delta}_{\pm}
	&= \frac{d_{\pm}}{2} \begin{pmatrix} 1 & \pm i \\ \mp i & 1 \end{pmatrix}.
\end{align}

To summarize, we find that the leading superconducting states for electron-driven (atomistic) pairing in TBG naturally host both intra- and inter-eigenband pairing, with distinct $d$-wave characters. Collecting the non-vanishing normal state and pairing terms we next construct a minimal Bogoliubov--de Gennes (BdG) Hamiltonian for the moir\'e flat bands. With valley being a preserved quantity in the normal state and only inter-valley pairing, we constrain ourselves to particles with $\eta=+$ and holes with $\eta^\prime=-$, corresponding to the Nambu spinor $\Psi_{\mathbf{k}} = ( \psi^{+ +}_{ \mathbf{k} \uparrow}, \psi^{- +}_{ \mathbf{k} \uparrow}, \psi^{+ - \dagger}_{ -\mathbf{k} \downarrow}, \psi^{- - \dagger}_{ -\mathbf{k} \downarrow } )^T$, where $\psi^{{b} \eta}_{\mathbf{k} \sigma}$ is the annihilation operator for the state with eigenband ${b}$, valley $\eta$, momentum $\mathbf{k}$, and spin $\sigma$. The BdG Hamiltonian is then
\begin{align} \label{eq:mapping:bdg_band}
	{H}_{\text{BdG}}(\mathbf{k}) &= 
	\begin{pmatrix} 
		{h}_{+}(\mathbf{k}) & {\Delta}(\mathbf{k}) \\ 
		{\Delta}^{\dagger}(\mathbf{k}) & -{h}_{-}^*(-\mathbf{k})
	\end{pmatrix},
\end{align}
which we numerically verify reproduces both the nodal points of the nematic states and the subgap bands of the chiral states discussed in Sec.~\ref{sec:atomistic:polarization}.

\subsection{Projection onto Chern basis}
\label{sec:mapping:chern}
In the moir\'e flat-band basis above, the normal-state contribution $h_{\eta}$ in the BdG Hamiltonian Eq.~\eqref{eq:mapping:bdg_band} is diagonal, whereas the pairing term $\Delta$ is not. In conventional superconductors, the band basis provides a natural starting point as it facilitates a projection onto the Fermi surface and inter-band pairing terms are typically neglected since they are off-shell in energy. However, in TBG and other flat-band systems, interaction strengths can easily exceed the bandwidth. As a result, inter-band pairing terms cannot be ignored \cite{Christos2023, Liu2025, Putzer2025}, while, at the same time, normal-state terms play a secondary role. We therefore adopt a basis in which $\Delta$ is diagonal instead, by performing the unitary transformation $\tilde{\Psi}_{\mathbf{k}} = U \, \Psi_{\mathbf{k}}$, with
\begin{align} \label{eq:mapping:chern_transform}
	U
	&= \begin{pmatrix} u & 0 \\ 0 & u \end{pmatrix},
	\quad
	u = \frac{1}{\sqrt{2}} \begin{pmatrix} 1 & i \\ 1 & -i \end{pmatrix}.
\end{align}

\begin{figure} 
	\includegraphics[width=\columnwidth]{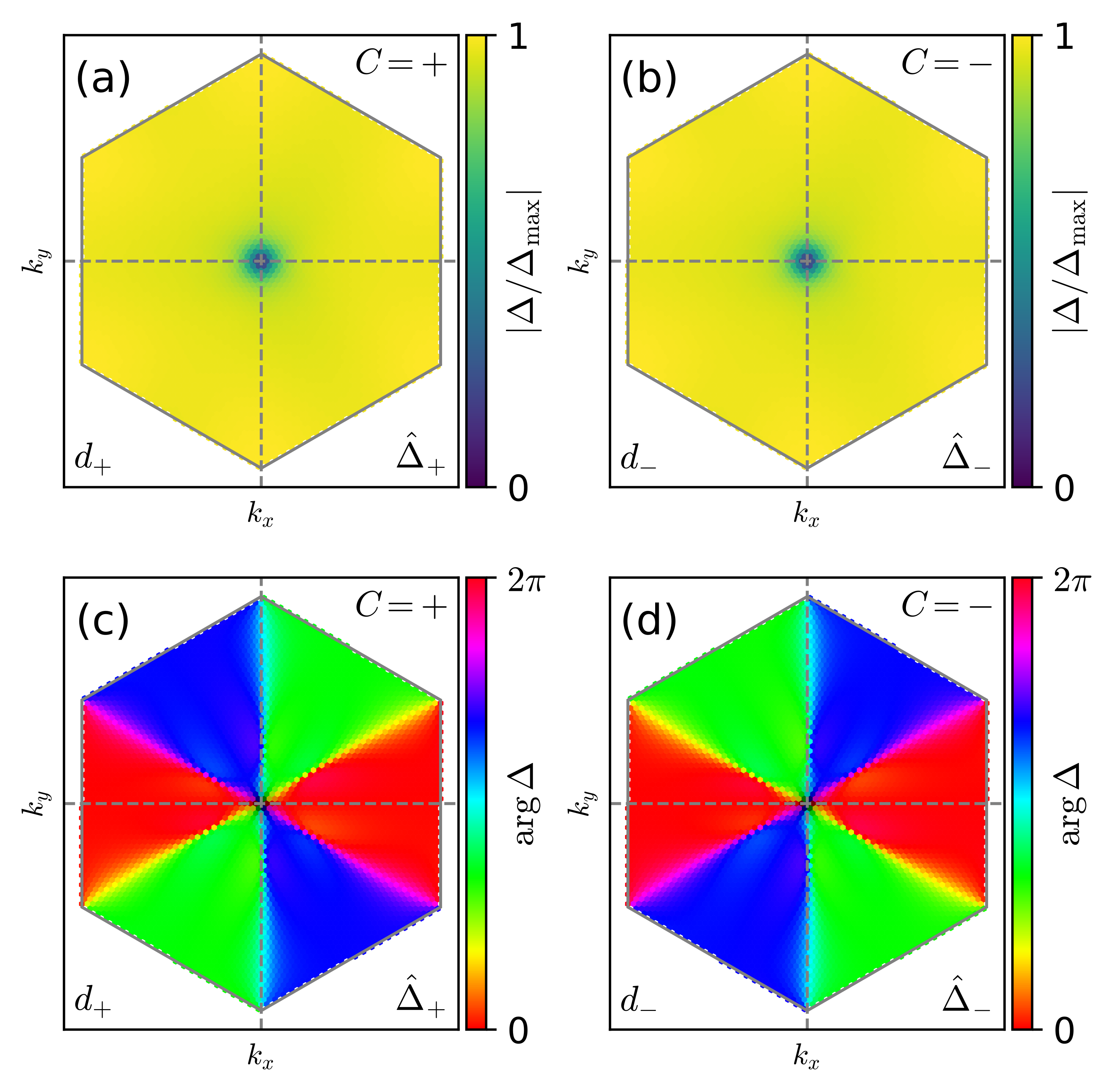}
	\caption{Amplitude (top) and phase (bottom) of the chiral form factors $d_{+}$ (left) and $d_{-}$ (right), extracted in each Chern sector $C=\pm$ for each chiral state $\hat{\Delta}_{\pm}$.}
	\label{fig:mapping:chiral}
\end{figure}

The pairing term transforms to $\tilde{\Delta}= u \, {\Delta} \, u^\dagger$, so that the chiral states in Eq.~\eqref{eq:mapping:chiral_band} become
\begin{align} \label{eq:mapping:chiral_chern}
	\tilde{\Delta}_{+} = \begin{pmatrix} d_{+} & 0 \\ 0 & 0 \end{pmatrix}, 
	\quad \tilde{\Delta}_{-} = \begin{pmatrix} 0 & 0 \\ 0 & d_{-} \end{pmatrix}.
\end{align}
We confirm this structure holds numerically by directly projecting and transforming the chiral states $\hat{\Delta}_{\pm}$, see SM. The resulting form factors $d_{\pm}$ obtained numerically in the Chern basis are depicted in Fig.~\ref{fig:mapping:chiral}. Their amplitude (a,b) is essentially constant across the Brillouin zone except for a small region around $\boldsymbol{\Gamma}$, while the phases (c,d) wind a total of $4\pi$ in opposite directions around $\boldsymbol{\Gamma}$, with the direction set by the chirality.

We next define the momentum-dependent pairing strength $\tilde{\Delta}_0(\mathbf{k}) = \tfrac{1}{2} \Delta_0 \, \abs{d_{+}(\mathbf{k})} $ and phase $\phi_{\mathbf{k}} = \phi + \delta \phi_{\mathbf{k}}$, where $\delta \phi_{\mathbf{k}} = \arg(d_-(\mathbf{k})) - \arg(d_+(\mathbf{k}))$. Its negative, $-\delta \phi_{\mathbf{k}}$, is the locally preferred nematic direction shown in Fig.~\ref{fig:introduction:diagram}(c), see Sec.~\ref{sec:competition:threemode}. Finally, noting that $d_{-}(\mathbf{k}) = d_{+}^*(\mathbf{k})$ and plugging the transformed chiral matrices in Eq.~\eqref{eq:mapping:chiral_chern} into Eq.~\eqref{eq:atomistic:parametrization}, we obtain the diagonal form of the pairing matrix for all superconducting states within the leading $E$ manifold
\begin{align} \label{eq:mapping:chern_pairing}
	\tilde{\Delta}(\mathbf{k}) = 2
	\tilde{\Delta}_0(\mathbf{k})
	\begin{pmatrix} 
		\cos \frac{\alpha}{2} \, e^{- i \phi_{\mathbf{k}}/2} & 0 \\ 
		0 & \sin \frac{\alpha}{2} \, e^{+ i \phi_{\mathbf{k}} / 2} 
	\end{pmatrix}.
\end{align}

We next calculate the transformed BdG Hamiltonian $\tilde{H}_{\text{BdG}} = U {H}_{\text{BdG}} U^\dagger$. For the normal-state term we have $\tilde{h}_{\eta} = u \, h_{\eta} \, u^\dagger$. We then define the energy splitting $\tilde{\epsilon}_{\eta} = \tfrac{1}{2} [\epsilon_{+, \eta}(\mathbf{k}) - \epsilon_{-, \eta}(\mathbf{k})]$ and the energy offset $\tilde{\mu}_{\eta} = \mu - \tfrac{1}{2} [\epsilon_{+, \eta}(\mathbf{k}) + \epsilon_{-, \eta}(\mathbf{k})] $ between the two flat bands in the same valley, as illustrated in Fig.~\ref{fig:atomistic:bands}(b). This allows us to rewrite $h_{\eta} = \tilde{\epsilon}_{\eta} \, \zeta_z - \tilde{\mu}_{\eta} \, \zeta_0$, where $\zeta_0$ and $\zeta_{x,y,z}$ are the identity and Pauli matrices in eigenband space, respectively. Noting that $u \, \zeta_z \, u^\dagger = \zeta_x$, we find
\begin{align} \label{eq:mapping:chern_normal}
	\tilde{h}_{\eta}(\mathbf{k}) = \tilde{\epsilon}_{\eta}(\mathbf{k}) \, \zeta_x - \tilde{\mu}_{\eta}(\mathbf{k}) \, \zeta_0.
\end{align}
Since $\epsilon_{b, \eta} (\mathbf{k}) = \epsilon_{b, -\eta} (-\mathbf{k})$, the effective parameters $\tilde{\epsilon}$ and $\tilde{\mu}$ inherit the same symmetry, as does the Hamiltonian $\tilde{h}$. Moreover, since Eq.~\eqref{eq:mapping:chern_normal} is real-valued, the hole sector of the BdG Hamiltonian transforms to $-\tilde{h}_{-}^*(-\mathbf{k}) = -\tilde{h}_{+}(\mathbf{k})$. Collecting the results, the transformed BdG Hamiltonian is
\begin{align} \label{eq:mapping:bdg_chern}
	\tilde{H}_{\text{BdG}}(\mathbf{k}) = 
	\begin{pmatrix} 
		\tilde{h}_{+}(\mathbf{k}) & \tilde{\Delta}(\mathbf{k}) \\ 
		\tilde{\Delta}^{\dagger}(\mathbf{k}) & -\tilde{h}_{+}(\mathbf{k})
	\end{pmatrix}.
\end{align}

The basis resulting from the transformation in Eq.~\eqref{eq:mapping:chern_transform} is in fact the well-known Chern basis \cite{Tarnopolsky2019, TBG_III}. The resulting Chern bands correspond to a superposition of the top and bottom eigenbands within each valley sector, such that they each host a finite Chern number $C=\pm 1$. This reflects the fact that within each valley the two Dirac points in the moir\'e Brillouin zone possess the same chirality \cite{Zou2018b}. From Eq.~\eqref{eq:mapping:chern_pairing}, we immediately find that pairing between different Chern sectors is always vanishing. Thus, by performing a projection onto the moir\'e flat-band subspace, along with a basis change and using symmetry properties of the normal state, we find that electron-driven nearest-neighbor pairing in TBG is well described by a minimal model where only pairing between bands with the same Chern number is present, so-called {\it intra-Chern pairing}.

\section{Intra-Chern pairing}
\label{sec:intra}

Having established that all leading $d$-wave nearest-neighbor pairing possibilities in TBG map to $d$-wave inter-valley and intra-Chern pairing, we next discuss the properties of this pairing. We are particularly interested in how it explains the spectral properties discussed in Sec.~\ref{sec:atomistic:polarization} and its relation to recent works focusing on electron-phonon pairing \cite{Liu2024, Liu2025, Wang2024}.

\subsection{Properties of intra-Chern pairing}
\label{sec:intra:properties}

The angle $\alpha$, which characterizes each superconducting state through Eq.~\eqref{eq:atomistic:parametrization}, or Eq.~\eqref{eq:mapping:chern_pairing} in the Chern basis, determines the relative pairing amplitude in the two Chern sectors. The nematic states, corresponding to $\alpha=\pi/2$, have equal magnitude pairing in both sectors, as schematically illustrated in Fig.~\ref{fig:introduction:diagram}(a). The angle $\phi$, which sets the (global) nematic direction breaking the $C_{3z}$ symmetry, appears in the Chern basis through the momentum-dependent relative phase $\phi_{\mathbf{k}}$ between the two Chern bands. Because of this phase, the Chern basis pairing matrix Eq.~\eqref{eq:mapping:chern_pairing} is generally complex, even for the nematic states preserving time-reversal symmetry. Time-reversal symmetry is instead manifest in the equal amplitude of both intra-Chern components. 
For the chiral states, corresponding to $\alpha=0,\pi$, we find that pairing vanishes within one of the Chern sectors, leaving the pairing matrix Eq.~\eqref{eq:mapping:chiral_chern} rank-deficient. Notably, this rank-deficiency is dictated by the chiral state preserving the $C_{3z}$ symmetry of the lattice and is thus an inherent and necessary property of the chiral state. Consequently, one set of Chern bands must always be left completely unpaired in the chiral state, as schematically illustrated in Fig.~\ref{fig:introduction:diagram}(b), a configuration to which we refer as (full) Chern-polarized pairing. Note that the nematic states, due to time-reversal symmetry, are instead guaranteed to have zero Chern polarization.

We next establish that the Chern polarization explains the sublattice-valley polarized subgap bands appearing in the chiral state in Fig.~\ref{fig:atomistic:polarization}(d). To understand this, we first consider the so-called chiral limit of the normal state, obtained by artificially suppressing all intra-sublattice hopping processes in Eq.~\eqref{eq:atomistic:Hamiltonian}. At the magic angle of this limit, and within the continuum approximation, all Chern bands become exact zero-energy eigenstates of the normal state, thus degenerate and absolutely flat \cite{Tarnopolsky2019}. There they also show perfect, valley-dependent sublattice polarization $P$, given in  Eq.~\eqref{eq:atomistic:polarization}~\cite{Bultinck2020, Wang2021}. 
By next introducing chiral pairing $\tilde{\Delta}_{\pm}$, we pair states in opposite valleys and in a single Chern sector, set by the chirality. Then, at each momentum, a spectral gap proportional to $\tilde{\Delta}_0(\mathbf{k})$ immediately opens up in that Chern sector, due to the absence of kinetic energy. By contrast, the normal-state bands from the unpaired Chern sector remain entirely unaffected by superconductivity.  Also, because only bands from opposite valleys and the same Chern number pair up, the pairs retain the sublattice-valley polarization of the normal-state bands. 

When moving away from the chiral limit by reinstating intra-sublattice hopping in Eq.~\eqref{eq:atomistic:Hamiltonian}, Chern bands within the same valley start to hybridize, leading to a partial suppression of their sublattice-valley polarization. In doing so, their energy dispersion also changes, such that the unpaired bands are no longer perfectly flat at the Fermi level. Still, for a pairing strength $\tilde{\Delta}_0$ sufficiently large compared to the normal-state hybridization energy $\tilde{\epsilon}$, the unpaired bands remain as partially polarized bands inside the superconducting gap opened by the paired bands. This is precisely what we see in Fig.~\ref{fig:atomistic:polarization}(d). Moreover, to first approximation, tuning the chemical potential $\mu$ shifts the subgap bands in energy, while keeping the gap unchanged. This means the subgap bands cross the superconducting gap as the chemical potential is tuned across charge neutrality, as we also find in Fig.~\ref{fig:atomistic:polarization}(b). 

We note that for the nematic states there is no unpaired Chern sector and thus no subgap states appear, as also evident in Fig.~\ref{fig:atomistic:polarization}(c). Moreover, their equal pairing in both Chern sectors explains the absence of sublattice-valley polarization. The intra-Chern minimal model Eq.~\eqref{eq:mapping:bdg_chern} thus not only reproduces, but also explains the low-energy spectral and polarization properties for electron-driven nearest-neighbor pairing. As such, it also explains the overall weakness of the chiral state compared to the nematic state: the subgap bands inherent to the chiral states are typically more detrimental to condensation energy than the narrow nodal points of the nematic state. This detrimental effect becomes less pronounced with doping, but is always finite, such that the picture above alone cannot account for a chiral ground state. Therefore, the nematic-to-chiral transition observed at large doping in Fig.~\ref{fig:atomistic:freeenergy}(a) suggests the existence of an additional mechanism suppressing the nematic state in that regime, a momentum-space frustration which we uncover in Sec.~\ref{sec:competition}.

\subsection{Alternative route to intra-Chern pairing}
\label{sec:intra:comparison}

Our effective model Eq.~\eqref{eq:mapping:bdg_chern} for electron-driven pairing closely resembles the $d$-wave intra-Chern model found as the subleading instability from electron-phonon interactions in Ref.~\cite{Liu2024}. We here compare the two. Remarkably, the two models only differ in the detailed momentum dependence of their Hamiltonian matrix elements. In Ref.~\cite{Liu2024}, the order-parameter form factor within each Chern sector takes the analytic form $d^{\text{ph}}_{\pm} = (k_x \pm i k_y)^2/(\mathbf{k}^2 + k_0^2)$, with $k_0$ a constant, for electron-phonon superconductivity. In our results, $d_{\pm}$ is instead obtained numerically but exhibits the same essential features characteristic of chiral $d$-wave superconductivity: a node at $\boldsymbol{\Gamma}$ surrounded by a $4\pi$ phase winding. The existence of nodes in the intra-Chern pairing matrix elements is guaranteed by the non-trivial Euler class in the flat bands responsible for their topological obstruction \cite{Yu2023, Ahn2019, Song2019b, TBG_II}.

In addition, our numerical results in Figs.~\ref{fig:mapping:nematic} and~\ref{fig:mapping:chiral} exhibit momentum-space periodicity as a direct consequence of the lattice model, while $d^{\text{ph}}_\pm$ lacks this property and should therefore only be interpreted as an expansion around the $\boldsymbol{\Gamma}$ point. Crucially, aside from a subtle point regarding the existence of exact nodes in the nematic state (discussed in the SM), these minor mathematical differences are not expected to significantly affect the physical pairing behavior.

Despite describing equivalent low-energy physics, Eq.~\eqref{eq:mapping:bdg_chern} and the intra-Chern model of Ref.~\cite{Liu2024} differ dramatically in origin. The latter employed a continuum low-energy formalism to study the coupling of the TBG flat bands to monolayer phonon modes, finding that only $K$-phonon modes can generate the scattering necessary for inter-valley pairing. Inter-Chern and intra-Chern channels were then found as the leading and subleading solutions to the linearized gap equation, respectively. Since the inter-Chern solution showed extended $s$-wave character, it had to be artificially neglected in order to reproduce the nematicity seen in experiments. 
This suggests that, despite experimental evidence of strong coupling to $K$-phonons \cite{Chen2024}, and results suggesting that such phonons are necessary to appropriately describe the phase diagram \cite{Kwan2024, Shi2025}, phonon interaction alone cannot explain superconductivity in TBG. Calculations taking into account the effects of Coulomb interaction further support this idea \cite{Wagner2024, Braz2024, ZiweiWang2025}. In contrast, in this work we generate the same intra-Chern pairing, and with no competing inter-Chern pairing, by considering electron-driven pairing on nearest-neighbor carbon bonds. Thus, these two mechanisms can reinforce each other constructively in the intra-Chern channel to generate nematic superconductivity.

As mentioned above, Refs.~\cite{Liu2024, Liu2025} identified intra-Chern pairing as a subleading instability in the bare bands of TBG, with extended $s$-wave inter-Chern pairing emerging as the dominant channel. The energetic competition between these two solutions is nevertheless very close, suggesting that relatively weak external perturbations can readily alter their ordering. In this regard, inter-valley inter-Chern pairing is predominantly intra-sublattice because of the sublattice polarization of the Chern bands. Consequently, repulsive intra-sublattice interactions are expected to suppress the inter-Chern channel. Consistent with this picture, Ref.~\cite{Wang2024} showed that incorporating a large on-site Hubbard interaction $U$ into a phonon-mediated pairing model stabilizes a $d$-wave ground state with signatures consistent with intra-Chern pairing. Likewise, nearest-neighbor $d$-wave pairing arises from spin fluctuations generated by a strong on-site interaction \cite{Fischer2021}. Together, these results suggest that strong on-site repulsion plays a central role in both phonon- and electron-mediated pairing in TBG. Rather than favoring a single microscopic mechanism, superconductivity may instead emerge from the cooperative interplay between phonons and electronic interactions, with both channels, in the presence of sufficiently strong on-site repulsion, promoting the same intra-Chern pairing state.

\section{Competition within the intra-Chern model}
\label{sec:competition}

The mapping in Sec.~\ref{sec:mapping} establishes intra-Chern pairing as a unifying framework for TBG superconductivity, finally clarifying connections between previously distinct models that both predict nematic and chiral superconductivity \cite{Fischer2021, Lothman2022, Liu2024, Liu2025, Wang2024}. The pressing question we address in this section is to explain the competition between these superconducting states in a symmetric parent state, including why and when a phase transition between them occurs. We do so aiming to find agreement with experimental findings of nematic signatures in the superconducting state in TBG  \cite{Cao2021a, Oh2021, DiBattista2022}.

Earlier results have argued \cite{Liu2025} that the disadvantage posed by the nodal quasiparticles in the nematic state is minimized by a quick recovery of the gap away from these momenta due to inter-eigenband pairing. On the other hand, we show through Eq.~\eqref{eq:mapping:chiral_band} that the chiral state also has large inter-eigenband pairing. At the same time, in Sec.~\ref{sec:intra} we establish the existence of subgap bands largely untouched by superconductivity in the chiral state, which are clearly unfavorable for the superconducting condensation energy. As a result, analyzing either the energy spectrum or the existence of inter-eigenband pairing is not sufficient to understand the competition between chiral and nematic superconductivity in TBG.

To investigate the nematic-chiral competition, we compare the condensation energies $E_c(\Delta_0,\alpha,\phi)$ of different superconducting states by constructing the mean-field Hamiltonian of the intra-Chern model. Starting from Eq.~\eqref{eq:atomistic:Hamiltonian} and performing the mapping in Sec.~\ref{sec:mapping}, while recalling that $\Delta_0 = |\hat{\Delta}|$ and introducing the renormalized interaction $\tilde{J} = J \abs{d_{+}(\mathbf{k})}^2$, we arrive at the 
minimal intra-Chern pairing model
\begin{align} \label{eq:competition:chern_meanfield}
	\tilde{H}
	&= \sum_{\mathbf{k}} \left[ \tfrac{1}{2}
	\tilde{\Psi}_{\mathbf{k}}^\dagger \tilde{H}_{\text{BdG}}(\mathbf{k}) \tilde{\Psi}_{\mathbf{k}}
	+ \tilde{\Delta}_0^2 / \tilde{J}
	\right],
\end{align}
where we omit an irrelevant constant energy shift from normal-state terms. Following Eq.~\eqref{eq:atomistic:condensation}, we then extract the condensation energy for a pairing configuration specified by $\Delta_0$, $\alpha$, and $\phi$. Notably, this expresses the condensation energy as a sum over momentum contributions. We next focus on a small set of representative momenta, where we can make analytical comparisons between the nematic and chiral condensation energies.

\subsection{Single-mode limit}
\label{sec:competition:singlemode}

We begin by restricting the sum in Eq.~\eqref{eq:competition:chern_meanfield} to a single momentum mode. In this \emph{single-mode limit}, the $\mathbf{k}$-dependence of $\phi_\mathbf{k}$ in Eq.~\eqref{eq:mapping:chern_pairing} becomes irrelevant and we set $\delta \phi_\mathbf{k} = 0$. This means the phase between the Chern sectors $\phi_\mathbf{k}$ is then simply equal to the global nematic direction, $\phi_\mathbf{k} = \phi$. Diagonalizing the intra-Chern Hamiltonian Eq.~\eqref{eq:mapping:bdg_chern} gives the eigenvalues
\begin{align} \label{eq:competition:singlemode_eigenvalues}
	\varepsilon_{\pm}^2
	&= \tilde{\epsilon}^2 + \tilde{\mu}^2 + 2\tilde{\Delta}_0^2 \nonumber \\
	& \pm 2 \tilde{\Delta}_0^2 \sqrt{\cos^2\alpha + \frac{\tilde{\epsilon}^2}{\tilde{\Delta}_0^2} \left(1 - \cos\phi \sin\alpha + \frac{\tilde{\mu}^2}{\tilde{\Delta}_0^2}\right)}.
\end{align}
Thus the single-mode contribution to the condensation energy of a given pairing configuration becomes
\begin{align} \label{eq:competition:singlemode_condensation}
	E^{(1)}_{c} (\alpha, \phi)
	&= \tfrac{1}{2} (\varepsilon_{+} + \varepsilon_{-}) - \max(|\tilde{\mu}|, |\tilde{\epsilon}|) - \tilde{\Delta}_0^2 / \tilde{J},
\end{align}
where we omitted the explicit dependence on $\tilde{\Delta}_0$ to focus instead on the behavior with respect to the angles $\alpha$ and $\phi$.

Immediately, we find that the chiral states $\tilde{\Delta}_{\pm}$ ($\alpha = 0, \pi$) are degenerate, and their energies also do not depend on $\phi$, as expected. We next directly compare the condensation energy of the nematic states ($\alpha = \pi/2$) to the chiral ones through 
\begin{align} \label{eq:competition:singlemode_cond_diff}
	\delta E^{(1)}_{c} (\phi)
	&\equiv E^{(1)}_{c} (\pi/2, \phi) - E^{(1)}_{c} (0,0).
\end{align}
In particular, for the state $\tilde{\Delta}_x$ ($\phi=0$) we find
\begin{align}
	\delta E^{(1)}_{c} (0)
	&= g\left( |\tilde{\epsilon} \, \tilde{\mu}| \right) - g \left( \sqrt{ \tilde{\Delta}_0^4 + \tilde{\epsilon}^2 \tilde{\Delta}_0^2 + \tilde{\epsilon}^2 \tilde{\mu}^2} \right),
\end{align}
where we define the auxiliary function $g(t) = \tfrac{1}{2}  \sqrt{\tilde{\epsilon}^2 + \tilde{\mu}^2 + 2\tilde{\Delta}_0^2 - 2t} + \tfrac{1}{2}  \sqrt{\tilde{\epsilon}^2 + \tilde{\mu}^2 + 2\tilde{\Delta}_0^2 + 2t}$. Since $g(t)$ is strictly decreasing for $t>0$, it follows that $\delta E^{(1)}_{c} > 0$ for all $\tilde{\Delta}_0 > 0$, implying that the nematic state $\tilde{\Delta}_x$ always has a higher condensation energy. The chiral state's weakness originates from the $\cos^2\alpha$ term inside the square root in Eq.~\eqref{eq:competition:singlemode_eigenvalues}, which produces a finite splitting between the two eigenvalue branches. This keeps the lower branch near its normal-state value, suppressing condensation. The lower branches from both valley sectors are precisely the unpaired subgap bands always present in the chiral state, see Figs.~\ref{fig:atomistic:polarization}(b,d), whose physical origin is the Chern polarization of the chiral state.

\begin{figure*}
	\includegraphics[width=\textwidth]{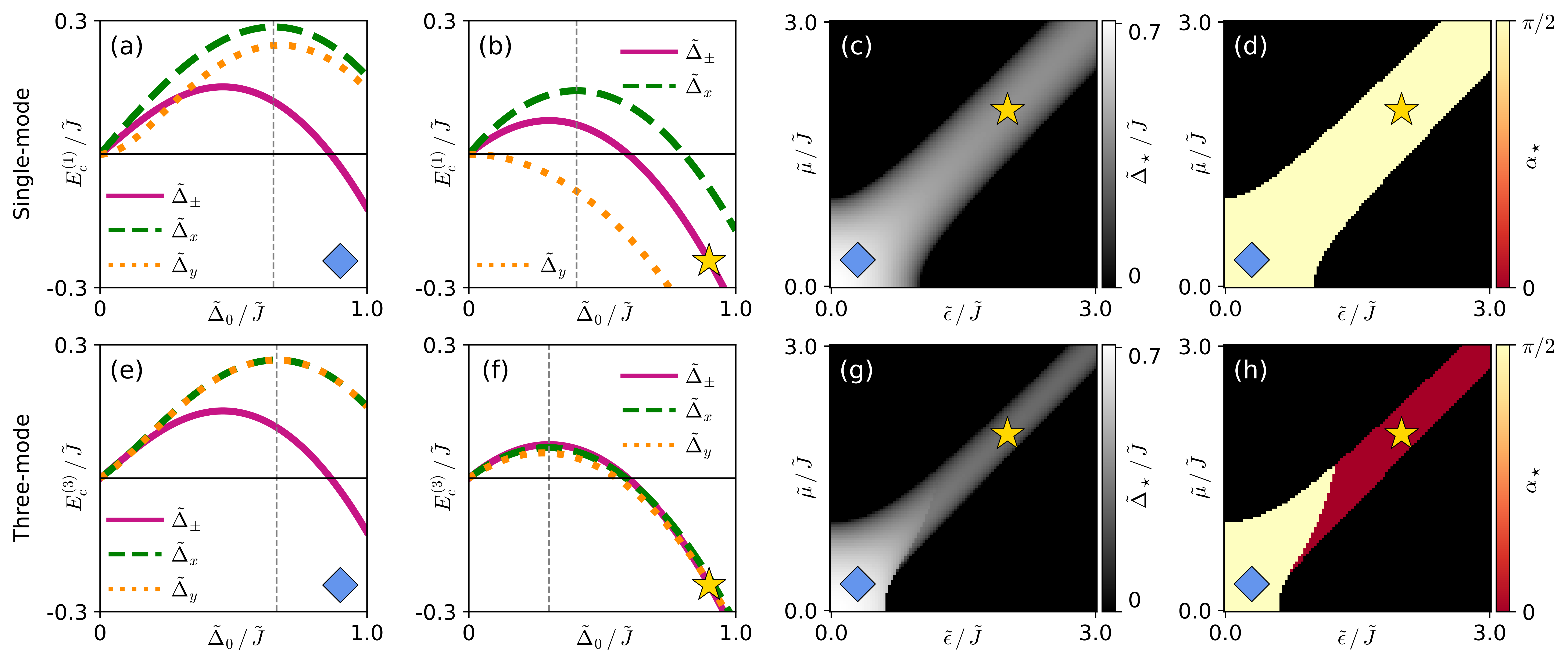}
	\caption{\textit{Top:} Single-mode intra-Chern model. Condensation energy $E^{(1)}_c$ as a function of $\tilde{\Delta}_0$ at (a) small energy splitting or large interactions $\tilde{\epsilon} = \tilde{\mu} = 0.3\tilde{J}$ and (b) large energy splitting or small interactions $\tilde{\epsilon}=\tilde{\mu}=2\tilde{J}$. (c) Optimal pairing strength $\tilde{\Delta}_\star$ as a function of $\tilde{\epsilon}$ and $\tilde{\mu}$ and (d) corresponding phase diagram for $\alpha_\star$ with normal phase, defined as $\tilde{\Delta}_0/\tilde{J} < 0.01$, masked in black. Blue diamond and yellow star denote parameters used in (a,b), respectively. \textit{Bottom:} Three-mode intra-Chern model. (e,f) Condensation energy $E^{(3)}_c$ and (g,h) phase diagram, for the same parameters as above. Competition between $C_{3z}$-symmetry related momenta frustrates the nematic state (pale yellow), stabilizing the chiral state (red). }
	\label{fig:competition:phasediagramcombined}
\end{figure*}

The favorability of $\tilde{\Delta}_x$ over the chiral state does not, however, extend to all nematic states. We first consider the limit of absolutely flat bands, $\tilde{\epsilon} \to 0$, where the eigenvalues $\varepsilon_{\pm}$ become independent of $\phi$ at any $\alpha$. This leaves the nematic direction $\phi$ as a global $U(1)$ symmetry for all superconducting states. We next lift this degeneracy by taking $\tilde{\epsilon} > 0$. Since we expect superconductivity to be stable only near $|\tilde{\mu}| = \tilde{\epsilon}$, where states cross the Fermi level in the normal state, we enforce this equality for now. Two regimes emerge depending on the ratio of pairing strength $\tilde{\Delta}_0$ to the normal-state energy splitting $\tilde{\epsilon}$. In the small energy splitting, or large interactions, regime $\tilde{\epsilon} \ll \tilde{\Delta}_0$, a first-order expansion of Eq.~\eqref{eq:competition:singlemode_cond_diff} gives
\begin{align} \label{eq:competition:singlemode_largeDelta}
	\delta E^{(1)}_{c} (\phi)
	&= \tilde{\Delta}_0 \left[ (\sqrt{2} - 1)  - \tfrac{1}{2} \tfrac{\tilde{\epsilon}}{\tilde{\Delta}_0}  + \mathcal{O}\!\left(\tfrac{\tilde{\epsilon}^2}{\tilde{\Delta}_0^2}\right) \right],
\end{align}
which is positive, indicating that all nematic states are favorable over the chiral states. In the opposite, large energy splitting, or small interactions, regime $\tilde{\epsilon} \gg \tilde{\Delta}_0$, we obtain
\begin{align} \label{eq:competition:singlemode_smallDelta}
	\delta E^{(1)}_{c} (\phi)
	&= \tfrac{1}{2} \left( \sqrt{2} \, \abs{\cos \tfrac{\phi}{2}} - 1 \right) \tilde{\Delta}_0 + \mathcal{O}\!\left(\tfrac{\tilde{\Delta}_0^2}{\tilde{\epsilon}}\right).
\end{align}
Here, the $\phi$-dependence plays a crucial role, with the nematic condensation energy reaching a maximum at $\phi = 0$ but dropping below that of the chiral state near $\phi = \pi$, i.e.~for the $\tilde{\Delta}_y$ state. The weakness of $\tilde{\Delta}_y$ in this regime arises from constructive contributions of terms inside the square root in Eq.~\eqref{eq:competition:singlemode_eigenvalues}, which increases the splitting of quasiparticle energies and reduces condensation energy. This splitting scales with $\tilde{\epsilon}$ and can thus exceed that of the chiral state, despite its subgap bands, explaining why the chiral state can still achieve a higher condensation energy.

Next, we numerically illustrate our results in the top row of Fig.~\ref{fig:competition:phasediagramcombined}, by evaluating Eq.~\eqref{eq:competition:singlemode_condensation} without approximations. We present variables in units of the effective interaction strength $\tilde{J}$, since it sets the energy scale. In Fig.~\ref{fig:competition:phasediagramcombined}(a,b) we plot $E^{(1)}_{c}$ as a function of $\tilde{\Delta}_0$ in two different regimes. In Fig.~\ref{fig:competition:phasediagramcombined}(a) we set $\tilde{\epsilon} = \tilde{\mu} = 0.3\tilde{J}$, finding that the maximal condensation energy is obtained by the nematic state $\tilde{\Delta}_x$ (green) at an optimal pairing strength $\tilde{\Delta}_\star$ (dashed gray line). The nematic state $\tilde{\Delta}_y$ (orange) closely follows with a slightly lower $E^{(1)}_{c}$, while the chiral states (magenta) perform worse. Here $\tilde{\Delta}_\star$ is large compared to $\tilde{\epsilon}$, corresponding to the small energy splitting regime of Eq.~\eqref{eq:competition:singlemode_largeDelta}. By contrast, in Fig.~\ref{fig:competition:phasediagramcombined}(b) we consider $\tilde{\epsilon} = \tilde{\mu} = 2\tilde{J}$, where the pairing $\tilde{\Delta}_\star$ is now much smaller than $\tilde{\epsilon}$, such that the large energy splitting regime of Eq.~\eqref{eq:competition:singlemode_smallDelta} applies. Although $\tilde{\Delta}_x$ remains the dominant state with the highest condensation energy, the chiral states are now preferred over the nematic $\tilde{\Delta}_y$ state.

We finally consider the whole phase diagram in Figs.~\ref{fig:competition:phasediagramcombined}(c,d) as a function of $\tilde{\epsilon}$ and $\tilde{\mu}$ by determining the ground state with respect to $\tilde{\Delta}_0$, $\alpha$, and $\phi$. Figure~\ref{fig:competition:phasediagramcombined}(c) shows the optimal pairing strength $\tilde{\Delta}_\star$. 
It is largest near $\tilde{\epsilon} = \tilde{\mu} = 0$ (white) but is still appreciable (gray) for all values in a window around $|\tilde{\mu}| \approx \tilde{\epsilon}$. We attribute the reduction away from the origin to suppression of inter-eigenband pairing. Blue diamond and yellow star denote the parameters used in Fig.~\ref{fig:competition:phasediagramcombined}(a,b), respectively.  Figure~\ref{fig:competition:phasediagramcombined}(d) shows the optimal $\alpha_\star$, corresponding to the value of $\alpha$ at the maximum of $E^{(1)}_{c}$. Across all probed parameters, the ground state is either a nematic state (pale yellow) or the normal state (black). We further verify that the $\tilde{\Delta}_x$ state ($\phi=0$) is always favored. Thus we find that at individual momenta, a nematic superconducting state is always more favorable than the chiral state, which we connect to the existence of the subgap bands of the chiral state. However, not all nematic states are more favorable. This points to the importance of considering more momentum modes.

\subsection{Three-mode limit}
\label{sec:competition:threemode}

In the single-mode analysis above, we neglected the momentum-dependent phase shift $\delta \phi_{\mathbf{k}}$ in Eq.~\eqref{eq:mapping:chern_pairing}. However, $d$-wave symmetry implies that $d_{\pm}$, and thus $\delta \phi_{\mathbf{k}}$, exhibit a nontrivial rotation pattern in reciprocal space, associated with the Cooper-pair angular momentum. Specifically, $e^{i\delta \phi_{\mathbf{k}}}$ transforms with eigenvalue $e^{-i\omega}$ under $C_{3z}$, where $\omega = 2\pi/3$. For the nematic states, which spontaneously break rotation symmetry, including this phase shift modifies the value of $\phi$ that maximizes the condensation energy to $\phi_{\star} = -\delta \phi_{\mathbf{k}}$, plotted in Fig.~\ref{fig:introduction:diagram}(c). Consequently, the value $\phi_{\star}$ that optimizes the condensation energy at a given momentum $\mathbf{k}_0$ is incompatible with the values preferred by the $C_{3z}$-related modes $\mathbf{k}_{\pm}$ (black markers), leading to a lowering of the overall condensation energy. This motivates restricting the momentum summation in Eq.~\eqref{eq:competition:chern_meanfield} to symmetry-related triplets of momenta, defining a \emph{three-mode limit}. The condensation energy averaged over three $C_{3z}$-related modes then becomes
\begin{align} \label{eq:competition:threemode_condensation}
	E^{(3)}_{c}(\alpha, \phi)
	&= \sum_{n=0}^{2} \tfrac{1}{3}E^{(1)}_{c}(\alpha, \phi + n \, \omega).
\end{align}

The three-mode condensation energy $E^{(3)}_{c}$ coincides with $E^{(1)}_{c}$ for chiral states, but differs significantly for nematic ones, as we illustrate in the bottom row of Fig.~\ref{fig:competition:phasediagramcombined}. There, we repeat the top-row calculations with the same parameters, but now for the three-mode case. In the small energy splitting regime Fig.~\ref{fig:competition:phasediagramcombined}(e), the condensation energies remain largely unchanged compared to the single-mode values. In the large energy splitting regime in Fig.~\ref{fig:competition:phasediagramcombined}(f), the previously optimal nematic state (green) is notably weakened, and both nematic states now have less condensation energy than the chiral state (magenta). We attribute this change to phase frustration induced by the spontaneously broken rotation symmetry. Choosing $\phi=0$ (corresponding to $\tilde{\Delta}_x$) maximizes $E^{(1)}_{c}$ at a specific momentum $\mathbf{k}_0$, but the phases that maximize $E^{(1)}_{c}$ at the $C_{3z}$-related momenta $\mathbf{k}_{\pm}$ are instead $\phi = \mp \omega$. As a result, the condensation energy of $\tilde{\Delta}_x$ is reduced in the three-mode case, $E^{(3)}_{c}(\tfrac{\pi}{2},0) \leq E^{(1)}_{c}(\tfrac{\pi}{2},0)$. This frustration across momentum modes is stronger in the large energy splitting regime, which exhibits larger differences in $E^{(1)}_{c}$ between nematic states as seen in Fig.~\ref{fig:competition:phasediagramcombined}(b). For sufficiently strong frustration, the chiral state can even become the ground state, which we verify in Fig.~\ref{fig:competition:phasediagramcombined}(g,h) by maximizing $E_{c}^{(3)}$ over $\tilde{\Delta}_0$, $\alpha$, and $\phi$, while keeping $\tilde{\epsilon}$ and $\tilde{\mu}$ fixed. The resulting optimal $\tilde{\Delta}_\star$ shows a narrower but still finite strip of superconductivity (white/gray) compared to the single-mode case. More importantly, in Fig.~\ref{fig:competition:phasediagramcombined}(h), the optimal $\alpha_\star$ reveals a region $ \abs{\tilde{\mu}} \approx \tilde{\epsilon} \gtrsim \tilde{J}$ with a chiral ground state. We thus conclude that the nematic state suffers from momentum-space frustration imposed by the normal-state $C_{3z}$ symmetry, such that sufficiently large flat-band energy splitting and/or weak interactions favor a chiral ground state. We note that this momentum-space phase frustration is generally present for any pairing configuration with a finite pairing amplitude in both Chern sectors. However, it is possible to show that, in the large energy splitting regime $ \tilde{\epsilon} \gg \tilde{\Delta}_0 $, the nematic states yield the largest variation of $E^{(1)}_{c}$ with respect to $\phi$, and hence are the most susceptible to phase frustration. This reflects the fact that these states host equal amplitudes of pairing in both Chern channels due to their preserved time reversal symmetry.

\begin{figure}
	\includegraphics[width=\columnwidth]{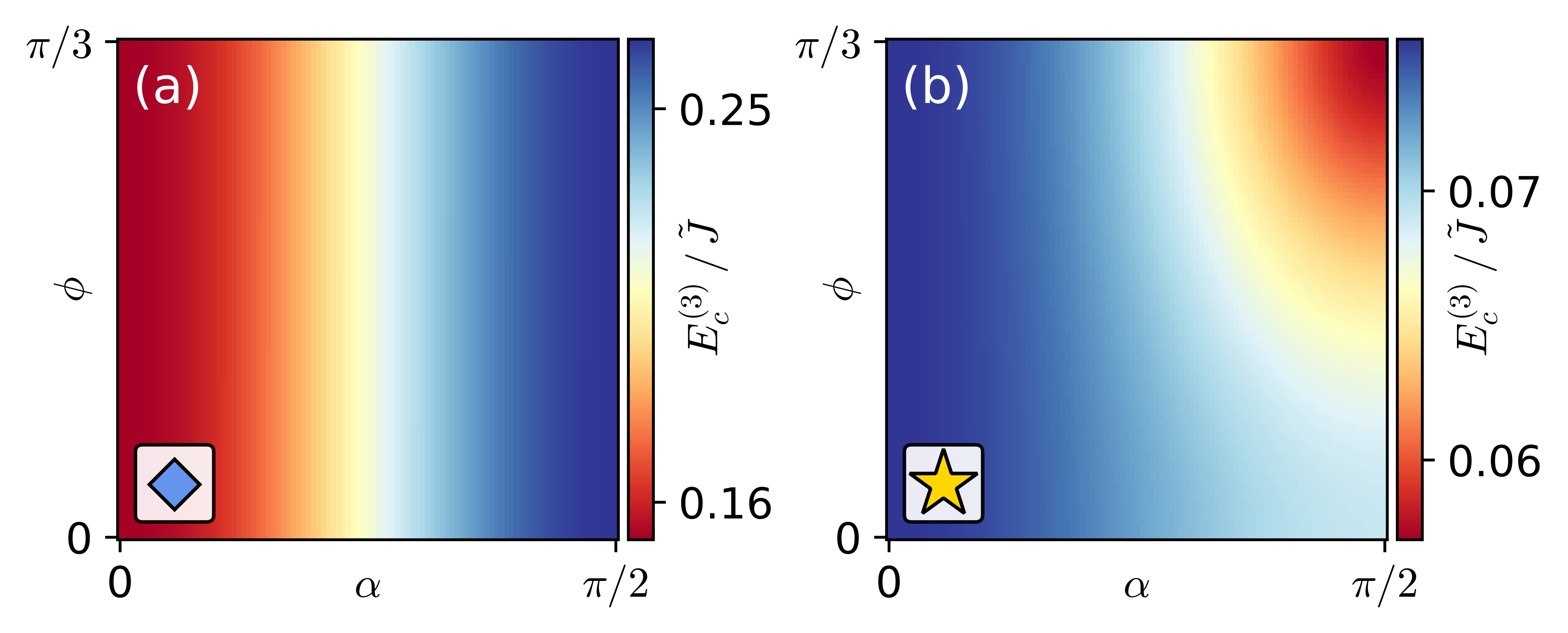}
	\caption{Three-mode condensation energy $E^{(3)}_{c}$ as a function of $(\alpha, \phi)$ at (a) $\tilde{\epsilon}=\tilde{\mu}=0.3\tilde{J}$ (blue diamond in Fig.~\ref{fig:competition:phasediagramcombined}) and (b) $\tilde{\epsilon}=\tilde{\mu}=2.0\tilde{J}$ (yellow star in Fig.~\ref{fig:competition:phasediagramcombined}). Red (blue) shows regions of lower (higher) condensation energy, with deepest blue indicating the superconducting ground state.}
	\label{fig:competition:freelandscape_minimal}
\end{figure}

To obtain a complete view of the energetic landscape within the three-mode model, we show in Fig.~\ref{fig:competition:freelandscape_minimal} how $E^{(3)}_{c}(\alpha,\phi)$ varies as a function of $\alpha$ and $\phi$ for fixed doping and interactions. We again consider representative small and large energy splitting parameters, $\tilde{\epsilon} = \tilde{\mu} = 0.3 \tilde{J}$ and $\tilde{\epsilon} = \tilde{\mu} = 2.0 \tilde{J}$, marked by the blue diamond and yellow star in Fig.~\ref{fig:competition:phasediagramcombined}, respectively. In the small energy splitting regime Fig.~\ref{fig:competition:freelandscape_minimal}(a), the nematic states ($\alpha = \pi/2$) have highest condensation energies and remain nearly degenerate with respect to $\phi$. In the large energy splitting regime in Fig.~\ref{fig:competition:freelandscape_minimal}(b), the chiral state ($\alpha = 0$) instead becomes the ground state.

\subsection{Full momentum integration}
\label{sec:competition:full}

The striking similarity between Fig.~\ref{fig:competition:freelandscape_minimal} and Fig.~\ref{fig:atomistic:freeenergy}(b,c) reveals that the three-mode intra-Chern pairing model is able to qualitatively capture the physics even at the atomistic level, including the nematic-to-chiral transition. To make the comparison quantitative, in the SM we also study the intra-Chern pairing model on the same momentum-space mesh (and then finer) as Fig.~\ref{fig:atomistic:freeenergy}, with both the normal-state spectrum and the pairing matrix extracted numerically from the atomistic model. By doing so, we find that the remote bands of TBG, i.e.~all those not included in the intra-Chern model, contribute significantly to total condensation energy, but in a way that is notably independent of pairing symmetry and which can be almost perfectly accounted for by simply rescaling the interaction strength $J$, see SM. Our analysis therefore exhaustively characterizes the intra-Chern pairing model in TBG: the nematic state is most often favored, but for weak interactions, or equivalently, large band splittings the chiral state may still emerge as the ground state.

\section{Symmetry-broken parent states}
\label{sec:parentstate}

So far we studied the energetics of intra-Chern pairing between the symmetric normal-state flat bands of TBG. Experimental data, however, suggests that the parent state developing superconductivity might already be symmetry broken. This includes external symmetry breaking due to the presence of lattice deformations, and spontaneous symmetry breaking due to electronic correlations, both of which we now investigate. The implementation of such parent states within an atomistic framework is challenging. We thus primarily adopt a hybrid approach, where we use the intra-Chern pairing structure obtained from our atomistic calculations with a symmetric parent state, but extract the symmetry-broken normal-state dispersion from the well-established topological heavy fermion (THF) model \cite{Song2022, Calugaru2023}, with parameters fitted to the Kang--Vafek (KV) model \cite{Kang2023, Vafek2023, HerzogArbeitman2025b} (see SM). We complement this with fully atomistic calculations for an inter-valley coherent parent state.

\begin{figure}
	\includegraphics[width=\columnwidth]{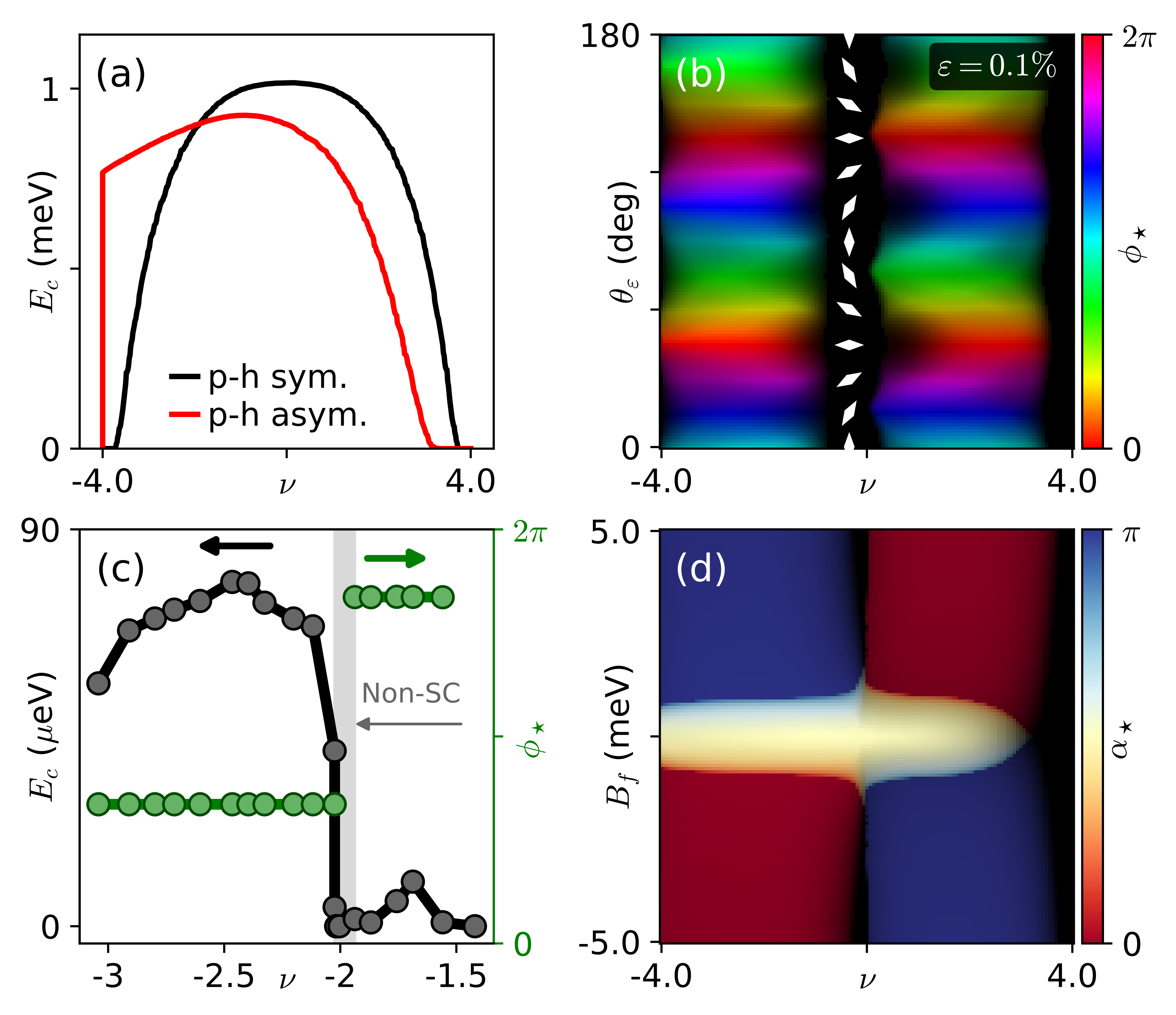}
	\caption{Intra-Chern pairing under symmetry-broken parent states. (a) Condensation energy $E_c$ versus filling $\nu$ in the presence (red) versus absence (black) of particle-hole symmetry breaking terms. (b) Optimal nematic direction $\phi_\star$ (hue) in the presence of finite heterostrain $\varepsilon=0.1\%$, as a function of $\nu$ and the heterostrain axis $\theta_{\varepsilon}$. White lozenges show the twofold axis of the real-space pairing pattern and highlight the $4\pi$ winding of $\phi_\star$. (c) $E_c$ (black) and $\phi_\star$ (green) versus $\nu$ in the presence of a finite IVC amplitude $U_{\textrm{IVC}}=27$~meV. Shaded area is non-superconducting. (d) Intra-Chern pairing angle $\alpha_\star$ versus $\nu$ and a Chern-polarizing perturbation $B_f$. In (b,d) brightness indicates magnitude of condensation energy, and black regions correspond to $E_c < 1~\mu$eV. In (a,b,d) the intra-Chern pairing obtained from the atomistic calculations is combined with the symmetry-broken normal-state spectrum obtained from the THF model using KV parameters at $1.05^\circ$ \cite{HerzogArbeitman2025b} (see SM), including particle-hole symmetry breaking terms unless otherwise noted and using a $65\times65$ momentum space sampling. In (c) we use the fully atomistic model on a $4\times4$ grid. $J=1.35~$eV is used throughout.}
	\label{fig:parentstate}
\end{figure}

\subsection{Lattice relaxation and heterostrain}
\label{sec:parentstate:lattice}
The atomistic model for TBG in Eq.~\eqref{eq:atomistic:Hamiltonian} is constructed for rigid graphene sheets, although the lattice is expected to undergo relaxation in an experimental setting \cite{Nam2017, YWLiu2020}. Relaxation primarily reduces the effective size of the $AA$ region, which increases the localization of the flat bands and thus we expect it to rescale the pairing amplitude. Other relaxation effects include a renormalization of the magic angle and of the gap between the flat and dispersive bands \cite{Carr2018, Carr2019, Cantele2020}, and also breaking of the approximate particle-hole symmetry that is otherwise present at small angles \cite{Song2019b, Kang2023}. In Fig.~\ref{fig:parentstate}(a) we show the effect of the latter for the intra-Chern pairing condensation energy as a function of filling $\nu$ in the absence (black) and presence (red) of particle-hole symmetry breaking terms. This mainly introduces an asymmetry in the superconducting dome, but preserves the overall $E_c$, meaning intra-Chern pairing stays robust.

More interesting is heterostrain, which is also present in typical TBG samples \cite{Huder2018, Qiao2018, Kazmierczak2021, Benschop2021}, and which results in increased splitting between the flat bands \cite{Mesple2021} and broken $C_{3z}$ rotation symmetry in the quasiparticle spectrum \cite{Bi2019}. For intra-Chern pairing, we find that such rotation symmetry breaking disrupts the momentum-space frustration of nematic pairing, such that heterostrain generally favors nematic superconductivity. At the same time, the increased flat-band splitting strongly reduces the condensation energy and, already at a small heterostrain $\varepsilon=0.1\%$, splits the superconducting dome into two around charge neutrality. We illustrate this in Fig.~\ref{fig:parentstate}(b), by plotting the superconducting phase diagram as a function of filling $\nu$ and the heterostrain principal axis $\theta_{\varepsilon}$. Bright regions support a large $E_c$, while black denotes vanishing superconductivity. For each $\theta_{\varepsilon}$ we generally find there is a single favored nematic direction $\phi_\star$ (hue), in contrast to the symmetric parent state which yields a threefold degenerate ground state. Notably, $\phi_\star$ winds twice over its space of physically distinct heterostrain directions (white lozenges), which in the SM we show derives from the $d$-wave nature of pairing. Our results thus predict that heterostrain along two perpendicular axes selects essentially the same nematic direction $\phi_{\star}$, albeit with different condensation energies.

\subsection{Inter-valley coherence}
\label{sec:parentstate:ivc}

In TBG, the most prominent superconducting domes flank the correlated insulator at $\nu = -2$ \cite{Cao2018b, Yankowitz2019, Lu2019, Codecido2019, Polshyn2019, Cao2020, Jaoui2022, Saito2020, Stepanov2020, Stepanov2021, Liu2021, Cao2021a, Oh2021, RodanLegrain2021, DiBattista2022, Nuckolls2023, Tian2023, Tanaka2025}. Superconductivity does not require this insulator: it persists when moderate screening suppresses the insulating state \cite{Saito2020, Stepanov2020, Liu2021}, although very strong screening suppresses superconductivity as well \cite{Barrier2026, Gao2026}. In samples where the two coexist, the superconducting state may nevertheless emerge from the doped correlated insulator. STM measurements indicate that this insulator is inter-valley coherent (IVC) \cite{Nuckolls2023}, and we therefore also study superconductivity emerging from an IVC parent state.

In the presence of heterostrain, inter-valley coherence is believed to take the form of an incommensurate Kekul\'e spiral (IKS) \cite{Kwan2021, Wagner2022, Herzog-Arbeitman2025, ZiweiWang2025}, where both intra- and inter-Chern correlations exist. An IVC state $|\xi\rangle$ composed of equal amplitudes in all Chern and valley sectors can be generally parametrized by angles $\phi_{\textrm{valley}}$, $\phi_{\textrm{Chern}}$, and $\phi_{\textrm{sublattice}}$, capturing phase differences between the denoted flat-band sectors (see SM). $C_{2z} \, \mathcal{T}$ symmetry, where $C_{2z}$ is a twofold in-plane rotation, is thought to be important for the emergence of the IKS \cite{Herzog-Arbeitman2025}. The IKS also preserves time-reversal symmetry, which requires $\phi_{\textrm{sublattice}}=0 \bmod \pi$ (see SM) \cite{Kwan2021}. The angle $\phi_{\textrm{valley}}$ corresponds to graphene-scale translations of the Kekul\'e pattern and is chosen spontaneously: since $C_{2z}\mathcal{T}$ does not commute with valley rotations, it fixes $\phi_{\textrm{valley}}$ only relative to the rotation center \cite{Kwan2021}. By contrast, $\phi_{\textrm{Chern}}$ encodes rotations of the Kekul\'e order and is believed to be determined by the heterostrain axis. The Coulomb interaction lowers $|\xi\rangle$ in energy, although $C_{2z} \, \mathcal{T}$-protected Dirac point crossings with the other bands still remain \cite{Herzog-Arbeitman2025}. Finally, a gap can open via a valley-dependent momentum boost \cite{Herzog-Arbeitman2025}. Such a boost breaks the moir\'e translation symmetry, making an atomistic formulation computationally demanding. We therefore focus on the IKS-precursor $|\xi\rangle$ to capture an IVC parent state. As the gap between the flat and remote bands closes for the IVC state \cite{Herzog-Arbeitman2025}, projection of pairing to the flat bands alone is not well justified. We therefore incorporate IVC physics within the fully atomistic treatment in Eq.~\eqref{eq:atomistic:Hamiltonian}. We achieve this by deriving the necessary Kekul\'e-modulated on-site, nearest-neighbor, and next-nearest-neighbor terms to implement the IKS-precursor $|\xi\rangle$, see SM. This accounts for both intra- and inter-Chern correlations known to be important for the development of the IKS \cite{Kwan2021}. The IVC strength $U_{\textrm{IVC}}$ controls the separation of the $|\xi\rangle$ band from the other flat bands. 

For concreteness, we choose $\phi_{\textrm{sublattice}} = 0$, as required by time-reversal symmetry, and $\phi_{\textrm{Chern}} = 0$ up to a $C_{3z}$ rotation (see SM). We then solve self-consistently for superconducting pairing, which we find is generally suppressed with increasing $U_{\textrm{IVC}}$ due to the decreased low-energy DOS. This results in loss of superconductivity around $\nu=-2$ due to the emergence of an IVC-driven suppression of low-energy DOS. We illustrate this in Fig.~\ref{fig:parentstate}(c), where $E_c$ (black, left axis) reveals a significant superconducting dome near $\nu \approx -2.5$ and a smaller one at $\nu \approx -1.7$, resembling experimental observations \cite{Cao2018b, Yankowitz2019, Cao2020, Cao2021a}. We find that the pairing is still strongly intra-Chern, with only small contributions distributed among other channels, see SM. Within the intra-Chern channel we find that the ground state is nematic, with the favored nematic direction $\phi_\star$ at each $\nu$ (green, right axis) undergoing a large shift between the two domes. This surprising shift is reminiscent of the filling-dependent change of nematic direction in experiments \cite{Cao2021a}. Our results thus pose an IVC parent state as a possible microscopic origin for this phenomenon.

\subsection{Chern polarization}
\label{sec:parentstate:chern}

Our results show that nematic superconductivity is favored in TBG because it makes full use of both Chern sectors for pairing, and also survives heterostrain and IVC parent states. On the other hand, chiral $d$-wave superconductivity has been proposed as a platform for topological quantum computing \cite{Margalit2022, Li2023, Brosco2024, Confalone2025}, motivating searches for mechanisms that stabilize chiral superconductivity. We next propose that Chern polarization of the parent state provides a natural route to achieve this. Notably, interaction-induced Chern-insulator states have been observed at $\nu=+1$ at zero magnetic field in hBN-non-aligned samples that also host superconducting domes \cite{Stepanov2021}, providing a possible origin of Chern polarization.

To demonstrate the effect of Chern polarization in the parent state, we turn again to the intra-Chern pairing model. Using the THF framework fitted to the KV model, we introduce Chern polarization by adding an energy splitting $B_f$ on the $f$-orbitals of the THF model, with the sign depending on the product of their valley and sublattice polarization. The resulting phase diagram is shown in Fig.~\ref{fig:parentstate}(d), with color denoting optimal $\alpha_\star$ as a function of $\nu$ and $B_f$, and brightness proportional to $E_c$. For small $|B_f|$ the ground state remains nematic (yellow), but as this magnitude increases the system evolves to a chiral configuration (blue/red). This is because at a fixed filling $|\nu|\neq 0, 4$ only Chern bands with Chern number $C=\sgn(B_f \, \nu)$ are near the Fermi level and are thus able to participate in pairing effectively. Superconductivity is strongly suppressed only close to $\nu = 0$, where a gap opens between the two Chern sectors for $|B_f| \gtrsim 1$~meV, and it is absent near the upper band edge, beyond the particle-hole asymmetric dome of Fig.~\ref{fig:parentstate}(a). Importantly, at large $|B_f|$ the chirality of the superconducting ground state is dictated by the active Chern sector.

We note that the locking between Chern polarization and Cooper-pair chirality in Fig.~\ref{fig:parentstate}(d) is reminiscent of recent results obtained in the Berry ``trashcan'' model for rhombohedral graphene \cite{Bernevig2025BerryTrashcanModel, Li2025BerryTrashcanPairing}, although in that case pairing is $p$-wave, intra-valley, and spin-triplet.

\section{Discussion}
\label{sec:discussion}
We study superconductivity in twisted bilayer graphene near the magic angle, starting from an electron-driven pairing model on the full carbon lattice. Solving self-consistently for the superconducting ground state with realistic attraction strength, and then projecting onto the Chern basis of the moir\'e flat bands, we find that the pairing is $d$-wave, spin singlet, inter-valley, and intra-Chern. This result unifies previously disparate proposals for superconductivity in TBG and establishes that both electron- and phonon-driven mechanisms \cite{BlackSchaffer2007, Fischer2021, Lothman2022, Liu2024, Wang2024, Liu2025} produce intra-Chern pairing. Thus, contrary to the usual expectation that these superconducting mechanisms compete or, at most, coexist, our results suggest that TBG provides a setting in which they naturally cooperate. This clearly enhances their appeal relative to other proposed pairing mechanisms, such as skyrmionic pairing \cite{Khalaf2021} or fluctuations associated with nearby correlated states \cite{You2019d, WangY2021, Christos2023}.

Importantly, we also provide a unifying framework for understanding why nematic superconductivity dominates in TBG. In particular, the rotation symmetry of chiral pairing requires fully Chern-polarized pairing, leaving an effectively unpaired band that reduces the superconducting condensation energy. As a result, nematic order is generally favored locally in reciprocal space. However, the momentum dependence of the preferred nematic direction generates frustration, 
which lowers its condensation energy. This is most pronounced for weak interactions or large band splittings, allowing the chiral phase to become favorable in this regime. Such behavior has been found in earlier calculations when tuning the interaction strengths \cite{Su2018, Kozii2019, Chichinadze2020, Julku2020, Fischer2021, Wang2024}, and our results explain this behavior and also establish the possibility of a filling-induced transition. We further establish that nematic superconductivity is preferred for parent states with finite heterostrain or inter-valley coherence. Notably, intra-Chern pairing persists for these parent states, while in the IVC case the nematic direction becomes filling dependent, which is reminiscent of experimental observations \cite{Cao2021a}.

We also comment on the assumptions made within our atomistic modeling. First, the approximately satisfied Pauli limit \cite{Cao2018b} cannot be taken as conclusive evidence for the spin-singlet pairing we assume, as it may instead signal a large orbital magnetization arising from a valley-polarized parent state \cite{Antebi2022}. However, we stress that spin-singlet pairing is still most consistent with available experimental evidence, which is suggestive of a spin-unpolarized parent state \cite{Cao2018, Cao2018b, Yankowitz2019, Hoke2024}, while equal-spin triplet pairing has been deemed incompatible with Andreev spectra \cite{Oh2021}. Recent theoretical developments further point to a spin-unpolarized, inter-valley coherent (i.e.~not valley-polarized) parent state \cite{Herzog-Arbeitman2025, ZiweiWang2025, Crippa2026}. Moreover, spin-singlet pairing is consistent with both a spin-fluctuation mechanism \cite{Fischer2021} and $K$-phonon superconductivity \cite{Wu2018, Wu2019, Liu2024, Liu2025}. Mixed-spin triplet pairing has, however, recently been suggested to arise from intra-valley interactions with Kekul\'e superconducting order carrying finite momentum on the moir\'e lattice scale \cite{Wang2026, Wang2026Stiffness}. It would in the future be interesting to extend our work to also include finite moir\'e momentum pairing. 

Second, we assume pairing on nearest-neighbor bonds, while spin fluctuations give weaker attraction also on longer-range inter-sublattice neighbors \cite{Fischer2021}, owing to an antiferromagnetic ordering tendency \cite{Klebl2019}. The inter-valley, intra-Chern pairing emerging from nearest-neighbor interactions can be related to the valley-dependent sublattice-polarization of the Chern bands. It is thus expected that longer-ranged inter-sublattice pairing terms will also favor this inter-valley, intra-Chern channel. Since our analysis of intra-Chern pairing relies only on rotation symmetry and distinct multiband structure properties that are insensitive to finer details, we conjecture that our main conclusions are robust to longer-ranged attraction. Moreover, longer-range pairing would normally add more nodal points in momentum-space, making such terms less energetically favorable. Conversely, we expect other putative mechanisms, if they favor intra-sublattice pairing, to yield inter-Chern superconductivity instead.

In terms of disorder effects, twist-angle disorder is known to be particularly pervasive in twisted graphene heterostructures \cite{Beechem2014, Uri2020, Kazmierczak2021, Benschop2021}. Recent work has shown that such disorder can obscure the characteristic signatures of gapped and nodal superconductivity \cite{Maccari2026}, complicating the interpretation of experimental data \cite{Tanaka2025}. It would be interesting to investigate how the intrinsically intra- and inter-band nature of intra-Chern pairing changes these results. Other relevant sources of disorder include vacancies and adatoms, which have been shown to produce pronounced moir\'e-scale responses even at extremely low concentrations \cite{Ramires2019, LopezBezanilla2019b, Baldo2023a}. Such impurities, as well as buckling of the graphene layers \cite{vanPoppelen2025}, can strongly couple to the sublattice degree of freedom, and the response of superconductivity to these sources of disorder is an interesting direction for future research. The absence of superconductivity in hBN-aligned samples \cite{Oh2021, Chen2024} in fact suggests that sublattice polarization may be detrimental to superconductivity in TBG. The inter-sublattice nature of intra-Chern pairing is compatible with such suppression.

Finally, our approach of projecting pairing on the atomic level onto the flat-band subspace can be easily extended to other moir\'e systems, including other multilayer twisted graphene structures. This may be particularly interesting in twisted trilayer graphene (TTG), which also hosts unconventional superconductivity \cite{Kim2022}, and for which atomistic descriptions of pairing have already been proposed \cite{Fischer2022}. There, the presence of a symmetry-broken parent state might be even more important, given reports of Pauli limit violation \cite{Cao2021PauliLimit} and of a double gap feature attributed to particle-hole and particle-particle correlations, respectively \cite{Park2025, Kim2026}. In fact, the superconducting anisotropy in TTG has been found to follow the nematic axis of the normal state \cite{Zhang2026}, in line with our finding that a rotation-symmetry-breaking parent state sets the nematic direction. Extremely recent experiments have further found a filling-dependent nematic axis in superconducting TTG \cite{Sun2026}, a system in which IVC order has also been imaged \cite{Kim2023}, making TTG a natural testing ground for our proposed parent-state control of the nematic direction.

\begin{acknowledgments}  
We thank Tomas L{\"o}thman, Quentin Marsal, Aaron Dunbrack and Jose L.~Lado for valuable discussions. This work was supported by the Swedish Research Council (Vetenskapsr\aa det) Grant No.~2022-03963 and the European Research Council (ERC) under the European Union's Horizon Europe research and innovation programme (ERC-2022-CoG, Grant agreement no.~101087096). Views and opinions expressed are, however, those of the author(s) only and do not necessarily reflect those of the European Union or the European Research Council Executive Agency. Neither the European Union nor the granting authority can be held responsible for them. Numerical calculations were enabled by resources provided by the National Academic Infrastructure for Supercomputing in Sweden (NAISS) at UPPMAX, partially funded by the Swedish Research Council through grant agreement no.~2022-06725.
\end{acknowledgments}

\bigskip

\section*{Data availability}
The data that support the findings of this article are openly available \cite{Baldo2026_Zenodo}.

\section*{Code availability}
The scripts that generate all figures from the raw data, together with the code for the flat-band projection of the atomistic results and the code for the self-consistent calculations in the flat-band-projected model, are openly available \cite{Baldo2026_Zenodo}. The code for the atomistic model is available from the corresponding author upon reasonable request, because it is part of a larger codebase developed together with researchers who were not involved in this work.

\bibliography{bibliography}

@misc{Sun2026,
      title={Nodal nematic superconductivity in moir\'e graphene}, 
      author={Shuwen Sun and Grgur Palle and Kenji Watanabe and Takashi Taniguchi and Rafael M. Fernandes and Pablo Jarillo-Herrero and Jeong Min Park},
      year={2026},
      eprint={2610.20605},
      archivePrefix={arXiv},
      primaryClass={cond-mat.supr-con},
      url={https://arxiv.org/abs/2610.20605}, 
}

@misc{Mahalingam2026,
      title={Cell Natural Orbitals in Quantum Materials}, 
      author={Harshitra Mahalingam and Nishchhal Verma and Daniel Muñoz-Segovia and Raquel Queiroz},
      year={2026},
      eprint={2608.10059},
      archivePrefix={arXiv},
      primaryClass={cond-mat.mtrl-sci},
      url={https://arxiv.org/abs/2608.10059}, 
}

@article{Solovyev2007,
  title = {Construction of Wannier functions from localized atomiclike orbitals},
  author = {Solovyev, I. V. and Pchelkina, Z. V. and Anisimov, V. I.},
  journal = {Phys. Rev. B},
  volume = {75},
  issue = {4},
  pages = {045110},
  numpages = {10},
  year = {2007},
  month = {Jan},
  publisher = {American Physical Society},
  doi = {10.1103/PhysRevB.75.045110},
  url = {https://link.aps.org/doi/10.1103/PhysRevB.75.045110}
}

@article{Vafek2023,
  title = {Continuum effective Hamiltonian for graphene bilayers for an arbitrary smooth lattice deformation from microscopic theories},
  author = {Vafek, Oskar and Kang, Jian},
  journal = {Phys. Rev. B},
  volume = {107},
  issue = {7},
  pages = {075123},
  numpages = {11},
  year = {2023},
  month = {Feb},
  publisher = {American Physical Society},
  doi = {10.1103/PhysRevB.107.075123},
  url = {https://link.aps.org/doi/10.1103/PhysRevB.107.075123}
}

@article{Kang2023,
  title = {Pseudomagnetic fields, particle-hole asymmetry, and microscopic effective continuum Hamiltonians of twisted bilayer graphene},
  author = {Kang, Jian and Vafek, Oskar},
  journal = {Phys. Rev. B},
  volume = {107},
  issue = {7},
  pages = {075408},
  numpages = {34},
  year = {2023},
  month = {Feb},
  publisher = {American Physical Society},
  doi = {10.1103/PhysRevB.107.075408},
  url = {https://link.aps.org/doi/10.1103/PhysRevB.107.075408}
}

@article{Choi2019,
	author = {Youngjoon Choi and Jeannette Kemmer and Yang Peng and Alex Thomson and Harpreet Arora and Robert Polski and Yiran Zhang and Hechen Ren and Jason Alicea and Gil Refael and Felix von Oppen and Kenji Watanabe and Takashi Taniguchi and Stevan Nadj-Perge},
	doi = {10.1038/s41567-019-0606-5},
	issn = {1745-2473},
	issue = {11},
	journal = {Nat. Phys.},
	month = {8},
	pages = {1174-1180},
	publisher = {Springer US},
	title = {Electronic correlations in twisted bilayer graphene near the magic angle},
	volume = {15},
	url = {https://www.nature.com/articles/s41567-019-0606-5},
	year = {2019}
}

@article{Xie2019,
	author = {Yonglong Xie and Biao Lian and Berthold Jäck and Xiaomeng Liu and Cheng-Li Chiu and Kenji Watanabe and Takashi Taniguchi and B. Andrei Bernevig and Ali Yazdani},
	doi = {10.1038/s41586-019-1422-x},
	issn = {14764687},
	issue = {7767},
	journal = {Nature},
	month = {8},
	pages = {101-105},
	publisher = {Nature Publishing Group},
	title = {Spectroscopic signatures of many-body correlations in magic-angle twisted bilayer graphene},
	volume = {572},
	url = {https://www.nature.com/articles/s41586-019-1422-x},
	year = {2019}
}

@article{Sharpe2019,
	author = {Aaron L. Sharpe and Eli J. Fox and Arthur W. Barnard and Joe Finney and Kenji Watanabe and Takashi Taniguchi and M. A. Kastner and David Goldhaber-Gordon},
	doi = {10.1126/science.aaw3780},
	issn = {0036-8075},
	issue = {6453},
	journal = {Science},
	month = {8},
	pages = {605-608},
	pmid = {31346139},
	title = {Emergent ferromagnetism near three-quarters filling in twisted bilayer graphene},
	volume = {365},
	url = {https://www.science.org/doi/10.1126/science.aaw3780},
	year = {2019}
}

@article{Serlin2020,
	author = {M. Serlin and C. L. Tschirhart and H. Polshyn and Y. Zhang and J. Zhu and K. Watanabe and T. Taniguchi and L. Balents and A. F. Young},
	doi = {10.1126/science.aay5533},
	issn = {0036-8075},
	issue = {6480},
	journal = {Science},
	month = {2},
	pages = {900-903},
	pmid = {31857492},
	title = {Intrinsic quantized anomalous Hall effect in a moiré heterostructure},
	volume = {367},
	url = {https://www.science.org/doi/10.1126/science.aay5533},
	year = {2020}
}

@article{Zondiner2020,
	author = {Uri Zondiner and Asaf Rozen and Daniel Rodan-Legrain and Yuan Cao and Raquel Queiroz and Takashi Taniguchi and Kenji Watanabe and Yuval Oreg and Felix von Oppen and Ady Stern and Erez Berg and Pablo Jarillo-Herrero and Shahal Ilani},
	doi = {10.1038/s41586-020-2373-y},
	issn = {14764687},
	issue = {7811},
	journal = {Nature},
	month = {6},
	pages = {203-208},
	pmid = {32528091},
	publisher = {Nature Research},
	title = {Cascade of phase transitions and Dirac revivals in magic-angle graphene},
	volume = {582},
	url = {https://www.nature.com/articles/s41586-020-2373-y},
	year = {2020}
}

@article{Choi2021,
	author = {Youngjoon Choi and Hyunjin Kim and Yang Peng and Alex Thomson and Cyprian Lewandowski and Robert Polski and Yiran Zhang and Harpreet Singh Arora and Kenji Watanabe and Takashi Taniguchi and Jason Alicea and Stevan Nadj-Perge},
	doi = {10.1038/s41586-020-03159-7},
	issn = {14764687},
	issue = {7843},
	journal = {Nature},
	month = {1},
	pages = {536-541},
	pmid = {33462504},
	publisher = {Nature Research},
	title = {Correlation-driven topological phases in magic-angle twisted bilayer graphene},
	volume = {589},
	url = {https://www.nature.com/articles/s41586-020-03159-7},
	year = {2021}
}

@article{Das2021,
	author = {Ipsita Das and Xiaobo Lu and Jonah Herzog-Arbeitman and Zhi-Da Song and Kenji Watanabe and Takashi Taniguchi and B. Andrei Bernevig and Dmitri K. Efetov},
	doi = {10.1038/s41567-021-01186-3},
	issn = {1745-2473},
	issue = {6},
	journal = {Nat. Phys.},
	month = {3},
	pages = {710-714},
	publisher = {Springer US},
	title = {Symmetry-broken Chern insulators and Rashba-like Landau-level crossings in magic-angle bilayer graphene},
	volume = {17},
	url = {https://www.nature.com/articles/s41567-021-01186-3},
	year = {2021}
}

@article{Saito2021,
	author = {Yu Saito and Fangyuan Yang and Jingyuan Ge and Xiaoxue Liu and Takashi Taniguchi and Kenji Watanabe and J. I. A. Li and Erez Berg and Andrea F. Young},
	doi = {10.1038/s41586-021-03409-2},
	issn = {14764687},
	issue = {7853},
	journal = {Nature},
	month = {4},
	pages = {220-224},
	pmid = {33828322},
	publisher = {Nature Research},
	title = {Isospin Pomeranchuk effect in twisted bilayer graphene},
	volume = {592},
	url = {https://www.nature.com/articles/s41586-021-03409-2},
	year = {2021}
}

@article{Rozen2021,
	author = {Asaf Rozen and Jeong Min Park and Uri Zondiner and Yuan Cao and Daniel Rodan-Legrain and Takashi Taniguchi and Kenji Watanabe and Yuval Oreg and Ady Stern and Erez Berg and Pablo Jarillo-Herrero and Shahal Ilani},
	doi = {10.1038/s41586-021-03319-3},
	issn = {14764687},
	issue = {7853},
	journal = {Nature},
	month = {4},
	pages = {214-219},
	pmid = {33828314},
	publisher = {Nature Research},
	title = {Entropic evidence for a Pomeranchuk effect in magic-angle graphene},
	volume = {592},
	url = {https://www.nature.com/articles/s41586-021-03319-3},
	year = {2021}
}

@article{Park2021,
	author = {Jeong Min Park and Yuan Cao and Kenji Watanabe and Takashi Taniguchi and Pablo Jarillo-Herrero},
	doi = {10.1038/s41586-021-03366-w},
	issn = {14764687},
	issue = {7852},
	journal = {Nature},
	month = {4},
	pages = {43-48},
	pmid = {33790447},
	publisher = {Nature Research},
	title = {Flavour Hund's coupling, Chern gaps and charge diffusivity in moiré graphene},
	volume = {592},
	url = {https://www.nature.com/articles/s41586-021-03366-w},
	year = {2021}
}

@article{Grover2022,
	author = {Sameer Grover and Matan Bocarsly and Aviram Uri and Petr Stepanov and Giorgio Di Battista and Indranil Roy and Jiewen Xiao and Alexander Y. Meltzer and Yuri Myasoedov and Keshav Pareek and Kenji Watanabe and Takashi Taniguchi and Binghai Yan and Ady Stern and Erez Berg and Dmitri K. Efetov and Eli Zeldov},
	doi = {10.1038/s41567-022-01635-7},
	issn = {1745-2473},
	issue = {8},
	journal = {Nat. Phys.},
	month = {6},
	pages = {885-892},
	publisher = {Springer US},
	title = {Chern mosaic and Berry-curvature magnetism in magic-angle graphene},
	volume = {18},
	url = {https://www.nature.com/articles/s41567-022-01635-7},
	year = {2022}
}

@misc{Bernevig2025BerryTrashcanModel,
      title={"Berry Trashcan" Model of Interacting Electrons in Rhombohedral Graphene}, 
      author={B. Andrei Bernevig and Yves H. Kwan},
      year={2025},
      eprint={2503.09692},
      archivePrefix={arXiv},
      primaryClass={cond-mat.str-el},
      url={https://arxiv.org/abs/2503.09692}, 
}

@misc{Li2025BerryTrashcanPairing,
      title={Berry Trashcan With Short Range Attraction:Exact $p_x+i p_y$ Superconductivity in Rhombohedral Graphene}, 
      author={Ming-Rui Li and Yves H. Kwan and Hong Yao and B. Andrei Bernevig},
      year={2025},
      eprint={2509.16312},
      archivePrefix={arXiv},
      primaryClass={cond-mat.supr-con},
      url={https://arxiv.org/abs/2509.16312}, 
}

@article{Antebi2022,
  title = {In-plane orbital magnetization as a probe for symmetry breaking in strained twisted bilayer graphene},
  author = {Antebi, Ohad and Stern, Ady and Berg, Erez},
  journal = {Phys. Rev. B},
  volume = {105},
  issue = {10},
  pages = {104423},
  numpages = {15},
  year = {2022},
  month = {Mar},
  publisher = {American Physical Society},
  doi = {10.1103/PhysRevB.105.104423},
  url = {https://link.aps.org/doi/10.1103/PhysRevB.105.104423}
}

@article{Beechem2014,
	title = {Rotational {Disorder} in {Twisted} {Bilayer} {Graphene}},
	volume = {8},
	issn = {1936-0851, 1936-086X},
	url = {https://pubs.acs.org/doi/10.1021/nn405999z},
	doi = {10.1021/nn405999z},
	number = {2},
	urldate = {2026-08-20},
	journal = {ACS Nano},
	author = {Beechem, Thomas E. and Ohta, Taisuke and Diaconescu, Bogdan and Robinson, Jeremy T.},
	month = feb,
	year = {2014},
	pages = {1655--1663},
}

@article{Uri2020,
	title = {Mapping the twist-angle disorder and {Landau} levels in magic-angle graphene},
	volume = {581},
	issn = {0028-0836, 1476-4687},
	url = {https://www.nature.com/articles/s41586-020-2255-3},
	doi = {10.1038/s41586-020-2255-3},
	number = {7806},
	urldate = {2026-08-20},
	journal = {Nature},
	author = {Uri, A. and Grover, S. and Cao, Y. and Crosse, J. A. and Bagani, K. and Rodan-Legrain, D. and Myasoedov, Y. and Watanabe, K. and Taniguchi, T. and Moon, P. and Koshino, M. and Jarillo-Herrero, P. and Zeldov, E.},
	month = may,
	year = {2020},
	pages = {47--52},
}

@misc{Maccari2026,
      title={How twist angle inhomogeneity masks the BKT transition and the order parameter symmetry}, 
      author={Ilaria Maccari and Louk Rademaker and Giulia Venditti},
      year={2026},
      eprint={2606.17191},
      archivePrefix={arXiv},
      primaryClass={cond-mat.supr-con},
      url={https://arxiv.org/abs/2606.17191}, 
}

@article{Klebl2019,
  title = {Inherited and flatband-induced ordering in twisted graphene bilayers},
  author = {Klebl, Lennart and Honerkamp, Carsten},
  journal = {Phys. Rev. B},
  volume = {100},
  issue = {15},
  pages = {155145},
  numpages = {8},
  year = {2019},
  month = {Oct},
  publisher = {American Physical Society},
  doi = {10.1103/PhysRevB.100.155145},
  url = {https://link.aps.org/doi/10.1103/PhysRevB.100.155145}
}

@article{MacDonald1988,
  title = {$\frac{t}{U}$ expansion for the Hubbard model},
  author = {MacDonald, A. H. and Girvin, S. M. and Yoshioka, D.},
  journal = {Phys. Rev. B},
  volume = {37},
  issue = {16},
  pages = {9753--9756},
  numpages = {0},
  year = {1988},
  month = {Jun},
  publisher = {American Physical Society},
  doi = {10.1103/PhysRevB.37.9753},
  url = {https://link.aps.org/doi/10.1103/PhysRevB.37.9753}
}

@article{Scalapino1995,
	title = {{The case for $d_{x^2 - y^2}$ pairing in the cuprate superconductors}},
	volume = {250},
	issn = {0370-1573},
	url = {https://www.sciencedirect.com/science/article/pii/037015739400086I},
	doi = {https://doi.org/10.1016/0370-1573(94)00086-I},
	number = {6},
	journal = {Phys. Rep.},
	author = {Scalapino, D. J.},
	year = {1995},
	pages = {329--365},
}

@article{Gonzalez1994,
title = {Non-Fermi liquid behavior of electrons in the half-filled honeycomb lattice (A renormalization group approach)},
journal = {Nucl. Phys. B},
volume = {424},
number = {3},
pages = {595-618},
year = {1994},
issn = {0550-3213},
doi = {https://doi.org/10.1016/0550-3213(94)90410-3},
url = {https://www.sciencedirect.com/science/article/pii/0550321394904103},
author = {J. González and F. Guinea and M.A.H. Vozmediano},
}

@article{Sheehy2007,
  title = {Quantum Critical Scaling in Graphene},
  author = {Sheehy, Daniel E. and Schmalian, J\"org},
  journal = {Phys. Rev. Lett.},
  volume = {99},
  issue = {22},
  pages = {226803},
  numpages = {4},
  year = {2007},
  month = {Nov},
  publisher = {American Physical Society},
  doi = {10.1103/PhysRevLett.99.226803},
  url = {https://link.aps.org/doi/10.1103/PhysRevLett.99.226803}
}

@article{Elias2011,
   title={Dirac cones reshaped by interaction effects in suspended graphene},
   volume={7},
   ISSN={1745-2481},
   url={http://dx.doi.org/10.1038/nphys2049},
   DOI={10.1038/nphys2049},
   number={9},
   journal={Nat. Phys.},
   publisher={Springer Science and Business Media LLC},
   author={Elias, D. C. and Gorbachev, R. V. and Mayorov, A. S. and Morozov, S. V. and Zhukov, A. A. and Blake, P. and Ponomarenko, L. A. and Grigorieva, I. V. and Novoselov, K. S. and Guinea, F. and Geim, A. K.},
   year={2011},
   pages={701–704} }

@article{Lee2026,
    author = {Lee, M. and Das, I. and Herzog-Arbeitman, J. and Papp, J. and Li, J. and Daschner, M. and Zhou, Z. and Bhatt, M. and Currle, M. and Yu, J. and Jiang, Y. and Becherer, M. and Mittermeier, R. and Altpeter, P. and Obermayer, C. and Lorenz, H. and Chavez, G. and Le, B. T. and Williams, J. and Watanabe, K. and Taniguchi, T. and Bernevig, B. A. and Efetov, D. K.},
    title = {Revealing Electron–Electron Interactions in Graphene at Room Temperature with a Quantum Twisting Microscope},
    journal = {Nano Letters},
    volume = {26},
    number = {12},
    pages = {4046-4052},
    year = {2026},
    month = {03},
    issn = {1530-6984},
    doi = {10.1021/acs.nanolett.5c05015},
    url = {https://doi.org/10.1021/acs.nanolett.5c05015},
    eprint = {https://pubs.acs.org/nalefd/article-pdf/26/12/4046/64800122/nl5c05015.pdf},
}

@misc{Fang2019arXiv,
      title={Angle-Dependent {\it Ab initio} Low-Energy Hamiltonians for a Relaxed Twisted Bilayer Graphene Heterostructure}, 
      author={Shiang Fang and Stephen Carr and Ziyan Zhu and Daniel Massatt and Efthimios Kaxiras},
      year={2019},
      eprint={1908.00058},
      archivePrefix={arXiv},
      primaryClass={cond-mat.mes-hall},
      url={https://arxiv.org/abs/1908.00058}, 
}

@article{Nam2017,
  title = {Lattice relaxation and energy band modulation in twisted bilayer graphene},
  author = {Nam, Nguyen N. T. and Koshino, Mikito},
  journal = {Phys. Rev. B},
  volume = {96},
  issue = {7},
  pages = {075311},
  numpages = {12},
  year = {2017},
  month = {Aug},
  publisher = {American Physical Society},
  doi = {10.1103/PhysRevB.96.075311},
  url = {https://link.aps.org/doi/10.1103/PhysRevB.96.075311}
}

@article{Carr2018,
  title = {Relaxation and domain formation in incommensurate two-dimensional heterostructures},
  author = {Carr, Stephen and Massatt, Daniel and Torrisi, Steven B. and Cazeaux, Paul and Luskin, Mitchell and Kaxiras, Efthimios},
  journal = {Phys. Rev. B},
  volume = {98},
  issue = {22},
  pages = {224102},
  numpages = {7},
  year = {2018},
  month = {Dec},
  publisher = {American Physical Society},
  doi = {10.1103/PhysRevB.98.224102},
  url = {https://link.aps.org/doi/10.1103/PhysRevB.98.224102}
}

@article{Carr2019,
  title = {Exact continuum model for low-energy electronic states of twisted bilayer graphene},
  author = {Carr, Stephen and Fang, Shiang and Zhu, Ziyan and Kaxiras, Efthimios},
  journal = {Phys. Rev. Res.},
  volume = {1},
  issue = {1},
  pages = {013001},
  numpages = {6},
  year = {2019},
  month = {Aug},
  publisher = {American Physical Society},
  doi = {10.1103/PhysRevResearch.1.013001},
  url = {https://link.aps.org/doi/10.1103/PhysRevResearch.1.013001}
}

@article{YWLiu2020,
  title = {Tunable Lattice Reconstruction, Triangular Network of Chiral One-Dimensional States, and Bandwidth of Flat Bands in Magic Angle Twisted Bilayer Graphene},
  author = {Liu, Yi-Wen and Su, Ying and Zhou, Xiao-Feng and Yin, Long-Jing and Yan, Chao and Li, Si-Yu and Yan, Wei and Han, Sheng and Fu, Zhong-Qiu and Zhang, Yu and Yang, Qian and Ren, Ya-Ning and He, Lin},
  journal = {Phys. Rev. Lett.},
  volume = {125},
  issue = {23},
  pages = {236102},
  numpages = {6},
  year = {2020},
  month = {Dec},
  publisher = {American Physical Society},
  doi = {10.1103/PhysRevLett.125.236102},
  url = {https://link.aps.org/doi/10.1103/PhysRevLett.125.236102}
}

@article{Cantele2020,
  title = {Structural relaxation and low-energy properties of twisted bilayer graphene},
  author = {Cantele, Giovanni and Alf\`e, Dario and Conte, Felice and Cataudella, Vittorio and Ninno, Domenico and Lucignano, Procolo},
  journal = {Phys. Rev. Res.},
  volume = {2},
  issue = {4},
  pages = {043127},
  numpages = {10},
  year = {2020},
  month = {Oct},
  publisher = {American Physical Society},
  doi = {10.1103/PhysRevResearch.2.043127},
  url = {https://link.aps.org/doi/10.1103/PhysRevResearch.2.043127}
}

@article{Woo2016,
  title = {Poisson's ratio in layered two-dimensional crystals},
  author = {Woo, Sungjong and Park, Hee Chul and Son, Young-Woo},
  journal = {Phys. Rev. B},
  volume = {93},
  issue = {7},
  pages = {075420},
  numpages = {5},
  year = {2016},
  month = {Feb},
  publisher = {American Physical Society},
  doi = {10.1103/PhysRevB.93.075420},
  url = {https://link.aps.org/doi/10.1103/PhysRevB.93.075420}
}

@article{Escudero2026,
  title = {Straintronics and twistronics in bilayer graphene},
  author = {Escudero, Federico and Wang, Dong and Pantale\'on, Pierre A. and Yuan, Shengjun and Guinea, Francisco and Zhan, Zhen},
  journal = {Phys. Rev. Res.},
  volume = {8},
  issue = {2},
  pages = {023225},
  numpages = {19},
  year = {2026},
  month = {May},
  publisher = {American Physical Society},
  doi = {10.1103/r8sv-6xmp},
  url = {https://link.aps.org/doi/10.1103/r8sv-6xmp}
}

@article{Song2022,
  title = {Magic-Angle Twisted Bilayer Graphene as a Topological Heavy Fermion Problem},
  author = {Song, Zhi-Da and Bernevig, B. Andrei},
  journal = {Phys. Rev. Lett.},
  volume = {129},
  issue = {4},
  pages = {047601},
  numpages = {10},
  year = {2022},
  month = {Jul},
  publisher = {American Physical Society},
  doi = {10.1103/PhysRevLett.129.047601},
  url = {https://link.aps.org/doi/10.1103/PhysRevLett.129.047601}
}

@article{Calugaru2023,
    author = {Călugăru, Dumitru and Borovkov, Maksim and Lau, Liam L. H. and Coleman, Piers and Song, Zhi-Da and Bernevig, B. Andrei},
    title = {Twisted bilayer graphene as topological heavy fermion: II. Analytical approximations of the model parameters},
    journal = {Low Temp. Phys.},
    volume = {49},
    number = {6},
    pages = {640-654},
    year = {2023},
    month = {06},
    issn = {1063-777X},
    doi = {10.1063/10.0019421},
    url = {https://doi.org/10.1063/10.0019421},
}

@article{YiZhang2020,
  title = {Correlated insulating phases of twisted bilayer graphene at commensurate filling fractions: A Hartree-Fock study},
  author = {Zhang, Yi and Jiang, Kun and Wang, Ziqiang and Zhang, Fuchun},
  journal = {Phys. Rev. B},
  volume = {102},
  issue = {3},
  pages = {035136},
  numpages = {9},
  year = {2020},
  month = {Jul},
  publisher = {American Physical Society},
  doi = {10.1103/PhysRevB.102.035136},
  url = {https://link.aps.org/doi/10.1103/PhysRevB.102.035136}
}

@article{DanielEParker2021,
  title = {Strain-Induced Quantum Phase Transitions in Magic-Angle Graphene},
  author = {Parker, Daniel E. and Soejima, Tomohiro and Hauschild, Johannes and Zaletel, Michael P. and Bultinck, Nick},
  journal = {Phys. Rev. Lett.},
  volume = {127},
  issue = {2},
  pages = {027601},
  numpages = {7},
  year = {2021},
  month = {Jul},
  publisher = {American Physical Society},
  doi = {10.1103/PhysRevLett.127.027601},
  url = {https://link.aps.org/doi/10.1103/PhysRevLett.127.027601}
}

@article{Calugaru2022,
  title = {Spectroscopy of Twisted Bilayer Graphene Correlated Insulators},
  author = {C\ifmmode \u{a}\else \u{a}\fi{}lug\ifmmode \u{a}\else \u{a}\fi{}ru, Dumitru and Regnault, Nicolas and Oh, Myungchul and Nuckolls, Kevin P. and Wong, Dillon and Lee, Ryan L. and Yazdani, Ali and Vafek, Oskar and Bernevig, B. Andrei},
  journal = {Phys. Rev. Lett.},
  volume = {129},
  issue = {11},
  pages = {117602},
  numpages = {8},
  year = {2022},
  month = {Sep},
  publisher = {American Physical Society},
  doi = {10.1103/PhysRevLett.129.117602},
  url = {https://link.aps.org/doi/10.1103/PhysRevLett.129.117602}
}

@article{Hoke2024,
   title={Uncovering the spin ordering in magic-angle graphene via edge state equilibration},
   volume={15},
   ISSN={2041-1723},
   url={http://dx.doi.org/10.1038/s41467-024-48385-z},
   DOI={10.1038/s41467-024-48385-z},
   number={1},
   journal={Nature Communications},
   publisher={Springer Science and Business Media LLC},
   author={Hoke, Jesse C. and Li, Yifan and May-Mann, Julian and Watanabe, Kenji and Taniguchi, Takashi and Bradlyn, Barry and Hughes, Taylor L. and Feldman, Benjamin E.},
   year={2024},
   month=May 
}

@article{Angeli2019,
  title = {Valley Jahn-Teller Effect in Twisted Bilayer Graphene},
  author = {Angeli, M. and Tosatti, E. and Fabrizio, M.},
  journal = {Phys. Rev. X},
  volume = {9},
  issue = {4},
  pages = {041010},
  numpages = {17},
  year = {2019},
  month = {Oct},
  publisher = {American Physical Society},
  doi = {10.1103/PhysRevX.9.041010},
  url = {https://link.aps.org/doi/10.1103/PhysRevX.9.041010}
}

@article{Angeli2020,
	author = {Mattia Angeli and Michele Fabrizio},
	doi = {10.1140/epjp/s13360-020-00647-7},
	issn = {21905444},
	issue = {8},
	journal = {Eur. Phys. J. Plus},
	month = {8},
	publisher = {Springer Science and Business Media Deutschland GmbH},
	title = {Jahn–Teller coupling to moiré phonons in the continuum model formalism for small-angle twisted bilayer graphene},
	volume = {135},
	year = {2020}
}

@article{Kwan2024,
  title = {Electron-phonon coupling and competing Kekul\'e orders in twisted bilayer graphene},
  author = {Kwan, Yves H. and Wagner, Glenn and Bultinck, Nick and Simon, Steven H. and Berg, Erez and Parameswaran, S. A.},
  journal = {Phys. Rev. B},
  volume = {110},
  issue = {8},
  pages = {085160},
  numpages = {9},
  year = {2024},
  month = {Aug},
  publisher = {American Physical Society},
  doi = {10.1103/PhysRevB.110.085160},
  url = {https://link.aps.org/doi/10.1103/PhysRevB.110.085160}
}

@article{Shi2025,
	author = {Hao Shi and Wangqian Miao and Xi Dai},
	doi = {10.1103/PhysRevB.111.155126},
	title = {{Moir\'e optical phonons coupled to heavy electrons in magic-angle twisted bilayer graphene}},
	issn = {24699969},
	issue = {15},
	pages = {155126},
	journal = {Phys. Rev. B},
	month = {4},
	publisher = {American Physical Society},
	volume = {111},
	url = {https://doi.org/10.1103/PhysRevB.111.155126},
	year = {2025},
}

@article{Wang2025,
  title = {Electron-phonon coupling in the topological heavy fermion model of twisted bilayer graphene},
  author = {Wang, Yi-Jie and Zhou, Geng-Dong and Lian, Biao and Song, Zhi-Da},
  journal = {Phys. Rev. B},
  volume = {111},
  issue = {3},
  pages = {035110},
  numpages = {47},
  year = {2025},
  month = {Jan},
  publisher = {American Physical Society},
  doi = {10.1103/PhysRevB.111.035110},
  url = {https://link.aps.org/doi/10.1103/PhysRevB.111.035110}
}

@article{Huder2018,
  title = {Electronic Spectrum of Twisted Graphene Layers under Heterostrain},
  author = {Huder, Lo\"{\i}c and Artaud, Alexandre and Le Quang, Toai and de Laissardi\`ere, Guy Trambly and Jansen, Aloysius G. M. and Lapertot, G\'erard and Chapelier, Claude and Renard, Vincent T.},
  journal = {Phys. Rev. Lett.},
  volume = {120},
  issue = {15},
  pages = {156405},
  numpages = {5},
  year = {2018},
  month = {Apr},
  publisher = {American Physical Society},
  doi = {10.1103/PhysRevLett.120.156405},
  url = {https://link.aps.org/doi/10.1103/PhysRevLett.120.156405}
}

@article{Qiao2018,
  title = {Twisted graphene bilayer around the first magic angle engineered by heterostrain},
  author = {Qiao, Jia-Bin and Yin, Long-Jing and He, Lin},
  journal = {Phys. Rev. B},
  volume = {98},
  issue = {23},
  pages = {235402},
  numpages = {8},
  year = {2018},
  month = {Dec},
  publisher = {American Physical Society},
  doi = {10.1103/PhysRevB.98.235402},
  url = {https://link.aps.org/doi/10.1103/PhysRevB.98.235402}
}

@article{Bi2019,
  title = {Designing flat bands by strain},
  author = {Bi, Zhen and Yuan, Noah F. Q. and Fu, Liang},
  journal = {Phys. Rev. B},
  volume = {100},
  issue = {3},
  pages = {035448},
  numpages = {9},
  year = {2019},
  month = {Jul},
  publisher = {American Physical Society},
  doi = {10.1103/PhysRevB.100.035448},
  url = {https://link.aps.org/doi/10.1103/PhysRevB.100.035448}
}

@article{Kazmierczak2021,
   title={Strain fields in twisted bilayer graphene},
   volume={20},
   ISSN={1476-4660},
   url={http://dx.doi.org/10.1038/s41563-021-00973-w},
   DOI={10.1038/s41563-021-00973-w},
   number={7},
   journal={Nat. Mater.},
   publisher={Springer Science and Business Media LLC},
   author={Kazmierczak, Nathanael P. and Van Winkle, Madeline and Ophus, Colin and Bustillo, Karen C. and Carr, Stephen and Brown, Hamish G. and Ciston, Jim and Taniguchi, Takashi and Watanabe, Kenji and Bediako, D. Kwabena},
   year={2021},
   month=Apr, pages={956–963} }

@article{Benschop2021,
  title = {Measuring local moir\'e lattice heterogeneity of twisted bilayer graphene},
  author = {Benschop, Tjerk and de Jong, Tobias A. and Stepanov, Petr and Lu, Xiaobo and Stalman, Vincent and van der Molen, Sense Jan and Efetov, Dmitri K. and Allan, Milan P.},
  journal = {Phys. Rev. Res.},
  volume = {3},
  issue = {1},
  pages = {013153},
  numpages = {14},
  year = {2021},
  month = {Feb},
  publisher = {American Physical Society},
  doi = {10.1103/PhysRevResearch.3.013153},
  url = {https://link.aps.org/doi/10.1103/PhysRevResearch.3.013153}
}

@article{Mesple2021,
  title = {Heterostrain Determines Flat Bands in Magic-Angle Twisted Graphene Layers},
  author = {Mesple, Florie and Missaoui, Ahmed and Cea, Tommaso and Huder, Loic and Guinea, Francisco and Trambly de Laissardi\`ere, Guy and Chapelier, Claude and Renard, Vincent T.},
  journal = {Phys. Rev. Lett.},
  volume = {127},
  issue = {12},
  pages = {126405},
  numpages = {6},
  year = {2021},
  month = {Sep},
  publisher = {American Physical Society},
  doi = {10.1103/PhysRevLett.127.126405},
  url = {https://link.aps.org/doi/10.1103/PhysRevLett.127.126405}
}

@article{HerzogArbeitman2025b,
  title = {Topological heavy fermion model as an efficient representation of atomistic strain and relaxation in twisted bilayer graphene},
  author = {Herzog-Arbeitman, Jonah and Yu, Jiabin and C\ifmmode \u{a}\else \u{a}\fi{}lug\ifmmode \u{a}\else \u{a}\fi{}ru, Dumitru and Hu, Haoyu and Regnault, Nicolas and Vafek, Oskar and Kang, Jian and Bernevig, B. Andrei},
  journal = {Phys. Rev. B},
  volume = {112},
  issue = {12},
  pages = {125128},
  numpages = {18},
  year = {2025},
  month = {Sep},
  publisher = {American Physical Society},
  doi = {10.1103/xv3m-vtlr},
  url = {https://link.aps.org/doi/10.1103/xv3m-vtlr}
}

@article{Kwan2021,
  title = {Kekul\'e Spiral Order at All Nonzero Integer Fillings in Twisted Bilayer Graphene},
  author = {Kwan, Y. H. and Wagner, G. and Soejima, T. and Zaletel, M. P. and Simon, S. H. and Parameswaran, S. A. and Bultinck, N.},
  journal = {Phys. Rev. X},
  volume = {11},
  issue = {4},
  pages = {041063},
  numpages = {23},
  year = {2021},
  month = {Dec},
  publisher = {American Physical Society},
  doi = {10.1103/PhysRevX.11.041063},
  url = {https://link.aps.org/doi/10.1103/PhysRevX.11.041063}
}

@article{Wagner2022,
  title = {Global Phase Diagram of the Normal State of Twisted Bilayer Graphene},
  author = {Wagner, Glenn and Kwan, Yves H. and Bultinck, Nick and Simon, Steven H. and Parameswaran, S. A.},
  journal = {Phys. Rev. Lett.},
  volume = {128},
  issue = {15},
  pages = {156401},
  numpages = {7},
  year = {2022},
  month = {Apr},
  publisher = {American Physical Society},
  doi = {10.1103/PhysRevLett.128.156401},
  url = {https://link.aps.org/doi/10.1103/PhysRevLett.128.156401}
}

@misc{ZiweiWang2025,
      title={Putting a new spin on the incommensurate Kekul\'e spiral: from spin-valley locking and collective modes to fermiology and implications for superconductivity}, 
      author={Ziwei Wang and Glenn Wagner and Yves H. Kwan and Nick Bultinck and Steven H. Simon and S. A. Parameswaran},
      year={2025},
      eprint={2509.12320},
      archivePrefix={arXiv},
      primaryClass={cond-mat.str-el},
      url={https://arxiv.org/abs/2509.12320}, 
}

@article{Crippa2026,
  title = {Interplay between many-body correlations, strain, and lattice relaxation in twisted bilayer graphene},
  author = {Crippa, Lorenzo and Rai, Gautam and C\ifmmode \u{a}\else \u{a}\fi{}lug\ifmmode \u{a}\else \u{a}\fi{}ru, Dumitru and Hu, Haoyu and Herzog-Arbeitman, Jonah and Bernevig, B. Andrei and Valent\'{\i}, Roser and Sangiovanni, Giorgio and Wehling, Tim},
  journal = {Phys. Rev. B},
  volume = {113},
  issue = {24},
  pages = {L241112},
  numpages = {7},
  year = {2026},
  month = {Jun},
  publisher = {American Physical Society},
  doi = {10.1103/x2mz-hm8y},
  url = {https://link.aps.org/doi/10.1103/x2mz-hm8y}
}

@article{Herzog-Arbeitman2025,
  title = {Kekul\'e spiral order from strained topological heavy fermions},
  author = {Herzog-Arbeitman, Jonah and C\ifmmode \u{a}\else \u{a}\fi{}lug\ifmmode \u{a}\else \u{a}\fi{}ru, Dumitru and Hu, Haoyu and Yu, Jiabin and Regnault, Nicolas and Kang, Jian and Bernevig, B. Andrei and Vafek, Oskar},
  journal = {Phys. Rev. B},
  volume = {112},
  issue = {12},
  pages = {125129},
  numpages = {34},
  year = {2025},
  month = {Sep},
  publisher = {American Physical Society},
  doi = {10.1103/rr5g-3js8},
  url = {https://link.aps.org/doi/10.1103/rr5g-3js8}
}

@article{Putzer2025,
	title = {Eliashberg theory and superfluid stiffness of band-off-diagonal pairing in twisted graphene},
	author = {Putzer, Bernhard and Scheurer, Mathias S.},
	journal = {Phys. Rev. B},
	volume = {111},
	issue = {14},
	pages = {144513},
	numpages = {16},
	year = {2025},
	month = {Apr},
	publisher = {American Physical Society},
	doi = {10.1103/PhysRevB.111.144513},
	url = {https://link.aps.org/doi/10.1103/PhysRevB.111.144513}
}

@article{Christos2023,
	author = {Maine Christos and Subir Sachdev and Mathias S. Scheurer},
	doi = {10.1038/s41467-023-42471-4},
	issn = {2041-1723},
	issue = {1},
	journal = {Nat. Commun.},
	month = {11},
	pages = {7134},
	pmid = {37932262},
	publisher = {Nature Research},
	title = {Nodal band-off-diagonal superconductivity in twisted graphene superlattices},
	volume = {14},
	url = {https://www.nature.com/articles/s41467-023-42471-4},
	year = {2023}
}

@article{Liu2025,
	author = {Chao-Xing Liu and B. Andrei Bernevig},
	doi = {10.1103/PhysRevB.111.L020502},
	issn = {2469-9950},
	issue = {2},
	journal = {Phys. Rev. B},
	month = {1},
	pages = {L020502},
	title = {Nodal nematic superconductivity in multiple flat-band systems},
	volume = {111},
	url = {https://link.aps.org/doi/10.1103/PhysRevB.111.L020502},
	year = {2025}
}

@article{Tarnopolsky2019,
	author = {Grigory Tarnopolsky and Alex Jura Kruchkov and Ashvin Vishwanath},
	doi = {10.1103/PhysRevLett.122.106405},
	issn = {0031-9007},
	issue = {10},
	journal = {Phys. Rev. Lett.},
	month = {3},
	pages = {106405},
	pmid = {30932657},
	publisher = {American Physical Society},
	title = {Origin of Magic Angles in Twisted Bilayer Graphene},
	volume = {122},
	url = {https://link.aps.org/doi/10.1103/PhysRevLett.122.106405},
	year = {2019}
}

@article{Wang2021,
	author = {Jie Wang and Yunqin Zheng and Andrew J. Millis and Jennifer Cano},
	doi = {10.1103/PhysRevResearch.3.023155},
	issn = {2643-1564},
	issue = {2},
	journal = {Phys. Rev. Res.},
	month = {5},
	pages = {023155},
	publisher = {American Physical Society},
	title = {Chiral approximation to twisted bilayer graphene: Exact intravalley inversion symmetry, nodal structure, and implications for higher magic angles},
	volume = {3},
	url = {https://link.aps.org/doi/10.1103/PhysRevResearch.3.023155},
	year = {2021}
}

@article{Lothman2022,
	author = {Tomas Löthman and Johann Schmidt and Fariborz Parhizgar and Annica M. Black-Schaffer},
	doi = {10.1038/s42005-022-00860-z},
	issn = {2399-3650},
	issue = {1},
	journal = {Commun. Phys.},
	pages = {92},
	title = {Nematic superconductivity in magic-angle twisted bilayer graphene from atomistic modeling},
	volume = {5},
	url = {https://www.nature.com/articles/s42005-022-00860-z},
	year = {2022}
}

@article{Wu2019,
	author = {Fengcheng Wu},
	doi = {10.1103/PhysRevB.99.195114},
	issn = {24699969},
	issue = {19},
	journal = {Phys. Rev. B},
	pages = {1-9},
	publisher = {American Physical Society},
	title = {Topological chiral superconductivity with spontaneous vortices and supercurrent in twisted bilayer graphene},
	volume = {99},
	year = {2019}
}

@article{Fischer2021,
	author = {Ammon Fischer and Lennart Klebl and Carsten Honerkamp and Dante M. Kennes},
	doi = {10.1103/PhysRevB.103.L041103},
	issn = {24699969},
	issue = {4},
	journal = {Phys. Rev. B},
	pages = {1-6},
	publisher = {American Physical Society},
	title = {Spin-fluctuation-induced pairing in twisted bilayer graphene},
	volume = {103},
	year = {2021}
}

@article{Fischer2022,
	title = {Unconventional superconductivity in magic-angle twisted trilayer graphene},
	volume = {7},
	issn = {2397-4648},
	url = {https://www.nature.com/articles/s41535-021-00410-w},
	doi = {10.1038/s41535-021-00410-w},
	number = {1},
	urldate = {2026-08-19},
	journal = {npj Quantum Mater.},
	author = {Fischer, Ammon and Goodwin, Zachary A. H. and Mostofi, Arash A. and Lischner, Johannes and Kennes, Dante M. and Klebl, Lennart},
	month = jan,
	year = {2022},
	pages = {5},
}

@article{Gonzalez2023,
	title = {Ising superconductivity induced from spin-selective valley symmetry breaking in twisted trilayer graphene},
	volume = {14},
	issn = {2041-1723},
	url = {https://www.nature.com/articles/s41467-023-38250-w},
	doi = {10.1038/s41467-023-38250-w},
	number = {1},
	urldate = {2026-09-08},
	journal = {Nat. Commun.},
	author = {González, J. and Stauber, T.},
	month = may,
	year = {2023},
	pages = {2746},
}

@article{Lee2006,
	author = {Patrick A. Lee and Naoto Nagaosa and Xiao-Gang Wen},
	doi = {10.1103/RevModPhys.78.17},
	issn = {0034-6861},
	issue = {1},
	journal = {Rev. Mod. Phys.},
	month = {1},
	pages = {17-85},
	publisher = {American Physical Society},
	title = {Doping a Mott insulator: Physics of high-temperature superconductivity},
	volume = {78},
	url = {https://link.aps.org/doi/10.1103/RevModPhys.78.17},
	year = {2006}
}

@article{BlackSchaffer2014a,
	author = {Annica M Black-Schaffer and Carsten Honerkamp},
	doi = {10.1088/0953-8984/26/42/423201},
	issn = {0953-8984},
	issue = {42},
	journal = {J. Phys. Condens. Matter},
	month = {10},
	pages = {423201},
	title = {Chiral d -wave superconductivity in doped graphene},
	volume = {26},
	url = {https://iopscience.iop.org/article/10.1088/0953-8984/26/42/423201},
	year = {2014}
}

@article{Su2018,
	author = {Ying Su and Shi-Zeng Lin},
	doi = {10.1103/PhysRevB.98.195101},
	issn = {2469-9950},
	issue = {19},
	journal = {Phys. Rev. B},
	month = {11},
	pages = {195101},
	publisher = {American Physical Society},
	title = {Pairing symmetry and spontaneous vortex-antivortex lattice in superconducting twisted-bilayer graphene: Bogoliubov-de Gennes approach},
	volume = {98},
	url = {https://link.aps.org/doi/10.1103/PhysRevB.98.195101},
	year = {2018}
}

@article{Fang2019,
	author = {Shi Chao Fang and Guang Kun Liu and Hai Qing Lin and Zhong Bing Huang},
	doi = {10.1103/PhysRevB.100.115135},
	issn = {24699969},
	issue = {11},
	journal = {Phys. Rev. B},
	pages = {115135},
	publisher = {American Physical Society},
	title = {{Quantum Monte Carlo study of magnetic ordering and superconducting pairing symmetry in twisted bilayer graphene}},
	volume = {100},
	url = {https://doi.org/10.1103/PhysRevB.100.115135},
	year = {2019}
}

@article{Ramires2018,
	author = {Aline Ramires and Jose L. Lado},
	doi = {10.1103/PhysRevLett.121.146801},
	issn = {0031-9007},
	issue = {14},
	journal = {Phys. Rev. Lett.},
	month = {10},
	pages = {146801},
	pmid = {30339453},
	publisher = {American Physical Society},
	title = {Electrically Tunable Gauge Fields in Tiny-Angle Twisted Bilayer Graphene},
	volume = {121},
	url = {https://link.aps.org/doi/10.1103/PhysRevLett.121.146801},
	year = {2018}
}

@article{Colomes2018,
	author = {E. Colomés and M. Franz},
	doi = {10.1103/PhysRevLett.120.086603},
	issn = {10797114},
	issue = {8},
	journal = {Phys. Rev. Lett.},
	pages = {1-5},
	pmid = {29542989},
	title = {Antichiral Edge States in a Modified Haldane Nanoribbon},
	volume = {120},
	year = {2018}
}

@article{Ramires2019,
	author = {Aline Ramires and Jose L. Lado},
	doi = {10.1103/PhysRevB.99.245118},
	issn = {2469-9950},
	issue = {24},
	journal = {Phys. Rev. B},
	month = {6},
	pages = {245118},
	publisher = {American Physical Society},
	title = {Impurity-induced triple point fermions in twisted bilayer graphene},
	volume = {99},
	url = {https://link.aps.org/doi/10.1103/PhysRevB.99.245118},
	year = {2019}
}

@article{LopezBezanilla2019b,
	author = {Alejandro Lopez-Bezanilla and J. L. Lado},
	doi = {10.1103/PhysRevMaterials.3.084003},
	issn = {2475-9953},
	issue = {8},
	journal = {Phys. Rev. Mater.},
	month = {8},
	pages = {084003},
	publisher = {American Physical Society},
	title = {Defect-induced magnetism and Yu-Shiba-Rusinov states in twisted bilayer graphene},
	volume = {3},
	url = {https://link.aps.org/doi/10.1103/PhysRevMaterials.3.084003},
	year = {2019}
}

@article{Liu2024,
	author = {Chao-Xing Liu and Yulin Chen and Ali Yazdani and B. Andrei Bernevig},
	doi = {10.1103/PhysRevB.110.045133},
	issn = {2469-9950},
	issue = {4},
	journal = {Phys. Rev. B},
	month = {7},
	pages = {045133},
	publisher = {American Physical Society},
	title = {{Electron-$K$-phonon interaction in twisted bilayer graphene}},
	volume = {110},
	url = {https://link.aps.org/doi/10.1103/PhysRevB.110.045133},
	year = {2024}
}

@article{Peng2024,
   title={Theoretical determination of the effect of a screening gate on plasmon-induced superconductivity in twisted bilayer graphene},
   volume={109},
   ISSN={2469-9969},
   url={http://dx.doi.org/10.1103/PhysRevB.109.045404},
   DOI={10.1103/physrevb.109.045404},
   number={4},
   journal={Phys. Rev. B},
   publisher={American Physical Society (APS)},
   author={Peng, Liangtao and Yudhistira, Indra and Vignale, Giovanni and Adam, Shaffique},
   year={2024},
   month=Jan }

@article{Bistritzer2011a,
	author = {Rafi Bistritzer and Allan H. MacDonald},
	doi = {10.1073/pnas.1108174108},
	isbn = {1108174108},
	issn = {0027-8424},
	issue = {30},
	journal = {Proc. Natl. Acad. Sci. U.S.A.},
	month = {7},
	pages = {12233-12237},
	pmid = {21730173},
	title = {Moiré bands in twisted double-layer graphene},
	volume = {108},
	url = {https://pnas.org/doi/full/10.1073/pnas.1108174108},
	year = {2011}
}

@article{Shallcross2010,
	title = {Electronic structure of turbostratic graphene},
	author = {Shallcross, S. and Sharma, S. and Kandelaki, E. and Pankratov, O. A.},
	journal = {Phys. Rev. B},
	volume = {81},
	issue = {16},
	pages = {165105},
	numpages = {15},
	year = {2010},
	month = {Apr},
	publisher = {American Physical Society},
	doi = {10.1103/PhysRevB.81.165105},
	url = {https://link.aps.org/doi/10.1103/PhysRevB.81.165105}
}

@article{TBG_II,
	author = {Zhi-Da Song and Biao Lian and Nicolas Regnault and B. Andrei Bernevig},
	doi = {10.1103/PhysRevB.103.205412},
	issn = {2469-9950},
	issue = {20},
	journal = {Phys. Rev. B},
	month = {5},
	pages = {205412},
	publisher = {American Physical Society},
	title = {Twisted bilayer graphene. II. Stable symmetry anomaly},
	volume = {103},
	url = {https://link.aps.org/doi/10.1103/PhysRevB.103.205412},
	year = {2021}
}

@article{TBG_III,
  title = {Twisted bilayer graphene. III. Interacting Hamiltonian and exact symmetries},
  author = {Bernevig, B. Andrei and Song, Zhi-Da and Regnault, Nicolas and Lian, Biao},
  journal = {Phys. Rev. B},
  volume = {103},
  issue = {20},
  pages = {205413},
  numpages = {34},
  year = {2021},
  month = {May},
  publisher = {American Physical Society},
  doi = {10.1103/PhysRevB.103.205413},
  url = {https://link.aps.org/doi/10.1103/PhysRevB.103.205413}
}

@article{Po2018b,
	author = {Hoi Chun Po and Liujun Zou and Ashvin Vishwanath and T. Senthil},
	doi = {10.1103/PhysRevX.8.031089},
	issn = {2160-3308},
	issue = {3},
	journal = {Phys. Rev. X},
	month = {9},
	pages = {031089},
	publisher = {American Physical Society},
	title = {Origin of Mott Insulating Behavior and Superconductivity in Twisted Bilayer Graphene},
	volume = {8},
	url = {https://doi.org/10.1103/PhysRevX.8.031089 https://link.aps.org/doi/10.1103/PhysRevX.8.031089},
	year = {2018}
}

@article{Kang2019,
	author = {Jian Kang and Oskar Vafek},
	doi = {10.1103/PhysRevLett.122.246401},
	issn = {0031-9007},
	issue = {24},
	journal = {Phys. Rev. Lett.},
	month = {6},
	pages = {246401},
	pmid = {31322361},
	publisher = {American Physical Society},
	title = {Strong Coupling Phases of Partially Filled Twisted Bilayer Graphene Narrow Bands},
	volume = {122},
	url = {https://link.aps.org/doi/10.1103/PhysRevLett.122.246401},
	year = {2019}
}

@article{Vafek2020,
	author = {Oskar Vafek and Jian Kang},
	doi = {10.1103/PhysRevLett.125.257602},
	issn = {0031-9007},
	issue = {25},
	journal = {Phys. Rev. Lett.},
	month = {12},
	pages = {257602},
	pmid = {33416368},
	publisher = {American Physical Society},
	title = {Renormalization Group Study of Hidden Symmetry in Twisted Bilayer Graphene with Coulomb Interactions},
	volume = {125},
	url = {https://link.aps.org/doi/10.1103/PhysRevLett.125.257602},
	year = {2020}
}

@article{Xie2020Insulator,
	author = {Ming Xie and Allan H. MacDonald},
	doi = {10.1103/PhysRevLett.124.097601},
	issn = {0031-9007},
	issue = {9},
	journal = {Phys. Rev. Lett.},
	month = {3},
	pages = {097601},
	pmid = {32202880},
	publisher = {American Physical Society},
	title = {Nature of the Correlated Insulator States in Twisted Bilayer Graphene},
	volume = {124},
	url = {https://link.aps.org/doi/10.1103/PhysRevLett.124.097601},
	year = {2020}
}

@article{TramblydeLaissardiere2010,
	author = {G. Trambly de Laissardière and D. Mayou and L. Magaud},
	doi = {10.1021/nl902948m},
	issn = {1530-6984},
	issue = {3},
	journal = {Nano Lett.},
	month = {3},
	pages = {804-808},
	pmid = {20121163},
	title = {Localization of Dirac Electrons in Rotated Graphene Bilayers},
	volume = {10},
	url = {https://pubs.acs.org/doi/10.1021/nl902948m},
	year = {2010}
}

@article{Song2019b,
	author = {Zhida Song and Zhijun Wang and Wujun Shi and Gang Li and Chen Fang and B. Andrei Bernevig},
		title = {All Magic Angles in Twisted Bilayer Graphene are Topological},
	doi = {10.1103/PhysRevLett.123.036401},
	issn = {0031-9007},
	issue = {3},
	journal = {Phys. Rev. Lett.},
	month = {7},
	pages = {036401},
	pmid = {31386469},
	publisher = {American Physical Society},
	volume = {123},
	url = {https://doi.org/10.1103/PhysRevLett.123.036401 https://link.aps.org/doi/10.1103/PhysRevLett.123.036401},
	year = {2019}
}

@article{Ahn2019,
	author = {Junyeong Ahn and Sungjoon Park and Bohm-Jung Yang},
	doi = {10.1103/PhysRevX.9.021013},
	issn = {2160-3308},
	issue = {2},
	journal = {Phys. Rev. X},
	month = {4},
	pages = {021013},
	publisher = {American Physical Society},
	title = {Failure of Nielsen-Ninomiya Theorem and Fragile Topology in Two-Dimensional Systems with Space-Time Inversion Symmetry: Application to Twisted Bilayer Graphene at Magic Angle},
	volume = {9},
	url = {https://link.aps.org/doi/10.1103/PhysRevX.9.021013},
	year = {2019}
}

@article{Yu2023,
	author = {Jiabin Yu and Ming Xie and Fengcheng Wu and Sankar Das Sarma},
	doi = {10.1103/PhysRevB.107.L201106},
	issn = {2469-9950},
	issue = {20},
	journal = {Phys. Rev. B},
	month = {5},
	pages = {L201106},
	publisher = {American Physical Society},
	title = {Euler-obstructed nematic nodal superconductivity in twisted bilayer graphene},
	volume = {107},
	url = {https://link.aps.org/doi/10.1103/PhysRevB.107.L201106},
	year = {2023}
}

@article{Wang2024,
	author = {Yi-Jie Wang and Geng-Dong Zhou and Shi-Yu Peng and Biao Lian and Zhi-Da Song},
	doi = {10.1103/PhysRevLett.133.146001},
	issn = {0031-9007},
	issue = {14},
	journal = {Phys. Rev. Lett.},
	month = {9},
	pages = {146001},
	pmid = {39423412},
	publisher = {American Physical Society},
	title = {Molecular Pairing in Twisted Bilayer Graphene Superconductivity},
	volume = {133},
	url = {https://link.aps.org/doi/10.1103/PhysRevLett.133.146001},
	year = {2024}
}

@article{Gu2020,
	author = {Xingyu Gu and Chuan Chen and Jia Ning Leaw and Evan Laksono and Vitor M. Pereira and Giovanni Vignale and Shaffique Adam},
	doi = {10.1103/PhysRevB.101.180506},
	issn = {2469-9950},
	issue = {18},
	journal = {Phys. Rev. B},
	month = {5},
	pages = {180506},
	publisher = {American Physical Society},
	title = {{Antiferromagnetism and chiral $d$--wave superconductivity from an effective $t$--$J$--$D$ model for twisted bilayer graphene}},
	volume = {101},
	url = {https://link.aps.org/doi/10.1103/PhysRevB.101.180506},
	year = {2020}
}

@article{Nandkishore2012b,
	author = {Rahul Nandkishore and Andrey V. Chubukov},
	doi = {10.1103/PhysRevB.86.115426},
	issn = {1098-0121},
	issue = {11},
	journal = {Phys. Rev. B},
	month = {9},
	pages = {115426},
	title = {Interplay of superconductivity and spin-density-wave order in doped graphene},
	volume = {86},
	url = {https://link.aps.org/doi/10.1103/PhysRevB.86.115426},
	year = {2012}
}

@article{Chichinadze2020,
	author = {Dmitry V. Chichinadze and Laura Classen and Andrey V. Chubukov},
   doi = {10.1103/PhysRevB.101.224513},
   issn = {2469-9950},
   issue = {22},
   journal = {Phys. Rev. B},
   month = {6},
   pages = {224513},
   publisher = {American Physical Society},
   title = {Nematic superconductivity in twisted bilayer graphene},
   volume = {101},
   url = {https://link.aps.org/doi/10.1103/PhysRevB.101.224513},
   year = {2020}
}

@article{Yu2021,
	author = {Tao Yu and Dante M. Kennes and Angel Rubio and Michael A. Sentef},
	doi = {10.1103/PhysRevLett.127.127001},
	issn = {0031-9007},
	issue = {12},
	journal = {Phys. Rev. Lett.},
	month = {9},
	pages = {127001},
	pmid = {34597086},
	publisher = {American Physical Society},
	title = {Nematicity Arising from a Chiral Superconducting Ground State in Magic-Angle Twisted Bilayer Graphene under In-Plane Magnetic Fields},
	volume = {127},
	url = {https://link.aps.org/doi/10.1103/PhysRevLett.127.127001},
	year = {2021}
}

@article{Wehling2011,
	author = {T. O. Wehling and E. Şaşıoğlu and C. Friedrich and A. I. Lichtenstein and M. I. Katsnelson and S. Blügel},
	doi = {10.1103/PhysRevLett.106.236805},
	issn = {0031-9007},
	issue = {23},
	journal = {Phys. Rev. Lett.},
	month = {6},
	pages = {236805},
	title = {Strength of Effective Coulomb Interactions in Graphene and Graphite},
	volume = {106},
	url = {https://link.aps.org/doi/10.1103/PhysRevLett.106.236805},
	year = {2011}
}

@article{Schler2013,
	author = {M. Schüler and M. Rösner and T. O. Wehling and A. I. Lichtenstein and M. I. Katsnelson},
	doi = {10.1103/PhysRevLett.111.036601},
	issn = {0031-9007},
	issue = {3},
	journal = {Phys. Rev. Lett.},
	month = {7},
	pages = {036601},
	title = {Optimal Hubbard Models for Materials with Nonlocal Coulomb Interactions: Graphene, Silicene, and Benzene},
	volume = {111},
	url = {https://link.aps.org/doi/10.1103/PhysRevLett.111.036601},
	year = {2013}
}

@article{Rosner2015,
  title = {Wannier function approach to realistic Coulomb interactions in layered materials and heterostructures},
  author = {R\"osner, M. and \ifmmode \mbox{\c{S}}\else \c{S}\fi{}a\ifmmode \mbox{\c{s}}\else \c{s}\fi{}\ifmmode \imath \else \i \fi{}o\ifmmode \breve{g}\else \u{g}\fi{}lu, E. and Friedrich, C. and Bl\"ugel, S. and Wehling, T. O.},
  journal = {Phys. Rev. B},
  volume = {92},
  issue = {8},
  pages = {085102},
  numpages = {10},
  year = {2015},
  month = {Aug},
  publisher = {American Physical Society},
  doi = {10.1103/PhysRevB.92.085102},
  url = {https://link.aps.org/doi/10.1103/PhysRevB.92.085102}
}

@article{DLee2025,
  title = {Self-consistent tight-binding calculations with extended Hubbard interactions in rhombohedral multilayer graphene},
  author = {Lee, Dongkyu and Yang, Wooil and Son, Young-Woo and Jung, Jeil},
  journal = {Phys. Rev. B},
  volume = {112},
  issue = {23},
  pages = {235160},
  numpages = {12},
  year = {2025},
  month = {Dec},
  publisher = {American Physical Society},
  doi = {10.1103/dn2n-5prq},
  url = {https://link.aps.org/doi/10.1103/dn2n-5prq}
}

@article{Zou2018b,
	author = {Liujun Zou and Hoi Chun Po and Ashvin Vishwanath and T. Senthil},
	doi = {10.1103/PhysRevB.98.085435},
	issn = {2469-9950},
	issue = {8},
	journal = {Phys. Rev. B},
	month = {8},
	pages = {085435},
	publisher = {American Physical Society},
	title = {Band structure of twisted bilayer graphene: Emergent symmetries, commensurate approximants, and Wannier obstructions},
	volume = {98},
	url = {https://link.aps.org/doi/10.1103/PhysRevB.98.085435},
	year = {2018}
}

@article{You2019d,
	author = {Yi-Zhuang You and Ashvin Vishwanath},
	doi = {10.1038/s41535-019-0153-4},
	issn = {2397-4648},
	issue = {1},
	journal = {npj Quantum Mater.},
	month = {4},
	pages = {16},
	publisher = {Springer US},
	title = {Superconductivity from valley fluctuations and approximate SO(4) symmetry in a weak coupling theory of twisted bilayer graphene},
	volume = {4},
	url = {https://www.nature.com/articles/s41535-019-0153-4},
	year = {2019}
}

@article{Lin2018,
	author = {Yu-Ping Lin and Rahul M. Nandkishore},
	doi = {10.1103/PhysRevB.98.214521},
	issn = {2469-9950},
	issue = {21},
	journal = {Phys. Rev. B},
	month = {12},
	pages = {214521},
	publisher = {American Physical Society},
	title = {Kohn-Luttinger superconductivity on two orbital honeycomb lattice},
	volume = {98},
	url = {https://link.aps.org/doi/10.1103/PhysRevB.98.214521},
	year = {2018}
}

@article{Lin2019,
	author = {Yu-Ping Lin and Rahul M. Nandkishore},
	doi = {10.1103/PhysRevB.100.085136},
	issn = {2469-9950},
	issue = {8},
	journal = {Phys. Rev. B},
	month = {8},
	pages = {085136},
	publisher = {American Physical Society},
	title = {Chiral twist on the high-   T c   phase diagram in moiré heterostructures},
	volume = {100},
	url = {https://link.aps.org/doi/10.1103/PhysRevB.100.085136},
	year = {2019}
}

@article{Wu2019c,
	author = {Fengcheng Wu and Euyheon Hwang and Sankar Das Sarma},
	doi = {10.1103/PhysRevB.99.165112},
	issn = {2469-9950},
	issue = {16},
	journal = {Phys. Rev. B},
	month = {4},
	pages = {165112},
	publisher = {American Physical Society},
	title = {Phonon-induced giant linear-in-  T  resistivity in magic angle twisted bilayer graphene: Ordinary strangeness and exotic superconductivity},
	volume = {99},
	url = {https://link.aps.org/doi/10.1103/PhysRevB.99.165112},
	year = {2019}
}

@article{Lian2019,
	author = {Biao Lian and Zhijun Wang and B. Andrei Bernevig},
	doi = {10.1103/PhysRevLett.122.257002},
	issn = {0031-9007},
	issue = {25},
	journal = {Phys. Rev. Lett.},
	month = {6},
	pages = {257002},
	pmid = {31347876},
	publisher = {American Physical Society},
	title = {Twisted Bilayer Graphene: A Phonon-Driven Superconductor},
	volume = {122},
	url = {https://link.aps.org/doi/10.1103/PhysRevLett.122.257002},
	year = {2019}
}

@article{Wu2018,
	author = {Fengcheng Wu and A. H. MacDonald and Ivar Martin},
	doi = {10.1103/PhysRevLett.121.257001},
	issn = {0031-9007},
	issue = {25},
	journal = {Phys. Rev. Lett.},
	month = {12},
	pages = {257001},
	pmid = {30608789},
	publisher = {American Physical Society},
	title = {Theory of Phonon-Mediated Superconductivity in Twisted Bilayer Graphene},
	volume = {121},
	url = {https://doi.org/10.1103/PhysRevLett.121.257001 https://link.aps.org/doi/10.1103/PhysRevLett.121.257001},
	year = {2018}
}

@article{Fernandes2021c,
	author = {Rafael M. Fernandes and Liang Fu},
	doi = {10.1103/PhysRevLett.127.047001},
	issn = {10797114},
	issue = {4},
	journal = {Phys. Rev. Lett.},
	pages = {47001},
	pmid = {34355931},
	publisher = {American Physical Society},
	title = {Charge- 4e Superconductivity from Multicomponent Nematic Pairing: Application to Twisted Bilayer Graphene},
	volume = {127},
	url = {https://doi.org/10.1103/PhysRevLett.127.047001},
	year = {2021}
}

@article{WangY2021,
	author = {Yuxuan Wang and Jian Kang and Rafael M. Fernandes},
	doi = {10.1103/PhysRevB.103.024506},
	issn = {2469-9950},
	issue = {2},
	journal = {Phys. Rev. B},
	month = {1},
	pages = {024506},
	publisher = {American Physical Society},
	title = {Topological and nematic superconductivity mediated by ferro-SU(4) fluctuations in twisted bilayer graphene},
	volume = {103},
	url = {https://link.aps.org/doi/10.1103/PhysRevB.103.024506},
	year = {2021}
}

@article{Sharma2020,
	author = {Gargee Sharma and Maxim Trushin and Oleg P. Sushkov and Giovanni Vignale and Shaffique Adam},
	doi = {10.1103/PhysRevResearch.2.022040},
	issn = {2643-1564},
	issue = {2},
	journal = {Phys. Rev. Res.},
	month = {5},
	pages = {022040},
	publisher = {American Physical Society},
	title = {Superconductivity from collective excitations in magic-angle twisted bilayer graphene},
	volume = {2},
	url = {https://link.aps.org/doi/10.1103/PhysRevResearch.2.022040},
	year = {2020}
}

@article{Isobe2018c,
	author = {Hiroki Isobe and Noah F.Q. Yuan and Liang Fu},
	doi = {10.1103/PhysRevX.8.041041},
	issn = {21603308},
	issue = {4},
	journal = {Phys. Rev. X},
	pages = {41041},
	publisher = {American Physical Society},
	title = {Unconventional Superconductivity and Density Waves in Twisted Bilayer Graphene},
	volume = {8},
	url = {https://doi.org/10.1103/PhysRevX.8.041041},
	year = {2018}
}

@article{Kennes2018,
	author = {Dante M. Kennes and Johannes Lischner and Christoph Karrasch},
	doi = {10.1103/PhysRevB.98.241407},
	issn = {2469-9950},
	issue = {24},
	journal = {Phys. Rev. B},
	month = {12},
	pages = {241407},
	publisher = {American Physical Society},
	title = {Strong correlations and   d + id   superconductivity in twisted bilayer graphene},
	volume = {98},
	url = {https://link.aps.org/doi/10.1103/PhysRevB.98.241407},
	year = {2018}
}

@article{Gonzalez2019,
	author = {J. González and T. Stauber},
	doi = {10.1103/PhysRevLett.122.026801},
	issn = {10797114},
	issue = {2},
	journal = {Phys. Rev. Lett.},
	pages = {26801},
	pmid = {30720323},
	publisher = {American Physical Society},
	title = {Kohn-Luttinger Superconductivity in Twisted Bilayer Graphene},
	volume = {122},
	url = {https://doi.org/10.1103/PhysRevLett.122.026801},
	year = {2019}
}

@article{Cea2021,
	author = {Tommaso Cea and Francisco Guinea},
	doi = {10.1073/pnas.2107874118},
	issn = {0027-8424},
	issue = {32},
	journal = {Proc. Natl. Acad. Sci. U.S.A.},
	month = {8},
	pages = {e2107874118},
	title = {Coulomb interaction, phonons, and superconductivity in twisted bilayer graphene},
	volume = {118},
	url = {https://pnas.org/doi/full/10.1073/pnas.2107874118},
	year = {2021}
}

@article{Shavit2021c,
	author = {Gal Shavit and Erez Berg and Ady Stern and Yuval Oreg},
	doi = {10.1103/PhysRevLett.127.247703},
	issn = {10797114},
	issue = {24},
	journal = {Phys. Rev. Lett.},
	pages = {247703},
	pmid = {34951791},
	publisher = {American Physical Society},
	title = {Theory of Correlated Insulators and Superconductivity in Twisted Bilayer Graphene},
	volume = {127},
	url = {https://doi.org/10.1103/PhysRevLett.127.247703},
	year = {2021}
}

@article{Islam2023,
	author = {SK Firoz Islam and A. Yu. Zyuzin and Alexander A. Zyuzin},
	doi = {10.1103/PhysRevB.107.L060503},
	issn = {2469-9950},
	issue = {6},
	journal = {Phys. Rev. B},
	month = {2},
	pages = {L060503},
	publisher = {American Physical Society},
	title = {Unconventional superconductivity with preformed pairs in twisted bilayer graphene},
	volume = {107},
	url = {https://link.aps.org/doi/10.1103/PhysRevB.107.L060503},
	year = {2023}
}

@article{MacCari2023,
	author = {I. Maccari and J. Carlström and E. Babaev},
	doi = {10.1103/PhysRevB.107.064501},
	issn = {2469-9950},
	issue = {6},
	journal = {Phys. Rev. B},
	month = {2},
	pages = {064501},
	publisher = {American Physical Society},
	title = {Prediction of time-reversal-symmetry breaking fermionic quadrupling condensate in twisted bilayer graphene},
	volume = {107},
	url = {https://link.aps.org/doi/10.1103/PhysRevB.107.064501},
	year = {2023}
}

@article{Braz2024,
	author = {Lauro B. Braz and George B. Martins and Luis G. G. V. Dias da Silva},
	doi = {10.1103/PhysRevB.109.184502},
	issn = {2469-9950},
	issue = {18},
	journal = {Phys. Rev. B},
	month = {5},
	pages = {184502},
	publisher = {American Physical Society},
	title = {Superconductivity from spin fluctuations and long-range interactions in magic-angle twisted bilayer graphene},
	volume = {109},
	url = {https://link.aps.org/doi/10.1103/PhysRevB.109.184502},
	year = {2024}
}

@article{Wagner2024,
	author = {Glenn Wagner and Yves H. Kwan and Nick Bultinck and Steven H. Simon and S. A. Parameswaran},
	doi = {10.1103/PhysRevB.109.104504},
	issn = {2469-9950},
	issue = {10},
	journal = {Phys. Rev. B},
	month = {3},
	pages = {104504},
	publisher = {American Physical Society},
	title = {{Coulomb-driven band unflattening suppresses $K$-phonon pairing in moir\'e graphene}},
	volume = {109},
	url = {https://link.aps.org/doi/10.1103/PhysRevB.109.104504},
	year = {2024}
}

@article{Chou2019,
	author = {Yang-Zhi Chou and Yu-Ping Lin and Sankar Das Sarma and Rahul M. Nandkishore},
	doi = {10.1103/PhysRevB.100.115128},
	issn = {2469-9950},
	issue = {11},
	journal = {Phys. Rev. B},
	month = {9},
	pages = {115128},
	publisher = {American Physical Society},
	title = {Superconductor versus insulator in twisted bilayer graphene},
	volume = {100},
	url = {https://link.aps.org/doi/10.1103/PhysRevB.100.115128},
	year = {2019}
}

@article{Fidrysiak2018,
	author = {M. Fidrysiak and M. Zegrodnik and J. Spałek},
	doi = {10.1103/PhysRevB.98.085436},
	issn = {2469-9950},
	issue = {8},
	journal = {Phys. Rev. B},
	month = {8},
	pages = {085436},
	publisher = {American Physical Society},
	title = {Unconventional topological superconductivity and phase diagram for an effective two-orbital model as applied to twisted bilayer graphene},
	volume = {98},
	url = {https://link.aps.org/doi/10.1103/PhysRevB.98.085436},
	year = {2018}
}

@article{Kozii2019,
	author = {Vladyslav Kozii and Hiroki Isobe and Jörn W.F. Venderbos and Liang Fu},
	doi = {10.1103/PhysRevB.99.144507},
	issn = {24699969},
	issue = {14},
	journal = {Phys. Rev. B},
	month = {4},
	pages = {144507},
	publisher = {American Physical Society},
	title = {Nematic superconductivity stabilized by density wave fluctuations: Possible application to twisted bilayer graphene},
	volume = {99},
	url = {https://link.aps.org/doi/10.1103/PhysRevB.99.144507},
	year = {2019}
}

@article{Liu2018,
	author = {Cheng-Cheng Liu and Li-Da Zhang and Wei-Qiang Chen and Fan Yang},
	issn = {0031-9007},
	issue = {21},
	journal = {Phys. Rev. Lett.},
	month = {11},
	pages = {217001},
	doi = {10.1103/PhysRevLett.121.217001},
	url = {https://link.aps.org/doi/10.1103/PhysRevLett.121.217001},
	pmid = {30517799},
	publisher = {American Physical Society},
	title = {Chiral Spin Density Wave and   d + i d   Superconductivity in the Magic-Angle-Twisted Bilayer Graphene},
	volume = {121},
	year = {2018}
}

@article{Khalaf2021,
	author = {Eslam Khalaf and Shubhayu Chatterjee and Nick Bultinck and Michael P. Zaletel and Ashvin Vishwanath},
	doi = {10.1126/sciadv.abf5299},
	issn = {2375-2548},
	issue = {19},
	journal = {Sci. Adv.},
	month = {5},
	pages = {1-11},
	pmid = {33952523},
	title = {Charged skyrmions and topological origin of superconductivity in magic-angle graphene},
	volume = {7},
	url = {https://www.science.org/doi/10.1126/sciadv.abf5299},
	year = {2021}
}

@article{Khalaf2022,
	author = {Eslam Khalaf and Patrick Ledwith and Ashvin Vishwanath},
	doi = {10.1103/PhysRevB.105.224508},
	issn = {24699969},
	issue = {22},
	journal = {Phys. Rev. B},
	pages = {1-12},
	publisher = {American Physical Society},
	title = {Symmetry constraints on superconductivity in twisted bilayer graphene: Fractional vortices, 4e condensates, or nonunitary pairing},
	volume = {105},
	year = {2022}
}

@article{Bultinck2020,
	author = {Nick Bultinck and Eslam Khalaf and Shang Liu and Shubhayu Chatterjee and Ashvin Vishwanath and Michael P. Zaletel},
	doi = {10.1103/PhysRevX.10.031034},
	issn = {2160-3308},
	issue = {3},
	journal = {Phys. Rev. X},
	month = {8},
	pages = {031034},
	publisher = {American Physical Society},
	title = {Ground State and Hidden Symmetry of Magic-Angle Graphene at Even Integer Filling},
	volume = {10},
	url = {https://link.aps.org/doi/10.1103/PhysRevX.10.031034},
	year = {2020}
}

@article{BlackSchaffer2014b,
	author = {Annica M. Black-Schaffer and Wei Wu and Karyn Le Hur},
	doi = {10.1103/PhysRevB.90.054521},
	issn = {1098-0121},
	issue = {5},
	journal = {Phys. Rev. B},
	month = {8},
	pages = {054521},
	title = {{Chiral $d$-wave superconductivity on the honeycomb lattice close to the Mott state}},
	volume = {90},
	url = {https://link.aps.org/doi/10.1103/PhysRevB.90.054521},
	year = {2014}
}

@article{BlackSchaffer2007,
	author = {Annica M. Black-Schaffer and Sebastian Doniach},
	doi = {10.1103/PhysRevB.75.134512},
	issn = {1098-0121},
	issue = {13},
	journal = {Phys. Rev. B},
	month = {4},
	pages = {134512},
	title = {{Resonating valence bonds and mean-field  $d$-wave superconductivity in graphite}},
	volume = {75},
	url = {https://link.aps.org/doi/10.1103/PhysRevB.75.134512},
	year = {2007}
}

@article{Stepanov2020,
	author = {Petr Stepanov and Ipsita Das and Xiaobo Lu and Ali Fahimniya and Kenji Watanabe and Takashi Taniguchi and Frank H.L. Koppens and Johannes Lischner and Leonid Levitov and Dmitri K. Efetov},
	doi = {10.1038/s41586-020-2459-6},
	issn = {14764687},
	issue = {7816},
	journal = {Nature},
	month = {7},
	pages = {375-378},
	pmid = {32632215},
	publisher = {Nature Research},
	title = {Untying the insulating and superconducting orders in magic-angle graphene},
	volume = {583},
	year = {2020}
}

@article{Baldo2023a,
	author = {Lucas Baldo and Tomas Löthman and Patric Holmvall and Annica M. Black-Schaffer},
	doi = {10.1103/PhysRevB.108.125141},
	issn = {2469-9950},
	issue = {12},
	journal = {Phys. Rev. B},
	month = {9},
	pages = {125141},
	publisher = {American Physical Society},
	title = {Defect-induced band restructuring and length scales in twisted bilayer graphene},
	volume = {108},
	url = {https://link.aps.org/doi/10.1103/PhysRevB.108.125141},
	year = {2023}
}

@misc{Kakoi2025,
	title={Spin-fluctuation-mediated chiral $d+id'$-wave superconductivity in the $\alpha$-$\mathcal{T}_3$ lattice with an incipient flat band}, 
	author={Masataka Kakoi and Kazuhiko Kuroki},
	year={2025},
	eprint={2512.14379},
	archivePrefix={arXiv},
	primaryClass={cond-mat.supr-con},
	url={https://arxiv.org/abs/2512.14379}, 
}

@misc{vanPoppelen2025,
	title={Buckling and flat bands in twisted bilayer graphene}, 
	author={Jannes van Poppelen and Annica M. Black-Schaffer},
	year={2025},
	eprint={2510.13471},
	archivePrefix={arXiv},
	primaryClass={cond-mat.mes-hall},
	url={https://arxiv.org/abs/2510.13471}, 
}

@article{Chamon2000,
  title = {Solitons in carbon nanotubes},
  author = {Chamon, C.},
  journal = {Phys. Rev. B},
  volume = {62},
  pages = {2806},
  year = {2000},
  publisher = {American Physical Society},
  doi = {10.1103/PhysRevB.62.2806},
  url = {https://link.aps.org/doi/10.1103/PhysRevB.62.2806}
}

@article{Hou2007,
  title = {Electron Fractionalization in Two-Dimensional Graphenelike Structures},
  author = {Hou, Chang-Yu and Chamon, Claudio and Mudry, Christopher},
  journal = {Phys. Rev. Lett.},
  volume = {98},
  pages = {186809},
  year = {2007},
  publisher = {American Physical Society},
  doi = {10.1103/PhysRevLett.98.186809},
  url = {https://link.aps.org/doi/10.1103/PhysRevLett.98.186809}
}

@article{Gutierrez2016,
  title = {Imaging chiral symmetry breaking from Kekul\'e bond order in graphene},
  author = {Guti\'errez, Christopher and Kim, Cheol-Joo and Brown, Lola and Schiros, Theanne and Nordlund, Dennis and Lochocki, Edward B. and Shen, Kyle M. and Park, Jiwoong and Pasupathy, Abhay N.},
  journal = {Nat. Phys.},
  volume = {12},
  pages = {950--958},
  year = {2016},
  publisher = {Springer Science and Business Media LLC},
  doi = {10.1038/nphys3776},
  url = {https://www.nature.com/articles/nphys3776}
}

@article{Blason2022,
  title = {Local Kekul\'e distortion turns twisted bilayer graphene into topological Mott insulators and superconductors},
  author = {Blason, Andrea and Fabrizio, Michele},
  journal = {Phys. Rev. B},
  volume = {106},
  issue = {23},
  pages = {235112},
  numpages = {8},
  year = {2022},
  month = {Dec},
  publisher = {American Physical Society},
  doi = {10.1103/PhysRevB.106.235112},
  url = {https://link.aps.org/doi/10.1103/PhysRevB.106.235112}
}

@article{Wang2026,
   title={Kekul\'e superconductivity in twisted magic angle bilayer graphene},
   volume={17},
   ISSN={2041-1723},
   url={http://dx.doi.org/10.1038/s41467-026-74845-9},
   DOI={10.1038/s41467-026-74845-9},
   number={1},
   journal={Nat. Commun.},
   publisher={Springer Science and Business Media LLC},
   author={Wang, Ke and Levin, K.},
   year={2026},}

@misc{Wang2026Stiffness,
      title={Geometric superfluid stiffness of Kekul\'e superconductivity in magic-angle twisted bilayer graphene}, 
      author={Ke Wang and Qijin Chen and Rufus Boyack and K. Levin},
      year={2026},
      eprint={2603.24766},
      archivePrefix={arXiv},
      primaryClass={cond-mat.supr-con},
      url={https://arxiv.org/abs/2603.24766}, 
}

@article{Julku2020,
	author = {A. Julku and T. J. Peltonen and L. Liang and T. T. Heikkilä and P. Törmä},
	doi = {10.1103/PhysRevB.101.060505},
	issn = {2469-9950},
	issue = {6},
	journal = {Phys. Rev. B},
	month = {2},
	pages = {060505},
	publisher = {American Physical Society},
	title = {Superfluid weight and Berezinskii-Kosterlitz-Thouless transition temperature of twisted bilayer graphene},
	volume = {101},
	url = {https://link.aps.org/doi/10.1103/PhysRevB.101.060505},
	year = {2020}
}

@article{Cao2018,
	author = {Yuan Cao and Valla Fatemi and Ahmet Demir and Shiang Fang and Spencer L. Tomarken and Jason Y. Luo and Javier D. Sanchez-Yamagishi and Kenji Watanabe and Takashi Taniguchi and Efthimios Kaxiras and Ray C. Ashoori and Pablo Jarillo-Herrero},
	doi = {10.1038/nature26154},
	issn = {0028-0836},
	issue = {7699},
	journal = {Nature},
	month = {4},
	pages = {80-84},
	pmid = {29512654},
	publisher = {Nature Publishing Group},
	title = {Correlated insulator behaviour at half-filling in magic-angle graphene superlattices},
	volume = {556},
	url = {https://www.nature.com/articles/nature26154},
	year = {2018}
}

@article{Cao2018b,
	author = {Yuan Cao and Valla Fatemi and Shiang Fang and Kenji Watanabe and Takashi Taniguchi and Efthimios Kaxiras and Pablo Jarillo-Herrero},
	doi = {10.1038/nature26160},
	issn = {0028-0836},
	issue = {7699},
	journal = {Nature},
	month = {4},
	pages = {43-50},
	pmid = {29512651},
	publisher = {Nature Publishing Group},
	title = {Unconventional superconductivity in magic-angle graphene superlattices},
	volume = {556},
	url = {http://www.nature.com/articles/nature26160},
	year = {2018}
}

@article{Codecido2019,
	author = {Emilio Codecido and Qiyue Wang and Ryan Koester and Shi Che and Haidong Tian and Rui Lv and Son Tran and Kenji Watanabe and Takashi Taniguchi and Fan Zhang and Marc Bockrath and Chun Ning Lau},
	doi = {10.1126/sciadv.aaw9770},
	issn = {2375-2548},
	issue = {9},
	journal = {Sci. Adv.},
	month = {9},
	pages = {eaaw9770},
	title = {Correlated insulating and superconducting states in twisted bilayer graphene below the magic angle},
	volume = {5},
	url = {https://www.science.org/doi/10.1126/sciadv.aaw9770},
	year = {2019}
}

@article{Yankowitz2019,
	author = {Matthew Yankowitz and Shaowen Chen and Hryhoriy Polshyn and Yuxuan Zhang and K. Watanabe and T. Taniguchi and David Graf and Andrea F. Young and Cory R. Dean},
	doi = {10.1126/science.aav1910},
	issn = {0036-8075},
	issue = {6431},
	journal = {Science},
	month = {3},
	pages = {1059-1064},
	pmid = {30679385},
	title = {Tuning superconductivity in twisted bilayer graphene},
	volume = {363},
	url = {https://www.science.org/doi/10.1126/science.aav1910},
	year = {2019}
}

@article{Lu2019,
	author = {Xiaobo Lu and Petr Stepanov and Wei Yang and Ming Xie and Mohammed Ali Aamir and Ipsita Das and Carles Urgell and Kenji Watanabe and Takashi Taniguchi and Guangyu Zhang and Adrian Bachtold and Allan H. MacDonald and Dmitri K. Efetov},
	title = {Superconductors, orbital magnets and correlated states in magic-angle bilayer graphene},
	doi = {10.1038/s41586-019-1695-0},
	isbn = {4158601916},
	issn = {0028-0836},
	issue = {7780},
	journal = {Nature},
	month = {10},
	pages = {653-657},
	pmid = {31666722},
	publisher = {Springer US},
	volume = {574},
	url = {http://dx.doi.org/10.1038/s41586-019-1695-0 http://www.nature.com/articles/s41586-019-1695-0 https://www.nature.com/articles/s41586-019-1695-0},
	year = {2019}
}

@article{Jiang2019,
	author = {Yuhang Jiang and Xinyuan Lai and Kenji Watanabe and Takashi Taniguchi and Kristjan Haule and Jinhai Mao and Eva Y. Andrei},
	doi = {10.1038/s41586-019-1460-4},
	issn = {14764687},
	issue = {7772},
	journal = {Nature},
	month = {9},
	pages = {91-95},
	pmid = {31365921},
	publisher = {Nature Publishing Group},
	title = {Charge order and broken rotational symmetry in magic-angle twisted bilayer graphene},
	volume = {573},
	year = {2019}
}

@article{Polshyn2019,
	author = {Hryhoriy Polshyn and Matthew Yankowitz and Shaowen Chen and Yuxuan Zhang and K. Watanabe and T. Taniguchi and Cory R. Dean and Andrea F. Young},
	doi = {10.1038/s41567-019-0596-3},
	issn = {1745-2473},
	issue = {10},
	journal = {Nat. Phys.},
	month = {10},
	pages = {1011-1016},
	publisher = {Nature Publishing Group},
	title = {Large linear-in-temperature resistivity in twisted bilayer graphene},
	volume = {15},
	url = {https://www.nature.com/articles/s41567-019-0596-3},
	year = {2019}
}

@article{Kerelsky2019,
	author = {Alexander Kerelsky and Leo J. McGilly and Dante M. Kennes and Lede Xian and Matthew Yankowitz and Shaowen Chen and K. Watanabe and T. Taniguchi and James Hone and Cory Dean and Angel Rubio and Abhay N. Pasupathy},
	doi = {10.1038/s41586-019-1431-9},
	issn = {14764687},
	issue = {7767},
	journal = {Nature},
	month = {8},
	pages = {95-100},
	pmid = {31367030},
	publisher = {Nature Publishing Group},
	title = {Maximized electron interactions at the magic angle in twisted bilayer graphene},
	volume = {572},
	year = {2019}
}

@article{Nuckolls2020,
   title={Strongly correlated Chern insulators in magic-angle twisted bilayer graphene},
   volume={588},
   ISSN={1476-4687},
   url={http://dx.doi.org/10.1038/s41586-020-3028-8},
   DOI={10.1038/s41586-020-3028-8},
   number={7839},
   journal={Nature},
   publisher={Springer Science and Business Media LLC},
   author={Nuckolls, Kevin P. and Oh, Myungchul and Wong, Dillon and Lian, Biao and Watanabe, Kenji and Taniguchi, Takashi and Bernevig, B. Andrei and Yazdani, Ali},
   year={2020},
   month=Dec, pages={610–615} }

@article{Stepanov2021,
	author = {Petr Stepanov and Ming Xie and Takashi Taniguchi and Kenji Watanabe and Xiaobo Lu and Allan H. MacDonald and B. Andrei Bernevig and Dmitri K. Efetov},
	doi = {10.1103/PhysRevLett.127.197701},
	issn = {0031-9007},
	issue = {19},
	journal = {Physical Review Letters},
	month = {11},
	pages = {197701},
	pmid = {34797145},
	publisher = {American Physical Society},
	title = {Competing Zero-Field Chern Insulators in Superconducting Twisted Bilayer Graphene},
	volume = {127},
	url = {https://link.aps.org/doi/10.1103/PhysRevLett.127.197701},
	year = {2021}
}

@article{Cao2020,
	author = {Yuan Cao and Debanjan Chowdhury and Daniel Rodan-Legrain and Oriol Rubies-Bigorda and Kenji Watanabe and Takashi Taniguchi and T. Senthil and Pablo Jarillo-Herrero},
	doi = {10.1103/PhysRevLett.124.076801},
	issn = {0031-9007},
	issue = {7},
	journal = {Phys. Rev. Lett.},
	month = {2},
	pages = {076801},
	pmid = {32142336},
	publisher = {American Physical Society},
	title = {Strange Metal in Magic-Angle Graphene with near Planckian Dissipation},
	volume = {124},
	url = {https://link.aps.org/doi/10.1103/PhysRevLett.124.076801},
	year = {2020}
}

@article{Wong2020,
	author = {Dillon Wong and Kevin P. Nuckolls and Myungchul Oh and Biao Lian and Yonglong Xie and Sangjun Jeon and Kenji Watanabe and Takashi Taniguchi and B. Andrei Bernevig and Ali Yazdani},
	doi = {10.1038/s41586-020-2339-0},
	issn = {14764687},
	issue = {7811},
	journal = {Nature},
	month = {6},
	pages = {198-202},
	pmid = {32528095},
	publisher = {Nature Research},
	title = {Cascade of electronic transitions in magic-angle twisted bilayer graphene},
	volume = {582},
	year = {2020}
}

@article{Saito2020,
	author = {Yu Saito and Jingyuan Ge and Kenji Watanabe and Takashi Taniguchi and Andrea F. Young},
	doi = {10.1038/s41567-020-0928-3},
	issn = {1745-2473},
	issue = {9},
	journal = {Nat. Phys.},
	month = {9},
	pages = {926-930},
	publisher = {Springer US},
	title = {Independent superconductors and correlated insulators in twisted bilayer graphene},
	volume = {16},
	url = {http://dx.doi.org/10.1038/s41567-020-0928-3 https://www.nature.com/articles/s41567-020-0928-3},
	year = {2020}
}

@article{Oh2021,
	author = {Myungchul Oh and Kevin P. Nuckolls and Dillon Wong and Ryan L. Lee and Xiaomeng Liu and Kenji Watanabe and Takashi Taniguchi and Ali Yazdani},
	doi = {10.1038/s41586-021-04121-x},
	issn = {0028-0836},
	issue = {7888},
	journal = {Nature},
	month = {12},
	pages = {240-245},
	pmid = {34670267},
	publisher = {Springer US},
	title = {Evidence for unconventional superconductivity in twisted bilayer graphene},
	volume = {600},
	url = {https://www.nature.com/articles/s41586-021-04121-x},
	year = {2021}
}

@article{Cao2021a,
	author = {Yuan Cao and Daniel Rodan-Legrain and Jeong Min Park and Noah F. Q. Yuan and Kenji Watanabe and Takashi Taniguchi and Rafael M. Fernandes and Liang Fu and Pablo Jarillo-Herrero},
	doi = {10.1126/science.abc2836},
	issn = {0036-8075},
	issue = {6539},
	journal = {Science},
	month = {4},
	pages = {264-271},
	pmid = {33859029},
	title = {Nematicity and competing orders in superconducting magic-angle graphene},
	volume = {372},
	url = {https://www.science.org/doi/10.1126/science.abc2836},
	year = {2021}
}

@article{Liu2021,
	author = {Xiaoxue Liu and Zhi Wang and K. Watanabe and T. Taniguchi and Oskar Vafek and J. I. A. Li},
	doi = {10.1126/science.abb8754},
	issn = {0036-8075},
	issue = {6535},
	journal = {Science},
	month = {3},
	pages = {1261-1265},
	pmid = {33737488},
	title = {Tuning electron correlation in magic-angle twisted bilayer graphene using Coulomb screening},
	volume = {371},
	url = {https://www.science.org/doi/10.1126/science.abb8754},
	year = {2021}
}

@article{RodanLegrain2021,
	title = {Highly tunable junctions and non-local {Josephson} effect in magic-angle graphene tunnelling devices},
	volume = {16},
	issn = {1748-3387, 1748-3395},
	url = {https://www.nature.com/articles/s41565-021-00894-4},
	doi = {10.1038/s41565-021-00894-4},
	number = {7},
	urldate = {2026-09-28},
	journal = {Nat. Nanotechnol.},
	author = {Rodan-Legrain, Daniel and Cao, Yuan and Park, Jeong Min and De La Barrera, Sergio C. and Randeria, Mallika T. and Watanabe, Kenji and Taniguchi, Takashi and Jarillo-Herrero, Pablo},
	month = jul,
	year = {2021},
	pages = {769--775},
}

@article{DiBattista2022,
	author = {Giorgio Di Battista and Paul Seifert and Kenji Watanabe and Takashi Taniguchi and Kin Chung Fong and Alessandro Principi and Dmitri K. Efetov},
	doi = {10.1021/acs.nanolett.1c04512},
	issn = {15306992},
	issue = {16},
	journal = {Nano Lett.},
	month = {8},
	pages = {6465-6470},
	pmid = {35917225},
	publisher = {American Chemical Society},
	title = {Revealing the Thermal Properties of Superconducting Magic-Angle Twisted Bilayer Graphene},
	volume = {22},
	year = {2022}
}

@article{Jaoui2022,
	author = {Alexandre Jaoui and Ipsita Das and Giorgio Di Battista and Jaime Díez-Mérida and Xiaobo Lu and Kenji Watanabe and Takashi Taniguchi and Hiroaki Ishizuka and Leonid Levitov and Dmitri K. Efetov},
	doi = {10.1038/s41567-022-01556-5},
	issn = {17452481},
	issue = {6},
	journal = {Nat. Phys.},
	month = {6},
	pages = {633-638},
	publisher = {Nature Research},
	title = {Quantum critical behaviour in magic-angle twisted bilayer graphene},
	volume = {18},
	year = {2022}
}

@article{Tian2023,
	author = {Haidong Tian and Xueshi Gao and Yuxin Zhang and Shi Che and Tianyi Xu and Patrick Cheung and Kenji Watanabe and Takashi Taniguchi and Mohit Randeria and Fan Zhang and Chun Ning Lau and Marc W. Bockrath},
	doi = {10.1038/s41586-022-05576-2},
	issn = {14764687},
	issue = {7948},
	journal = {Nature},
	pages = {440-444},
	pmid = {36792742},
	publisher = {Springer US},
	title = {Evidence for Dirac flat band superconductivity enabled by quantum geometry},
	volume = {614},
	year = {2023}
}

@article{Nuckolls2023,
	author = {Kevin P. Nuckolls and Ryan L. Lee and Myungchul Oh and Dillon Wong and Tomohiro Soejima and Jung Pyo Hong and Dumitru Călugăru and Jonah Herzog-Arbeitman and B. Andrei Bernevig and Kenji Watanabe and Takashi Taniguchi and Nicolas Regnault and Michael P. Zaletel and Ali Yazdani},
	doi = {10.1038/s41586-023-06226-x},
	issn = {0028-0836},
	issue = {7974},
	journal = {Nature},
	month = {2},
	pages = {525-532},
	pmid = {37587297},
	publisher = {Springer US},
	title = {Quantum textures of the many-body wavefunctions in magic-angle graphene},
	volume = {620},
	url = {https://www.nature.com/articles/s41586-023-06226-x http://arxiv.org/abs/2303.00024 http://dx.doi.org/10.1038/s41586-023-06226-x},
	year = {2023}
}

@article{Chen2024,
	author = {Cheng Chen and Kevin P. Nuckolls and Shuhan Ding and Wangqian Miao and Dillon Wong and Myungchul Oh and Ryan L. Lee and Shanmei He and Cheng Peng and Ding Pei and Yiwei Li and Chenyue Hao and Haoran Yan and Hanbo Xiao and Han Gao and Qiao Li and Shihao Zhang and Jianpeng Liu and Lin He and Kenji Watanabe and Takashi Taniguchi and Chris Jozwiak and Aaron Bostwick and Eli Rotenberg and Chu Li and Xu Han and Ding Pan and Zhongkai Liu and Xi Dai and Chaoxing Liu and B. Andrei Bernevig and Yao Wang and Ali Yazdani and Yulin Chen},
	doi = {10.1038/s41586-024-08227-w},
	issn = {14764687},
	issue = {8042},
	journal = {Nature},
	month = {12},
	pages = {342-347},
	pmid = {39663492},
	publisher = {Nature Research},
	title = {Strong electron–phonon coupling in magic-angle twisted bilayer graphene},
	volume = {636},
	year = {2024}
}

@article{Park2025,
	author = {Jeong Min Park and Shuwen Sun and Kenji Watanabe and Takashi Taniguchi and Pablo Jarillo-Herrero},
	doi = {10.1126/science.adv8376},
	issn = {0036-8075},
	issue = {6780},
	volume = {391},
	journal = {Science},
	month = {11},
	pages = {eadv8376},
	title = {Experimental evidence for nodal superconducting gap in moiré graphene},
	url = {https://www.science.org/doi/10.1126/science.adv8376},
	year = {2025}
}

@article{Kim2022,
	author = {Hyunjin Kim and Youngjoon Choi and Cyprian Lewandowski and Alex Thomson and Yiran Zhang and Robert Polski and Kenji Watanabe and Takashi Taniguchi and Jason Alicea and Stevan Nadj-Perge},
	doi = {10.1038/s41586-022-04715-z},
	issn = {0028-0836},
	issue = {7914},
	journal = {Nature},
	month = {6},
	pages = {494-500},
	pmid = {35705819},
	publisher = {Springer US},
	title = {Evidence for unconventional superconductivity in twisted trilayer graphene},
	volume = {606},
	url = {https://www.nature.com/articles/s41586-022-04715-z},
	year = {2022}
}

@article{Kim2023,
	author = {Hyunjin Kim and Youngjoon Choi and Étienne Lantagne-Hurtubise and Cyprian Lewandowski and Alex Thomson and Lingyuan Kong and Haoxin Zhou and Eli Baum and Yiran Zhang and Ludwig Holleis and Kenji Watanabe and Takashi Taniguchi and Andrea F. Young and Jason Alicea and Stevan Nadj-Perge},
	doi = {10.1038/s41586-023-06663-8},
	issn = {0028-0836},
	issue = {7989},
	journal = {Nature},
	month = {11},
	pages = {942-948},
	pmid = {37968401},
	title = {Imaging inter-valley coherent order in magic-angle twisted trilayer graphene},
	volume = {623},
	url = {https://www.nature.com/articles/s41586-023-06663-8},
	year = {2023}
}

@article{Zhang2026,
	title = {Angular interplay of nematicity, superconductivity and strange metallicity in magic-angle twisted trilayer graphene},
	issn = {1745-2473, 1745-2481},
	url = {https://www.nature.com/articles/s41567-026-03202-w},
	doi = {10.1038/s41567-026-03202-w},
	urldate = {2026-04-14},
	journal = {Nat. Phys.},
	author = {Zhang, Naiyuan J. and Nosov, Pavel A. and Sommer, Ophelia Evelyn and Wang, Yibang and Watanabe, Kenji and Taniguchi, Takashi and Khalaf, Eslam and Li, J.I.A.},
	month = {mar},
	year = {2026},
}

@article{Cao2021PauliLimit,
	title = {Pauli-limit violation and re-entrant superconductivity in moiré graphene},
	volume = {595},
	issn = {0028-0836, 1476-4687},
	url = {https://www.nature.com/articles/s41586-021-03685-y},
	doi = {10.1038/s41586-021-03685-y},
	number = {7868},
	urldate = {2026-08-19},
	journal = {Nature},
	author = {Cao, Yuan and Park, Jeong Min and Watanabe, Kenji and Taniguchi, Takashi and Jarillo-Herrero, Pablo},
	month = jul,
	year = {2021},
	pages = {526--531},
}

@article{Kim2026,
	author = {Hyunjin Kim and Gautam Rai and Lorenzo Crippa and Dumitru Călugăru and Haoyu Hu and Youngjoon Choi and Lingyuan Kong and Eli Baum and Yiran Zhang and Ludwig Holleis and Kenji Watanabe and Takashi Taniguchi and Andrea F. Young and B. Andrei Bernevig and Roser Valentí and Giorgio Sangiovanni and Tim Wehling and Stevan Nadj-Perge},
	doi = {10.1038/s41586-025-10067-1},
	issn = {0028-0836},
	journal = {Nature},
	month = {2},
	title = {{Resolving intervalley gaps and many-body resonances in moir\'e superconductors}},
	url = {https://www.nature.com/articles/s41586-025-10067-1},
	year = {2026},
	pages = {}
}

@article{Tanaka2025,
	author = {Miuko Tanaka and Joel Î.j. Wang and Thao H. Dinh and Daniel Rodan-Legrain and Sameia Zaman and Max Hays and Aziza Almanakly and Bharath Kannan and David K. Kim and Bethany M. Niedzielski and Kyle Serniak and Mollie E. Schwartz and Kenji Watanabe and Takashi Taniguchi and Terry P. Orlando and Simon Gustavsson and Jeffrey A. Grover and Pablo Jarillo-Herrero and William D. Oliver},
	doi = {10.1038/s41586-024-08494-7},
	issn = {14764687},
	issue = {8049},
	journal = {Nature},
	month = {2},
	pages = {99-105},
	pmid = {39910388},
	publisher = {Nature Research},
	title = {Superfluid stiffness of magic-angle twisted bilayer graphene},
	volume = {638},
	year = {2025}
}

@article{Uchoa2007,
	author = {Bruno Uchoa and A. H. Castro Neto},
	doi = {10.1103/PhysRevLett.98.146801},
	issn = {00319007},
	issue = {14},
	journal = {Phys. Rev. Lett.},
	month = {4},
	title = {Superconducting states of pure and doped graphene},
	volume = {98},
	year = {2007}
}

@article{Honerkamp2008,
	author = {Carsten Honerkamp},
	doi = {10.1103/PhysRevLett.100.146404},
	issn = {0031-9007},
	issue = {14},
	journal = {Phys. Rev. Lett.},
	month = {4},
	pages = {146404},
	title = {Density Waves and Cooper Pairing on the Honeycomb Lattice},
	volume = {100},
	url = {https://link.aps.org/doi/10.1103/PhysRevLett.100.146404},
	year = {2008}
}

@article{Nandkishore2012,
	author = {Rahul Nandkishore and L. S. Levitov and A. V. Chubukov},
	doi = {10.1038/nphys2208},
	issn = {1745-2473},
	issue = {2},
	journal = {Nat. Phys.},
	month = {2},
	pages = {158-163},
	publisher = {Nature Publishing Group},
	title = {Chiral superconductivity from repulsive interactions in doped graphene},
	volume = {8},
	url = {https://www.nature.com/articles/nphys2208},
	year = {2012}
}

@article{Pathak2010,
	author = {Sandeep Pathak and Vijay B. Shenoy and G. Baskaran},
	doi = {10.1103/PhysRevB.81.085431},
	issn = {10980121},
	issue = {8},
	journal = {Phys. Rev. B},
	month = {2},
	title = {{Possible high-temperature superconducting state with a $d+id$ pairing symmetry in doped graphene}},
	pages = {085431},
	volume = {81},
	year = {2010}
}

@article{Ma2011,
	author = {Tianxing Ma and Zhongbing Huang and Feiming Hu and Hai Qing Lin},
	doi = {10.1103/PhysRevB.84.121410},
	issn = {10980121},
	issue = {12},
	journal = {Phys. Rev. B},
	month = {9},
	title = {{Pairing in graphene: A quantum Monte Carlo study}},
	volume = {84},
	year = {2011},
	pages = {085431},
}

@article{Kiesel2012,
	author = {Maximilian L. Kiesel and Christian Platt and Werner Hanke and Dmitry A. Abanin and Ronny Thomale},
	doi = {10.1103/PhysRevB.86.020507},
	issn = {10980121},
	issue = {2},
	journal = {Phys. Rev. B},
	month = {7},
	title = {Competing many-body instabilities and unconventional superconductivity in graphene},
	volume = {86},
	year = {2012},
	pages = {020507},
}

@article{Wang2012,
	author = {Wan Sheng Wang and Yuan Yuan Xiang and Qiang Hua Wang and Fa Wang and Fan Yang and Dung Hai Lee},
	doi = {10.1103/PhysRevB.85.035414},
	issn = {10980121},
	issue = {3},
	journal = {Phys. Rev. B},
	month = {1},
	title = {{Functional renormalization group and variational Monte Carlo studies of the electronic instabilities in graphene near $\tfrac{1}{4}$ doping}},
	volume = {85},
	pages = {035414},
	year = {2012}
}

@article{Lewandowski2021Plasmon,
   title={Pairing in magic-angle twisted bilayer graphene: Role of phonon and plasmon umklapp},
   volume={103},
   ISSN={2469-9969},
   url={http://dx.doi.org/10.1103/PhysRevB.103.235401},
   DOI={10.1103/physrevb.103.235401},
   number={23},
   journal={Physical Review B},
   publisher={American Physical Society (APS)},
   author={Lewandowski, Cyprian and Chowdhury, Debanjan and Ruhman, Jonathan},
   year={2021},}

@article{Kallin2016,
	title={Chiral superconductors},
	volume={79},
	ISSN={1361-6633},
	url={http://dx.doi.org/10.1088/0034-4885/79/5/054502},
	DOI={10.1088/0034-4885/79/5/054502},
	number={5},
	journal={Rep. Prog. Phys.},
	publisher={IOP Publishing},
	author={Kallin, Catherine and Berlinsky, John},
	year={2016},
	month=apr, pages={054502} }

@article{Sigrist1991,
	author = {Manfred Sigrist and Kazuo Ueda},
	doi = {10.1103/RevModPhys.63.239},
	issn = {0034-6861},
	issue = {2},
	journal = {Reviews of Modern Physics},
	month = {4},
	pages = {239-311},
	title = {Phenomenological theory of unconventional superconductivity},
	volume = {63},
	url = {https://link.aps.org/doi/10.1103/RevModPhys.63.239},
	year = {1991}
}

@article{Margalit2022,
	author = {Gilad Margalit and Binghai Yan and Marcel Franz and Yuval Oreg},
	doi = {10.1103/PhysRevB.106.205424},
	issn = {2469-9950},
	issue = {20},
	journal = {Phys. Rev. B},
	month = {11},
	pages = {205424},
	publisher = {American Physical Society},
	title = {Chiral Majorana modes via proximity to a twisted cuprate bilayer},
	volume = {106},
	url = {https://link.aps.org/doi/10.1103/PhysRevB.106.205424},
	year = {2022}
}

@article{Li2023,
	author = {Yu-Xuan Li and Cheng-Cheng Liu},
	doi = {10.1103/PhysRevB.107.235125},
	issn = {2469-9950},
	issue = {23},
	journal = {Phys. Rev. B},
	month = {6},
	pages = {235125},
	publisher = {American Physical Society},
 	title = {High-temperature Majorana corner modes in a $d+i{d}^{\ensuremath{'}}$ superconductor heterostructure: Application to twisted bilayer cuprate superconductors},
	volume = {107},
	url = {https://link.aps.org/doi/10.1103/PhysRevB.107.235125},
	year = {2023}
}

@article{Brosco2024,
	author = {Valentina Brosco and Giuseppe Serpico and Valerii Vinokur and Nicola Poccia and Uri Vool},
	doi = {10.1103/PhysRevLett.132.017003},
	issn = {0031-9007},
	issue = {1},
	journal = {Phys. Rev. Letters},
	month = {1},
	pages = {017003},
	pmid = {38242651},
	publisher = {American Physical Society},
	title = {Superconducting Qubit Based on Twisted Cuprate Van der Waals Heterostructures},
	volume = {132},
	url = {https://link.aps.org/doi/10.1103/PhysRevLett.132.017003},
	year = {2024}
}

@article{Confalone2025,
	author = {Tommaso Confalone and Flavia Lo Sardo and Yejin Lee and Sanaz Shokri and Giuseppe Serpico and Alessandro Coppo and Luca Chirolli and Valerii M. Vinokur and Valentina Brosco and Uri Vool and Domenico Montemurro and Francesco Tafuri and Kornelius Nielsch and Golam Haider and Nicola Poccia},
	doi = {10.1002/qute.202500203},
	issn = {2511-9044},
	issue = {11},
	journal = {Adv. Quantum Technol.},
	month = {11},
	publisher = {John Wiley and Sons Inc},
	title = {Cuprate Twistronics for Quantum Hardware},
	volume = {8},
	url = {https://advanced.onlinelibrary.wiley.com/doi/10.1002/qute.202500203},
	year = {2025},
	pages = {2500203}
}

@misc{Baldo2026_Zenodo,
	author = {Baldo, Lucas},
	title  = {{Data and scripts for manuscript ``Unifying description of competing chiral and nematic superconducting states in twisted bilayer graphene''}},
	howpublished = {Zenodo \href{https://doi.org/10.5281/zenodo.18783903}{https://doi.org/10.5281/zenodo.18783903}},
	year   = {2026},
}

@article{Barrier2026,
  title = {Coulomb Screening of Superconductivity in Magic-Angle Graphene},
  author = {Barrier, Julien and Peng, Liangtao and Xu, Shuigang and De Beule, Christophe and Fal'ko, V. I. and Watanabe, K. and Tanigushi, T. and Geim, A. K. and Adam, Shaffique and Berdyugin, Alexey I.},
  journal = {Phys. Rev. X},
  volume = {16},
  issue = {3},
  pages = {031040},
  numpages = {8},
  year = {2026},
  month = {Aug},
  publisher = {American Physical Society},
  doi = {10.1103/z9qg-287y},
  url = {https://link.aps.org/doi/10.1103/z9qg-287y}
}

@article{Gao2026,
	title = {Double-edged role of interactions in superconducting twisted bilayer graphene},
	volume = {22},
	issn = {1745-2473, 1745-2481},
	url = {https://www.nature.com/articles/s41567-026-03243-1},
	doi = {10.1038/s41567-026-03243-1},
	number = {5},
	urldate = {2026-09-08},
	journal = {Nat. Phys.},
	author = {Gao, Xueshi and Jimeno-Pozo, Alejandro and Pantaleon, Pierre A. and Rajesh, Aatmaj and Codecido, Emilio and Sharifi, Daria L. and Zhang, Zheneng and Liu, Youwei and Watanabe, Kenji and Taniguchi, Takashi and Bockrath, Marc W. and Guinea, Francisco and Lau, Chun Ning},
	month = may,
	year = {2026},
	pages = {692--697},
}
\end{document}


\title{Supplemental Material to ``Unifying description of competing chiral and nematic superconducting states in twisted bilayer graphene''}

\author{Lucas Baldo\,\orcidlink{0009-0002-7612-8521}}
\affiliation{Department of Physics and Astronomy, Uppsala University, Box 516, SE-752 37 Uppsala, Sweden}

\author{Patric Holmvall\,\orcidlink{0000-0002-1866-2788}}
\affiliation{Department of Physics and Astronomy, Uppsala University, Box 516, SE-752 37 Uppsala, Sweden}

\author{Annica M. Black-Schaffer\,\orcidlink{0000-0002-4726-5247}}
\affiliation{Department of Physics and Astronomy, Uppsala University, Box 516, SE-752 37 Uppsala, Sweden}

\date{\today}
\maketitle
\tableofcontents

\section*{Notation}
In this work, we mainly consider models expressed in three distinct bases: the atomistic orbital basis, the eigenband basis, and the Chern basis. To clarify the basis in which a matrix is represented, we adopt a system of accents: a hat denotes the atomic basis ($\hat{M}$), no accent denotes the eigenband basis ($M$), and a tilde denotes the Chern basis ($\tilde{M}$). This convention also applies to scalar quantities and Fock operators.

\section{Atomistic modeling}
In this section we provide details on the atomistic model used in the main text.

\subsection{Normal state}

A commensurate twisted structure for TBG is obtained by starting from an $AA$ stacked graphene bilayer and rotating the top and bottom layers respectively by $\pm\theta/2$. The rotation axis is taken to be the out-of-plane axis that passes through sites in both layers, which are denoted as $A$ sites, yielding a so-called type~I structure \cite{Zou2018b}. The commensurate angle is parametrized by positive integers $p,q$ so that \cite{Shallcross2010}
\begin{align} \label{eq:atomistic:normal:angle}
	\theta &= \cos^{-1}\left( \frac{3q^2 - p^2}{3q^2 + p^2} \right).
\end{align}
For the calculations presented in this work, we use $(p,q)=(1,55)$, giving $\theta\approx 1.2^\circ$. For each structure, the crystal is comprised of moir\'e unit cells with origins at positions $\mathbf{R} = m_1 \boldsymbol{a}_1 + m_2 \boldsymbol{a}_2$, where the integers $m_1, m_2$ label the unit cell and $\boldsymbol{a}_{1,2}$ are the moir\'e lattice basis vectors. The unit cell consists of a finite number $N_c$ of carbon sites (or equivalently carbon $p_z$ orbitals) at positions $\mathbf{r}_i$ with respect to the unit cell origin. Then, $\hat{c}_{i \sigma} (\mathbf{R})$ is the annihilation operator for an electron at that site with spin $\sigma$. We define the Fourier transformed operators
\begin{align} \label{eq:atomistic:normal:fourier}
	\hat{c}_{i \boldsymbol{k} \sigma}
	&= \frac{1}{\sqrt{N_k}} \sum_{\mathbf{R}} e^{i \boldsymbol{k} \cdot \mathbf{R} } \, \hat{c}_{i \sigma} (\mathbf{R}),
\end{align}
where $N_k$ is the number of moir\'e unit cells in the crystal. The normal-state Hamiltonian is a sum of terms
\begin{align}
	\hat{H}_{0} (\mathbf{k})
	&= \sum_{\sigma} \sum_{i,j} [\hat{h}_0 (\mathbf{k})]_{i j} \, \hat{c}^{\dagger}_{i \boldsymbol{k} \sigma} \hat{c}_{j \boldsymbol{k} \sigma}.
\end{align}
The matrix elements $[\hat{h}_0 (\boldsymbol{k})]_{i j}$ are the momentum-space Slater--Koster overlap integrals between $p_z$ orbitals,
\begin{align} \label{eq:atomistic:normal:overlap}
	[\hat{h}_0 (\boldsymbol{k})]_{i j}
	&= -\mu \delta_{i j} - \sum_{\delta \mathbf{R}} e^{i \boldsymbol{k} \cdot \delta \mathbf{R}} \left[ t_{\pi} e^{(a_c - r_{i j})/\lambda} (1 - (\boldsymbol{\hat{r}}_{i j} \cdot \boldsymbol{\hat{z}})^2 ) + t_{\sigma} e^{(d_0 - r_{i j})/\lambda} (\boldsymbol{\hat{r}}_{i j} \cdot \boldsymbol{\hat{z}})^2 \right],
\end{align}
where the distance between two carbon sites is given by $\mathbf{r}_{i j} (\delta \mathbf{R}) = \mathbf{r}_{i} - \mathbf{r}_{j} + \delta\mathbf{R}$ with $\delta \mathbf{R}$ an element of the moir\'e lattice, and further $r = \abs{\mathbf{r}}$ and $\hat{\mathbf{r}} = \mathbf{r}/r$ are the norm and unit vector, respectively. The complex exponential is a Bloch wrapping factor. The chemical potential $\mu$ is taken as a tuning parameter for the filling. The parameters $t_{\pi} = 2.7$~eV, $t_{\sigma} = -0.48$~eV, $a_c=1.42$~\AA{}, $d_0=3.35$~\AA{}, and $\lambda = 0.45$~\AA{} are the in-plane and out-of-plane hopping strengths, intra-layer carbon to carbon separation, inter-layer distance, and decay length, respectively, all estimated from \emph{ab initio} calculations \cite{TramblydeLaissardiere2010}. For computational purposes, we introduce a cutoff for the sum over sites. For inter-layer terms, we keep all terms with $r < 6 a$, where $a = \sqrt{3} a_c$ is the monolayer lattice constant, verifying convergence of the band structure. For intra-layer terms we find it sufficient to keep only the nearest-neighbor terms.

For a given chemical potential $\mu$, we calculate the normal-state filling factor $\nu$ at zero temperature by counting the occupied flat-band states per moir\'e unit cell relative to charge neutrality. This is expressed in terms of the normal-state flat-band density of states (DOS) as
\begin{align} \label{eq:atomistic:filling}
    \nu = -4 + 2 \int_{\textrm{occ}} d\epsilon \, \textrm{DOS}(\epsilon).
\end{align}
The factor of two above comes from spin degeneracy and the offset means we measure $\nu$ relative to charge neutrality. DOS and $\nu$ are calculated using a $24\times24$ grid or finer to ensure convergence. Throughout our work, $\nu$ refers to the normal-state filling, also when the superconducting state at the same $\mu$ is considered.

\subsection{Symmetries of the normal state} \label{SM:sec:model:symmetries}
 
With the construction of the normal state as described above, the lattice exhibits an exact $D_3$ point-group symmetry, generated by $C_{3z}$ operations (threefold rotations about the same out-of-plane axis as the twist) and $C_{2y}$ operations (twofold rotations about axes parallel to one of the moir\'e lattice vectors). Furthermore, for any two carbon sites $i,j$, the overlap matrix in Eq.~\eqref{eq:atomistic:normal:overlap} is real, so the normal-state Hamiltonian possesses an exact spinless time-reversal symmetry
\begin{align} \label{eq:atomistic:symmetries:TRS}
	{\mathcal{T}}^{-1} \hat{h}_0(\mathbf{k}) {\mathcal{T}} = \hat{h}_0^*(-\mathbf{k}) = \hat{h}_0(\mathbf{k}).
\end{align}

Besides these exact time-reversal and lattice symmetries, the system we consider has an additional set of emergent, or approximate, symmetries, which we now review. We start by noting that the low-energy description of each layer is dominated by two Dirac cones, corresponding to the two valleys $\eta = \pm$. Interlayer coupling hybridizes the cones from opposite layers, but for small twist angles inter-valley scattering is negligible, such that $\eta$ remains a good quantum number, associated with an approximate $U(1)$ charge-conservation symmetry in each valley. In the atomistic framework, the valley $\eta$ of an energy eigenstate can be computed as the sign of the expectation value of the valley operator \cite{Ramires2019},
\begin{align} \label{eq:atomistic:symmetries:valley}
	\hat{\mathcal{V}} &= \hat{\mathcal{V}}_1 - \hat{\mathcal{V}}_2,
	\quad \hat{\mathcal{V}}_\ell = \frac{i}{3 \sqrt{3}} \sum_{\sigma} \sum_{\langle \langle i,j \rangle \rangle \in \ell} \eta_{ij} \, s_{i} \, \hat{c}^{\dagger}_{i \sigma} \hat{c}_{j \sigma},
\end{align}
where $\langle \langle i,j \rangle \rangle \in \ell$ denotes next-nearest-neighbor pairs in layer $\ell$, $\eta_{ij} = \pm$ indicates whether the path between these sites winds clockwise or anticlockwise, and $s_{i}$ is a sublattice-dependent sign factor. This operator is essentially a sublattice-dependent Haldane term \cite{Colomes2018, Ramires2018}, which distinguishes Bloch states of different chirality. Because it is purely imaginary, the valley operator in Eq.~\eqref{eq:atomistic:symmetries:valley} anticommutes with time-reversal, ${\mathcal{T}}^{-1} \hat{\mathcal{V}} {\mathcal{T}} = -\hat{\mathcal{V}}$, meaning that $\mathcal{T}$ flips the value of $\eta$. This fact becomes important when we project the pairing matrix onto the flat bands in Sec.~\ref{sec:pairing-basis:projection}.

If we instead consider type~II structures of TBG, where the twist axis passes through a carbon ring (hexagon) center rather than a carbon site \cite{Zou2018b}, there is an additional invariance under a lattice transformation $C_{2z}$, consisting of a twofold rotation centered on the central carbon ring of the $AA$ region. This symmetry inverts the momentum, valley, and sublattice quantum numbers. Although it is not a lattice symmetry of the type~I structures we consider, both structures are known to share the same low-energy description at small angles \cite{Zou2018b}, so $C_{2z}$ remains a spectral symmetry to very good approximation also for type~I structures. The operation $C_{2z} \mathcal{T}$ is therefore also an approximate symmetry, and is in fact responsible for protecting the topology of the flat bands \cite{Ahn2019, Song2019b, TBG_II}.

Finally, we consider two symmetries of idealized models of TBG. Neglecting intra-sublattice hopping terms renders the system bipartite, giving rise to a chiral symmetry (not to be confused with the chiral superconducting ground state). In this so-called \emph{chiral limit}, the chiral symmetry operator $\mathcal{C}$, corresponding to a $\pi$ phase shift on one sublattice, anticommutes with the normal-state Hamiltonian \cite{Tarnopolsky2019, Wang2021}. In the eigenband basis, this flips the eigenband number $b$, while preserving all other quantum numbers, and in the Chern basis its presence results in perfectly sublattice-polarized wavefunctions \cite{Tarnopolsky2019, Wang2021}. In realistic models of TBG, chiral symmetry is substantially broken \cite{TBG_II}, but its remnants persist as a finite sublattice polarization of the Chern bands \cite{Bultinck2020}. In the atomistic model we use, interlayer hopping naturally includes intra-sublattice terms that break the chiral symmetry, yet the sublattice polarization remains, becoming apparent through the intra-Chern pairing discussed in the main text.

The second such symmetry is (normal-state) particle-hole symmetry $\mathcal{P}$, which flips both the eigenband and momentum quantum numbers, but preserves the valley number \cite{Song2019b}. This symmetry becomes exact at small twist angles under the assumption of local interlayer tunneling \cite{TBG_II}. Particle-hole symmetry breaking is visible in the spectrum of the atomistic model we use as a consequence of including also nonlocal interlayer tunneling, but it remains weak in the eigenstates, owing to the rigid lattice and the isotropic Slater--Koster hopping. Including lattice relaxation and anisotropy in the Slater--Koster hopping would lead to stronger particle-hole breaking \cite{Fang2019arXiv, Carr2019, Kang2023, HerzogArbeitman2025b}. The combined transformation $\mathcal{C} \mathcal{P}$, which relates states of the same eigenband and valley at opposite momenta, is likewise generally broken. We discuss the importance of these points to the projected pairing matrix symmetries in Sec.~\ref{SM:sec:multiband:symmetries}.

\subsection{Nearest-neighbor pairing}

In order to study $d$-wave superconductivity in TBG from an atomistic perspective we consider the simplest term beyond on-site, namely nearest-neighbor spin-singlet pairing. Such pairing has been shown to arise via spin fluctuations induced by strong on-site Coulomb repulsion away from integer fillings \cite{Fischer2021}. The related interactions are well-captured by an effective $t$-$J$ model without projecting out doubly occupied sites, with antiferromagnetic interaction $J$ between nearest neighbors within each layer, where we assume the interaction strength $J$ to be constant throughout the unit cell. Implementing the nearest-neighbor constraint through $\delta_{\langle \mathbf{r}, \mathbf{r}^\prime \rangle}$, the interacting term of the Hamiltonian is
\begin{align} \label{eq:atomistic:pairing:HJ}
	\hat{H}_J
	&= J \sum_{i,j} \sum_{\mathbf{R} \mathbf{R}^\prime} \delta_{\langle \mathbf{r}_{i} + \mathbf{R}, \mathbf{r}_{j} + \mathbf{R}^\prime \rangle} \left( \hat{\mathbf{S}}_{i \mathbf{R}} \cdot \hat{\mathbf{S}}_{j \mathbf{R}^\prime} - \tfrac{1}{4} \hat{n}_{i \mathbf{R}} \hat{n}_{j \mathbf{R}^\prime} \right).
\end{align}
The spin operators are $\hat{\mathbf{S}}_{i \mathbf{R}}^{\mu} = \tfrac{1}{2} \sum_{\sigma_1, \sigma_2} \hat{c}^\dagger_{i \sigma_1} (\mathbf{R}) [\sigma_\mu]_{\sigma_1 \sigma_2} \hat{c}_{i \sigma_2} (\mathbf{R})$, where $\mu = x,y,z$ and $\sigma_\mu$ are the Pauli matrices in spin space. Similarly, the number operator is $\hat{n}_{i \mathbf{R}} = \sum_{\sigma_1, \sigma_2} \hat{c}^\dagger_{i \sigma_1} (\mathbf{R}) [\sigma_0]_{\sigma_1 \sigma_2} \hat{c}_{i \sigma_2} (\mathbf{R})$, where $\sigma_0$ is the identity matrix. The ordered sum in Eq.~\eqref{eq:atomistic:pairing:HJ} counts each bond twice, so with $\hat{\mathbf{S}}_i\cdot\hat{\mathbf{S}}_j-\tfrac14\hat{n}_{i} \hat{n}_{j}=-\tfrac{1}{2} \hat{s}^{\dagger}_{ij} \hat{s}_{ij}$ it equals $-J \sum_{\langle i,j\rangle} \hat{s}^{\dagger}_{ij} \hat{s}_{ij}$. We then Fourier transform the operators and perform a mean-field decoupling in the Cooper pairing channel, assuming the ground state preserves moir\'e translational symmetry, $\langle \hat{c}^\dagger_{i \mathbf{k} \sigma} \hat{c}^\dagger_{j \mathbf{k}^\prime \bar{\sigma}} \rangle \propto \delta_{\mathbf{k}, -\mathbf{k}^\prime}$. Defining the spin-singlet pair operator $\hat{s}^\dagger_{ij}(\mathbf{k}) = \hat{c}^\dagger_{i\mathbf{k}\uparrow} \hat{c}^\dagger_{j -\mathbf{k}\downarrow} - \hat{c}^\dagger_{i\mathbf{k}\downarrow} \hat{c}^\dagger_{j -\mathbf{k}\uparrow}$, the mean-field Hamiltonian becomes
\begin{align} \label{eq:transition:Hamiltonian}
	\hat{H}
	&= \sum_{\mathbf{k}} \left[ \hat{H}_0(\mathbf{k}) + \sum_{\langle i,j \rangle} \left( \hat{\Delta}_{ij}(\mathbf{k}) \hat{s}_{ij}^\dagger(\mathbf{k}) + \mathrm{H.c.} + \frac{|\hat{\Delta}_{ij}|^2}{J}  \right) \right],
\end{align}
where the order parameter is
\begin{align} \label{eq:atomistic:pairing:Delta}
	\hat{\Delta}_{i j}^{} (\mathbf{k}) 
	&= -\frac{J}{N_k} \sum_{\mathbf{k}^\prime} \varphi_{ij}(\mathbf{k}^\prime - \mathbf{k}) \, \langle \hat{s}_{ij}(\mathbf{k}^\prime) \rangle,
	\quad 
	\varphi_{ij}(\mathbf{k}^\prime - \mathbf{k}) 
	= \sum_{\delta \mathbf{R}} e^{i [\mathbf{k}^\prime - \mathbf{k}] \cdot \delta \mathbf{R}} \, \delta_{\langle \mathbf{r}_i, \mathbf{r}_{j} + \delta \mathbf{R} \rangle}.
\end{align}
It satisfies $\hat{\Delta}_{j i}^{} (\mathbf{k}) = \hat{\Delta}_{i j}^{} (-\mathbf{k})$, as required by spin-singlet pairing. At the angle $\theta\approx 1.2^\circ$ we consider, there are $13{,}614$ distinct intra-layer bonds within the moir\'e unit cell, each corresponding to an independent pairing matrix value $\hat{\Delta}_{i j}$ which needs to be solved for self-consistently.

We rewrite Eq.~\eqref{eq:transition:Hamiltonian} in the Bogoliubov--de Gennes (BdG) formalism by defining the Nambu spinor $\mathbf{\hat{c}}_{\mathbf{k} \uparrow} = ( \{\hat{c}_{i \mathbf{k} \uparrow}\}, \{\hat{c}^\dagger_{i -\mathbf{k} \downarrow}\})^T$. The resulting BdG Hamiltonian in this sector is
\begin{align} \label{eq:atomistic:pairing:BdG}
	\hat{H}_{\text{BdG}} (\mathbf{k})
	&= \begin{pmatrix} \hat{h}_0(\mathbf{k}) & \hat{\Delta}^{}(\mathbf{k}) \\ \hat{\Delta}^{\dagger}(\mathbf{k}) & - \hat{h}_0^*(-\mathbf{k}) \end{pmatrix}.
\end{align} 
We numerically diagonalize the BdG Hamiltonian in Eq.~\eqref{eq:atomistic:pairing:BdG}, resulting in the diagonal matrix $\hat{D}(\mathbf{k}) = {U}_{\mathbf{k}} \hat{H}_{\text{BdG}} (\mathbf{k}) {U}^{\dagger}_{\mathbf{k}} = \textrm{diag}(\hat{E}_\mathbf{k}, -\hat{E}_{-\mathbf{k}})$ containing the energies $\hat{E}_\mathbf{k} \geq 0$. The unitary matrix has the form
\begin{align} \label{eq:atomistic:pairing:unitary}
	{U}_{\mathbf{k}}
	&= 
	\begin{pmatrix} {u}_{\mathbf{k} \uparrow} & {v}_{\mathbf{k} \uparrow} \\ {v}^*_{-\mathbf{k} \downarrow} & {u}^*_{-\mathbf{k} \downarrow} \end{pmatrix}.
\end{align}
To solve self-consistently for superconductivity we calculate the expectation values as
\begin{align} \label{eq:atomistic:pairing:expectation}
	\langle \hat{s}_{ij}(\mathbf{k}) \rangle 
	= 2 \, \langle \hat{c}_{j -\mathbf{k} \downarrow} \hat{c}_{i \mathbf{k} \uparrow} \rangle 
	&= 2 \, [ v^T_{-\mathbf{k} \downarrow} (1 - F^T_{-\mathbf{k}}) u^*_{-\mathbf{k} \downarrow} + u^\dagger_{\mathbf{k} \uparrow} F_{\mathbf{k}} v_{\mathbf{k} \uparrow} ]_{i j},
\end{align}
where $F_{\mathbf{k}} = f(\hat{E}_\mathbf{k})$, with $f(\epsilon) = (e^{\beta \epsilon} + 1)^{-1}$ the Fermi-Dirac distribution and $\beta$ the inverse temperature.

The linearized form of Eq.~\eqref{eq:atomistic:pairing:Delta} yields a 2D manifold of solutions with the highest $T_c$, corresponding to an $E$ irreducible representation (irrep) of the crystal symmetry group $D_{3}$ \cite{Lothman2022}. Below $T_c$, Eqs.~\eqref{eq:atomistic:pairing:Delta}-\eqref{eq:atomistic:pairing:expectation} define a non-linear self-consistency loop. We focus on zero temperature, where $F_{\mathbf{k}} = 0$. To study specific pairing configurations we can modify this self-consistency loop by projecting $\hat{\Delta}$ at each step into one of the solutions in the $E$ irrep, while keeping the overall norm $\Delta_0$ as a free parameter. This yields the optimal $\Delta_\star$ minimizing the free energy (thereby maximizing the condensation energy) for the chosen configuration. This method allows us to make a direct comparison between superconducting solutions with different symmetries. In all our atomistic self-consistent calculations we sample the moir\'e Brillouin zone in a $4\times4$ grid, which has been shown to well capture the nematic ground state. 

For a given pairing configuration and chemical potential $\mu$, we calculate the electronic density of states (DOS) at zero temperature as the DOS obtained from the BdG spectrum in a finer $24\times24$ momentum space grid, and weighted by the electronic part of the wavefunctions. This is how we obtain the DOS shown in Fig.~5(a,b) of the main text.

\subsection{Interaction energy scales}
\label{sec:atomistic:scales}

In our atomistic model, we treat the interaction strength $J$ in Eq.~\eqref{eq:atomistic:pairing:HJ} as a de facto tuning parameter, which we set to $J=1.35$~eV in order to approximately reproduce the experimentally observed pairing gap and critical temperature. We here further justify this choice, and clarify a point of possible confusion regarding its magnitude relative to other energy scales discussed in the literature. 

Ab initio estimates of the underlying on-site Hubbard repulsion $U$ in graphene systems are known to vary substantially with system configuration, in particular whether the system is singly or doubly screened and the associated screening length, as well as more general modeling choices \cite{Wehling2011, Schler2013, Rosner2015, DLee2025, Saito2020, Stepanov2020, Liu2021}. Early constrained RPA calculations reported $U\approx9.3$~eV for monolayer graphene \cite{Wehling2011}, while double metallic encapsulation was subsequently shown to reduce this to $U\approx7.3$~eV in Bernal bilayer graphene \cite{Rosner2015}. More recent self-consistent DFT+U+V calculations found $U\approx6.2$~eV for $n$-layer rhombohedral multilayer graphene, which encompasses both the monolayer and Bernal bilayer cases \cite{DLee2025}. A commonly employed estimate relates the on-site repulsion to the nearest-neighbor exchange coupling via $2J \approx 4t^2/U$ \cite{MacDonald1988, Scalapino1995, Lee2006}, where the factor of two arises because Eq.~\eqref{eq:atomistic:pairing:HJ} counts each nearest-neighbor bond twice. The range of $U$ values above then gives $J\approx1.6$--$2.4$~eV, somewhat above our value $J=1.35$~eV. An RPA analysis that derives the attraction directly from the bare Hubbard repulsion yields a nearest-neighbor attraction that depends on $U$ and is of order $1$~eV \cite{Fischer2021}. Our value of $J$ is thus compatible with literature estimates.

 
Despite the solid physical arguments presented above, a value of $J\sim1$~eV may appear anomalously large when compared against the energy scales commonly invoked in discussions of correlated and superconducting states within the flat bands of twisted bilayer graphene (TBG), which are instead typically framed on the order of meV, rather than eV. This apparent mismatch has in fact already led to unfortunate comparisons in the literature. For instance, Ref.~\cite{Wang2026} found that intra-valley pairing in TBG is stabilized already by attractive interactions in the sub-meV range, and contrasted this with the much larger, eV-scale couplings required to stabilize nearest-neighbor pairing in earlier works \cite{Julku2020, Fischer2021, Lothman2022}. We show below that this discrepancy is actually not physical, but is instead an artifact of comparing interaction strengths defined in a fully local atomic basis as in our work, to those defined in a delocalized plane-wave basis, which is the natural starting point for continuum models of TBG. In fact, these two bases carry different natural energy scales, so making this distinction explicit is essential for correctly interpreting and comparing interaction strengths across different models.

To proceed, we use a slightly different notation to describe the same TBG geometry as before. Starting from two $AA$ stacked graphene layers, we define a supercell containing $N_{\textrm{g}} = N^{\textrm{g}}_{1} \times N^{\textrm{g}}_{2}$ graphene unit cells. We then introduce the moir\'e lattice positions $\mathbf{R} = m^{\textrm{M}}_{1} \mathbf{a}^{\textrm{M}}_{1} + m^{\textrm{M}}_{2} \mathbf{a}^{\textrm{M}}_{2}$ and graphene lattice positions $\mathbf{r} = m^{\textrm{g}}_{1} \mathbf{a}^{\textrm{g}}_{1} + m^{\textrm{g}}_{2} \mathbf{a}^{\textrm{g}}_{2}$, where $\mathbf{a}^{M,g}_{1,2}$ are the moir\'e and (non-rotated) graphene lattice vectors, respectively. The integers $m_{i}^{M,g} = 0,...,N_i^{\textrm{M,g}}-1$ count the moir\'e unit cells in the TBG lattice and the graphene unit cells within the moir\'e unit cell, respectively, in each crystalline direction. Each layer $l$ is then rotated according to the rotation matrix $R^{(l)}$. The physical site positions on the TBG lattice are then $R^{(l)} \, (\mathbf{R} + \mathbf{r} + \delta_{s,-} \, \boldsymbol{\delta} ) + \delta_{l,1} \, d_0 \, \hat{\mathbf{z}}$, where $\delta_{s,-}$ selects the $B$ sites, $\boldsymbol{\delta}$ is a vector connecting nearest-neighbor atoms in the unrotated layer, $d_{0}$ is the inter-layer spacing, and $\hat{\mathbf{z}}$ is the unit vector in the out-of-plane direction. Further, we write the annihilation operator at such a lattice site as $\hat{c}_{l, s, \sigma} (\mathbf{R}, \mathbf{r})$, where $s=\pm$ denotes the $A,B$ sublattice and $\sigma$ is the spin. We use the spin and number operators of Eq.~\eqref{eq:atomistic:pairing:HJ}, written in this notation as
\begin{align}
    \hat{S}^{\mu}_{l, s}(\mathbf{R}, \mathbf{r}) &= \tfrac{1}{2} \sum_{\sigma_1, \sigma_2} \hat{c}_{l, s, \sigma_1}^{\dagger} (\mathbf{R}, \mathbf{r}) \, [\sigma_\mu]_{\sigma_1 \sigma_2} \, \hat{c}_{l, s, \sigma_2} (\mathbf{R}, \mathbf{r}),
\end{align}
with $\mu = 0, x, y, z$ and $\sigma_0$ the identity, so that $\hat{S}^{x,y,z}_{l,s}$ are the spin operators and $\hat{S}^{0}_{l,s} = \tfrac{1}{2}\hat{n}_{l,s}$. We then write the most general interaction Hamiltonian with separable spin and spatial degrees of freedom,
\begin{align} \label{eq:atomistic:scales:interactions}
\hat{H}_{\mathrm{int}}
&=
\sum_{l_{1} l_{2}} \sum_{s_{1} s_{2}} \sum_{\mathbf{R}_{1} \mathbf{R}_{2}} \sum_{\mathbf{r}_{1} \mathbf{r}_{2}}
V_{s_{1} s_{2}}^{l_{1} l_{2}}
(\mathbf{R}_{1}, \mathbf{r}_{1}; \mathbf{R}_{2}, \mathbf{r}_{2}) \sum_{\mu} S_{\mu} \, \hat{S}^{\mu}_{l_{1}, s_{1}}(\mathbf{R}_{1}, \mathbf{r}_{1}) \, \hat{S}^{\mu}_{l_{2}, s_{2}}(\mathbf{R}_{2}, \mathbf{r}_{2}).
\end{align}
The coefficients $S_{\mu}$ set the spin structure of the interactions. For example, the $S_{0}$ term describes pure charge-charge interaction terms, such as the ones obtained from phonons \cite{Wu2018, Lian2019, Cea2021, Liu2024} or other pairing mechanisms \cite{Gonzalez2019, Lewandowski2021Plasmon, Blason2022, Gonzalez2023, Peng2024}. The $\mu\neq0$ terms instead describe spin-spin interaction terms, such that the Heisenberg exchange interaction of Eq.~\eqref{eq:atomistic:pairing:HJ} corresponds to $S_{0} = -1$ and $S_{x} = S_{y} = S_{z} = 1$. 

Regardless of the spin structure $S_{\mu}$, the Hamiltonian Eq.~\eqref{eq:atomistic:scales:interactions} describes interactions between localized states, and its natural energy scale is thus on the order of eV, since the kinetic Hamiltonian in such a basis is dominated by the nearest-neighbor hopping $t_{\pi} \approx 2.7$~eV. We next show that by changing to a delocalized, plane-wave, basis over the moir\'{e} scale, the interaction form factor $V$ is automatically rescaled by the number $N_{\textrm{g}}$ of graphene unit cells within the moir\'e unit cell. To obtain such a basis we expand the electron operators in terms of Fourier components in both the moir\'e and graphene coordinates,
\begin{align} \label{eq:atomistic:scales:fourier-both-scales}
\hat{c}_{ l s \sigma}(\mathbf{R}, \mathbf{r})
&=
\frac{1}{\sqrt{N_k}} \frac{1}{\sqrt{N_{\textrm{g}}}}
\sum_{\mathbf{k}} \sum_{\mathbf{p}}
e^{i \mathbf{k} \cdot \mathbf{R}} \, e^{i \mathbf{p} \cdot \mathbf{r}} \,
\hat{c}_{l s \sigma}(\mathbf{k},\mathbf{p}),
\end{align}
where $N_k$ is again the number of moir\'e unit cells. For simplicity, we assume the interaction to be translation invariant on both the moir\'e and graphene scales, $V(\mathbf{R}_{1}, \mathbf{r}_{1}; \mathbf{R}_{2}, \mathbf{r}_{2}) = V(\mathbf{R}_{1}-\mathbf{R}_{2}, \mathbf{r}_{1}-\mathbf{r}_{2}; 0, 0)$, although our results can be generalized to include modulation of the interactions in either scale. We then define the Fourier-space interaction tensors
\begin{align} \label{eq:atomistic:scales:fourier-form-factors}
V_{s_{1} s_{2}}^{l_{1} l_{2}} (\mathbf{Q},\mathbf{q})
&=
\sum_{\delta \mathbf{R}} \sum_{\delta \mathbf{r}}
e^{-2i \mathbf{Q} \cdot \delta \mathbf{R} } \, e^{-2i \mathbf{q} \cdot \delta \mathbf{r} } \,
V_{s_{1} s_{2}}^{l_{1} l_{2}} (\delta \mathbf{R}, \delta \mathbf{r}; 0, 0),
\end{align}
and the same operators in momentum space
\begin{align} \label{eq:atomistic:scales:fourier-density}
    \hat{S}^{\mu}_{l, s}(\mathbf{k}, \mathbf{p}; \mathbf{Q}, \mathbf{q}) &= \tfrac{1}{2} \sum_{\sigma_1, \sigma_2} \hat{c}_{l, s, \sigma_1, \mathbf{k} + \mathbf{Q}, \mathbf{p} + \mathbf{q}}^{\dagger} \, [\sigma_\mu]_{\sigma_1 \sigma_2} \, \hat{c}_{l, s, \sigma_2, \mathbf{k} - \mathbf{Q}, \mathbf{p} - \mathbf{q}}.
\end{align}
Combining Eqs.~\eqref{eq:atomistic:scales:fourier-both-scales}, \eqref{eq:atomistic:scales:fourier-form-factors}, and \eqref{eq:atomistic:scales:fourier-density} into \eqref{eq:atomistic:scales:interactions}, we find 
\begin{align} \label{eq:atomistic:scales:interactions-momentum}
\hat{H}_{\mathrm{int}}
= 
\frac{1}{N_k} \frac{1}{N_{\textrm{g}}}
\sum_{l_{1} l_{2}} \sum_{s_{1} s_{2}}
\sum_{\mathbf{k} \mathbf{k}^{\prime} \mathbf{Q}} 
\sum_{\mathbf{p} \mathbf{p}^{\prime} \mathbf{q}} \,
V_{s_{1} s_{2}}^{l_{1} l_{2}} (\mathbf{Q},\mathbf{q})
\sum_{\mu} S_{\mu} \, 
\hat{S}^{\mu}_{l_{1}, s_{1}}(\mathbf{k}, \mathbf{p}; \mathbf{Q}, \mathbf{q}) \, 
\hat{S}^{\mu}_{l_{2}, s_{2}}(\mathbf{k}^{\prime}, \mathbf{p}^{\prime}; -\mathbf{Q}, -\mathbf{q}).
\end{align}

Notably, the prefactor $1/(N_k N_{\textrm{g}})$ in Eq.~\eqref{eq:atomistic:scales:interactions-momentum} is accompanied by independent sums over the moir\'e momenta $\mathbf{k}, \mathbf{k}^{\prime}, \mathbf{Q}$ and the graphene momenta $\mathbf{p}, \mathbf{p}^{\prime}, \mathbf{q}$, each running over $N_k$ and $N_{\textrm{g}}$ values, respectively. Whether either prefactor remains after summation therefore depends on how many of these momenta remain active once the interaction is projected onto the flat bands. The moir\'e momentum $\mathbf{k}$ retains its full range: the flat bands span the entire moir\'e Brillouin zone, so all $N_k$ values contribute, and the corresponding sum then compensates the $1/N_k$ prefactor. The graphene momentum $\mathbf{p}$ is different: the flat-band states are built almost entirely out of monolayer states near the graphene Dirac points, so only a small, $\mathbf{k}$-independent set of $\mathbf{p}$ values close to the Dirac points contributes appreciably to the projected interaction, rather than the full set of $N_{\textrm{g}}$ graphene momenta. Consequently, the sum over $\mathbf{p}$ does not compensate the $1/N_{\textrm{g}}$ prefactor, and the projected interaction is automatically suppressed by $N_{\textrm{g}}$. Thus, expressed in the delocalized plane-wave basis for the moir\'{e} flat bands, the effective interaction scale is set by $V / N_{\textrm{g}}$. For our nearest-neighbor coupling of $J=1.35$~eV, defined in the local atomic basis, this translates to an effective plane-wave-basis interaction of $J_{g}\approx0.595$~meV, which lies squarely within the same meV range as the intra-valley interactions recently identified as relevant in Refs.~\cite{Wang2026, Wang2026Stiffness}. The interaction strengths used in our work are therefore entirely consistent with, rather than in tension with, the interaction scales reported in the continuum-model literature: the two are simply expressed in different bases, each carrying its own natural energy scale.

\section{Spectral properties of the nematic state}
In this section we elaborate on the spectrum and especially the nodal points of the nematic state.

\subsection{Quasiparticle spectrum}

\begin{figure}
	\includegraphics[width=0.8\textwidth]{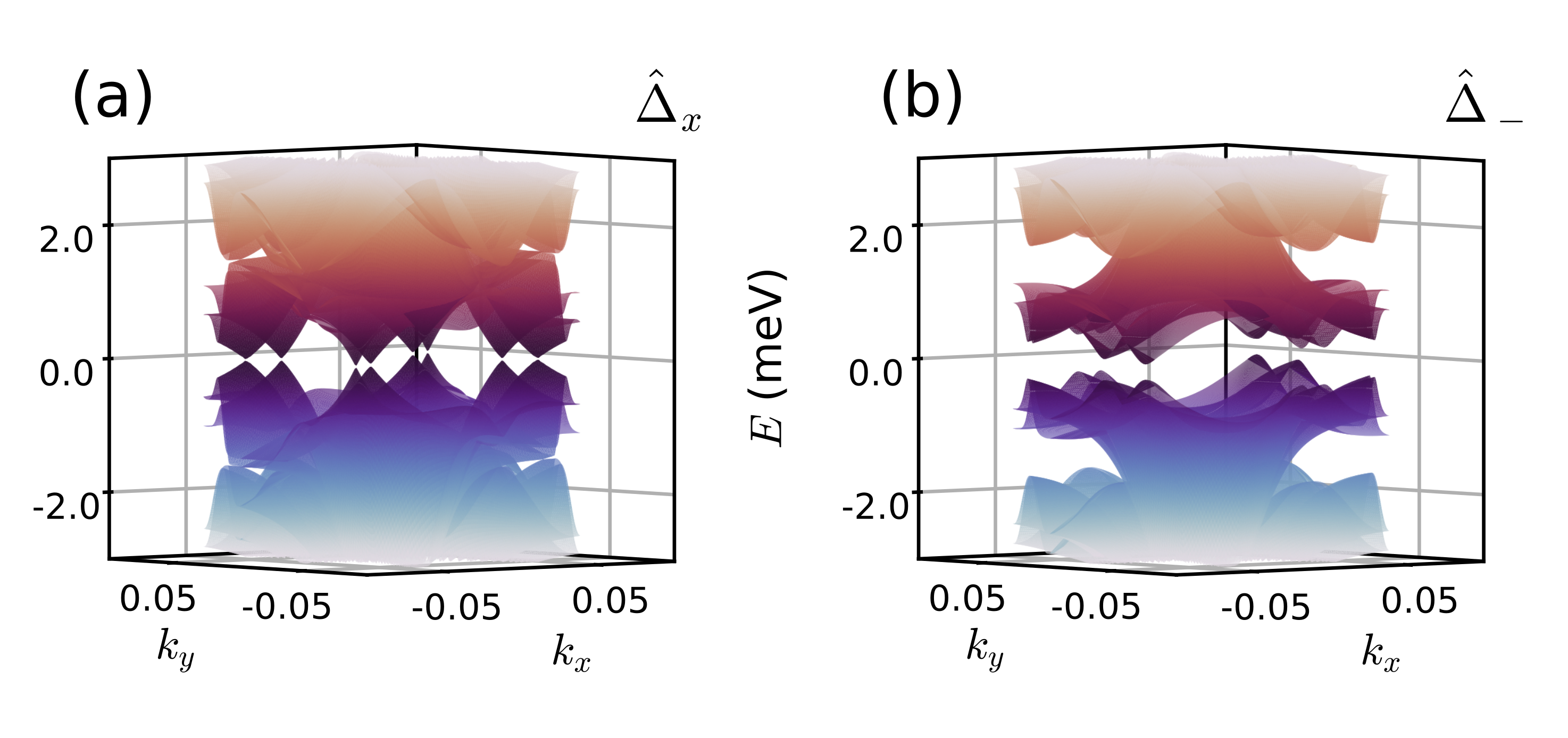}
	\caption{Momentum-resolved BdG energy spectra of the (a) nematic $\hat{\Delta}_x$ and (b) chiral $\hat{\Delta}_+$ pairing states in the two-dimensional Brillouin zone near $\boldsymbol{\Gamma}$, computed using the same parameters as in the main text, $\Delta_0 \approx 49$~meV and $\mu \approx 0.5$~meV. Momenta $k_{x,y}$ are in units of the reciprocal moir\'e lattice vectors.}
	\label{fig:spectrum}
\end{figure}

In the main text we mention that the nematic superconducting state has nodes but at the same time the low-energy density of states (DOS) seems to display a full gap. Here we provide a momentum-resolved view of the numerical BdG spectrum in order to clarify the precise nature of the low-energy spectrum. Figure~\ref{fig:spectrum} shows the BdG eigenvalues across a small region of the two-dimensional Brillouin zone near $\boldsymbol{\Gamma}$ for both the nematic $\hat{\Delta}_x$ (a) and chiral $\hat{\Delta}_+$ (b) states. Parameters used are the same as in the main text, namely $\Delta_0 \approx 49$~meV and $\mu \approx 0.5$~meV. In each panel, all moir\'e flat bands are visible and color is used solely to aid visual separation of the different bands.

For the nematic state in Fig.~\ref{fig:spectrum}(a) isolated nodal points are visible, where the lowest positive and highest negative BdG eigenvalues reach down towards zero energy, demonstrating that the nematic state is technically nodal. Importantly, the spectrum recovers very rapidly away from these nodes, with all bands quickly developing a sizable gap over the rest of the Brillouin zone. Consequently, the phase space of low-energy quasiparticles is strongly restricted, explaining why the DOS plotted in the main text appears effectively gapped despite these nodal points. In fact, we find that an extremely fine $192\times192$ sampling of this small region of the moir\'e Brillouin zone is needed in order to resolve the nodal features and that they happen away from symmetry lines, explaining why they were missed by previous works \cite{Lothman2022} and not visible in the main text. Narrow nodal regions have also been documented for phonon-driven superconductivity in the intra-Chern basis \cite{Liu2024, Liu2025}. For comparison, Fig.~\ref{fig:spectrum}(b) shows the energy spectrum for the chiral state $\hat{\Delta}_+$. In this case, the BdG spectrum is fully gapped: no band crosses zero energy anywhere in the Brillouin zone. However, two distinct sets of bands can be seen. Low-energy bands corresponding to the unpaired Chern sector discussed in the main text lie inside a larger superconducting gap associated with the paired sector, visible at higher energies.

\subsection{Protected nodes in the nematic spectrum}

We next demonstrate that the nodal points of the nematic state are robust features of its spectrum and also derive an analytical criterion for their existence. Beyond reinforcing the numerical results above, this analysis resolves an apparent discrepancy in the literature. Specifically, Ref.~\cite{Yu2023} showed that nematic intra-Chern pairing in TBG is Euler obstructed, implying the existence of nodal points protected by the effective $C_{2z}\mathcal{T}$ symmetry of the normal state. As discussed there, this obstruction originates from the nonzero Euler numbers of the TBG flat bands, which prevent the construction of symmetry-preserving, maximally localized Wannier orbitals from the flat bands alone \cite{Ahn2019, Song2019b}. By contrast, Refs.~\cite{Liu2024, Liu2025} reported a fully gapped nematic phase within the minimal intra-Chern model, seemingly at odds with the Euler obstruction.

The resolution lies in two observations. First, the argument of Ref.~\cite{Yu2023} applies only when the chemical potential lies within a model-dependent energy window. Second, the nematic intra-Chern model of Refs.~\cite{Liu2024, Liu2025} is obtained from a continuum expansion around the $\boldsymbol{\Gamma}$ point. As a result, the corresponding chiral form factors $d^{\mathrm{ph}}_{\pm}$ are not periodic over the Brillouin zone and consequently exhibit a net winding. By contrast, the form factors $d_{\pm}$ obtained from the atomistic model display a $4\pi$ winding around $\boldsymbol{\Gamma}$ that is exactly compensated by $2\pi$ counter-windings centered at the Dirac points $\mathbf{K}_{\pm}$. We next clarify the distinct physical origins of these vortex structures.

The vortex centered at $\boldsymbol{\Gamma}$ is a direct consequence of the Euler obstruction. More precisely, the obstruction guarantees the existence of two zeros of the pairing matrix, each carrying a $2\pi$ vortex, somewhere in the Brillouin zone. Their number and vorticity are fixed by the Euler number of the flat bands \cite{Yu2023}. Because of the irreducible representation structure of the intra-Chern pairing states under the $D_{3}$ lattice symmetry, these vortices are pinned to high-symmetry points in the Brillouin zone. Since $\boldsymbol{\Gamma}$ is the only momentum at which the pairing amplitude $|d_{\pm}|$ vanishes, the observed $4\pi$ winding around this point must correspond to the merger of the two obstruction-induced vortices. Thus, this winding is a direct manifestation of the topology of the symmetric flat bands within the intra-Chern channel.

The counter-windings at $\mathbf{K}_{\pm}$ have a different origin. They arise from the gauge choice that fixes the kinetic term $\tilde{\epsilon}$ to have a uniform global phase throughout the Brillouin zone. A different gauge may transfer these counter-winding vortices from the pairing matrix to the kinetic term without affecting any gauge-invariant observable, including the quasiparticle spectrum. We therefore focus on the quasiparticle energies and show that the compensating winding necessarily enforces spectral nodes in the nematic state, provided the chemical potential lies within the flat-band manifold.

We start by writing the quasiparticle energies for the nematic configuration
\begin{align}
	\varepsilon_{\pm}^2
	&= \tilde{\epsilon}^2 + \tilde{\mu}^2 + 2 \tilde{\Delta}_0^2 \pm 2 \sqrt{\tilde{\epsilon}^2 \left( \tilde{\mu}^2 + 2 \tilde{\Delta}_0^2 - \tilde{\Delta}_0^2 \delta^2 \right)},
\end{align}
where we define $\delta^2 = 1 + \cos\phi_\mathbf{k}$. The condition for a nodal point is $\varepsilon_{-}=0$, which leads to two possible solutions for $\tilde{\epsilon}$, which we denote by
\begin{align} \label{eq:cond_ev}
	\tilde{\epsilon}_{\pm}^2
	&= (\tilde{\mu}^2 + 2 \tilde{\Delta}_0^2) - 2 \tilde{\Delta}_0^2 \delta^2 \pm 2 \tilde{\Delta}_0 \delta \sqrt{\tilde{\Delta}_0^2 \delta^2 - (\tilde{\mu}^2 + 2 \tilde{\Delta}_0^2)}.
\end{align}
This expression is real-valued if and only if
\begin{align} \label{eq:cond0}
	\tilde{\Delta}_0^2 \delta^2 \geq \tilde{\mu}^2 + 2 \tilde{\Delta}_0^2 \quad \text{or} \quad \tilde{\Delta}_0 \delta = 0.
\end{align}
Since $0 \leq \delta^2 \leq 2$, the left condition of Eq.~\eqref{eq:cond0} is never fulfilled except at the fine-tuned case $\Tilde{\mu} = \phi_\mathbf{k} = 0$, which we disregard. We therefore focus on the condition on the right. Noting that $\tilde{\Delta}_0$ is always finite away from $\boldsymbol{\Gamma}$, it implies
\begin{align} \label{eq:cond1}
	\phi_\mathbf{k}=\pi.
\end{align}
At the same time, together with Eq.~\eqref{eq:cond_ev}, we have
\begin{align} \label{eq:cond2}
	\tilde{\epsilon}^2 = \tilde{\mu}^2 + 2 \tilde{\Delta}_0^2.
\end{align}

Equations~\eqref{eq:cond1}-\eqref{eq:cond2} are necessary and sufficient conditions for the existence of nodes in the spectrum of the nematic state. We next show that, as long as the chemical potential is tuned within the flat bands, there are always momenta $\mathbf{k}$ satisfying both of these equations. Eq.~\eqref{eq:cond1} is satisfied at momenta $\mathbf{k}$ where $d_+(\mathbf{k})$ and $d_-(\mathbf{k})$ satisfy the relation $[\arg(d_+) - \arg(d_-)] \bmod 2\pi = \pi$. Since these two functions wind twice ($4\pi$) in opposite directions when following a path around $\boldsymbol{\Gamma}$, there must be four momentum points satisfying the above relation along these paths. The collection of all such points forms four distinct, smooth curves $\gamma_1$ starting from $\boldsymbol{\Gamma}$, along which \eqref{eq:cond1} is satisfied. Importantly, such curves $\gamma_1$ end at $\mathbf{K}_{\pm}$, due to the $2\pi$ counter-winding there. 

At the same time, the condition Eq.~\eqref{eq:cond2} consists of the intersection between two surfaces, defined by $\tilde{\epsilon}^2$ and $\tilde{\mu}^2 + 2 \tilde{\Delta}_0^2$. We next assume that the band edges of the non-interacting flat bands are located at $\boldsymbol{\Gamma}$, which is typically the case. We then note that $\tilde{\epsilon}$ is vanishing at $\mathbf{K}_{\pm}$ and finite at $\boldsymbol{\Gamma}$, while the converse is true for $\tilde{\Delta}_0$. On the other hand, $\tilde{\mu}^2 + 2 \tilde{\Delta}_0^2$ is generally finite everywhere. Since $\tilde{\Delta}_0$ is finite at $\mathbf{K}_{\pm}$, we have $\tilde{\epsilon}^2 < \tilde{\mu}^2 + 2 \tilde{\Delta}_0^2$ there. Now, if $\tilde{\epsilon}^2 > \tilde{\mu}^2 + 2 \tilde{\Delta}_0^2$ at $\boldsymbol{\Gamma}$, then $\tilde{\epsilon}^2$ and $\tilde{\mu}^2 + 2 \tilde{\Delta}_0^2$ must intersect along curves $\gamma_2$ in the Brillouin zone, which encircle either $\boldsymbol{\Gamma}$ or $\mathbf{K}_{\pm}$. As a consequence, these curves $\gamma_2$ are guaranteed to intersect with each of the $\gamma_1$ curves connecting $\boldsymbol{\Gamma}$ and $\mathbf{K}_{\pm}$. At such intersections, both Eq.~\eqref{eq:cond1} and Eq.~\eqref{eq:cond2} are satisfied and the BdG Hamiltonian hosts a zero mode there, that is, its spectrum is gapless. Since there are four $\gamma_1$ curves, we expect four nodal points per inter-valley sector. Indeed, in Fig.~\ref{fig:spectrum} we observe eight nodal points in total, since in the atomistic model there are two inter-valley sectors.

If, on the other hand, $\tilde{\epsilon}^2 < \tilde{\mu}^2 + 2 \tilde{\Delta}_0^2$ at $\boldsymbol{\Gamma}$, there is no guarantee that $\tilde{\epsilon}^2$ and $\tilde{\mu}^2 + 2 \tilde{\Delta}_0^2$ cross somewhere in the Brillouin zone and the nodal points are no longer protected. The two scenarios discussed above are separated by the special case $\tilde{\epsilon}^2(\boldsymbol{\Gamma}) = \tilde{\mu}^2(\boldsymbol{\Gamma})$, where we used that $\tilde{\Delta}_0$ vanishes at $\boldsymbol{\Gamma}$. Recalling $\tilde{\epsilon} = \tfrac{1}{2} (\epsilon_{+,+} - \epsilon_{-,+})$ and $\tilde{\mu} = \mu - \tfrac{1}{2} (\epsilon_{+,+} + \epsilon_{-,+})$, we find that nodes in the nematic quasiparticle spectrum are guaranteed to appear for 
\begin{align}
	\epsilon_{-,+}(\boldsymbol{\Gamma}) < \mu < \epsilon_{+,+}(\boldsymbol{\Gamma}),
\end{align}
which is precisely the condition that the chemical potential is tuned inside the flat bands.

\section{Pairing in the moir\'e band basis}
\label{sec:pairing-basis}
In this section we further explain the procedure for obtaining the pairing structure in the low-energy moir\'e bands and report results supporting those in the main text. We then analyze the resulting pairing matrix elements in the context of the symmetries discussed in Sec.~\ref{SM:sec:model:symmetries}.

\subsection{Projection onto moir\'e flat bands}
\label{sec:pairing-basis:projection}

To achieve an effective low-energy description of the electron-electron driven pairing on the carbon lattice, we project the BdG Hamiltonian written in the atomic basis $\hat{H}_{\text{BdG}}$ in Eq.~\eqref{eq:atomistic:pairing:BdG} onto the moir\'e flat bands. At each $\mathbf{k}$, we first diagonalize the normal-state Hamiltonian, obtaining the unitary matrix $w_{\mathbf{k}}$, such that
\begin{align}
	w_{\mathbf{k}} \, \hat{h}_0(\mathbf{k}) \, w^{\dagger}_{\mathbf{k}} 
	&= \epsilon(\mathbf{k}),
\end{align}
where $\epsilon(\mathbf{k})$ is a diagonal matrix whose elements are the normal-state quasiparticle energies. We then use $w_{\mathbf{k}}$ to rewrite $\hat{H}_{\text{BdG}}$ in the band basis,
\begin{align}
	H_{\mathrm{BdG}}(\mathbf{k}) = {W}_{\mathbf{k}} \, \hat{H}_{\mathrm{BdG}}(\mathbf{k})\, {W}^{\dagger}_{\mathbf{k}}
	&= \begin{pmatrix} \epsilon(\mathbf{k}) & {\Delta}(\mathbf{k}) \\ {\Delta}^{\dagger}(\mathbf{k}) & -\epsilon(-\mathbf{k}) \end{pmatrix},
	\quad {W}_{\mathbf{k}} \equiv \begin{pmatrix} w_{\mathbf{k}} & 0 \\ 0 & w^{*}_{-\mathbf{k}} \end{pmatrix}.
\end{align}
In this new basis, the normal-state sectors are diagonal and all mixing between bands resides in the transformed pairing matrix
\begin{align}
	\Delta_{mn}(\mathbf{k})
	&= \sum_{i j} [w_{\mathbf{k}}]_{m i} \hat{\Delta}_{i j}(\mathbf{k}) [w^{T}_{-\mathbf{k}}]_{j n} = \langle \, m, \mathbf{k} \mid \hat{\Delta}(\mathbf{k}) \overline{\mid n, -\mathbf{k} \, \rangle},
\end{align}
where $\mid m, \mathbf{k} \, \rangle$ is the eigenvector corresponding to the $m$th eigenvalue of $\hat{h}_0(\mathbf{k})$. The bar denotes complex conjugation, which appears because the pairing matrix is an operator connecting the particle and hole sectors of the Hilbert space. The eigenvectors above are obtained numerically, hence their global phases are completely random. In order to achieve pairing matrices locally smooth in momentum space, we fix this gauge freedom within the flat-band subspace, as described below.

We next project the band-basis pairing matrices onto the moir\'e flat-band subspace. We do this by restricting the band indices $m,n$ to the subspace of the four bands in the middle of the spectrum. We then rename the states in this moir\'e flat-band subspace according to their valley ($\eta$) and eigenband ($b$) quantum numbers. The valley number is computed through the sign of the expectation value of the valley operator Eq.~\eqref{eq:atomistic:symmetries:valley}, while the eigenband index is assigned by ordering the states from a given valley with increasing energy, with $b=+$ ($b=-$) the upper (lower) band. We also note that a given $m$ yields opposite $\eta$ values at opposite momenta, as illustrated in Fig.~2 of the main text. This results in the basis $\mid b \, \eta \, \mathbf{k} \, \rangle$ used in Eqs.~(6)--(8) of the main text. The pairing matrix in this moir\'e flat-band basis is explicitly 
\begin{align}
	{\Delta}_{b b^\prime}^{\eta \eta^\prime}(\mathbf{k})
	&= \langle \, b \, \eta \, \mathbf{k} \mid \hat{\Delta}(\mathbf{k}) \overline{\mid  \, b^\prime \, \eta^\prime \, (-\mathbf{k}) \, \rangle}.
\end{align}
We next recall that complex conjugation implements spinless time-reversal symmetry ($\mathcal{T}$) within the atomistic model, which transforms the quantum numbers according to
\begin{align}
	\mathcal{T}: \quad b \rightarrow b,\quad \eta \rightarrow -\eta ,\quad \mathbf{k} \rightarrow -\mathbf{k}.
\end{align}
Using this we obtain
\begin{align} \label{eq:projection:finalOP}
	\Delta_{b b^\prime}^{\eta \eta^\prime}(\mathbf{k})
	&= \langle \, b \, \eta \, \mathbf{k} \mid \hat{\Delta}(\mathbf{k}) \mid  \, b^\prime \, (-\eta^\prime) \, \mathbf{k} \, \rangle.
\end{align}
This is exactly Eq.~(8) given in the main text. We thus find that the transformed pairing matrix at a given momentum $\mathbf{k}$ can be obtained by using only eigenvectors computed at that same $\mathbf{k}$, instead of also depending on quantities at $-\mathbf{k}$. We fix the gauge of the flat-band states through the Chern basis. In each valley $\eta$ we diagonalize the sublattice polarization $\hat n_A-\hat n_B$ within the two flat bands and label its eigenvectors by $C=\eta\,\mathrm{sgn}\langle\hat n_A-\hat n_B\rangle$. The (approximate) $C_{2z}\mathcal{T}$ symmetry relates the two Chern states but leaves their relative phase free. We fix it by requiring the kinetic matrix element $\langle C{=}+, \eta, \, \mathbf{k}|\hat h_0|C{=}- , \, \eta, \, \mathbf{k}\rangle$ to be real and positive, and the common phase by making the $C=+$ state real on the central site of the $AA$ region in the top layer. The states of the opposite valley are obtained by time reversal. The eigenband states then follow from Eq.~(13) of the main text, and coincide with the flat-band eigenstates of $\hat h_0$ to numerical precision. A single global phase, common to all states and pairing configurations, finally makes the nematic pairing matrices real.

We label each Chern band by its Chern number $C = \frac{1}{2\pi} \int_{\mathrm{BZ}} d^2k \, (\partial_{k_x} A_y -\partial_{k_y} A_x)$, with $\mathbf A = i\langle u_{\mathbf k}|\nabla_{\mathbf k}u_{\mathbf k}\rangle$, evaluated on the Chern states; the bands labeled $C=\pm$ carry Chern numbers $\pm1$ in both valleys. The Chern basis pairing matrices are then related to the eigenband basis pairing matrices via
\begin{align} \label{eq:projection:finalOP_chern}
	\tilde{\Delta}_{C C^\prime}^{\eta \eta^\prime}(\mathbf{k})
	&= [U \, \Delta(\mathbf{k}) \, U^\dagger]_{C C^\prime}^{\eta \eta^\prime},
\end{align}
where $U$ is defined in Eq.~(13) of the main text.

\subsection{Valley and multiband structure of pairing} \label{sec:multiband}

In the main text we consider the pairing matrix projected onto the moir\'e flat-band subspace, where we focus only on a single inter-valley sector. Here we provide the full structure of the superconducting pairing in both valley and eigenband index for the moir\'e flat bands. In particular, we show that all intra-valley terms have vanishing amplitude. 

\begin{figure} [b]
	\includegraphics[width=\columnwidth]{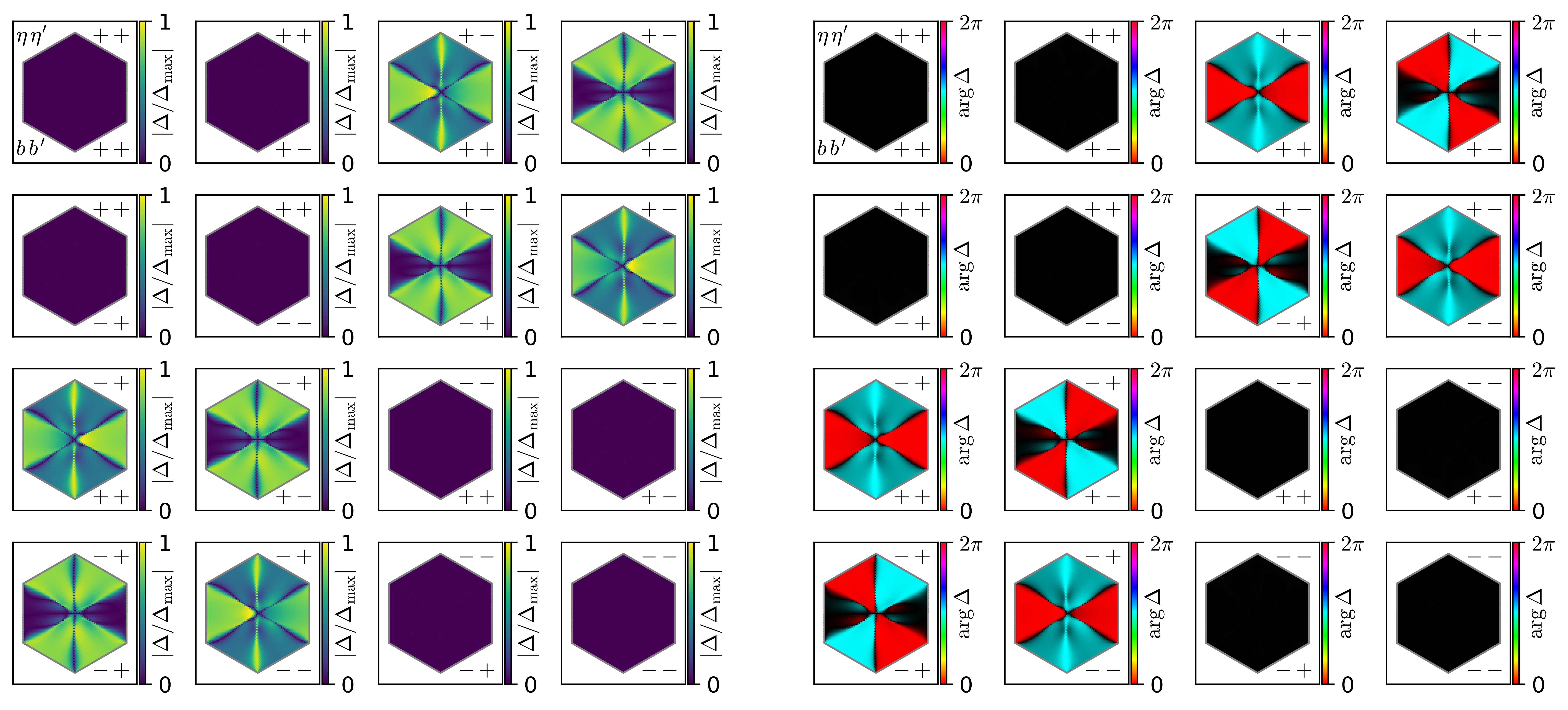}
	\caption{Multiband pairing structure of the nematic state $\hat{\Delta}_x$ within the flat-band subspace with amplitude (left) and phase (right) of the pairing matrix elements $\Delta_{b b^\prime}^{\eta \eta^\prime}$. Top left and bottom right blocks show vanishing amplitude, indicating pairing is inter-valley only.}
	\label{fig:multiband:Deltax}
\end{figure}

Starting with the nematic case $\hat{\Delta}_x$, i.e.~$(\alpha, \phi) = (\tfrac{\pi}{2}, 0)$, which corresponds to the ground state in the regime of small normal-state band splitting and/or large interactions, we show in Fig.~\ref{fig:multiband:Deltax} the pairing amplitude (left) and phase (right) of the elements $\Delta_{b b^\prime}^{\eta \eta^\prime}$ for every combination of eigenband and valley index. Each panel corresponds to a different matrix element and is appropriately labeled. We find that all intra-valley terms ($\eta = \eta^\prime$, main diagonal blocks) show vanishing pairing amplitude. Therefore only inter-valley terms ($\eta \neq \eta^\prime$, off-diagonal blocks) have finite pairing strength throughout the Brillouin zone and we further find that all such elements can be made real under an appropriate gauge choice. 


Most interestingly, we observe a similarity in the momentum-space structure of the inter-valley elements. In particular, all intra-eigenband terms $\Delta_{b b}^{\eta (-\eta)}$ show a very similar form, which is (approximately) even under the independent mirror transformations $k_x \rightarrow -k_x$ and $k_y \rightarrow -k_y$ (dashed gray lines in Fig.~6 of the main text). We here ignore symmetry breaking features, which we comment on below, and thus label this form factor by $d_{x^2-y^2}$. In the gauge we consider, all such terms have the same global phase. The inter-eigenband terms $\Delta_{b (-b)}^{\eta (-\eta)}$ also show approximately the same form among themselves, but which is instead (approximately) odd under the mirror symmetries above, so we label them by $d_{xy}$. These show an alternating global phase with either eigenband or valley index, $\Delta_{b (-b)}^{\eta (-\eta)} \approx -\Delta_{(-b) b}^{\eta (-\eta)} \approx -\Delta_{b (-b)}^{(-\eta) \eta}$. Taken together, this results in Eq.~(9) in the main text.

\begin{figure} [h]
	\includegraphics[width=\columnwidth]{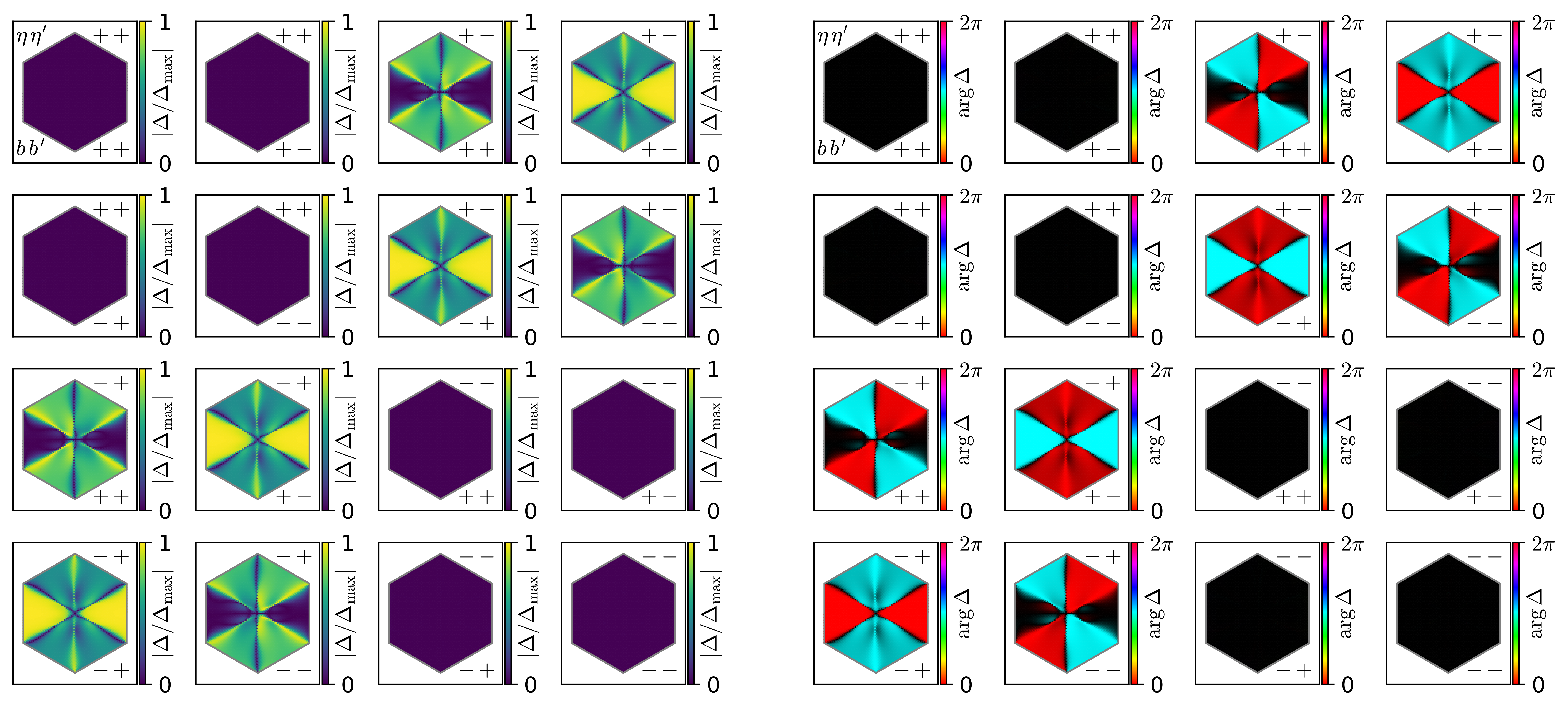}
	\caption{Same as Fig.~\ref{fig:multiband:Deltax} but for the nematic state $\hat{\Delta}_y$.}
	\label{fig:multiband:Deltay}
\end{figure}

We next consider the other, orthogonal (and suboptimal), nematic solution $\hat{\Delta}_y$, corresponding to $(\alpha, \phi) = (\tfrac{\pi}{2}, \pi)$, in Fig.~\ref{fig:multiband:Deltay}. Similar to $\hat{\Delta}_x$, we also here find vanishing intra-valley pairing. This time, however, the intra- and inter-eigenband terms have swapped approximate forms $d_{xy}$ and $d_{x^2-y^2}$, while the global phase structure is still even in the intra-eigenband case and odd in the inter-eigenband one. This results in Eq.~(10) in the main text.

\begin{figure} [h]
	\includegraphics[width=\columnwidth]{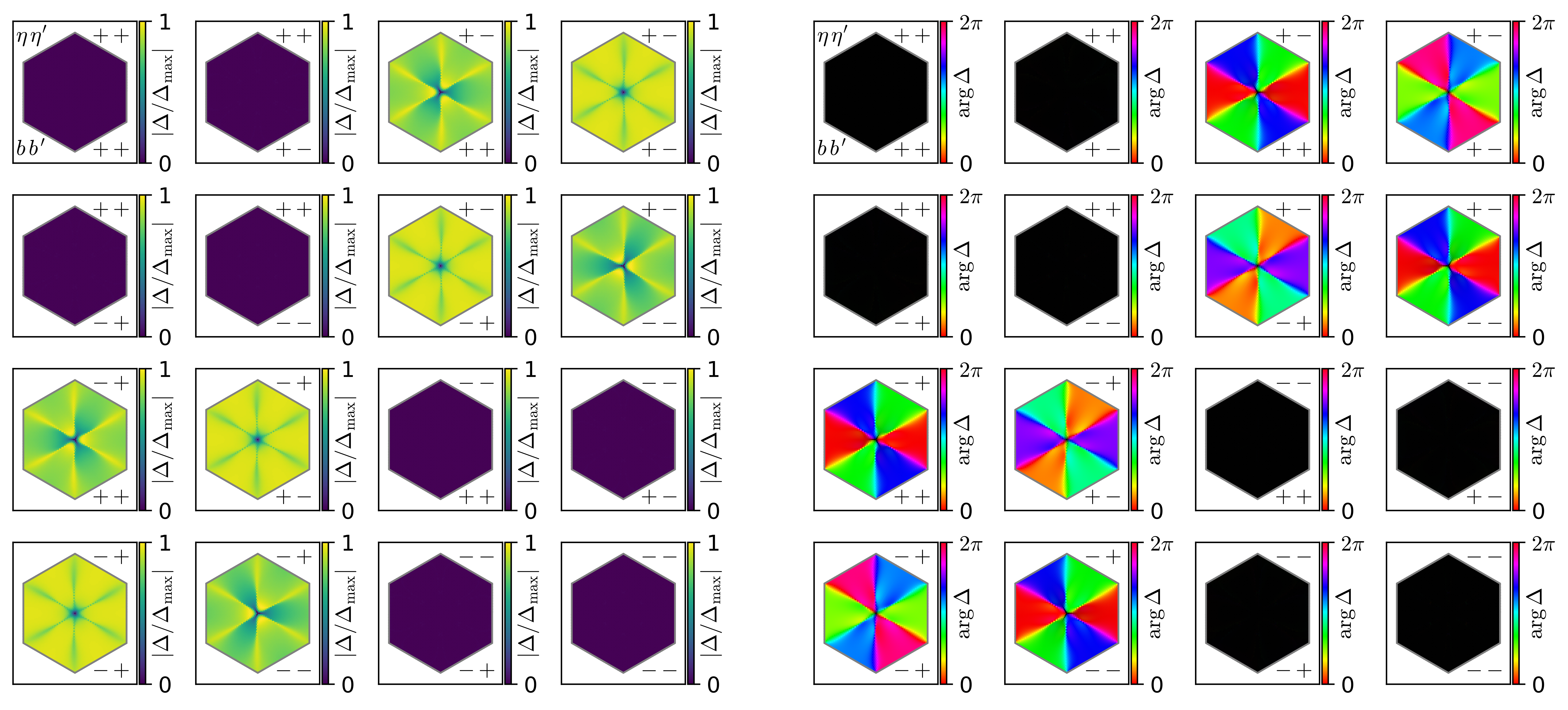}
	\caption{Same as Fig.~\ref{fig:multiband:Deltax} but for the chiral state $\hat{\Delta}_+$.}
	\label{fig:multiband:Deltadpid}
\end{figure}

\begin{figure}  [h]
	\includegraphics[width=\columnwidth]{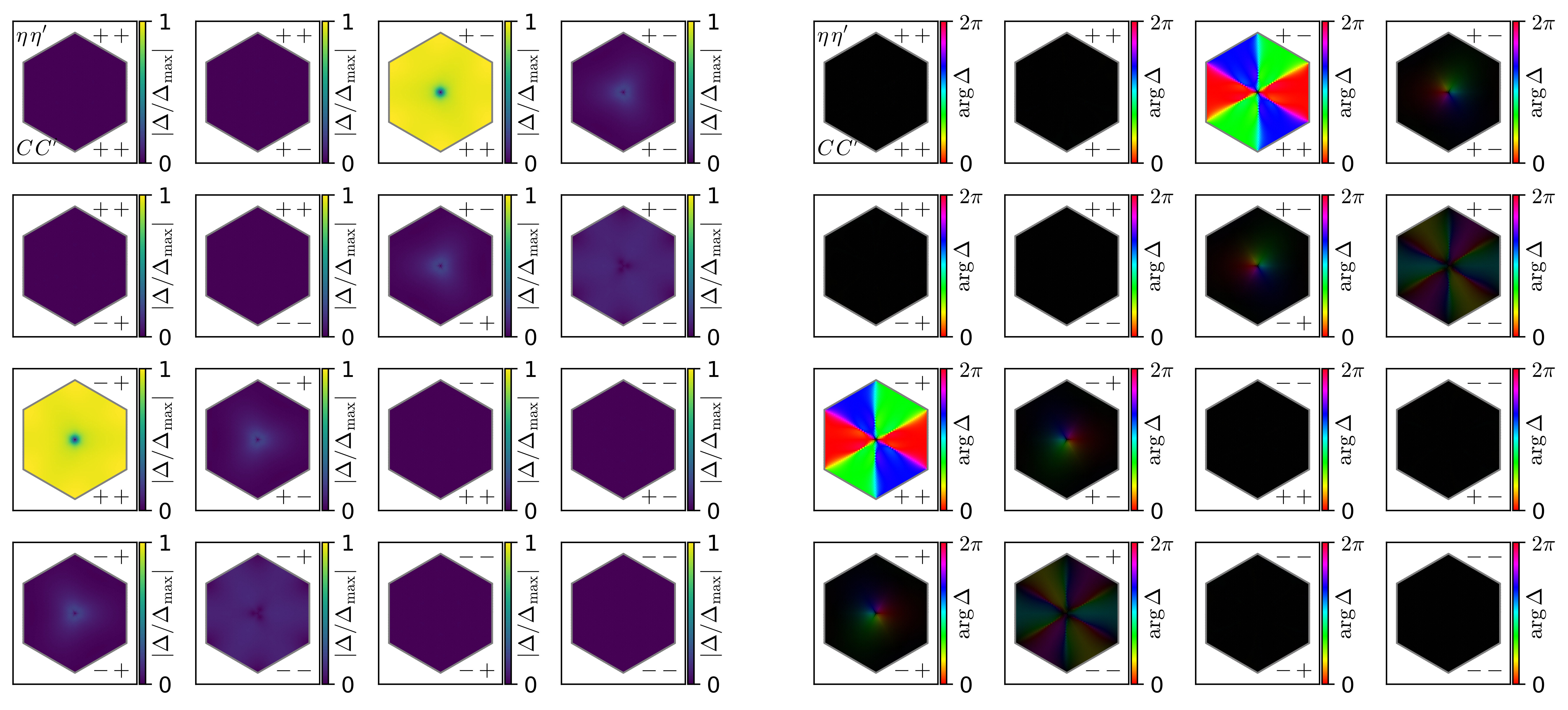}
	\caption{Multiband pairing structure of the chiral state $\hat{\Delta}_+$ within the flat-band subspace in the Chern basis. Amplitude (left) and phase (right) of the pairing matrix elements $\tilde{\Delta}_{C C^\prime}^{\eta \eta^\prime}$, where $C$, $C^\prime$ denote the Chern sectors.}
	\label{fig:multiband:Deltadpid_Chern}
\end{figure}

\begin{figure}  [h]
	\includegraphics[width=\columnwidth]{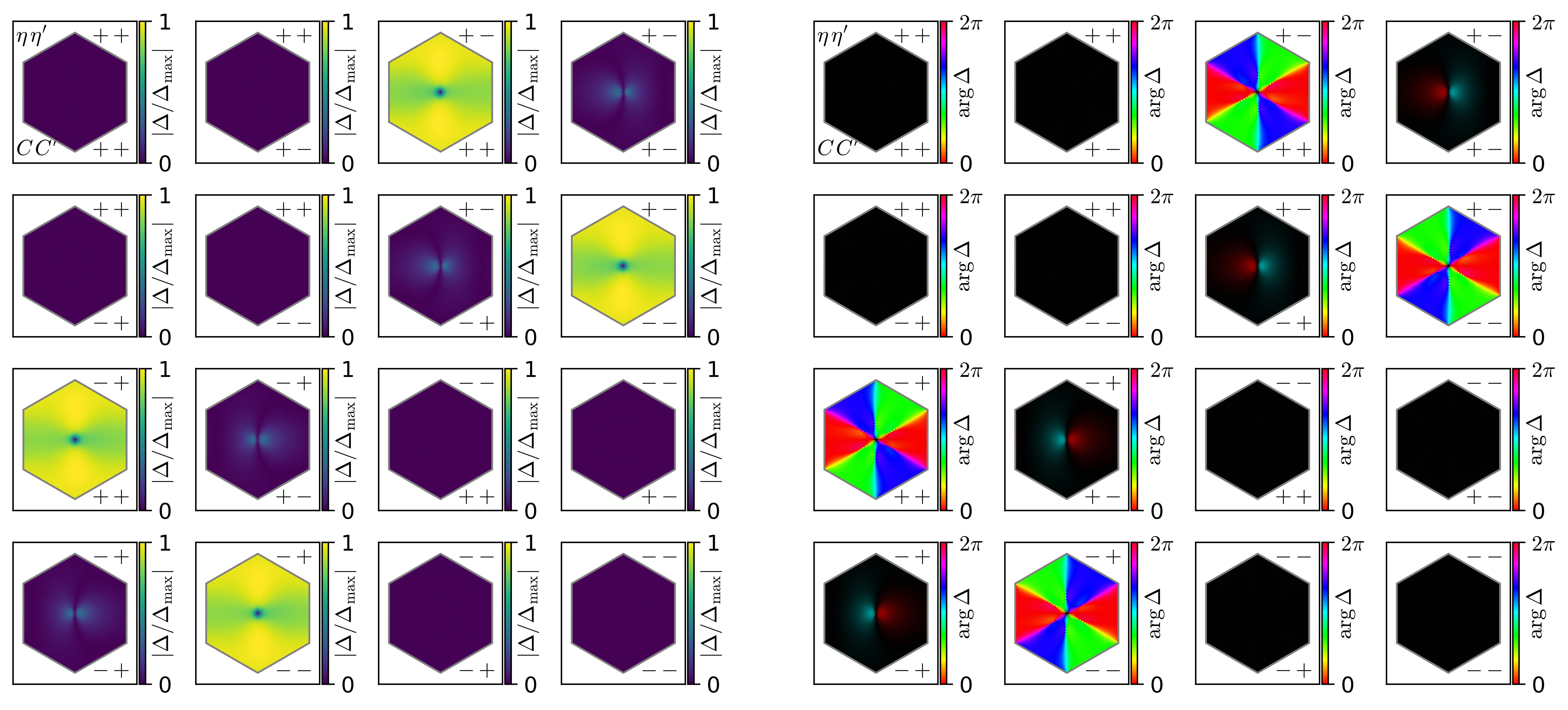}
	\caption{Same as Fig.~\ref{fig:multiband:Deltadpid_Chern} but for the nematic state $\hat{\Delta}_x$.}
	\label{fig:multiband:Deltadpid_Chern_nematic}
\end{figure}

Finally, we consider the chiral state, starting with the pairing matrix elements in the eigenband basis in Fig.~\ref{fig:multiband:Deltadpid}. Here, we again find inter-valley pairing only, and both intra- and inter-eigenband terms. The elements $\Delta_{b b^\prime}^{\eta (-\eta)}(\mathbf{k})$ cannot however be made real across the whole Brillouin zone, but instead show a $4\pi$ winding around $\boldsymbol{\Gamma}$, as described in the main text. In addition, a $\pi/2$ phase shift is present between the intra- and inter-eigenband terms. This results in the pairing structure of Eq.~(11) in the main text. Together, the observations allow us to constrain the pairing matrix of the nematic and chiral states to the form presented in the main text in Eqs.~(9)--(11). From there, the intra-Chern pairing is found through a basis change. We check that the chiral pairing matrix has indeed the Chern-polarized form Eq.~(14) by performing the basis change Eq.~\eqref{eq:projection:finalOP_chern} numerically, and show our results in Fig.~\ref{fig:multiband:Deltadpid_Chern}, where each panel corresponds to an element $\tilde{\Delta}_{C C^\prime}^{\eta \eta^\prime}$, where $C=\pm$ labels each Chern sector. In this basis, the only relevant elements of the inter-valley sectors $\tilde{\Delta}_{C C^\prime}^{\eta (-\eta)}$ correspond to $C=C^\prime=+$, that is, pairing is polarized to this Chern sector. Nematic configurations yield instead a pairing matrix with the same amplitude in both intra-Chern sectors $C=C^\prime=\pm$, as shown in Fig.~\ref{fig:multiband:Deltadpid_Chern_nematic} for $\hat{\Delta}_x$.

\clearpage

\clearpage

\clearpage

\subsection{Symmetries of the pairing matrix} 
\label{SM:sec:multiband:symmetries}

In the discussion above, as well as in the main text, we classify the pairing matrix elements according to a $d$-wave momentum-space symmetry. We here note that for the individual matrix elements, such symmetry is not exact, although it is a very good approximation away from $\boldsymbol{\Gamma}$. More precisely, a matrix element $\Delta_{b b^\prime}^{\eta \eta^\prime}(\mathbf{k})$ or $\tilde{\Delta}_{C C^\prime}^{\eta \eta^\prime}(\mathbf{k})$ is not generally invariant under momentum inversion $\mathbf{k} \rightarrow -\mathbf{k}$. To understand why, we recall the discussion on the normal-state symmetries in Sec.~\ref{SM:sec:model:symmetries}. Momentum inversion, while preserving $b$ and $\eta$, is achieved through the combination of chiral $\mathcal{C}$ and particle-hole $\mathcal{P}$ symmetries. By visually comparing two intra-eigenband matrix elements $\Delta_{b b}^{\eta \bar{\eta}}(\mathbf{k})$ and $\Delta_{\bar{b} \bar{b}}^{\eta \bar{\eta}}(-\mathbf{k})$, where the bar denotes the opposite index ($\bar{b} = -b$, $\bar{\eta} = -\eta$), we find that the pairing matrices show preserved particle-hole symmetry. However, as discussed in Sec.~\ref{SM:sec:model:symmetries}, $\mathcal{C}$ is broken by intra-sublattice hopping terms, which are significant in the atomistic model we use, thus explaining the lack of $\mathcal{C} \mathcal{P}$ symmetry in the pairing matrices. Such symmetry breaking is evidenced for example by peaks of pairing amplitude appearing close to $\boldsymbol{\Gamma}$ in Fig.~\ref{fig:multiband:Deltax}, but which have no counterpart under $\mathbf{k} \rightarrow -\mathbf{k}$. The fact that the symmetry seems better preserved in other regions of the Brillouin zone suggests that chiral symmetry breaking is strongest near $\boldsymbol{\Gamma}$. Finally, we note that this symmetry breaking could in principle be exactly accounted for by replacing the form factors in the lower row in the definition of the $\Delta_{x,y}$ matrices in Eqs.~(9) and (10) of the main text by their momentum conjugates. Since we do not expect such details to meaningfully modify the energetic competition between nematic and chiral states we focus on in the main text, we neglect them in our intra-Chern minimal model.

Finally, we note that any pairing matrix must have an additional exact symmetry stemming from the fermionic statistics of the electrons composing the Cooper pair: it must be anti-symmetric under the exchange of \emph{all} degrees of freedom. Since we consider spin-singlet pairing, we also have the constraint that the pairing matrix must be anti-symmetric with respect to the exchange of spin degrees of freedom alone. In turn, this means it must be symmetric under the exchange of the remaining degrees of freedom $\Delta_{b b^\prime}^{\eta \eta^\prime}(\mathbf{k}) = \Delta_{b^\prime b}^{\eta^\prime \eta}(-\mathbf{k})$. We have verified that all our pairing matrices satisfy this symmetry to numerical accuracy.

\subsection{Remote-band contribution to condensation energy}
\label{sec:pairing-basis:remote}

\begin{figure}
	\includegraphics[width=\textwidth]{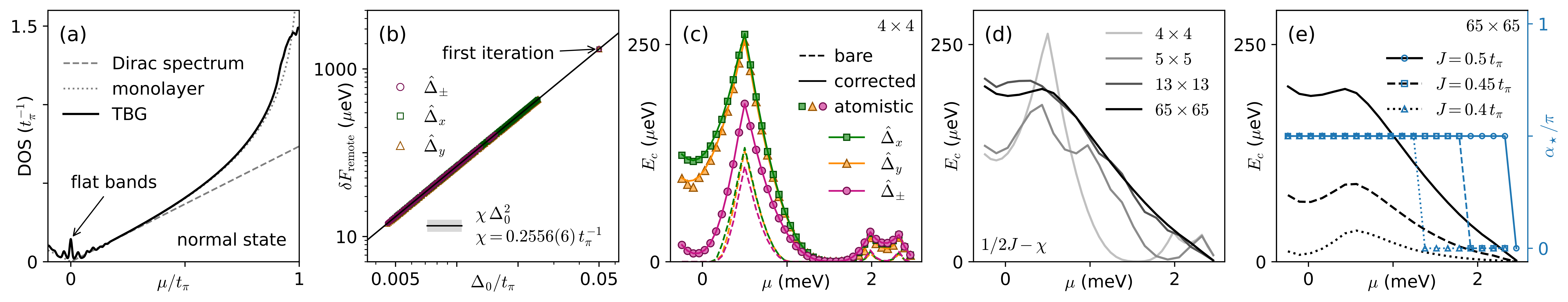}
	\caption{Calibration of the intra-Chern pairing model. (a) DOS versus chemical potential $\mu$ of the normal state in TBG (black) and two decoupled graphene monolayers (dotted gray) at large energy scales. Dashed line denotes the Dirac regime in monolayer graphene. (b) Remote-band contribution to condensation energy, $\delta F_{\textrm{remote}}$, as a function of pairing amplitude $\Delta_0 = |\hat{\Delta}|$. Open symbols, one shape per state as in Fig.~4(a) of the main text: all iterations of the self-consistency loops underlying that figure; isolated points at $\Delta_0 = 0.05\,t_{\pi}$ are the initial guesses. Line: $\chi \Delta_0^2$ fit, with $\chi = 0.2556(6) \, t_{\pi}^{-1}$. (c) Condensation energy at $J = 0.5\,t_{\pi}$ from the atomistic model (symbols) and from the intra-Chern pairing model without (dashed) and with (solid) correction in Eq.~\eqref{eq:remote:Jeff}, due to the remote bands, using a $4\times4$ momentum grid. (d) Condensation energy of the intra-Chern pairing ground state with remote-band corrections at $J = 0.5\,t_{\pi}$ using momentum grids from $4\times4$ to $65\times65$. (e) Remote-band corrected condensation energy (black, left axis) and $\alpha_\star$ (blue, right axis) with varying $J = 0.5$, $0.45$, and $0.4\,t_{\pi}$, using a $65\times65$ momentum grid. In all panels, $\mu$ is measured from the Dirac point and energies are per moir\'e unit cell and spin sector.}
	\label{fig:remote}
\end{figure}

The minimal intra-Chern pairing model, Eq.~(18) in the main text, retains only the four moir\'e flat bands, whereas the atomistic calculations in Fig.~4(a) of the main text include all, approximately $10^4$ bands. Here we quantify the contribution of the remaining, so-called remote, bands to the condensation energy $E_c$ per moir\'e unit cell and per spin sector and show that it amounts almost perfectly to a simple rescaling of the interaction $J$. All results are summarized in Fig.~\ref{fig:remote}. We start by rewriting the zero-temperature condensation energy of the atomistic model, Eq.~(4) in the main text, as $E_c (\hat{\Delta}) = \delta F_{\textrm{qp}}(\hat{\Delta}) - \Delta_0^2/2J$, where $\delta F_{\textrm{qp}}$ is the contribution from the change in quasiparticle energies with finite superconducting pairing and $\Delta_0 = |\hat{\Delta}|$ is the pairing strength. We then further denote by $\delta F_{\textrm{flat}}$ the part of the quasiparticle contribution coming solely from the moir\'e flat bands included in the intra-Chern model. The difference $\delta F_{\textrm{remote}} = \delta F_{\textrm{qp}} - \delta F_{\textrm{flat}}$ therefore measures the contribution of all remote bands to the condensation energy, which is missed by the flat-band projection performed to arrive at the intra-Chern pairing model. 

Performing a second order expansion on the quasiparticle energies of the remote bands in the presence of pairing we obtain the general energy $\epsilon + g \, \Delta_{0}^2/\epsilon$, where $\epsilon$ is the normal-state quasiparticle energy with respect to the Fermi level and $g$ is a coefficient encoding the strength with which the states at $\epsilon$ couple via pairing. We here neglect a possible dependence of $g$ on $\epsilon$. Within this approximation, the condensation energy contribution coming from the remote bands thus takes the form 
\begin{align} \label{eq:remote_dF}
	\delta F_{\textrm{remote}}
	&\propto \int_{W_-}^{W_+} d\epsilon \, \textrm{DOS}(\epsilon) \, g \, \frac{\Delta_0^2}{\abs{\epsilon}},
\end{align}
where $W_{+}<0$ is below the flat-band energy scale so that the energy interval $[W_-, W_+]$ contains only remote bands. As shown in Fig.~\ref{fig:remote}(a), sufficiently away from the flat bands, the normal-state DOS of small angle TBG (black) resembles that of two uncoupled graphene monolayers (dotted gray). This includes a linear increase at intermediate $\epsilon$ originating from the Dirac dispersion (dashed gray), followed by a peak at $|\mu|=t_{\pi}$ originating from the monolayer van Hove singularity. This DOS is thus able to overcome the factor $1/\epsilon$ in the integrand in Eq.~\eqref{eq:remote_dF}, which means the remote bands can make a sizable contribution to condensation energy. 

According to Eq.~\eqref{eq:remote_dF}, this contribution to the condensation energy is quadratic in $\Delta_0$. 
Verifying this quadratic dependence requires calculating $\delta F_{\textrm{remote}}$ for different pairing configurations $\hat{\Delta}$. This is computationally demanding, as it requires diagonalizing the full atomistic Hamiltonian. For convenience, we instead use the pairing configurations $\hat{\Delta}_{x}$, $\hat{\Delta}_{y}$ and $\hat{\Delta}_{+}$, for which $F_{\textrm{qp}}$ was already calculated during the self-consistency loops resulting in Fig.~4(a) of the main text. We plot in Fig.~\ref{fig:remote}(b) the resulting 
$\delta F_{\textrm{remote}}$ as a function of $\Delta_0$, confirming a clear quadratic behavior. Fitting the data yields
\begin{align} \label{eq:remote:chi}
	\delta F_{\textrm{remote}} 
    \approx \chi \, \Delta_0^2, \quad \chi = 0.2556(6) \, t_{\pi}^{-1}.
\end{align}
Notably, all sample points lie within about $1\%$ of this curve, the residual spread mainly reflecting a weak dependence on $\mu$. Importantly, this means $\delta F_{\textrm{remote}}$ is virtually independent of $\alpha$ and $\phi$. The competition between chiral and nematic states is therefore entirely decided by the flat bands, exactly as assumed in the intra-Chern pairing model of the main text. 

The quadratic nature of the remote-band condensation energy, extracted in Eq.~\eqref{eq:remote:chi}, means it can be absorbed into the mean-field term in Eq.~\eqref{eq:transition:Hamiltonian}. The condensation energy per spin then becomes
\begin{align} \label{eq:remote:Jeff}
	E_c = \delta F_{\textrm{flat}}(\hat{\Delta}) - \frac{\Delta_0^2}{2 J_{\textrm{eff}}},
	\quad
    J_{\textrm{eff}} = \frac{J}{1 - 2\chi J},
\end{align}
i.e.~the remote bands simply rescale the effective pairing interaction acting on the flat bands. The effective interaction $J_{\textrm{eff}}$ has a pole at $J_{\textrm{pole}} = 1/2\chi = 1.956(5) \,t_{\pi}$, which is close to the quantum critical point for nearest-neighbor pairing in monolayer graphene $J_{\textrm{QCP}}=1.91 \, t_{\pi}$ \cite{BlackSchaffer2007}. Thus for $J>J_{\textrm{pole}}$, the remote bands in TBG generate superconductivity by themselves.
The calculations we report in this work are well below this regime, such that superconductivity is only possible due to the TBG flat bands. For the value $J = 0.5\,t_{\pi} \approx 1.35$~eV used in the main text, we obtain $J_{\textrm{eff}} \approx 0.672\,t_{\pi}$, corresponding to $J_{\textrm{g}} \approx 0.800$~meV instead of $0.595$~meV in the plane-wave basis of Sec.~\ref{sec:atomistic:scales}. The minimal intra-Chern pairing model of the main text thus applies with $J$ replaced by $J_{\textrm{eff}}$ in $\tilde{J}$, when we also include the effects of the remote bands. Note that the axes in Figs.~8 and 9 of the main text are expressed in units of $\tilde{J}$ and their results are thus unaffected by this rescaling. 
A quantitative comparison between the atomistic and moir\'e flat-band-projected models, by contrast, is sensitive to this rescaling. Fig.~\ref{fig:remote}(c) compares the atomistic results of Fig.~4(a) of the main text with the moir\'e flat-band results on the same $4\times4$ momentum grid. Without rescaling, the flat-band description recovers only about half of the maximal atomistic $E_c$, and almost nothing near charge neutrality. 
Including the rescaling in Eq.~\eqref{eq:remote:Jeff} results in a near-perfect agreement instead.

The model only including the moir\'e flat bands can be solved on much finer momentum grids than the atomistic model, which allows us to test the $4\times4$ grid used in the main text. Fig.~\ref{fig:remote}(d) shows the ground state condensation energy of the intra-Chern pairing model at $J = 0.5\,t_{\pi}$ on grids from $4\times4$ to $65\times65$. The $4\times4$ grid captures the overall scale of $E_c$, but its pronounced maximum near $\mu \approx 0.5$~meV and its suppression near $\mu \approx 1.5$~meV are finite-size features that follow the sampled normal-state DOS, as noted in the main text. For denser momentum grids these features are washed out. We find that a $13\times13$ grid already gives approximately converged results. Finally, in Fig.~\ref{fig:remote}(e) we investigate the nature of the ground state on the converged $65\times65$ grid as a function of $\mu$ for different $J$. At $J = 0.5\,t_{\pi}$ (solid line) the ground state is nematic, $\alpha_\star = \pi/2$ (blue, right axis), wherever $E_c$ is appreciable (black, left axis). Reducing $J$ lowers $E_c$, while at the same time the chiral state ($\alpha_\star = 0$) takes over at the upper end of the doping range. Thus, the chiral phase appears at the edge of the superconducting dome, where it grows as $J$ decreases, in line with the preference for the chiral state at weak interactions and also a doping-induced nematic-to-chiral transition, established in Sec.~V of the main text.

\section{Hybrid atomistic-continuum approach for symmetry-broken parent states}
\label{sec:parentstate:hybrid}

Incorporating symmetry-broken parent states in atomistic simulations is challenging. Heterostrain, for example, changes the commensurability between the layers and thereby the construction of the moir\'e unit cell \cite{Escudero2026}, while lattice relaxation and spontaneous symmetry breaking require additional, computationally intensive calculations. Continuum formulations such as the topological heavy fermion (THF) model \cite{Song2022, Calugaru2023} offer a computationally much cheaper route, and parametrizations of lattice relaxation, non-local tunneling, and heterostrain within this model have recently become available \cite{HerzogArbeitman2025b}.

We therefore treat particle-hole asymmetry, heterostrain, and Chern polarization in the parent states of Fig.~10(a,b,d) of the main text within a hybrid approach. We keep the atomistically resolved $d$-wave pairing projected onto the Chern basis, Eq.~(15) of the main text, but determine its amplitude $\Delta_0$ and angles $\alpha$ and $\phi$ for each symmetry-broken parent state and chemical potential, by maximizing the condensation energy. For the normal state, instead of using Eq.~\eqref{eq:atomistic:normal:overlap} to obtain Eq.~(16) of the main text, we now use the THF Hamiltonian including the symmetry-breaking terms, projected onto a fixed Chern basis: that of the THF flat bands with particle-hole asymmetry but without heterostrain or Chern polarization (Sec.~\ref{sec:parentstate:hybrid:model}). In doing so, it is important to match the gauge of the two models so that the atomistic and THF Chern bases are compatible. This construction captures all modifications of the flat-band energies and the couplings between the Chern bands, but does not change the pairing form factors $d_{\pm}(\mathbf{k})$. This is justified as long as the symmetry-breaking and pairing terms are small compared with the separation between the flat and the remote bands. The inter-valley coherent parent state, which closes this band separation \cite{Herzog-Arbeitman2025}, is instead treated within the fully atomistic model in Sec.~\ref{sec:parentstate:ivc}, since the band-projected hybrid approach is not justified there. Below we first construct the hybrid model and then apply it to the particle-hole asymmetric, heterostrained, and Chern-polarized parent states.

\subsection{Hybrid atomistic-continuum model}
\label{sec:parentstate:hybrid:model}

To describe the normal state within our hybrid atomistic-continuum model approach, we need to first find the normal state within the THF model. The low-energy spectrum of pristine TBG in valley $\eta$ in the THF model is described by the Hamiltonian \cite{Song2022, Calugaru2023}
\begin{align} \label{eq:THF-pristine}
    h_{\eta}^{\textrm{THF}}(\mathbf{k})
    &= \begin{pmatrix}
        0 & v \mathbf{k} \cdot (\eta \sigma_0,i\sigma_z) & \gamma \sigma_0 + v^\prime \mathbf{k} \cdot (\eta \sigma_x, \sigma_y) \\
        v \mathbf{k} \cdot (\eta \sigma_0, -i\sigma_z) & M \sigma_x & 0 \\
        \gamma \sigma_0 + v^\prime \mathbf{k} \cdot (\eta \sigma_x, \sigma_y) & 0 & 0
    \end{pmatrix},
\end{align}
where for brevity we omit a damping factor, which is restored via $\gamma, v^{\prime} \rightarrow \gamma e^{-\lambda^2 k^2/2}, v^{\prime}e^{-\lambda^2 k^2/2}$. The Hamiltonian acts on six orbitals in each valley, with the Pauli matrices $\sigma_{0,x,y,z}$ acting in an orbital space tied to the sublattice degree of freedom. The upper $4\times4$ block describes four delocalized $c$ electrons, which disperse linearly with velocity $v$ around $\boldsymbol{\Gamma}$, where two of them are split by $M$. The lower $2\times2$ block describes two $f$ orbitals at zero energy, localized at the $AA$ sites on the scale $\lambda$ and hybridized with the $c$ electrons through $\gamma$ and $v^\prime$. We use the parameters fitted in Ref.~\cite{HerzogArbeitman2025b} to the Kang--Vafek (KV) model at $\theta = 1.05^\circ$ \cite{Kang2023, Vafek2023}, here collected in Table~\ref{tab:THF} together with those of the particle-hole asymmetric and heterostrain terms introduced below. Eq.~\eqref{eq:THF-pristine} is by itself not periodic in momentum space. Periodicity is restored by introducing copies of the $c$ electrons at $\mathbf{k}+\mathbf{G}$, for reciprocal lattice vectors $\mathbf{G}$, all coupled to the same $f$ orbitals through the off-diagonal blocks of Eq.~\eqref{eq:THF-pristine} evaluated at $\mathbf{k}+\mathbf{G}$ \cite{Song2022}. We keep the copies at $\mathbf{G}=0$ and at the $18$ shortest nonzero $\mathbf{G}$ (three shells), which converges the flat bands throughout the moir\'e Brillouin zone.

\begin{table}[t]
    \centering
    \[ \begin{array}{c|c|cccc|cccc|cccccc}
        \theta & \lambda & \gamma & M & v & v^\prime & \mu_1 & \mu_2 & v_1 & v_2 & c & c^\prime & c^{\prime \prime} & M_f & \gamma^\prime & M^\prime \\
        (^\circ) & (L_m) & (\textrm{meV}) & (\textrm{meV}) & (\textrm{eV Å}) & (\textrm{eV Å}) & (\textrm{meV}) & (\textrm{meV}) & (\textrm{eV Å}) & (\textrm{eV Å}) & (\textrm{meV}) & (\textrm{meV}) & (\textrm{meV}) & (\textrm{meV}) & (\textrm{meV}) & (\textrm{meV}) \\
        \hline
        1.05 & 0.3 & -14.5 & -8.5 & -3.8 & 1.2 & 23.8 & 9.8 & 0.2 & -0.5 & -6290 & 630 & 4990 & -7190 & -2810 & -3430
    \end{array} \]
    \caption{Parameter values for the THF model at twist angle $1.05^\circ$ extracted from Ref.~\cite{HerzogArbeitman2025b} and there fitted to the Kang--Vafek model \cite{Kang2023}.
    Parameters $\gamma$, $M$, $v$, and $v^\prime$ describe the fully symmetric low-energy dispersion, while $c$, $c^{\prime}$, $c^{\prime \prime}$, $M_f$, $\gamma^\prime$, and $M^\prime$ capture the effects of heterostrain. The particle-hole symmetry breaking terms $\mu_1$, $\mu_2$, $v_1$, $v_2$ incorporate the effects of non-local tunneling and lattice relaxation.}
    \label{tab:THF}
\end{table}

To combine the THF normal state with the atomistic pairing, we express the flat-band projection of the THF Hamiltonian in a Chern basis, which requires a gauge with the correct symmetry representations. Here and below, the flat bands are the two bands per valley closest to charge neutrality of the full THF Hamiltonian, including the $c$-electron copies; they are mixtures of $f$ and $c$ electrons, $f$-like except close to $\boldsymbol{\Gamma}$. In the basis of Eq.~\eqref{eq:THF-pristine}, $C_{2z}\mathcal{T}$ is represented by $D[C_{2z}\mathcal{T}] = (\sigma_x \oplus \sigma_x \oplus \sigma_x)\,\mathcal{K}$, where $\mathcal{K}$ is complex conjugation. It is an exact symmetry of $h^{\textrm{THF}}_{\eta}$, also with the particle-hole asymmetric terms of Sec.~\ref{sec:parentstate:phs} and with heterostrain \cite{HerzogArbeitman2025b}. At each $\mathbf{k}$ we choose the phase of each flat-band eigenstate $|b\,\eta\,\mathbf{k}\rangle_{\textrm{THF}}$, with $b=\pm$ the upper and lower band, such that $C_{2z}\mathcal{T}|b\,\eta\,\mathbf{k}\rangle_{\textrm{THF}} = |b\,\eta\,\mathbf{k}\rangle_{\textrm{THF}}$. This fixes each state up to a sign. We then define the Chern states as in Eq.~(13) of the main text,
\begin{align} \label{eq:THF-Chern-basis}
    |C \, \eta \, \mathbf{k} \rangle_{\textrm{THF}}^{\textrm{Chern}} = (|+ \, \eta \, \mathbf{k} \rangle_{\textrm{THF}} - i C \, |- \, \eta \, \mathbf{k} \rangle_{\textrm{THF}})/\sqrt{2}.
\end{align}
The states from the opposite valley are obtained by spinless time reversal $\mathcal{T}$. As discussed in Ref.~\cite{TBG_II}, the relative sign of the two band states leaves two solution branches, which differ in which combination carries which Chern number. We resolve this ambiguity by calculating the sublattice polarization of the Chern bands $S_{\textrm{THF}} = -(\sigma_z \oplus \sigma_z \oplus \sigma_z)$, and choosing the branch with $\sgn \langle S_{\textrm{THF}} \rangle = C\eta$. With this choice, $C$ is the sign of the Chern number of each band, and the Chern band $C$ has strongest weight in the $f$ orbital with $\langle \sigma_z \rangle = -C\eta$. For the parameters of Table~\ref{tab:THF}, we find $|\langle S_{\textrm{THF}}\rangle| \geq 0.8$ at all $\mathbf{k}$, so this assignment is unambiguous. In this Chern basis, the projection of the unperturbed THF Hamiltonian takes the same form as Eq.~(16) of the main text but with THF-derived parameters,
\begin{align} \label{eq:THF-Chern-normal}
    \tilde{h}^{\textrm{THF}}_{\eta}(\mathbf{k}) = \tilde{\epsilon}^{\textrm{THF}}_{\eta}(\mathbf{k}) \, \zeta_x - \tilde{\mu}^{\textrm{THF}}_{\eta}(\mathbf{k}) \, \zeta_0,
\end{align}
with $\tilde{\epsilon}^{\textrm{THF}}_{\eta} \geq 0$ being half the splitting of the two THF flat bands and $\tilde{\mu}^{\textrm{THF}}_{\eta}$ the chemical potential measured from their mean energy. Here and below, the superscript $\textrm{THF}$ marks quantities of the THF Hamiltonian without the heterostrain and Chern-polarizing perturbations of Secs.~\ref{sec:parentstate:heterostrain} and \ref{sec:parentstate:chern}, projected onto the Chern basis of Eq.~\eqref{eq:THF-Chern-basis}. The same symbols without superscript, as in Eqs.~(16) and (17) of the main text, refer to the atomistic model, and the normal state of the hybrid model, constructed below from $\tilde{h}^{\textrm{THF}}_{\eta}$, is denoted $\tilde{h}^{\textrm{hyb}}_{\eta}$.

To proceed, we note that within each model the relative phase between the Chern states with $C=+$ and $C=-$ is a gauge choice. Changing it rotates the kinetic and pairing terms together and leaves all observables unchanged. In the hybrid model, however, the kinetic term comes from the THF model and the pairing term from the atomistic model, each in its own gauge. The quasiparticle spectrum depends on the relative phase between the two terms, which is gauge invariant within one model, but which in the hybrid model is set by how the THF Chern states are identified with the atomistic ones. Fixing the gauge of the THF Chern basis such that it describes the same physical states as the atomistic one is therefore part of the construction of the hybrid model, not a gauge choice within it. The simplest identification, a real and positive inter-Chern kinetic element in both models, as in Eq.~(16) of the main text and Eq.~\eqref{eq:THF-Chern-normal}, fails for the parameters from the KV model. With these parameters, the THF flat bands host additional Dirac cones inside the moir\'e Brillouin zone, at which $[\tilde{h}^{\textrm{THF}}_{\eta}(\mathbf{k})]_{+-}$ vanishes and around which its phase winds by $\pm 2\pi$, whereas the atomistic flat bands have Dirac points only at the corners of the Brillouin zone. These cones are a property of the THF band structure for these parameters, not of the gauge, but matching the kinetic phases turns them into spurious vortices of the pairing phase relative to the normal state.


We therefore reference the relative phase in each model and valley $\eta$ to a pair of $AA$-localized, $f$-like orbitals $g_{\pm,\eta}$, one per Chern sector. The relative Chern phase is then $\rho_{\eta}(\mathbf{k}) = \arg \langle g_{+,\eta} | +\,\eta\,\mathbf{k}\rangle - \arg \langle g_{-,\eta} | -\,\eta\,\mathbf{k}\rangle$, which we evaluate in both models as $\rho_{\textrm{THF},\eta}$ and $\rho_{\textrm{atom},\eta}$. In the THF model, $g_{C,\eta}$ is the $f$ orbital with $\sigma_z = -\eta C$, which hosts most of the Chern band $C$. The atomistic model has no $f$ orbitals, so we construct $g_{\pm,\eta}$ from its two flat bands of valley $\eta$. We weight each site $\mathbf{r}$ by the momentum-integrated density $n^{AA}_{\eta}(\mathbf{r})$ of these bands within a disk of radius $0.3\,L_m$ around the $AA$ center, and set $n^{AA}_{\eta} = 0$ outside. Among the normalized superpositions of the flat-band Bloch states of valley $\eta$ with a given $C_{3z}$ eigenvalue, we then take the one that maximizes $\sum_{\mathbf{r}} n^{AA}_{\eta}(\mathbf{r})\,|g(\mathbf{r})|^2$. Here $C_{3z}$ rotates by $2\pi/3$, counterclockwise as seen from above the top layer, about the central carbon site of the $AA$ region. The eigenvalue $1$ gives $g_{\eta,\eta}$ and the eigenvalue $e^{-2\pi i\eta/3}$ gives $g_{-\eta,\eta}$, and each $g_{C,\eta}$ lies almost entirely ($99.9\%$) in the Chern sector $C$. These eigenvalues, $e^{i\pi(C-\eta)/3}$, differ from those of the THF $f$ orbitals, $e^{-2\pi i C/3}$, by the factor $e^{2\pi i\eta/3}$, because the atomistic $C_{3z}$ axis passes through a carbon site rather than a hexagon center. This construction is reminiscent of cell natural orbitals \cite{Mahalingam2026} and of earlier work \cite{Solovyev2007}.

The construction above leaves the phase of each $g_{C,\eta}$ free. We fix it by symmetry alone, requiring $g_{\pm,\eta}$ to transform as the THF $f$ orbitals do. The exact $C_{2y}\mathcal{T}$, with $C_{2y}$ the layer-exchanging twofold rotation of Fig.~3 of the main text, maps each $g_{C,\eta}$ onto itself and thus fixes it up to a sign. The approximate $C_{2z}\mathcal{T}$, about the center of the hexagon that contains the $AA$ site and lies on the $C_{2y}$ axis, maps $g_{+,\eta}$ onto $g_{-,\eta}$ and thus fixes their relative sign. The remaining common sign drops out of $\rho_{\eta}$, so neither $\rho_{\textrm{THF},\eta}$ nor $\rho_{\textrm{atom},\eta}$ involves a fitted parameter. Finally, we rotate the THF Chern states so that $\rho_{\textrm{THF},\eta}$ becomes $\rho_{\textrm{atom},\eta}$,
\begin{align} \label{eq:hybrid:regauge}
    |C\,\eta\,\mathbf{k}\rangle^{\textrm{Chern}}_{\textrm{THF}} \rightarrow e^{-iC\vartheta_{\eta, \mathbf{k}}/2}\,|C\,\eta\,\mathbf{k}\rangle^{\textrm{Chern}}_{\textrm{THF}},
    \quad
    \vartheta_{\eta, \mathbf{k}} = \rho_{\textrm{THF},\eta}(\mathbf{k}) - \rho_{\textrm{atom},\eta}(\mathbf{k}).
\end{align}
Since $\mathcal{T}$ maps the orbitals of one valley onto those of the other with $C \rightarrow -C$, it gives $\vartheta_{\eta,\mathbf{k}} = \vartheta_{-\eta,-\mathbf{k}}$, so it suffices to construct the orbitals in one valley.

The transformation Eq.~\eqref{eq:hybrid:regauge} preserves the Chern labels, the normal-state spectrum, $\mathcal{T}$, and $C_{2z}\mathcal{T}$. Since it acts on the THF Chern basis alone, it is a gauge transformation of the THF model but not of the hybrid model: it changes the BdG spectrum, by aligning the phase of the normal state with that of the atomistic pairing. In the rotated basis, any operator projected onto the Chern basis of Eq.~\eqref{eq:THF-Chern-basis} transforms with $U_{\eta}(\mathbf{k}) = e^{i\zeta_z\vartheta_{\eta, \mathbf{k}}/2}$. The normal state of the hybrid model is therefore
\begin{align} \label{eq:hybrid:regauge-h}
    \tilde{h}^{\textrm{hyb}}_{\eta}(\mathbf{k}) = U_{\eta}(\mathbf{k}) \left[ \tilde{h}^{\textrm{THF}}_{\eta}(\mathbf{k}) + \delta\tilde{h}_{\eta}(\mathbf{k}) \right] U^{\dagger}_{\eta}(\mathbf{k}),
\end{align}
where $\delta\tilde{h}_{\eta}$ is the projection of a perturbation, such as those of Secs.~\ref{sec:parentstate:heterostrain} and \ref{sec:parentstate:chern}, onto the basis of Eq.~\eqref{eq:THF-Chern-basis}. In particular, the unperturbed part becomes $U_{\eta}\tilde{h}^{\textrm{THF}}_{\eta}U^{\dagger}_{\eta} = \tilde{\epsilon}^{\textrm{THF}}_{\eta}(\cos\vartheta_{\eta, \mathbf{k}}\,\zeta_x - \sin\vartheta_{\eta, \mathbf{k}}\,\zeta_y) - \tilde{\mu}^{\textrm{THF}}_{\eta}\,\zeta_0$, i.e.~its inter-Chern element acquires the phase $e^{i\vartheta_{\eta, \mathbf{k}}}$ and the normal state is no longer real. For a rotation-symmetric parent state, this phase has little effect on the energetics. It becomes important, however, once the parent state breaks rotation symmetry, such as under heterostrain (Sec.~\ref{sec:parentstate:heterostrain}), where it is also crucial for obtaining the correct favored nematic direction.

Finally, we use $\tilde{h}^{\textrm{hyb}}_{\eta}$ of Eq.~\eqref{eq:hybrid:regauge-h} in place of the atomistic normal state $\tilde{h}_{\eta}$ in the intra-Chern BdG Hamiltonian, Eq.~(17) of the main text,
\begin{align} \label{eq:hybrid:bdg}
    \tilde{H}^{\textrm{hyb}}_{\textrm{BdG}}(\mathbf{k}) =
    \begin{pmatrix}
        \tilde{h}^{\textrm{hyb}}_{+}(\mathbf{k}) & \tilde{\Delta}(\mathbf{k}) \\
        \tilde{\Delta}^{\dagger}(\mathbf{k}) & -\tilde{h}^{\textrm{hyb}\,*}_{-}(-\mathbf{k})
    \end{pmatrix}.
\end{align}
Here $\tilde{\Delta}$ is the inter-valley sector of the projected pairing matrix obtained from $\hat{\Delta}$ in Eq.~(3) of the main text, expressed in the Chern basis. Its intra-Chern elements carry $99.5\%$ of the weight, so that it takes the form of Eq.~(15) of the main text to this accuracy. For parent states that preserve both $\mathcal{T}$ and $C_{2z}\mathcal{T}$, such as the particle-hole asymmetric and heterostrained ones, $\tilde{h}^{\textrm{hyb}}_{-}(-\mathbf{k}) = \tilde{h}^{\textrm{hyb}}_{+}(\mathbf{k})$, and the hole block thus becomes $-\tilde{h}^{\textrm{hyb}\,*}_{+}(\mathbf{k})$.

We also note that the superconducting condensation energy follows from the quasiparticle energies of Eq.~\eqref{eq:hybrid:bdg} as in Eq.~\eqref{eq:remote:Jeff}, i.e.~it is necessary to include the remote-band renormalization $J \rightarrow J_{\textrm{eff}}$ of the mean-field term, with $J = 0.5\,t_{\pi} \approx 1.35$~eV used in the main text. At each chemical potential $\mu$ we maximize the condensation energy with respect to $\Delta_0$, $\alpha$, and $\phi$ at zero temperature. All calculations within the hybrid model use a $65\times65$ momentum grid on which the atomistic pairing matrices are computed, which is closed under $\mathbf{k}\rightarrow-\mathbf{k}$ and $C_{3z}$. Technically, the atomistic ($\theta \approx 1.2^\circ$) and THF ($\theta = 1.05^\circ$) moir\'e Brillouin zones, both close to the magic angle of the respective model, are matched by expressing momenta in units of the respective reciprocal lattice vectors. The filling $\nu$ is computed as in Eq.~\eqref{eq:atomistic:filling}, from the four projected flat bands, including any parent-state perturbation.

\subsection{Symmetry-broken parent state: particle-hole asymmetry}
\label{sec:parentstate:phs}

At small twist angles, the continuum model of TBG with local interlayer tunneling has an approximate particle-hole symmetry $\mathcal{P}$ \cite{Song2019b, TBG_II}, which our atomistic model of rigid graphene layers also approximately inherits. However, non-local interlayer tunneling and lattice relaxation break this symmetry \cite{Kang2023, Vafek2023, HerzogArbeitman2025b}. In the THF model these effects enter through the additional terms \cite{HerzogArbeitman2025b}
\begin{align} \label{eq:THF-phs}
    \delta h_{\eta}^{\mathcal{P}}(\mathbf{k})
    &= \begin{pmatrix}
        \mu_{1} \sigma_0 + v_2 \mathbf{k} \cdot (\eta \sigma_x, \sigma_y) & v_1 \mathbf{k} \cdot (\eta \sigma_x, -\sigma_y) & 0 \\
        v_1 \mathbf{k} \cdot (\eta \sigma_x, -\sigma_y) & \mu_{2} \sigma_0 & 0 \\
        0 & 0 & 0
    \end{pmatrix},
\end{align}
which act only on the $c$ electrons. The constant terms $\mu_1$ and $\mu_2$ shift the two pairs of $c$ electrons by different energies, while $v_1$ and $v_2$ add velocity terms. These terms respect the crystalline symmetries, including $C_{2z}\mathcal{T}$, and $\mathcal{T}$, but break $\mathcal{P}$ \cite{HerzogArbeitman2025b}. Since $\mathcal{P}$ is the only (approximate) operation relating $\mathbf{k}$ and $-\mathbf{k}$ within one valley, breaking it lifts the relation $\epsilon_{+,\eta}(\mathbf{k}) = -\epsilon_{-,\eta}(-\mathbf{k})$, while time reversal still guarantees $\epsilon_{b\eta}(\mathbf{k}) = \epsilon_{b,-\eta}(-\mathbf{k})$. Since these terms are part of the parent state in all our hybrid calculations and preserve $C_{2z}\mathcal{T}$, we add them to the unperturbed Hamiltonian: the Chern basis, Eq.~\eqref{eq:THF-Chern-basis}, and $\tilde{h}^{\textrm{THF}}_{\eta}$, Eq.~\eqref{eq:THF-Chern-normal}, are then obtained from the flat bands of $h^{\textrm{THF}}_{\eta} + \delta h^{\mathcal{P}}_{\eta}$. Unless stated otherwise, all results from the hybrid model, including those for heterostrain and Chern polarization, include Eq.~\eqref{eq:THF-phs}.

With the parameters from the KV model in Table~\ref{tab:THF}, the terms Eq.~\eqref{eq:THF-phs} make the flat-band spectrum strongly asymmetric. Fig.~10(a) of the main text compares the condensation energy at zero strain with (red) and without (black) the terms of Eq.~\eqref{eq:THF-phs}. Without them, $E_c$ forms a single dome, symmetric around charge neutrality. Particle-hole asymmetry makes the dome lopsided, resulting in a large $E_c$ at the lower band edge that reflects an extreme flatness of the lower band. Notably, particle-hole asymmetry however does not change the overall magnitude of $E_c$, consistent with the robustness of intra-Chern pairing discussed in the main text.

\subsection{Symmetry-broken parent state: heterostrain}
\label{sec:parentstate:heterostrain}


Heterostrain, an asymmetric strain of the two layers \cite{Huder2018, Qiao2018}, is ubiquitous in TBG samples \cite{Kazmierczak2021}. It increases the splitting of the flat bands \cite{Bi2019, Mesple2021} and breaks their $C_{3z}$ symmetry \cite{Bi2019}. Following Refs.~\cite{Bi2019, HerzogArbeitman2025b}, we consider strain tensors $\mathcal{E}_{1} = -\mathcal{E}_{2}$ that are opposite in the two layers, with isotropic part $\varepsilon_{+}$ and anisotropic and shear parts combined into $\boldsymbol{\varepsilon}=(\varepsilon_{xy}, \, \varepsilon_{-}) = \varepsilon_{\textrm{an}} \, (\sin 2 \theta_{\varepsilon}, \, \cos 2 \theta_{\varepsilon})$. For a uniaxial heterostrain of strength $\varepsilon$ along the axis at angle $\theta_{\varepsilon}$ from the $x$-axis of Fig.~3 of the main text, the Poisson contraction across the axis gives $\varepsilon_{+} = (\nu_G - 1)\,\varepsilon/2$ and $\varepsilon_{\textrm{an}} = -(1 + \nu_G)\,\varepsilon/2$, with $\nu_G = 0.16$ \cite{Woo2016, Bi2019}. In the THF model, heterostrain enters through \cite{HerzogArbeitman2025b, Herzog-Arbeitman2025}
\begin{align} \label{eq:THF-perturbation}
    \delta h_{\eta}^{\varepsilon}
    &= \begin{pmatrix}
        c \boldsymbol{\varepsilon} \cdot (\sigma_x, \eta \sigma_y) & c^\prime \boldsymbol{\varepsilon} \cdot (\sigma_x, -\eta \sigma_y) & +i \eta \gamma^\prime \varepsilon_{+} \sigma_z \\
        c^\prime \boldsymbol{\varepsilon} \cdot (\sigma_x, -\eta \sigma_y) & M^\prime \varepsilon_{+} \eta \sigma_y & c^{\prime \prime} \boldsymbol{\varepsilon} \cdot (\sigma_0, -i \eta \sigma_z) \\
        -i \eta \gamma^\prime \varepsilon_{+} \sigma_z & c^{\prime \prime} \boldsymbol{\varepsilon} \cdot (\sigma_0, +i \eta \sigma_z) & M_f \boldsymbol{\varepsilon} \cdot (\sigma_x, \eta \sigma_y)
    \end{pmatrix}.
\end{align}
This term preserves $\mathcal{T}$ and $C_{2z}\mathcal{T}$, but breaks $C_{3z}$.



We include the particle-hole asymmetric terms of Sec.~\ref{sec:parentstate:phs} exactly, in the Hamiltonian $h^{\textrm{THF}}_{\eta} + \delta h^{\mathcal{P}}_{\eta}$ from which the flat bands and the Chern basis are obtained. Heterostrain, in contrast, we treat as a perturbation within the flat-band subspace of this unstrained Hamiltonian. We thus project $\delta h^{\varepsilon}_{\eta}$ onto the Chern states of Eq.~\eqref{eq:THF-Chern-basis}. This keeps a single reference basis for all strain strengths and axes, in which the pairing form factors and the phase reference of Sec.~\ref{sec:parentstate:hybrid:model} are fixed, and it makes the strain-induced coupling between the two Chern bands explicit. The projection neglects the strain-induced mixing of the flat bands with the other bands, which affects the flat-band energies only at second order in $\delta h^{\varepsilon}_{\eta}$, while the strain-induced couplings within the flat bands are kept exactly. This approach is justified as long as heterostrain does not change the flat-band subspace substantially. We checked this at $\varepsilon = 0.2\%$ and $\theta_{\varepsilon} = 0$, where the overlap of the flat-band subspaces with and without strain exceeds $90\%$ except close to $\boldsymbol{\Gamma}$, where intra-Chern pairing has a node. The projection of Eq.~\eqref{eq:THF-perturbation} onto the Chern basis takes the form
\begin{align} \label{eq:THF-projected}
    \delta \tilde{h}^{\varepsilon}_{\eta}(\mathbf{k}) =
    \begin{pmatrix}
        -\tilde{\mu}_{\varepsilon,\eta}(\mathbf{k}) & w_{\varepsilon,\eta}(\mathbf{k}) \\
        w^{*}_{\varepsilon,\eta}(\mathbf{k}) & -\tilde{\mu}_{\varepsilon,\eta}(\mathbf{k})
    \end{pmatrix},
\end{align}
where the intra-Chern elements are equal by $C_{2z}\mathcal{T}$ symmetry. The complex inter-Chern element $w_{\varepsilon,\eta}$ dominates; its root-mean-square value is about four times that of $\tilde{\mu}_{\varepsilon,\eta}$. It stems mainly from the $f$-orbital term $M_f$. The Chern band $C$ of valley $\eta$ resides mainly in the $f$ orbital with $\sigma_z = -C\eta$ ($86\%$ of its weight on average over the Brillouin zone). The factor $\eta$ of the $M_f$ term in Eq.~\eqref{eq:THF-perturbation} therefore drops out, and the phase of $w_{\varepsilon,\eta}$ turns approximately $e^{-2i\theta_{\varepsilon}}$ with the strain axis $\theta_{\varepsilon}$ in both valleys. The $c$--$f$ coupling $c^{\prime\prime}$ contributes a component $\propto e^{+2i\theta_{\varepsilon}}$ to $w_{\varepsilon,\eta}$ about three times weaker, but all other terms are smaller by an order of magnitude.

The effect of heterostrain on intra-Chern pairing is most transparent at a single momentum. By Eq.~\eqref{eq:hybrid:regauge-h}, the inter-Chern element of $\tilde{h}^{\textrm{hyb}}_{\eta}$ with heterostrain included is $w_{\eta}(\mathbf{k}) = e^{i\vartheta_{\eta, \mathbf{k}}}\,[\tilde{\epsilon}^{\textrm{THF}}_{\eta}(\mathbf{k}) + w_{\varepsilon,\eta}(\mathbf{k})]$, with $\mathcal{T}$ implying $w_{-\eta}(\mathbf{k}) = w_{\eta}(-\mathbf{k})$. Consider the Cooper pair formed by the electron in $(C,\eta,\mathbf{k})$ and its partner in $(C,-\eta,-\mathbf{k})$. A further rotation of the two Chern states of both partners by $e^{iC\arg w_{\eta}(\mathbf{k})/2}$ makes the inter-Chern element real and shifts the pairing phase of Eq.~(15) of the main text by $2\arg w_{\eta}(\mathbf{k})$. The single-mode results of the main text therefore hold for this pair with $\tilde{\epsilon} \rightarrow |w_{\eta}(\mathbf{k})|$, $\tilde{\mu} \rightarrow \tilde{\mu}^{\textrm{THF}}_{\eta} + \tilde{\mu}_{\varepsilon,\eta}$, and $\phi_{\mathbf{k}} \rightarrow \phi + \delta\phi_{\eta\mathbf{k}} + 2\arg w_{\eta}(\mathbf{k})$. In particular, its condensation energy depends on $\phi$ only through $\cos(\phi + \delta\phi_{\eta\mathbf{k}} + 2\arg w_{\eta})$ and is now largest for
\begin{align} \label{eq:heterostrain_star}
    \phi_{\star,\eta,\mathbf{k}} = - \left[ \delta\phi_{\eta\mathbf{k}} + 2 \arg w_{\eta}(\mathbf{k}) \right] \bmod{2\pi},
\end{align}
the nematic direction favored by this pair. Time-reversal symmetry implies $\phi_{\star,-,\mathbf{k}} = \phi_{\star,+,-\mathbf{k}}$. The two valleys thus favor the same direction at a given $\mathbf{k}$ only where $\phi_{\star,\eta,\mathbf{k}}$ is even in $\mathbf{k}$. Because $\mathcal{P}$ is the only operation relating $\mathbf{k}$ and $-\mathbf{k}$ within a valley, the particle-hole asymmetric terms make $\phi_{\star,\eta,\mathbf{k}}$ valley dependent, most strongly close to $\boldsymbol{\Gamma}$. The two valleys therefore do not cooperate perfectly in selecting a nematic direction, but we still find this valley effect to be small. At a single momentum, heterostrain thus mainly increases the band splitting $|w_{\eta}|$, which reduces the condensation energy, and shifts the favored nematic direction via Eq.~\eqref{eq:heterostrain_star}.

\begin{figure}
	\includegraphics[width=\columnwidth]{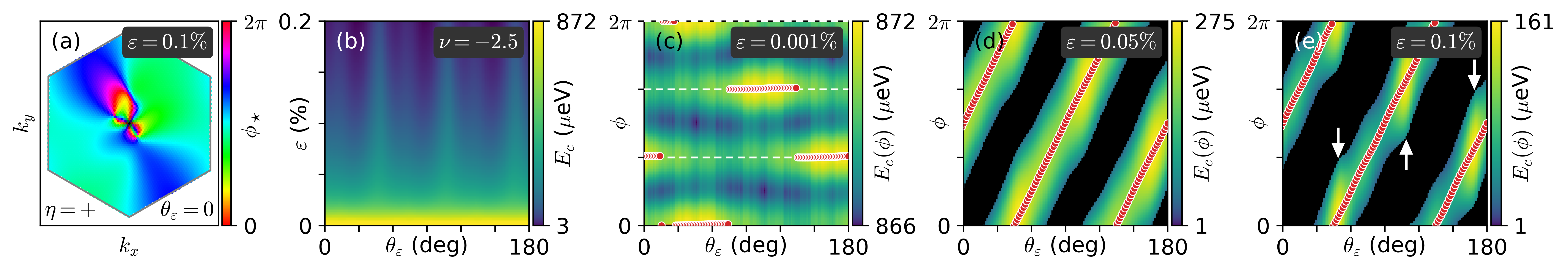}
	\caption{Intra-Chern pairing under heterostrain. (a) Favored nematic direction $\phi_{\star,+,\mathbf{k}}$, Eq.~\eqref{eq:heterostrain_star}, in the Brillouin zone for $\varepsilon=0.1\%$ and $\theta_{\varepsilon}=0$; it does not depend on the filling. (b) Ground-state condensation energy $E_c$ at $\nu = -2.5$ versus heterostrain strength $\varepsilon$ and axis $\theta_{\varepsilon}$, with a nematic ground state throughout. (c-e) $E_c$ of the nematic states at $\nu = -2.5$ versus $\phi$ and $\theta_{\varepsilon}$ for (c) $\varepsilon=0.001\%$, (d) $0.05\%$, and (e) $0.1\%$, with the ground state $\phi_\star$ (red dots). Black marks regions with $E_c < 1~\mu$eV. Dashed lines in (c) mark the three $C_{3z}$-related nematic states of the symmetric parent state and white arrows in (e) mark the maxima of $E_c$. Hybrid model of Sec.~\ref{sec:parentstate:hybrid} with $J=1.35$~eV.}
	\label{fig:heterostrain}
\end{figure}

In Fig.~\ref{fig:heterostrain}(a) we investigate the results above by plotting $\phi_{\star,+,\mathbf{k}}$ at $\varepsilon = 0.1\%$ and $\theta_{\varepsilon} = 0$. Without strain, the favored nematic direction differs strongly between $C_{3z}$-related momenta, see Fig.~1(c) of the main text, which frustrates the nematic state. At $\varepsilon = 0.1\%$, $w_{\varepsilon,+}$ already exceeds the kinetic energy at most momenta, and the momentum dependence of $2\arg w_{+}$ somewhat compensates that of $\delta\phi_{\mathbf{k}}$, to reduce the strong $C_{3z}$-frustration. Away from $\boldsymbol{\Gamma}$, $\phi_{\star,+,\mathbf{k}}$ in fact approaches $\pi$, i.e.~$\hat{\Delta}_y$. Full momentum integration reveals that $\hat{\Delta}_y$ is indeed the ground state at this strain and axis. The compensation is not exact, which leaves some residual variations that decrease at larger strain. We further note that close to $\boldsymbol{\Gamma}$, the flat bands lose their $f$ character, $w_{\varepsilon,+}$ vanishes together with it, and $\phi_{\star,+,\mathbf{k}}$ thus keeps following the winding of $\delta\phi_{\mathbf{k}}$. There, however, the pairing amplitude is small, so these momenta contribute little to the condensation energy.

We next consider full momentum space integration. In Fig.~\ref{fig:heterostrain}(b) we show the condensation energy $E_c$ as a function of the heterostrain strength $\varepsilon$ and axis $\theta_{\varepsilon}$. As $\varepsilon$ increases, $E_c$ is continuously suppressed and develops a modulation with period $60^\circ$ in $\theta_{\varepsilon}$. The ground state shows a $4\pi$ winding of the nematic direction with $\theta_{\varepsilon}$, also reported in Fig.~10(b) of the main text, whose origin we now explain. Without strain, $C_{3z}$ maps $\phi \rightarrow \phi + 2\pi/3 \bmod{2\pi}$, so that the three nematic states $\phi = 0, \, \pm 2\pi/3$ of the symmetric parent state are degenerate. Heterostrain lifts this degeneracy by coupling $\phi$ to the strain axis, and the form of this coupling follows from the $d$-wave nature of the pairing. Since the two chiral components carry angular momenta $\pm 2$, a rotation of the system by an angle $\beta$ changes their relative phase $\phi$ by $4\beta$, while it changes the strain axis $\theta_{\varepsilon}$ by $\beta$. Where the strain dominates the flat-band kinetic energy, it enters mainly through the $f$-orbital term $M_f$, which is covariant under all rotations. The dominant part of the coupling between $\phi$ and the strain axis is then rotation invariant and can depend only on $\phi - 4\theta_{\varepsilon}$. It locks the ground state to $\phi_\star = 4\theta_{\varepsilon} + \textrm{const}$, which winds by $4\pi$ as the strain axis turns by $\pi$. The single-mode picture above shows how this coupling arises. The favored direction of each mode, Eq.~\eqref{eq:heterostrain_star}, contains the phase $-2\arg w_{\eta}(\mathbf{k})$, with a factor $2$ since both electrons of a Cooper pair must hop between the Chern sectors to connect the pairing in one sector with that in the other. In this regime, $w_{\eta}(\mathbf{k})$ turns approximately as $e^{-2i\theta_{\varepsilon}}$ with the strain axis, so that the favored direction of each mode shifts by $4\theta_{\varepsilon}$, and this coupling grows as $\varepsilon^2$. The exact rotation symmetry of the lattice is only $C_{3z}$, which also allows a coupling through $\phi + 2\theta_{\varepsilon}$. This coupling is linear in $\varepsilon$ and stems mainly from the interference of the strain with the kinetic energy, but its coefficient is small, and the quadratic coupling exceeds it already for $\varepsilon \gtrsim 0.003\%$.

We finally illustrate the analysis above with numerical calculations. At extremely weak heterostrain, $\varepsilon = 0.001\%$, the landscape of $E_c$ versus $\phi$ and $\theta_{\varepsilon}$ in Fig.~\ref{fig:heterostrain}(c) is quite flat, with faint horizontal stripes at the ground states of the symmetric system (white dashed lines), signaling the weak $C_{3z}$-symmetric dependence on $\phi$. Heterostrain already makes these three states inequivalent, each being the ground state (red dots) in a different range of $\theta_{\varepsilon}$. Above $\varepsilon \approx 0.01\%$, well below typical sample values \cite{Kazmierczak2021, Mesple2021}, the quadratic coupling exceeds the $C_{3z}$-symmetric dependence on $\phi$, and $\phi_\star$ winds continuously. Fig.~\ref{fig:heterostrain}(d) shows this for $\varepsilon = 0.05\%$, where $\phi_\star \simeq 4\theta_{\varepsilon} + \pi$ (red dots) winds by $4\pi$ as the strain axis turns by $\pi$, in agreement with the analysis above. At $\varepsilon = 0.1\%$, in Fig.~\ref{fig:heterostrain}(e), an anisotropy in $\theta_{\varepsilon}$ develops, such that three heterostrain directions yield a higher $E_c$ (white arrows). Finally, in real space the twofold axis of the nematic bond pattern lies at $-\phi/2$ from the $x$-axis, i.e.~at $\theta_{\textrm{nem}} = \pi/2 - 2\theta_{\varepsilon}$. Strain along $x$ thus selects $\hat{\Delta}_y$, and the nematic pairing axis turns twice as fast as the strain axis, in the opposite sense. Since $\phi_\star$ changes by $2\pi$ when the strain axis turns by $\pi/2$, perpendicular strain axes select the same nematic state, but with different condensation energies, since they are not related by $C_{3z}$.

\subsection{Symmetry-broken parent state: Chern polarization}
\label{sec:parentstate:chern}

Interactions in the flat bands of TBG can favor Chern-polarized states, in which the two Chern sectors are unequally occupied \cite{Bultinck2020}, as observed in zero-field Chern insulators \cite{Sharpe2019, Serlin2020, Stepanov2021}. We model such a parent state phenomenologically by an energy splitting $B_f$ of the $f$ orbitals,
\begin{align} \label{eq:THF-Bf}
    \delta h_{\eta}^{B} = 0 \, \oplus \, 0 \, \oplus \, (-\eta B_f \sigma_z),
\end{align}
added to $h^{\textrm{THF}}_{\eta} + \delta h^{\mathcal{P}}_{\eta}$. Since the Chern band $C$ resides mainly in the $f$ orbital with $\sigma_z = -C\eta$, see Sec.~\ref{sec:parentstate:hybrid:model}, this term shifts the Chern band $C$ by $C B_f$ times its $f$-orbital polarization, with the same sign in both valleys. It breaks both $\mathcal{T}$ and $C_{2z}\mathcal{T}$, each of which maps the Chern band $C$ onto $-C$, as required for a net Chern polarization.

As for heterostrain, we project $\delta h^{B}_{\eta}$ onto the Chern basis constructed at $B_f = 0$, Eq.~\eqref{eq:THF-Chern-basis}. This is necessary, since $\delta h^{B}_{\eta}$ breaks the $C_{2z}\mathcal{T}$ symmetry on which the construction of the Chern basis from the THF bands relies. The projection is diagonal in Chern index and therefore unaffected by the rotation $U_{\eta}$ of Eq.~\eqref{eq:hybrid:regauge-h},
\begin{align} \label{eq:THF-Bf-projected}
    \delta\tilde{h}^{B}_{\eta}(\mathbf{k}) = B_f \, p_f(\mathbf{k}) \, \zeta_z,
\end{align}
where $p_f(\mathbf{k}) = |\langle C\,\eta\,\mathbf{k}|\,0\oplus0\oplus\sigma_z\,|C\,\eta\,\mathbf{k}\rangle|$ is the $f$-orbital polarization of the Chern bands. $p_f$ is the same for both Chern bands, by $C_{2z}\mathcal{T}$, and for the states at $\mathbf{k}$ in valley $\eta$ and at $-\mathbf{k}$ in valley $-\eta$, by $\mathcal{T}$. It vanishes at $\boldsymbol{\Gamma}$, where the flat bands have no $f$ character, and is close to one wherever the flat bands are $f$-like. In the BdG Hamiltonian, Eq.~\eqref{eq:hybrid:bdg}, the electron in $(C, +, \mathbf{k})$ and its Cooper partner in $(C, -, -\mathbf{k})$ are therefore shifted by the same energy $C B_f p_f(\mathbf{k})$. Chern polarization thus acts as a Chern-dependent chemical potential, $\tilde{\mu}^{\textrm{THF}}_{\eta} \rightarrow \tilde{\mu}^{\textrm{THF}}_{\eta} - C B_f p_f(\mathbf{k})$, which is not pair breaking for intra-Chern pairing but moves the two Chern sectors in opposite directions relative to the Fermi level. This term thus strongly favors a chiral superconducting ground state because it suppresses pairing in one of the Chern sectors by moving it away from the Fermi level.

Fig.~10(d) of the main text shows the resulting phase diagram. The ground state remains nematic up to $|B_f| \approx 0.2$--$0.9$~meV, depending on $\nu$, and becomes chiral above with a continuous crossover with intermediate $\alpha_\star$. The chiral ground state has $\alpha_\star = 0$ ($\hat{\Delta}_+$) for $B_f\nu>0$ and $\alpha_\star=\pi$ ($\hat{\Delta}_-$) for $B_f\nu<0$. At $\nu=0$, a gap opens between the two Chern sectors for $|B_f| \gtrsim 1.2$~meV and strongly suppresses superconductivity. At the lower band edge, $\nu=-4$, $E_c$ instead remains finite in the range of $B_f$ considered, owing to the extreme flatness of the lower band.

\section{Symmetry-broken parent state: inter-valley coherence}
\label{sec:parentstate:ivc}

Symmetry breaking of the parent state for superconductivity in TBG may occur via the formation of emergent particle-hole ordering at low temperatures \cite{Cao2018, Cao2018b, Yankowitz2019, Choi2019, Kerelsky2019, Sharpe2019, Xie2019, Jiang2019, Lu2019, Serlin2020, Wong2020, Zondiner2020, Stepanov2020, Saito2020, Nuckolls2020, Choi2021, Das2021, Park2021, Rozen2021, Saito2021, Cao2021a, Grover2022, Nuckolls2023}. The most prominent superconducting domes typically appear by doping away from a correlated insulator state at $\nu=-2$, which recent experiments point to being inter-valley-coherent (IVC) in both twisted bilayer and trilayer graphene \cite{Nuckolls2023, Kim2023, Kim2026}. How far such electronic order persists at fillings close to the superconducting domes, and whether its nature changes there, is still unclear, but superconductivity may well emerge from such an interaction-driven, symmetry-broken parent state. We here address this possibility by fully atomistic, self-consistent calculations of superconductivity in the presence of an inter-valley-coherent background. We first briefly review the current understanding of the IVC insulator in TBG, which is formulated in the language of the THF model of Sec.~\ref{sec:parentstate:hybrid:model} \cite{Herzog-Arbeitman2025}, and then construct its atomistic counterpart in Sec.~\ref{sec:parentstate:ivc-intra-Chern-atomistic}. A fully atomistic treatment is required here because the IVC parent state closes the energy gap between the moir\'e flat bands and the remote bands \cite{Herzog-Arbeitman2025}, so that the band-projected hybrid approach of Sec.~\ref{sec:parentstate:hybrid} is not justified.

It is by now understood that the symmetry-broken orders of TBG emerge following a hierarchy of interactions \cite{Po2018b, Kang2019, Xie2020Insulator, Bultinck2020, Vafek2020, TBG_III}, where the long-range Coulomb interaction corresponds to the highest energy scale. It was initially proposed that Coulomb interactions lead to a Kramers inter-valley-coherent (K-IVC) ground state at even integer fillings \cite{YiZhang2020, Bultinck2020}. Shortly thereafter, however, heterostrain was shown to destabilize such a state \cite{DanielEParker2021} and experimental data later revealed Kekul\'e distortions near $|\nu|=2$ \cite{Nuckolls2023} that are incompatible with the K-IVC state \cite{Calugaru2022}. In samples with ultra-low heterostrain, current data are instead suggestive of a time-reversal-symmetric inter-valley-coherent (T-IVC) state, which has been shown to be stabilized when coupling to $K$-phonons dominates over the inter-valley Coulomb scattering \cite{Angeli2019, Angeli2020, Kwan2024, Shi2025, Wang2025}. In typical samples, however, heterostrain is larger and experimental data is compatible with an incommensurate Kekul\'e spiral state (IKS) \cite{Nuckolls2023}, on which we therefore focus.

The IKS state was early on shown to be stabilized by heterostrain \cite{Kwan2021}, while more recent results have also revealed that its spin structure is dictated by the strength of the coupling to $K$-phonons \cite{ZiweiWang2025}. In the scenario of strong coupling, which is supported by experiments \cite{Chen2024}, the resulting IKS is in the charge sector (i.e.~it has no spin texture). We here assume such a scenario, which is also relevant for intra-Chern pairing, since our work reveals that phonon-mediated and spin-fluctuation mechanisms for superconductivity can cooperate in this channel (see main text). In this scenario, a single state $|\xi\rangle$ is selected in each spin sector, which in the language of the THF model reads \cite{Herzog-Arbeitman2025}
\begin{align} \label{eq:ivc}
    |\xi\rangle = \tfrac{1}{2} \sum_{s \eta} \exp[\tfrac{i}{2} (\eta \, \phi_{\textrm{valley}} + s \, \phi_{\textrm{sublattice}} + s \, \eta \, \phi_{\textrm{Chern}}) ] \, |f_{s \eta} \rangle_{\textrm{THF}},
\end{align}
characterized by the set of momentum-independent angles $\phi_{\textrm{valley}}$, $\phi_{\textrm{Chern}}$, $\phi_{\textrm{sublattice}}$. Here, $|f_{s \eta} \rangle_{\textrm{THF}}$ is the band formed by the THF $f$ orbital of Eq.~\eqref{eq:THF-pristine} with sublattice $s$ and valley $\eta$. Since this orbital belongs mainly to the Chern band $C = -s\eta$ (see Sec.~\ref{sec:parentstate:hybrid:model}), $|\xi\rangle$ has equal weight in all four Chern bands and captures both intra- and inter-Chern inter-valley correlations.

To understand how $|\xi\rangle$ emerges in the presence of heterostrain, recall that the latter acts on the $f$ orbitals in two ways: it splits them in energy within each valley, and it simultaneously establishes a global phase coherence between the two Chern sectors, i.e.~a definite value of $\phi_{\textrm{Chern}}$ set by the heterostrain axis. In the THF model, both effects come mainly from the $f$-orbital term $M_f\,\boldsymbol{\varepsilon} \cdot (\sigma_x, \eta \sigma_y)$ of Eq.~\eqref{eq:THF-perturbation}, which couples the two $f$ orbitals of each valley with a phase that turns with the strain axis. Together, these effects leave a single band per valley and spin at lower energy. It has recently been argued that the selection of the correlated insulator parent state in the presence of heterostrain is then dominated by the development of inter-valley coherence exclusively in this lower-energy manifold \cite{Herzog-Arbeitman2025}. In this process, the values of $\phi_{\textrm{valley}}$ and $\phi_{\textrm{sublattice}}$ become fixed through spontaneous symmetry breaking. Then the long-ranged part of the Coulomb interaction lowers the energy of the selected state $|\xi\rangle$ by an amount $U_1$ \cite{Herzog-Arbeitman2025}. This is captured by a term 
\begin{align} \label{eq:ivc_THF_hamiltonian}
    h_{\textrm{IVC}}^{\textrm{THF}} = -U_1 \, |\xi\rangle\langle\xi|
\end{align}
in the effective THF normal-state Hamiltonian of each spin sector. Finally, the many-body parent state is constructed by filling both spin flavors of $|\xi\rangle$, yielding a spin-unpolarized many-body ground state compatible with experimental observations \cite{Cao2018, Yankowitz2019, Oh2021, Hoke2024}.

Due to the $C_{2z} \, \mathcal{T}$-protected topology of the flat bands \cite{Ahn2019, Song2019b, TBG_II}, the resulting band structure is still gapless, with Dirac points connecting the lower moir\'e flat band to the other flat bands. Such gaplessness is typically energetically unfavorable. However, it was recently shown that the Dirac point protection can be circumvented, while still preserving the $C_{2z} \, \mathcal{T}$ symmetry, through a valley-dependent momentum boost $\mathbf{q}$ that opens a gap by performing a non-Abelian braiding of the Dirac points \cite{Herzog-Arbeitman2025}. Such a boost breaks the moir\'e translation symmetry with a generally incommensurate wave vector, which makes it unfeasible to describe within an atomistic model. To circumvent this, we focus on the translation-symmetric parent state described by $|\xi\rangle$. This is an approximation of the $\mathbf{q}=0$ limit of the heterostrain-induced IKS state, and for this reason we refer to it as the IKS-precursor. In the future, it would be interesting to extend our investigation to the true IKS case by using e.g.~the THF description \cite{Song2022, Calugaru2023, Liu2024, Liu2025} combined with a full Hartree--Fock--Bogoliubov treatment of the interactions.

We also note that the spin-unpolarized charge IVC has been argued to be the ground state for filling $-2 \geq \nu \geq \nu_c$, with $\nu_c \approx -2.6$. Below $\nu_c$ spin-valley locking is expected to occur \cite{ZiweiWang2025}. We therefore focus on fillings around $\nu = -2$ with $\nu > \nu_c$, which are also relevant for superconductivity in TBG \cite{Cao2018b, Stepanov2021, Cao2021a}. Results for $\nu < \nu_c$, included in Fig.~\ref{fig:ivc_atomistic_results}(e) and Fig.~10(c) of the main text for completeness, should be read with this caveat. We also set $\phi_{\textrm{valley}} = \phi_{\textrm{sublattice}} = 0$, such that $\mathcal{T}$ and $C_{2z}$ symmetries are individually preserved. In particular, $C_{2z}\mathcal{T}$, which is thought to be important for the emergence of the IKS \cite{Herzog-Arbeitman2025}, is then preserved as well. Of the two conditions, time reversal requires $\phi_{\textrm{sublattice}} = 0 \bmod \pi$, while $C_{2z}\mathcal{T}$ fixes $\phi_{\textrm{valley}}$ relative to the rotation center. Finite values of $\phi_{\textrm{valley}}$ amount to a global valley rotation, i.e.~an atomic-scale shift of the Kekul\'e pattern, under which the state stays $C_{2z}\mathcal{T}$-symmetric up to this rotation \cite{Kwan2021}. We can therefore set it to zero, while finite $\phi_{\textrm{Chern}}$ rotates the Kekul\'e pattern. Without strain, values of $\phi_{\textrm{Chern}}$ related by a $C_{3z}$ rotation are equivalent; in a strained sample, its value is set by the heterostrain axis. The value used in our atomistic calculations is given in Sec.~\ref{sec:parentstate:ivc-intra-Chern-atomistic}.

We here use the THF description only to characterize the IVC parent state, i.e.~to fix the structure of $|\xi\rangle$ and the role of the angles $\phi_{\textrm{X}}$, with $\textrm{X} \in \{\textrm{valley}, \textrm{Chern}, \textrm{sublattice}\}$. Thus, unlike for the parent states of Sec.~\ref{sec:parentstate:hybrid}, we do not combine it with the atomistic pairing in a hybrid model. Instead, all our results for an IVC parent state are obtained in the fully atomistic model. This yields the most accurate results by accounting for all bands, especially important as the IVC perturbation closes the gap to the remote bands, and requires us to  develop the implementation of Eq.~\eqref{eq:ivc_THF_hamiltonian} within an atomistic framework, which we do next in Sec.~\ref{sec:parentstate:ivc-intra-Chern-atomistic}.

\subsection{Atomistic construction of IVC parent state}
\label{sec:parentstate:ivc-intra-Chern-atomistic}

The IKS-precursor Hamiltonian Eq.~\eqref{eq:ivc_THF_hamiltonian} effectively lowers one of the moir\'e flat bands, such that it overlaps in energy with the remote valence bands \cite{Herzog-Arbeitman2025}. A projection of the pairing onto the flat bands alone, as in the hybrid model of Sec.~\ref{sec:parentstate:hybrid}, might thus yield unreliable results. We therefore do not use the hybrid model here, and instead incorporate the IVC parent state directly in the atomistic model, where the superconducting pairing involves all bands. We then solve for the superconducting pairing matrix self-consistently and without constraints. Our main results are summarized in Fig.~10(c) of the main text, while we here provide the technical details behind these calculations. In order to perform them, we develop, to the best of our knowledge, the first implementation of the IKS-precursor state characterized by $|\xi\rangle$ in Eq.~\eqref{eq:ivc} within a fully atomistic framework. This corresponds to an atomistic Hamiltonian perturbation term $\delta \hat{h}_{\textrm{IVC}}$ to be added to the normal-state Hamiltonian Eq.~\eqref{eq:atomistic:normal:overlap}. Our implementation accounts for both the intra- and inter-Chern correlations known to be important for the development of the incommensurate Kekul\'e spiral \cite{Kwan2021}.

\begin{figure}
	\includegraphics[width=\columnwidth]{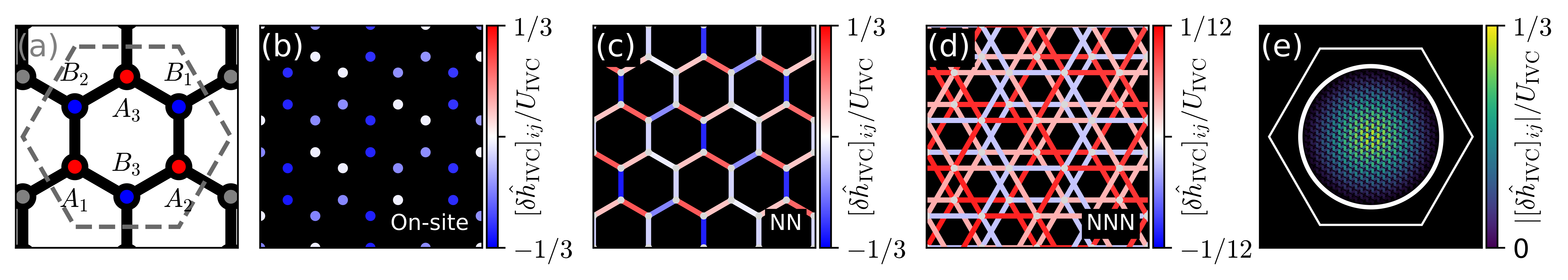}
	\caption{Atomistic construction of the IVC perturbation $\delta \hat{h}_{\textrm{IVC}}$, Eq.~\eqref{eq:IVC:TBGelements}. (a) Kekul\'e unit cell used to define the sublattice-polarized wavefunctions in Eq.~\eqref{eq:Kekule-wavefunctions}, with sites labeled $\kappa=A_{1,2,3}/B_{1,2,3}$ and dashed hexagon denoting the Kekul\'e cell boundary. (b-d) Top-layer real-space matrix elements $[\delta \hat{h}_{\textrm{IVC}}]_{ij}$ in the $AA$ region, separated into (b) on-site, (c) nearest-neighbor and (d) next-nearest-neighbor terms. Note the Kekul\'e-scale periodicity of all patterns. (e) Magnitude $|[\delta \hat{h}_{\textrm{IVC}}]_{ij}|$ across the top layer of the moir\'e unit cell. All non-vanishing elements are overlaid. Opacity matches intensity for better visualization of the relevant terms. White circle marks $r_{\textrm{max}}^{\textrm{IVC}} = \lambda_{\textrm{IVC}}$, beyond which the envelope vanishes, and the white hexagon marks the moir\'e unit-cell boundary, showing the perturbation is localized to the central $AA$ region. We use the angles of Sec.~\ref{sec:parentstate:ivc-intra-Chern-atomistic}; the matrix elements are real-valued since $\phi^{\textrm{g}}_{\textrm{sublattice}} = 0$. }
	\label{fig:ivc_atomistic_construction}
\end{figure}

To proceed we first note that the $f$ states in Eq.~\eqref{eq:ivc} are trivial (atomic-insulator) states built from Wannier orbitals peaked at the $AA$ regions of the moir\'e cell, with well-defined valley and sublattice character \cite{Song2022, Calugaru2023}. In each $AA$ region we therefore approximate their wavefunction in each layer by the corresponding monolayer state at the Dirac point. To describe inter-valley coherence we express these atomic-scale wavefunctions in the Kekul\'e basis~\cite{Chamon2000, Hou2007, Gutierrez2016}. The Kekul\'e description of graphene generates a tiling of the honeycomb lattice that corresponds to an enlarged monolayer unit cell containing three inequivalent sites per sublattice, labeled $\kappa=A/B_{1,2,3}$, as plotted in Fig.~\ref{fig:ivc_atomistic_construction}(a). Note that there are three inequivalent possible tilings, corresponding to different choices $A_{1,2,3}$ as the site at the origin. We find that in each layer $\ell$, the sublattice-polarized Kekul\'e wavefunctions $\psi_{s \eta}$ at the Dirac point are
\begin{align} \label{eq:Kekule-wavefunctions}
	\psi_{++} &= (1, \omega, \omega^*, 0, 0, 0)^T, &
	\psi_{+-} &= (1, \omega^*, \omega, 0, 0, 0)^T, &
	\psi_{-+} &= (0, 0, 0, 1, \omega^*, \omega)^T, &
	\psi_{--} &= (0, 0, 0, 1, \omega, \omega^*)^T.
\end{align}
We here use the basis $(A_1,A_2,A_3,B_1,B_2,B_3)^T$, where the subscripts $s=\pm$ label the sublattice $A/B$, $\eta=\pm$ label the valley, and with $\omega=e^{2\pi i/3}$. Now, let $|\xi(\ell)\rangle_{\textrm{g}}$ be the IVC state obtained by replacing $f\rightarrow\psi$ and $\phi_{\textrm{X}} \rightarrow \phi^{\textrm{g}}_{\textrm{X},\ell}$ in Eq.~\eqref{eq:ivc}, where the label $\textrm{g}$ indicates that the state and its angles are defined on the graphene monolayer, and $\ell$ denotes the layer. The angles $\phi^{\textrm{g}}_{\textrm{X},\ell}$ play the same roles as the THF angles $\phi_{\textrm{X}}$, but are related to them only through the phase reference of Sec.~\ref{sec:parentstate:hybrid:model}, as we show below. Its projector, $P_{\textrm{g}, \xi(\ell)} = |\xi(\ell)\rangle\langle\xi(\ell)|_{\textrm{g}}$, encodes the Kekul\'e structure within each layer. To turn $P_{\textrm{g}, \xi(\ell)}$ into the atomistic perturbation $\delta \hat{h}_{\textrm{IVC}}$, akin to Eq.~\eqref{eq:ivc_THF_hamiltonian}, on the full TBG lattice requires two further steps.

First, Eq.~\eqref{eq:Kekule-wavefunctions} only fixes the wavefunctions exactly at the monolayer Dirac point, so we must specify how the corresponding real-space matrix elements decay in order to extend $P_{\textrm{g}, \xi(\ell)} $ across the moir\'e Brillouin zone. Rather than a cumbersome extraction of this full decay from an atomistic Hartree-Fock treatment of the Coulomb interaction in the presence of heterostrain, we here retain only on-site, nearest-neighbor, and next-nearest-neighbor intra-layer terms for computational feasibility within an atomistic framework. Their amplitude is given by $[P_{\textrm{g}, \xi(\ell)}]_{\kappa  \kappa^\prime} / g_{\kappa \kappa^\prime}$, where $\kappa$, $\kappa^\prime$ denote the Kekul\'e site species and $g_{\kappa \kappa^\prime}$ is the multiplicity of the term: $g_{\kappa \kappa^\prime}=1$ for on-site and nearest-neighbor terms and $g_{\kappa \kappa^\prime}=3$ for next-nearest-neighbor terms. This truncation still reproduces $P_{\textrm{g}, \xi(\ell)}$ exactly at the Dirac point, and hence the correct valley and orbital structure, while also remaining Kekul\'e-periodic.

Second, the Kekul\'e tiling is incommensurate with the moir\'e unit cell, so imposing it uniformly on every site of the TBG lattice would require an enlarged supercell that is currently computationally unfeasible near the magic angle. We sidestep this by assigning each site's Kekul\'e index $\kappa_i$ according to the tiling of its nearest $AA$ center, and by suppressing the resulting domain walls between neighboring tilings with a phenomenological envelope function
\begin{align} \label{eq:IVC:envelope}
	\Theta_{ij} =
	\begin{cases}
		\tanh\!\left[ \left( r_{\textrm{max}}^{\textrm{IVC}} - \tfrac{1}{2}(r^{AA}_i + r^{AA}_j) \right) \, \lambda_{\textrm{IVC}}^{-1} \right]  \big/ \tanh\!\left[ r_{\textrm{max}}^{\textrm{IVC}} \, \lambda_{\textrm{IVC}}^{-1}  \right], & \tfrac{1}{2}(r^{AA}_i + r^{AA}_j) < r_{\textrm{max}}^{\textrm{IVC}}, \\[4pt]
		0, & \textrm{otherwise},
	\end{cases}
\end{align}
where $r^{AA}_i$ is the distance from site $i$ to the nearest $AA$ center. The envelope is maximal at the $AA$ center and decreases smoothly to zero at $r_{\textrm{max}}^{\textrm{IVC}}$, which avoids discontinuities between neighboring tilings and allows the problem to be treated within a single moir\'e unit cell. Since the flat bands are themselves strongly localized in the $AA$ regions, the IVC term acts mainly where the flat-band weight resides. We use $r_{\textrm{max}}^{\textrm{IVC}} = \lambda_{\textrm{IVC}} = 0.4\,L_m$, with $L_m$ the moir\'e length, for which $\Theta$ drops to $\approx 0.6$ at $0.2\,L_m$ from the $AA$ center, and check that the main spectral features are robust to small variations of these parameters.

Modulating the elements derived from $P_{\textrm{g}, \xi(\ell)}$ by $\Theta_{ij}$, and introducing the IVC amplitude $U_{\textrm{IVC}}$, gives the TBG lattice IVC perturbation for layer $\ell$. The matrix elements between sites $i,j \in \ell$ are non-zero only for on-site, nearest-neighbor and next-nearest-neighbor pairs, where they take the form
\begin{align} \label{eq:IVC:TBGelements}
	[\delta \hat{h}_{\textrm{IVC}}(\xi(\ell);\mathbf{k})]_{ij} = -U_{\textrm{IVC}} \, \frac{1}{g_{\kappa_{i} \kappa_{j}}} \sum_{\delta\mathbf{R}} e^{i\mathbf{k}\cdot\delta\mathbf{R}} \, [P_{\textrm{g}, \xi(\ell)}]_{\kappa_i \kappa_j} \, \Theta_{ij},
\end{align}
where the complex exponential is the same Bloch wrapping factor used in Eq.~\eqref{eq:atomistic:normal:overlap}. The atomic-scale structure of these matrix elements is most easily seen in the $AA$ region, where $\Theta_{ij}$ is approximately uniform, and we display them in Fig.~\ref{fig:ivc_atomistic_construction}(b-d). We also show the moir\'e-scale modulation induced by $\Theta_{ij}$ in Fig.~\ref{fig:ivc_atomistic_construction}(e). Adding Eq.~\eqref{eq:IVC:TBGelements} to the kinetic term $\hat{h}_0(\mathbf{k})$ in Eq.~\eqref{eq:atomistic:normal:overlap} yields the symmetry-broken, IVC normal-state Hamiltonian. Because of the truncation and the bond-by-bond envelope, $\delta\hat{h}_{\textrm{IVC}}$ is not strictly a projector, but like $-U_1|\xi\rangle\langle\xi|$ it acts only near the $AA$ centers and couples states of definite valley and sublattice character. 

\begin{figure}
	\includegraphics[width=\columnwidth]{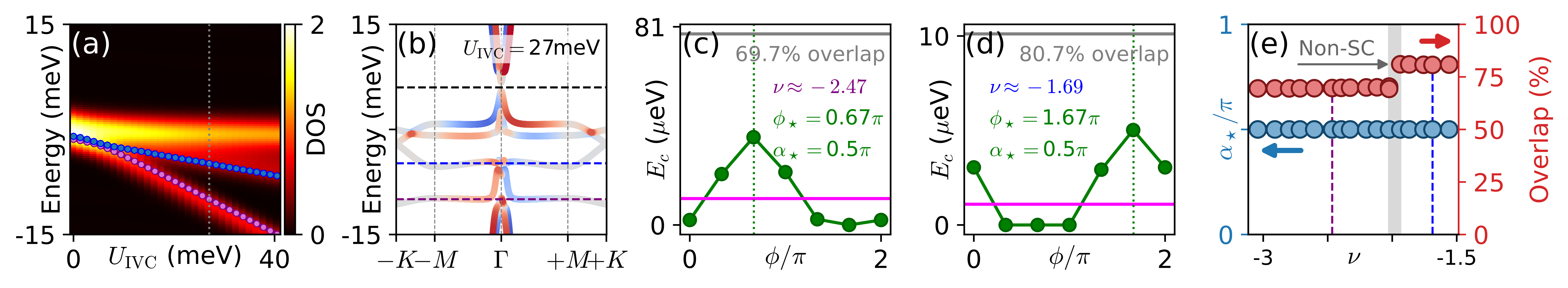}
	\caption{Normal-state and self-consistent pairing properties of the atomistic IVC background. (a) DOS as a function of energy and amplitude $U_{\textrm{IVC}}$. Blue and purple markers track fillings $\nu=-1.69$ and $\nu=-2.47$, respectively. Vertical dotted line denotes value $U_{\textrm{IVC}} = 27$~meV used in (b-e). (b) Band structure showing an IVC band (purple dashed) split off from the other three flat bands (blue dashed and above). Purple and blue dashed lines correspond to same-color markers in (a), and also mark the filling used in (c,d), respectively. (c,d) Self-consistent condensation energy $E_c$ for representative fillings (c) $\nu=-2.47$ and (d) $\nu=-1.69$. Green and magenta are solutions constrained to nematic and chiral intra-Chern pairing, respectively. Gray denotes a fully unconstrained solution (i.e.~the true ground state). The $x$-axis quantifies the value of $\phi$ of the different nematic solutions. Chiral and unconstrained solutions are $\phi$-independent. Dotted vertical line denotes the $\phi$ value extracted by projecting the unconstrained solution onto the original intra-Chern pairing subspace, and matches the best constrained nematic solution. Overlap between the unconstrained solution and original intra-Chern pairing subspace is annotated in gray. (e) Optimal $\alpha_\star$ (blue, left axis) obtained by projecting the unconstrained solution onto the original intra-Chern pairing subspace, as a function of filling $\nu$. We also show the overlap between the unconstrained solution and the intra-Chern subspace (red, right axis). Shaded area denotes region of vanishing condensation.}
	\label{fig:ivc_atomistic_results}
\end{figure}

The angles $\phi^{\textrm{g}}_{\textrm{X}, \ell}$ determining the IVC structure can in principle be chosen independently in each layer $\ell$, and we parametrize them as $\phi^{\textrm{g}}_{\textrm{X}, \ell} = \phi^{\textrm{g}}_{\textrm{X}} \pm \delta \phi^{\textrm{g}}_{\textrm{X}}/2$ in the top (bottom) layer, where $\delta \phi^{\textrm{g}}_{\textrm{X}}$ denotes the shift between layers. We set $\delta \phi^{\textrm{g}}_{\textrm{valley}} = \delta \phi^{\textrm{g}}_{\textrm{sublattice}}=0$ and $\delta \phi^{\textrm{g}}_{\textrm{Chern}} = \pi$, which reproduces the main features expected for the IVC parent state, namely a split-off flat band displaying strong inter-valley coherence. Comparing with the $f$-like orbitals $g_{\pm,\eta}$ of Sec.~\ref{sec:parentstate:hybrid:model}, we find that this layer shift makes both layers represent the same THF state $|\xi\rangle$, and that $\phi^{\textrm{g}}_{\textrm{Chern}} = \pi$, which we use below, corresponds to $\phi_{\textrm{Chern}} \approx 4\pi/3$ in the THF model, i.e.~to $\phi_{\textrm{Chern}} = 0$ up to a $C_{3z}$ rotation. In Fig.~\ref{fig:ivc_atomistic_results}(a) we plot the normal-state DOS as function of $U_{\rm IVC}$, which shows a quasiparticle branch that separates from the other flat bands and lowers in energy as $U_{\textrm{IVC}}$ increases. Purple dots tracks filling $\nu=-2.47$, where the Fermi level is tuned to this lowering branch, while blue dots tracks $\nu=-1.69$ corresponding to the flat band with second lowest energy. For concreteness, we set $U_{\textrm{IVC}} = 27$~meV (vertical dotted line). The amplitude $U_{\textrm{IVC}}$ plays the role of $U_1$ in Eq.~\eqref{eq:ivc_THF_hamiltonian}, and this chosen value reproduces qualitative features of the band structure of the IKS-precursor of Ref.~\cite{Herzog-Arbeitman2025}. The resulting band structure, plotted in Fig.~\ref{fig:ivc_atomistic_results}(b), indeed reveals a lower band (purple dashed line) which has strong inter-valley coherence, as indicated by the gray color away from $\boldsymbol{\Gamma}$. Inter-valley coherence is weak at $\boldsymbol{\Gamma}$ because the flat-band wavefunctions there are $c$-electron-like and overlap only weakly with $|\xi(\ell)\rangle$ \cite{Song2022}, as also observed in the vanishing $f$-orbital polarization $p_f$ at $\boldsymbol{\Gamma}$ of Sec.~\ref{sec:parentstate:chern}.

Setting the phases $\phi^{\textrm{g}}_{\textrm{valley}} = \phi^{\textrm{g}}_{\textrm{sublattice}} = 0$, $\phi^{\textrm{g}}_{\textrm{Chern}} = \pi$ and amplitude $U_{\textrm{IVC}} = 27$~meV, as discussed above, to model the IVC parent state within an atomistic framework, we can finally solve self-consistently for the nearest-neighbor pairing $\hat{\Delta}$, for a range of chemical potentials $\mu$ corresponding to fillings around $\nu=-2$. In Fig.~\ref{fig:ivc_atomistic_results}(c,d) we plot the condensation energies $E_{c}$ obtained at the representative fillings $\nu=-2.47$ and $\nu=-1.69$, respectively. In each case we first constrain the order parameter to lie along some direction $(\alpha, \phi)$ inside the intra-Chern pairing manifold obtained from the symmetric parent state. Here we find that at least some nematic configurations (green dots) achieve higher condensation energy $E_c$ than the chiral one (magenta line). Moreover, we find that the optimal nematic direction $\phi_\star$ (dotted vertical line) changes as the filling increases, approximately from $2\pi/3$ to $5\pi/3$ for the chosen IVC angles, consistent with Fig.~10(c) in the main text. We then verify these constrained results by performing fully unconstrained self-consistent calculations, where we do not constrain the order parameter at all. This yields the ground state $E_c$ (gray line), with a roughly twofold increase in $E_c$ compared to the constrained calculations. We then project this fully unconstrained self-consistent solution onto the intra-Chern pairing manifold and extract the corresponding $\alpha_{\star}$ and $\phi_{\star}$ values, which match those of the optimal constrained nematic solution. The overlap is high, but not perfect (gray numbers), consistent with the increase in $E_c$ for the fully unconstrained self-consistent calculation. In Fig.~\ref{fig:ivc_atomistic_results}(e) we verify the generality of these results by performing unconstrained calculations across $-3 \lesssim \nu \lesssim -1.5$. Projected onto the intra-Chern pairing manifold, these solutions give $\alpha_{\star} = \pi/2$ throughout (blue, left axis), i.e.~a nematic ground state wherever superconductivity is stabilized (outside the shaded area), with an overlap of $\gtrsim 70\%$ (red, right axis). The corresponding $E_c$ and $\phi_\star$ are shown in Fig.~10(c) of the main text.

\begin{figure} [b]
	\includegraphics[width=\columnwidth]{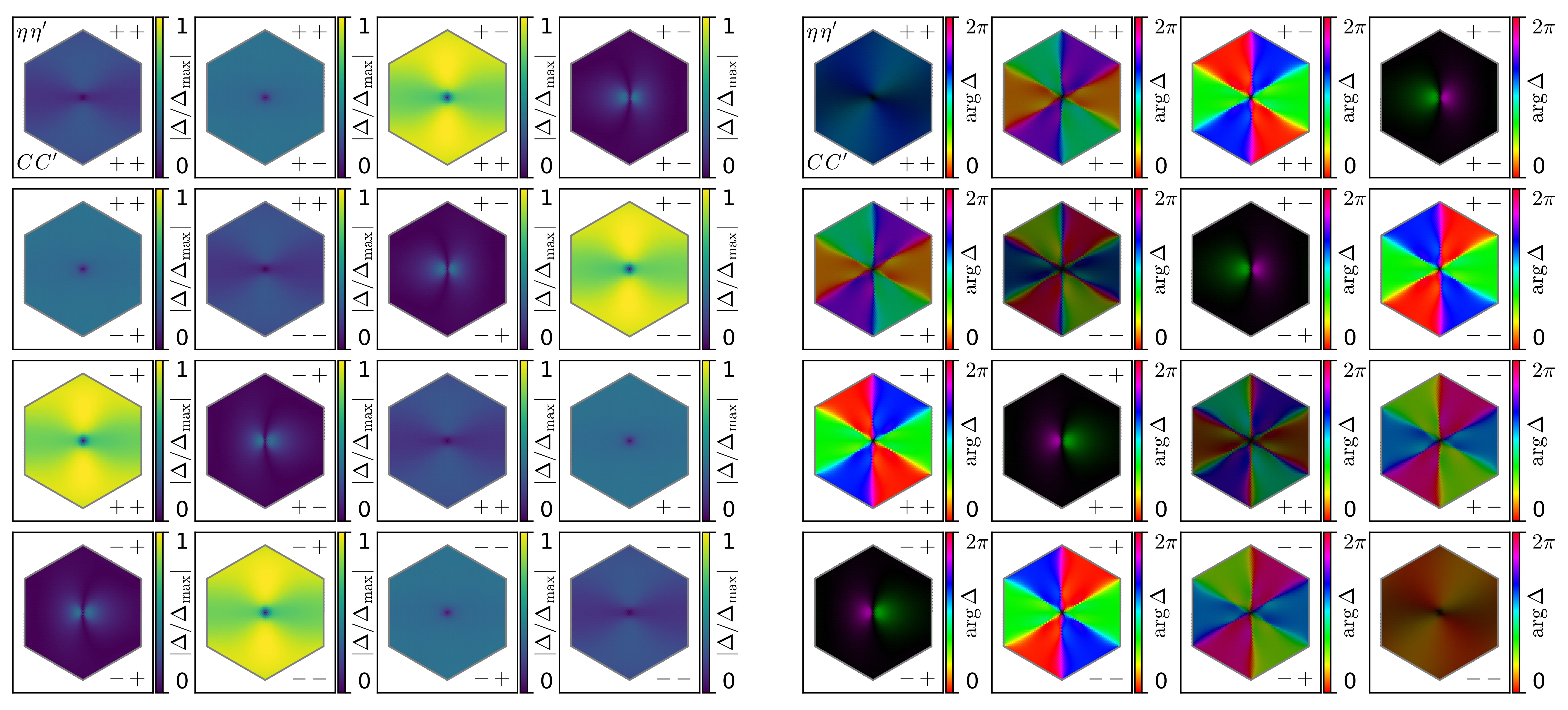}
	\caption{Multiband pairing structure of the superconducting ground state at $\nu=-2.47$ for an IVC parent state with $U_{\textrm{IVC}} = 27$~meV (one of the data points in Fig.~10(c) of the main text), projected onto the Chern basis of the unperturbed ($U_{\textrm{IVC}}=0$) flat bands. Amplitude (left) and phase (right) of the pairing matrix elements $\tilde{\Delta}_{C C^\prime}^{\eta \eta^\prime}$, where $C$, $C^\prime$ denote the Chern sectors.}
	\label{fig:ivc_atomistic_pairing_matrix}
\end{figure}

The results above reveal the large overlap between the unconstrained self-consistent solution and the original intra-Chern pairing channel. We further validate the intra-Chern nature of pairing under an IVC parent state by computing the projection of the unconstrained self-consistent pairing matrix onto the Chern basis of the symmetric flat bands. In Fig.~\ref{fig:ivc_atomistic_pairing_matrix} we plot the results for the unconstrained solution in Fig.~\ref{fig:ivc_atomistic_results}(c). Visual inspection reveals that although finite intra-valley and inter-Chern pairing are now technically finite, they remain small compared to the inter-valley intra-Chern contributions, and the pairing matrix overall remains very similar to that of the symmetric parent state in Fig.~\ref{fig:multiband:Deltadpid_Chern_nematic}. This visually confirms that the intra-Chern pairing model remains valid for an IVC parent state. This can further be quantified by computing the momentum-integrated components $\int_{\textrm{BZ}} |\tilde{\Delta}_{C C^\prime}^{\eta \eta^\prime}|^2 \, d^2k$, where the matrix elements $\tilde{\Delta}_{C C^\prime}^{\eta \eta^\prime}$ are obtained through Eq.~\eqref{eq:projection:finalOP_chern} using the symmetric Chern band basis. For the unconstrained solution in Fig.~\ref{fig:ivc_atomistic_results}(c), the inter-valley intra-Chern channel contribution corresponds to $81.7\%$ of the total pairing, while the rest is distributed among the remaining channels: intra-valley inter-Chern ($13.0\%$), intra-valley intra-Chern ($4.8\%$), and inter-valley inter-Chern ($0.5\%$). We note that this inter-valley intra-Chern weight is larger than the overlap of $69.7\%$ in Fig.~\ref{fig:ivc_atomistic_results}(c), which measures the overlap with the intra-Chern pairing manifold of the symmetric parent state, i.e.~with its specific form factors, rather than the weight of the whole channel. This establishes inter-valley intra-Chern pairing as the clearly dominant channel even in the presence of substantial inter-valley coherence in the parent state, thus validating the extracted $\phi_{\star}$ shown in Fig.~10(c) of the main text. We stress, however, the qualitative nature of this analysis. Simultaneous self-consistent treatment of both the Hartree-Fock and Cooper channels may be necessary to fully quantify intra-Chern pairing for an IVC parent state, which we leave for future work.

\clearpage
\bibliography{bibliography}